\documentclass[rmp,aps,nofootinbib,endfloatsamsmath,amsmath,amssymb,twocolumn]{revtex4}
\pdfoutput=1
\usepackage{graphicx}
\usepackage{dcolumn}
\usepackage{bm}
\usepackage{bbm}
\usepackage[usenames]{color}
\usepackage{color} 
\usepackage{amsmath}
\usepackage{empheq}

\usepackage{manfnt}
\usepackage{wasysym}
\usepackage{empheq}
\usepackage{paralist}

\newcommand{\ket}[1]{|{#1}\rangle}
\def\dblone{\hbox{$1\hskip -1.2pt\vrule depth 0pt height 1.6ex width 0.7pt \vrule depth 0pt height 0.3pt width 0.12em$}}
\definecolor{myblue}{rgb}{.9, .9, 1}

\begin{document}

\title{Compass and Kitaev models -- Theory and Physical Motivations}

\author{Zohar Nussinov}
 \affiliation{Department of Physics, Washington University, St. Louis, MO 63160, USA}
 
\author{Jeroen van den Brink}
\affiliation{Institute for Theoretical Solid State Physics, IFW Dresden, 01069 Dresden, Germany}
\affiliation{Department of Physics, Technical University Dresden, 01062 Dresden, Germany}

\date{\today}

\begin{abstract}
{\it Compass models} are theories of matter in which the couplings between the internal spin (or other relevant field) components are inherently spatially (typically, direction) dependent. A simple illustrative example is furnished by the ``90$^{\circ}$ compass" model on a square lattice in which only couplings of the form $\tau^{x}_{i} \tau^{x}_{j}$ (where $\{\tau^{a}_{i}\}_{a}$ denote Pauli operators at site $i$) are associated with nearest neighbor sites $i$ and $j$ separated along the $x$ axis of the lattice while $\tau^{y}_{i} \tau^{y}_{j}$ couplings appear for sites separated by a lattice constant along the $y$ axis. Such compass-type interactions appear in diverse physical systems including Mott insulators with orbital degrees of freedom (where interactions sensitively depend on the spatial orientation of the orbitals involved), the low energy effective theories of frustrated quantum magnets, systems with strong spin-orbit couplings (such as the iridates), vacancy centers, and cold atomic gases. Kitaev's models, in particular the compass variant on the honeycomb lattice, realize basic notions of topological quantum computing. The fundamental inter-dependence between internal (spin, orbital, or other) and external (i.e., spatial) degrees of freedom which underlies compass models generally leads to very rich behaviors including the frustration of (semi-)classical ordered states on non-frustrated lattices and to enhanced quantum effects prompting, in certain cases, the appearance of zero temperature quantum spin liquids.  As a consequence of these frustrations, new types of symmetries and their associated degeneracies may appear. These {\it intermediate symmetries} lie midway between the extremes of global symmetries and local gauge symmetries and lead to effective dimensional reductions. We review compass models in a unified manner, paying close attention to exact consequences of these symmetries, and to thermal and quantum fluctuations that stabilize orders via {\it order out of disorder} effects. We review non-trivial statistics and the appearance of  {\it topological quantum orders} in compass systems in which, by virtue of their intermediate symmetry standard orders do not arise. This is complemented by a survey of numerical results. Where appropriate theoretical and experimental results are compared. 
\end{abstract}

\maketitle
\tableofcontents

\section{Introduction \&  Outline}
\label{sec:intro}

\subsection{Introduction}
This article reviews compass models. The term "compass models"  refers to a family of closely related lattice models involving interacting quantum degrees of freedom (and their classical approximants). Members of this family appear in very different physical contexts. Already three decades ago they were first encountered as minimal models to describe interactions between orbital degrees of freedom in strongly correlated electron materials~\cite{Kugel82}. The name {\it orbital compass model} was coined at the time, but only in the past decade these models started to receive wide-spread attention to describe physical properties of materials with orbital degrees of freedom~\cite{Tokura00,Brink04,Khaliullin05}. 

In different guises, these models describe the phase variable in certain superconducting Josephson-junction arrays~\cite{Xu04,Nussinov05} and exchange interactions in ultra-cold atomic gasses~\cite{Duan03,Wu08}. Last but not least,  quantum compass models have recently made an entrance to the scene of quantum information theory as mathematical models for topological quantum computing~\cite{Kitaev03}: The much-studied Kitaev's honeycomb model has the structure of a compass model. It is interesting to note that the apparently different fields dealing with orbital degrees of freedom in complex oxides and dealing with models for quantum computing have compass models in common and can thereby in principle cross-fertilize. Kitaev's honeycomb model has, for instance, been put forward to describe the interactions between magnetic moments in certain iridium-oxide materials~\cite{Jackeli09}. 

Here we review the different incarnations of compass models, their physical motivations, symmetries, ordering and excitations. In doing so, we aim to highlight in particular the relation between orbital models and Kitaev's models for quantum computation. One should stress however that although the investigation of compass and Kitaev models has grown into a considerable area of research, this is an active field of research with still many interesting and open problems, as will become more explicit in the following. 

\subsection{Outline of the Review}

We start by introducing and defining, in Section \ref{sec:basic}, various compass models.  Next, in Section \ref{sec:generalized_compass}, we discuss viable extensions of more typical compass models including, e.g., ring-exchange and extensions to general spatial dimensions.  While the most common representation of compass models is that on a lattice, other representations are noteworthy. 

In Section \ref{sec:representations}, we put to the fore continuum representations that are suited for field theoretic treatments, introduce general momentum space representations and illustrate how it naturally suggests the presence of dimensional reductions in compass models. We furthermore discuss classical incommensurate ground states and the representation of a quantum compass model as an unusual {\it anisotropic} classical Ising model. In subsection \ref{d-eom}, the general equations of motion associated with compass theories are presented; these equations capture the quintessential anisotropic character of the compass models. 

Next, in Section \ref{sec:motivation}, we discuss the physical contexts that motivate compass models and derive them for special cases. This includes situations where the compass degrees of freedom represent orbital degrees of freedom [subsection \ref{sec:orbitals}]. We review how they emerge, how they  interact, and how they are described mathematically in terms of orbital  Hamiltonians. Most typical representations rely on SU(2) algebra but we also discuss SU(3) Gell-mann and other matrix forms that are better suited for the description of certain orbital systems. We conclude subsection \ref{sec:orbitals} by illustrating how strong spin-orbit effects can lead, within the subspace of low-energy locked orbit and spin states, to compass model hybrids, in particular to the so-called {\it Heisenberg-Kitaev} model of  pertinence to the iridates. A brief summary of how compass models arise vacancy center and trapped ion systems with effective dipolar interactions is provided in subsection \ref{VCsection}.  In subsection \ref{sec:cold_atoms} we proceed with a review of the realization of compass models in cold atomic systems. We conclude our general discussion of incarnations of compass models in general physical systems in subsection \ref{sec:chiral} where we review how the effective low energies theories in chiral frustrated magnets (such as the Kagome and triangular antiferromagnets) are of the compass model type. 

In Section \ref{sec:sym}, we turn to one of the most common unifying features of compass models: the {\it intermediate symmetries} that they exhibit. We review what these symmetries are and place in them in perspective to the two extremes of global and local gauge symmetries. We discuss precise consequences of these symmetries notably those concerning effective dimensional reductions, briefly allude to relations to topological quantum orders and illustrate how these symmetries arise in the various compass models. 

In Section \ref{new_theorem_on_bands}, we introduce a new result: an exact relation between intermediate symmetries and band structures. In particular, we illustrate how flat bands can arise and are protected by the existence of these symmetries and demonstrate how this is materialized in various compass models. One common and important consequence of intermediate symmetries is the presence of an exponentially large ground state degeneracy. We will discuss situations where this degeneracy is exact and ones in which it emerges in various limits. 

In Section \ref{sec:diso}, we review how low temperature orders in various compass models nevertheless appear and are stabilized by fluctuations or, as they are often termed, {\it order out of disorder} effects. Orders in classical compass models that we review are, rigorously, stabilized by thermal fluctuations. This ordering tendency is further bolstered by quantum zero point fluctuations.  Due an exact equivalence between the large $n$ and high temperature limits, the low temperature behavior of compass models is supplanted by exact results at high temperatures as review in Section \ref{sec:high_t_section}.

Following the review of these earlier analytic results concerning the limiting behaviors at both low and high temperatures, we turn in Section \ref{sec:compass_phase_diagram} to numerical results concerning the phases and transitions  in various compass model systems. In Section \ref{HKCS}, we present a discussion (containing both rigorous and numerical results) of the hybrid Heisenberg-Kitaev model and its possible connection to iridate compounds (along with a comparison between theoretical and experimental results). 

In the final part of this article, Section \ref{sec:kit}, we review Kitaev's honeycomb model and its context. This exactly solvable model was inspired by the ideas of topological quantum computing yet also exhibits many other notable features including spin liquid type ground states. Both these aspects we will present and review in a largely self-contained manner.


\section{Compass Model Overview}
\label{sec:basic}

\subsection{Definition of  Quantum Compass Models}
\label{sec:definition}
In order to define quantum compass models, we start by considering a lattice with sites on which quantum degrees of freedom live. Throughout this review the total number of lattice sites is denoted by  by $N$. When square (or cubic) lattices will be involved, these will be consider of dimension $N=L \times L$ (or $N= L \times L \times L$). On more general lattices, $L$ denotes the typical linear dimension (i.e., linear extent along one of the crystalline axis). We set the lattice constant to unity. The spatial dimensionality of the lattice is denoted by $D$ (e.g., $D=2$ for the square and honeycomb lattices, $D=3$ in cubic and pyrochlore lattices etc.). 

Depending on the problem at hand, we will refer to these degrees of freedom at the lattice sites as spins, pseudospins or orbitals. We denote these degrees of freedom by $\bm{\tau}_i$, where $i$ labels the lattice sites and $\bm{\tau}\equiv  \frac{1}{2}(\sigma^{x},\sigma^{y},\sigma^{z})$, where $\sigma^{x}$, $\sigma^{y}$ and $\sigma^{z}$ are the Pauli matrices. In terms of the creation ($c_\alpha$) and annihilation  ($c^\dagger_\alpha$) operator for an electron in state $\alpha$, the pseudospin operator  ${\bm \tau}$ can be expressed as ${\bm \tau}=\frac{1}{2} \sum_{\alpha \beta}c^\dagger_\alpha {\bm \sigma}_{\alpha \beta}c_\beta$, where the sum if over the two different possibilities for each $\alpha$ and $\beta$. Here $\bm \tau$ is the fundamental $T=1/2$ representation of SU(2), for $T>1/2$ we use $\bm T$.

A representation in terms of Pauli matrices is particularly useful for degrees of freedom that have two flavors, for instance two possible orientations of a spin (up or down) or two possible orbitals that an electron can occupy, as the Pauli matrices are generators of SU(2), the group of $2 \times 2$ matrices with determinant one. For degrees of freedom with $n$ flavors, it makes sense to use a representation in terms of the generators of SU($n$), which for the particular case of $n=3$ are the eight Gell-Mann matrices $\lambda_i$, with $i=1,8$ (see Appendix, Sec.~\ref{sec:appendix}).

The name that one chooses to bestow upon the degree of freedom (whether {\it spin}, {\it pseudospin}, {\it color}, {\it flavor} or {\it orbital}) is of course mathematically irrelevant. For SU(2) quantum compass models it is important that the components of $\bm{\tau}$ obey the well-known commutation relation $[\tau^x,\tau^y]=i \tau^z$, and its cyclic permutations and that $(\tau^{\gamma})^{2}=1/4$ for any component $\gamma=x,y,$ or $z$. In the case of SU(3), in the fundamental representation ${\bm \tau}$ is the eight component vector ${\bm \tau}=\frac{1}{2} \sum_{\alpha \beta}c^\dagger_\alpha {\bm \lambda}_{\alpha \beta}c_\beta$, with the commutation relations governed by those of the Gell-Mann matrices. 

Compass models are characterized by the specific form that the interaction between the degrees of freedom assumes: $(i)$ there is only an interaction between certain {\it vector components} of $\bm{\tau}$ and $(ii)$ on different bonds in the lattice, different vector components interact. When, for instance, a site $i$ is linked to nearest neighbor sites $j$ and $k$, the interaction along the lattice link $\langle ij \rangle$  can be of the type $\tau^x_i \tau^x_j$, whereas on the link $\langle ik \rangle$ it is  $\tau^y_i \tau^y_k$. In the following sections specific Hamiltonians corresponding to various quantum compass models are introduced, in particular the $90^\circ$ compass models, Kitaev's honeycomb model, $120^\circ$ compass models and a number of generalizations thereof.

\subsubsection{$90^\circ$ compass models} 
\begin{figure}
\centering
\includegraphics[width=.6\columnwidth]{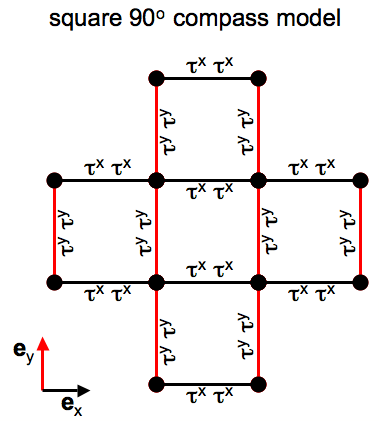}
\caption{The planar $90^\circ$ compass model on a square lattice: the interaction of (pseudo-)spin degrees of freedom ${\bm \tau}=(\tau^x,\tau^y)$ along horizontal bonds that are connected by the unit vector ${\bf e}_x$ is $\tau^x_i \tau^x_{i+{\bf e}_x}$. Along vertical bonds ${\bf e}_y$ it is $\tau^y_i \tau^y_{i+{\bf e}_y}$.
}
\label{fig:s90compass}
\end{figure}

A basic realization of a quantum compass model can be set up on a two-dimensional square lattice, where every site has two horizontal and two vertical bonds. If one defines the interaction along horizontal lattice links $\langle ij \rangle_H$ to be $J \tau^x_i \tau^x_j$ and along the vertical links $\langle ij \rangle_V$ to be $J \tau^y_i \tau^y_j$, we have constructed the so-called {\em{two-dimensional $90^\circ$ quantum compass model}} also known as the {\it planar $90^\circ$ orbital compass model}, see Fig.~\ref{fig:s90compass}. Its Hamiltonian is
\begin{eqnarray}
\label{2dpocm}
H_{\square}^{90^\circ}=  - J_{x} \sum_{\langle ij \rangle_H} \tau^x_i \tau^x_j - J_{y} \sum_{\langle ij \rangle_V} \tau^y_i \tau^y_j. 
\end{eqnarray}
The isotropic variant of this system has equal couplings along the vertical and horizontal directions ($J_{x}=J_{y}=J$). 
The minus signs that appear in this Hamiltonian were chosen such that the interactions between the pseudospins $\bf \tau$ tend to stabilize uniform ground states with "ferro" pseudospin order. (In $D=2$ the $90^\circ$ compass models with "ferro" and "antiferro" interactions are directly related by symmetry, see Section~\ref{sec:120_compass_models}). For clarity, we note that the isotropic two dimensional compass model is very different from the two-dimensional Ising model
\begin{eqnarray}
H^{Ising}_{\square}&=& - J \sum_{\langle ij \rangle_H} \tau^x_i \tau^x_j - J \sum_{\langle ij \rangle_V} \tau^x_i \tau^x_j \nonumber \\
&=&-J\sum_{\langle ij \rangle_H,\langle ij \rangle_V} \tau^x_i \tau^x_j , 
\label{eq:HIsing}
\end{eqnarray}
where on each horizontal and vertical vertex of the square lattice the interaction is the same and of the form $J \tau^x_i \tau^x_j$  -- it is also very different from the two-dimensional XY model
\begin{eqnarray}
H^{XY}_{\square}= - J \sum_{\langle ij \rangle_H,\langle ij \rangle_V} 
(\tau^x_i \tau^x_j + \tau^y_i \tau^y_j),
\label{eq:HXY}
\end{eqnarray}
because also in this case on all bonds the interaction terms in the Hamiltonian are of the same form. 

\begin{center}
\begin{table*}
\begin{tabular}{c|c|c|l|c }
\hline
 \multicolumn{5}{c}{Model Hamiltonian: ${\cal H}= - \sum_{i,\gamma} \tau^\gamma_i \tau^\gamma_{i+\bm{e}_\gamma}$}  \\  \hline
$\{\tau^\gamma\}$ &  $\{\bm{e}_\gamma\}$ &   model name & symbol & dimension\\   
\hline \hline
 $\{\tau^x\}$	& $\{\bm{e}_x\}$ & Ising chain &  $H_1^{Ising}$ & 1 \\
 $\{\tau^x,\tau^y\}$	& $\{\bm{e}_x,\bm{e}_x\}$ & XY chain & $H_1^{XY}$ & 1 \\
  $\{\tau^x,\tau^y,\tau^z\}$	& $\{\bm{e}_x,\bm{e}_x,\bm{e}_x\}$ & Heisenberg chain  & $H_1^{Heis}$ & 1 \\
 $\{\tau^x,\tau^x\}$	& $\{\bm{e}_x,\bm{e}_y\}$ & square Ising & $H_\square^{Ising}$ & 2\\
 $\{\tau^x,\tau^x,\tau^x\}$	 & $\{\bm{e}_x,\bm{e}_y,\bm{e}_z\}$ &  cubic Ising & $H_{3\square}^{Ising}$ & 3 \\
 $\{\tau^x,\tau^y,\tau^x,\tau^y\}$	 & $\{\bm{e}_x,\bm{e}_x,\bm{e}_y,\bm{e}_y\}$ &  square XY & $H_{\square}^{XY}$ & 2 \\
 $\{\tau^x,\tau^y,\tau^z,\tau^x,\tau^y,\tau^z\}$	 & $\{\bm{e}_x,\bm{e}_x,\bm{e}_x,\bm{e}_y,\bm{e}_y,\bm{e}_y\}$ &  square Heisenberg & $H_{\square}^{Heis}$ & 2 \\
 $\{\tau^x,\tau^y\}$	 & $\{\bm{e}_x,\bm{e}_y\}$ & 90$^\circ$ square compass & $H_{\square}^{90^\circ}$ & 2 \\
 $\{\tau^x,\tau^y,\tau^z\}$	 & $\{\bm{e}_x,\bm{e}_y,\bm{e}_z\}$ & 90$^\circ$ cubic compass & $H_{3\square}^{90^\circ}$ & 3 \\ 
 \hline \hline \multicolumn{2}{c}{With $\{\theta_\gamma\}=\{0,2\pi/3,4\pi/3\}$: }  & \multicolumn{2}{c}{}  & \\ \hline \hline
 $\{\tau^x,\tau^x,\tau^x\}$	& $ \bm{e}_x\cos{\theta_\gamma}  +  \bm{e}_y \sin{\theta_\gamma} $ & honeycomb Ising & $H_{\varhexagon}^{Ising}$ & 2 \\
$\{\tau^x,\tau^y,\tau^z\}$	& $ \bm{e}_x\cos{\theta_\gamma}  +  \bm{e}_y \sin{\theta_\gamma} $ & honeycomb Kitaev  &  $H_{\varhexagon}^{Kitaev}$ & 2 \\
$\{\tau^x,\tau^x,\tau^z\}$	& $ \bm{e}_x\cos{\theta_\gamma}  +  \bm{e}_y \sin{\theta_\gamma} $ & honeycomb XXZ  &  $H_{\varhexagon}^{XXZ}$ & 2 \\
 $\pi^\gamma=\tau^x\cos{\theta_\gamma}  +  \tau^y \sin{\theta_\gamma} $ &$\{\bm{e}_x,\bm{e}_y,\bm{e}_z\}$ & cubic 120$^\circ$ &$H_{3\square}^{120^\circ}$ & 3 \\
  $\pi^\gamma$ &$ \bm{e}_x\cos{\theta_\gamma}  +  \bm{e}_y \sin{\theta_\gamma} $ & honeycomb 120$^\circ$  &$H_{\varhexagon}^{120^\circ}$ & 2 \\
\hline \hline \multicolumn{2}{c}{With $\{\theta_\gamma\}=\{0,2\pi/3,4\pi/3\}$ and $\eta=\pm1$: }  & \multicolumn{2}{c}{}  & \\ \hline \hline
$\{\tau^x,\tau^y,\tau^z\}$ & $\eta \bm{e}_x\cos{\frac{\theta_\gamma}{2}}  +  \eta \bm{e}_y \sin{\frac{\theta_\gamma}{2}} $ & triangular Kitaev  &  $H^{Kitaev}_{\vartriangle}$ & 2 \\
$\pi^\gamma$ &$ \eta \bm{e}_x\cos{\frac{\theta_\gamma}{2}}  +  \eta \bm{e}_y \sin{\frac{\theta_\gamma}{2}} $ & triangular 120$^\circ$  &  $H^{120}_{\vartriangle}$ & 2 \\
\hline
\end{tabular}
\caption{Generalized notation that casts compass models and more well-known model Hamiltonians such as the Ising, XY or Heisenberg model on various lattices of different dimensions in the same form. When coupling constants $J_\gamma$ depend on the bond direction $\{\bm{e}_\gamma$, connecting sites $i$ and $j$, an additional spatial anisotropy is introduced which changes the strengths of the interaction on different links, but not the form of those interactions which is determined by how different components of the vectors ${\bm \tau}_i$ and ${\bm \tau}_j$ couple.}
\label{table1}
\end{table*}
\end{center}

One can rewrite the $90^\circ$ compass Hamiltonian in a more compact form by introducing unit vectors $\bm{e}_x$ and $\bm{e}_y$ that denote the bonds along the $x$- and $y$-direction in the 2D lattice, so that 
\begin{eqnarray}
\label{s90comp}
H_\square^{90^\circ}= - J \sum_{i} (\tau^x_i \tau^x_{i+\bm{e}_x} + \tau^y_i \tau^y_{i+\bm{e}_y}). 
\end{eqnarray}
With this notation the compass model Hamiltonian can be cast in the more general form
\begin{eqnarray} 
\label{s90eq}
H_\square^{90^\circ}= - J \sum_{i,\gamma}  \tau^\gamma_i \tau^\gamma_{i+\bm{e}_\gamma},
\end{eqnarray}
where for the $90^\circ$ square lattice compass model, $H_\square^{90^\circ}$, we have $\gamma =1,2$, $\{\tau^\gamma\}=\{\tau^1,\tau^2\}= \{\tau^x,\tau^y\}$ and $\{ \bm{e}_\gamma\}=\{ \bm{e}_1,\bm{e}_2\}=\{\bm{e}_x,\bm{e}_y\}$. 

This generalized notion allows for different compass models and the more well-known models such as the Ising or Heisenberg model to be cast in the same form, see Table~\ref{table1}. For instance the two-dimensional square lattice Ising model $H^{Ising}_{\square}$  corresponds to $\gamma =1,2$ with $\{\tau^\gamma\}= \{\tau^x,\tau^x\}$ and $\{\bm{e}_\gamma\}=\{\bm{e}_x,\bm{e}_y\}$. The Ising model on a three dimensional cubic lattice is then given by $\gamma =1...3$, $\{\tau^\gamma\}= \{\tau^x,\tau^x,\tau^x\}$ and $\{\bm{e}_\gamma\}=\{\bm{e}_x,\bm{e}_y,\bm{e}_z\}$. The XY model on a square lattice $H^{XY}_{\square}$ corresponds to $\gamma =1...4$, $\{\tau^\gamma\}= \{\tau^x,\tau^y,\tau^x,\tau^y\}$ and $\{\bm{e}_\gamma\}=\{\bm{e}_x,\bm{e}_x,\bm{e}_y,\bm{e}_y\}$. Another example is the square lattice Heisenberg model, where we have  $\gamma =1...6$, $\{\tau^\gamma\}= \{\tau^x,\tau^y,\tau^z,\tau^x,\tau^y,\tau^z\}$ and $\{\bm{e}_\gamma\}=\{\bm{e}_x,\bm{e}_x,\bm{e}_x,\bm{e}_y,\bm{e}_y,\bm{e}_y\}$, so that in this case $\sum_\gamma \tau^\gamma_i \tau^\gamma_{i+\bm{e}_\gamma}$ is equal to $\frac{1}{3}\sum_\gamma {\bm \tau}_i \cdot {\bm \tau}_{i+\bm{e}_\gamma}$.  

This class of compass models can be further generalized in a straightforward manner by allowing for a coupling strength $J_{\gamma}$ between the pseudospins $\tau^\gamma$ that depends on the direction of the bond $\gamma$  ({\it anisotropic compass models} \cite{Nussinov05}) and by adding a field $h_\gamma$ that couples to $\tau^\gamma$ linearly \cite{Nussinov08b, Scarola09}. This general class of compass models is then defined by the Hamiltonian
\begin{eqnarray} 
{\cal H}_{compass}=  -\sum_{i,\gamma} \left( J_\gamma \tau^\gamma_i \tau^\gamma_{i+\bm{e}_\gamma} +h_\gamma \tau^\gamma_i \right).
\label{eq:general_compass}
\end{eqnarray} 

\begin{figure}
\centering
\includegraphics[width=.7\columnwidth]{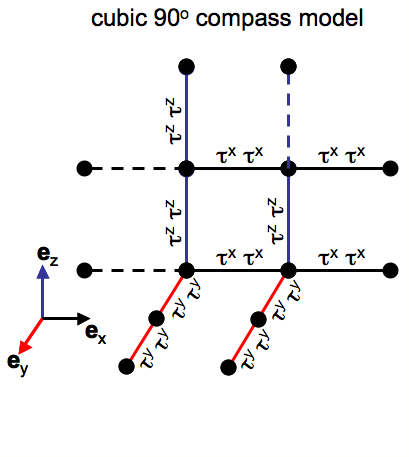}
\caption{The $90^\circ$ compass model on a cubic lattice: the interaction of (pseudo-)spin degrees of freedom ${\bm \tau}=(\tau^x,\tau^y,\tau^z)$ along horizontal bonds that are connected by the unit vector ${\bf e}_x$ is $J \tau^x_i \tau^x_{i+{\bf e}_x}$. On bonds connected by ${\bf e}_y$ it is $J \tau^y_i \tau^y_{i+{\bf e}_y}$ and along the vertical bonds it is $J \tau^z_i \tau^z_{i+{\bf e}_z}$.
}
\label{fig:c90compass}
\end{figure}

From a historical (as well as somewhat practical) viewpoint the {\em{three dimensional 90$^\circ$ compass model}} is particularly interesting. Denoted by $H_{3\square}^{90^\circ}$, it is customarily defined on a cubic lattice and given by $\cal H$ (Eq.~(\ref{eq:general_compass})) where $\gamma$ spans three Cartesian directions: $\gamma = 1...3$ with $\{\tau^\gamma\}= \{\tau^x,\tau^y,\tau^y\}$, $J_\gamma=J=1$, $h_\gamma=0$ and $\{\bm{e}_\gamma\}=\{\bm{e}_x,\bm{e}_y,\bm{e}_z\}$, so that
\begin{eqnarray}
\label{c90comp}
H_{3\square}^{90^\circ}=  - J \sum_{i} (\tau^x_i \tau^x_{i+\bm{e}_x} + \tau^y_i \tau^y_{i+\bm{e}_y} + \tau^z_i \tau^z_{i+\bm{e}_z}). 
\end{eqnarray}
Thus, the square lattice 90 degree compass model of Eq. (\ref{s90eq}) is trivially extended to three spatial dimensions by allowing $\gamma$ to assume values $\gamma=1,2,3$, 
Thus, with appropriate generalizations, in an arbitrary spatial dimension $D$ (which we will return to in later sections), $\gamma =1,2, ..., D$. 
The structure of this Hamiltonian is schematically represented in Fig.~\ref{fig:c90compass}. This compass model is actually the one that was originally proposed by~\cite{Kugel82} in the context of orbital ordering. At that time it was noted that even if the interaction on each individual bond is Ising-like, the overall symmetry of the model is considerably more complicated, as will be reviewed in Sec.~\ref{sec:orbitals}. 

In alternative notations for compass model Hamiltonians one introduces the unit vector $\bm n$ connecting neighboring lattice sites $i$ and $j$. Along the three Cartesian axes on a cubic lattice, for instance, ${\bm n}$ equals $\bm{e}_x=(1,0,0)$, $\bm{e}_y=(0,1,0)$ or $\bm{e}_z=(0,0,1)$. With this one can express $\tau^x$ as $\tau^x = \bm{\tau} \cdot \bm{e}_x$ and with this vector notation
\begin{eqnarray}
H_{3\square}^{90^\circ}=  -\sum_{i,\gamma} \tau^\gamma_i \tau^\gamma_{i+\bm{e}_\gamma}= -\sum_{ij} \left(\bm{\tau}_i \cdot{\bm n} \right) \left(\bm{\tau}_j \cdot{\bm n} \right).
\label{eq:90compass}
\end{eqnarray}
The Hamiltonian in vector form stresses the compass nature of the interactions between the pseudospins. The vector notation, however, not always generalizes naturally to cases with higher dimensions and/or different lattice geometries. All Hamiltonians in this review will therefore be given in terms $\tau^\gamma$ operators and be complemented by an expression in vector notation where appropriate.
 
\begin{figure}
\centering
\includegraphics[width=.7\columnwidth]{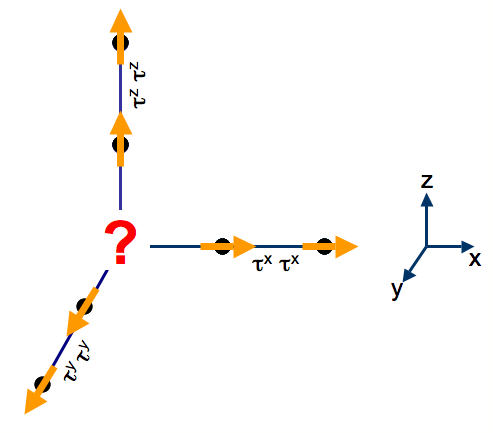}
\caption{Frustration in the $90^\circ$ compass model on a cubic lattice. The interactions between pseudospins $\bm \tau$ are such that the pseudospins tend to align their components $\tau^x$, $\tau^y$ and $\tau^z$ along the $x$, $y$ and $z$-axis, respectively. This causes mutually exclusive ordering patterns.}
\label{fig:frustration}
\end{figure}

It is typical for compass models that even the ground state structure is non-trivial. For a system governed by $H_{3\square}^{90^\circ}$, pairs of pseudospins on lattice links parallel to the $x$-axis, for instance, favor pointing their pseudospins $\bm{\tau}$ along $x$ so that the expectation value $\langle \tau^x \rangle \neq 0$, see Fig.~\ref{fig:frustration}. Similarly, on bonds parallel to the $y$-direction, it is advantageous for the pseudospins to align along the $y$ direction, so that $\langle \tau^y \rangle \neq 0$. It is clear that at a site the bonds along $x$, $y$ and $z$ cannot be satisfied at the same time, so that the interactions are in fact strongly frustrated. This situation bears resemblance to the dipole-dipole interactions between magnetic needles positioned on a lattice, and hence the Hamiltonian above was coined a {\it compass} model. 

Such a frustration of interactions is typical of compass models, but of course also appears in numerous other systems. Indeed, on a conceptual level, many of the ideas and results that will be discussed in this review such as renditions of thermal and quantum fluctuation-driven ordering effects, unusual symmetries and ground state sectors labeled by topological invariants have similar incarnations in frustrated spin, charge, cold atom and Josephson junction array systems. Although these similarities are mostly conceptual there are also instances where there are exact correspondences. For instance, the two dimensional 90$^\circ$ compass model is, in fact, dual to the Moore-Lee model describing Josephson coupling between superconducting grains in a square lattice~\cite{Xu05,Xu04,Moore04} that exhibits time reversal symmetry breaking~\cite{Nussinov05, cobanera}.

\subsubsection{Kitaev's honeycomb model} 
\label{sec:defineKHM}
\begin{figure}
\centering
\includegraphics[width=.9\columnwidth]{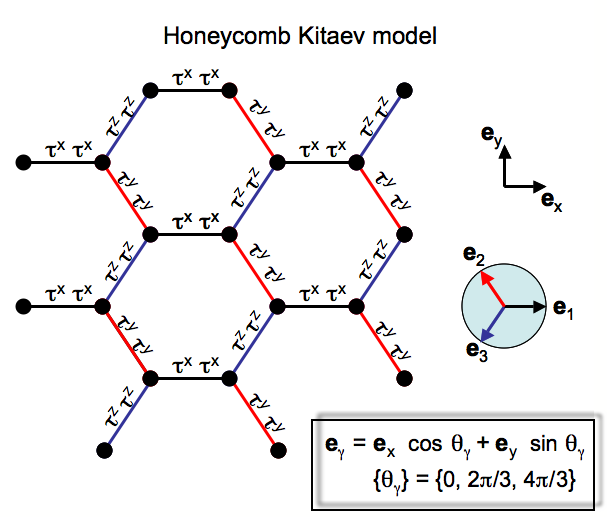}
\caption{Kitaev's compass model on a honeycomb lattice: the interaction of (pseudo-)spin degrees of freedom ${\bm \tau}=(\tau^x,\tau^y,\tau^z)$ along the three bonds that each site is connected to are $\tau^x_i \tau^x_{i+{\bf e}_1}$, $\tau^y_i \tau^y_{i+{\bf e}_2}$ and $\tau^z_i \tau^z_{i+{\bf e}_z}$, where the bond-vectors of the honeycomb lattice $\{{\bf e}_1,{\bf e}_2,{\bf e}_3\}$ are $\{{\bf e}_x,-{\bf e}_x/2+\sqrt{3}{\bf e}_y/2,-{\bf e}_x/2-\sqrt{3}{\bf e}_y/2\}$, respectively.
}
\label{fig:KitaevH}
\end{figure}

In 2006, Alexei Kitaev introduced a type of compass model that has interesting topological properties and excitations, which are relevant and much studied in the context of topological quantum computing~\cite{Kitaev06}. The model  is defined on a honeycomb lattice and is referred to either as {\it Kitaev's honeycomb model} or the {\it XYZ honeycomb compass model}. The lattice links on a honeycomb lattice may point along three different directions, see Fig.~\ref{fig:KitaevH}. One can label the bonds along these directions by $\bm{e}_1$, $\bm{e}_2$ and $\bm{e}_3$, where the angle between the three unit lattice vectors is $120^\circ$. With these preliminaries, the Kitaev's honeycomb model Hamiltonian $H_{\varhexagon}^{Kitaev}$ reads
\begin{eqnarray}
-J_x\sum_{ \begin{smallmatrix} \bm{e}_1-&\\ {\sf bonds}& \end{smallmatrix}}\tau^x_{i}\tau^x_{j} 
-J_y \sum_{\begin{smallmatrix} \bm{e}_2-&\\ {\sf bonds}& \end{smallmatrix}}\tau^y_{i}\tau^y_{j} 
-J_z \sum_{\begin{smallmatrix} \bm{e}_3-&\\ {\sf bonds}& \end{smallmatrix}}\tau^z_{i}\tau^z_{j}.\nonumber\end{eqnarray}
One can re-express this model in the form of $H_{compass}$ introduced above, where
\begin{eqnarray}
H_{\varhexagon}^{Kitaev}&=& - \sum_{i,\gamma} J_\gamma \tau^\gamma_i \tau^\gamma_{i+\bm{e}_\gamma} \nonumber \\
&&with
\left\{ 
\begin{array}{l}
  \{\tau^\gamma\}=\{\tau^x,\tau^y,\tau^z\} \\
    \{J_\gamma\}=\{J_x,J_y,J_z\} \\
  \bm{e}_\gamma=\bm{e}_x\cos{\theta_\gamma}  +  \bm{e}_y \sin{\theta_\gamma} \\
  \{\theta_\gamma\}=\{0,2\pi/3,4\pi/3\}
\end{array} 
\right.
\label{eq:HKT}
\end{eqnarray}

It was proven that for large $J_z$, the model Hamiltonian $H_{\varhexagon}^{Kitaev}$ maps onto a square lattice model  known as {\it Kitaev's toric code model}~\cite{Kitaev03}. We will return to these models of Kitaev in 
Sec.~\ref{sec:kit} and discuss there other related quantum computing models. Numerous other aspects of these models have been investigated in great depth. These include,
amongst others, issues pertaining to quench dynamics \cite{mondal2008,sengupta2008,Sen2010}. Related hybrid models (see sections \ref{hybrid}, \ref{sec:sporb}, \ref{HKCS}) were suggested to be of relevance to certain iridium oxide materials. 
To highlight the pertinent interactions and geometry of Kitaev's honeycomb model as a compass model, it may also be termed an {\it XYZ honeycomb compass model}. It suggests variants such as
the {\it XXZ honeycomb compass model} which we define next. 

\subsubsection{The XXZ honeycomb compass model}
\label{xxz:sec}
\begin{figure}
\centering
\includegraphics[width=.7\columnwidth]{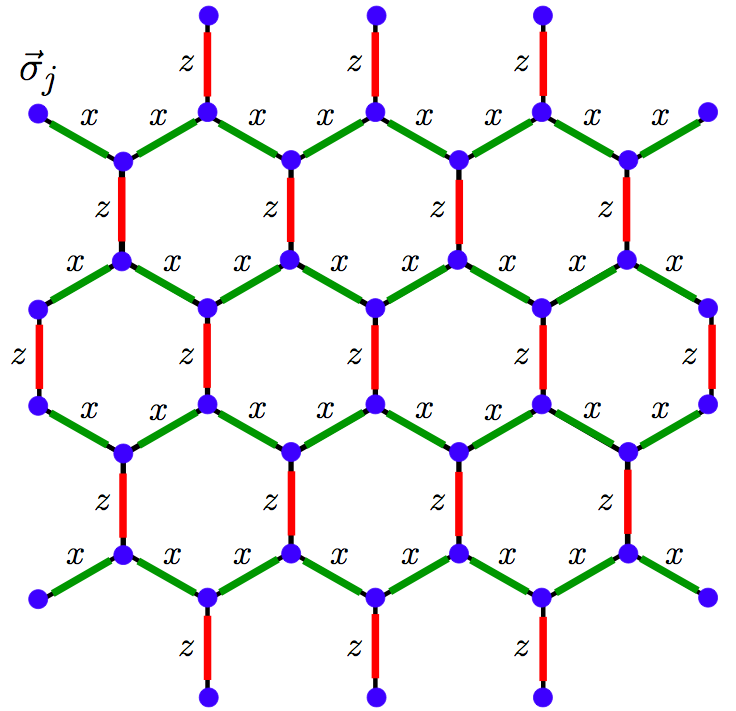}
\caption{Schematic representation of the XXZ honeycomb compass model \cite{Nussinov2012}.}
\label{fig:xxzhc}
\end{figure}

A variation of the Kitaev honeycomb compass Hamiltonian $H_{\varhexagon}^{Kitaev}$ in Eq. (\ref{eq:HKT}) is to consider a compass model where on bonds in two directions there is an $\tau^x \tau^x$-type interaction and in the third direction a $\tau^z \tau^z$ interaction. This model goes under the name of the {\it XXZ honeycomb compass model} \cite{Nussinov2012}. Explicitly, it is given by the Hamiltonian
\begin{eqnarray}
H_{\varhexagon}^{XXZ}&=& - \sum_{i,\gamma} J_\gamma \tau^\gamma_i \tau^\gamma_{i+\bm{e}_\gamma} \nonumber \\
&&with
\left\{ 
\begin{array}{l}
  \{\tau^\gamma\}=\{\tau^x,\tau^x,\tau^z\} \\
    \{J_\gamma\}=\{J_x,J_x,J_z\} \\
  \bm{e}_\gamma=\bm{e}_x\cos{\theta_\gamma}  +  \bm{e}_y \sin{\theta_\gamma} \\
  \{\theta_\gamma\}=\{0,2\pi/3,4\pi/3\}
\end{array} 
\right.
\label{xxzhc}
\end{eqnarray}
A schematic is provided in Fig. \ref{fig:xxzhc}.
The key defining feature of this Hamiltonian vis a vis the original Kitaev model of Section \ref{sec:defineKHM}- the interactions along both the diagonal (``zig-zag'') - ``x'' and ``y''-  directions of the honeycomb lattice are of the $ \tau^x \tau^x  $ type (as opposed to both $\tau^x \tau^x$ and $\tau^y \tau^y$ in Kitaev's model). Similar to Kitaev's honeycomb model, all interactions along the vertical (``z'' direction) are of the $\tau^{z} \tau^z$ type.  
While in Eq. (\ref{xxzhc}) only two couplings, $J_x$ and $J_z$, appear, the model can of course be further generalized to having three different couplings on the three different types of links (and more generally
to have non-uniform spatially dependent couplings), while the interactions retain their $XXZ$ form. In all of these cases, an exact duality to a corresponding Ising lattice gauge theory on a square lattice which we
will elaborate on in later in this review (Section \ref{sec:xxz}) exists.

\subsubsection{120$^\circ$ compass models} 
\label{sec:120_compass_models}
The {\it 120$^\circ$ compass model} has the form  of $H_{compass}$ (Eq.~(\ref{eq:general_compass})) and is defined on a  general lattice having three distinct lattice directions ${\bm e}_{\gamma}$ for nearest neighbor links. As for the other compass models on these lattice links different components of $\bm \tau$ interact. Its particularity is that the three components of $\bm \tau$ are not orthogonal. Along bond $\gamma$ the interaction is between the vector components $\tau^x  \cos{\theta} + \tau^y  \sin{\theta}$ of the two sites connected by the bond, where for the three different links of each site $\theta=0, 2\pi/3$ and $4\pi/3$ respectively.

\begin{figure}
\centering
\includegraphics[width=0.9\columnwidth]{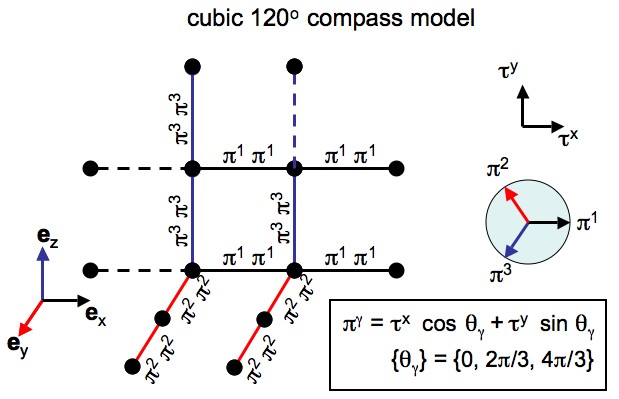}
\caption{The $120^\circ$ compass model on a cubic lattice: the interaction of (pseudo-)spin degrees of freedom ${\bm \tau}=(\tau^x,\tau^y,\tau^z)$ along the three bonds that each site is connected to are $\hat{\pi}^1_i \hat{\pi}^1_{i+{\bf e}_x}$, $\hat{\pi}^2_i \hat{\pi}^2_{i+{\bf e}_y}$ and $\hat{\pi}^3_i \hat{\pi}^3_{i+{\bf e}_z}$, where the different components  $\{\hat{\pi}^1,\hat{\pi}^2,\hat{\pi}^3\}$ of the vector $\hat{\bm \pi}=( \tau^x,(-\tau^x+\sqrt{3}\tau^y)/2,(-\tau^x-\sqrt{3}\tau^y)/2)$ interact along the different bonds $\{ {\bf e}_x,{\bf e}_y,{\bf e}_z\}$. }
\label{fig:120compass}
\end{figure}

The model was first studied on the cubic lattice \cite{Nussinov04,  Biskup05, Brink04} and later on the honeycomb \cite{Nasu08,Wu08,Zhao08} and pyrochlore lattice \cite{Chern10}. The general $120^\circ$ Hamiltonian can be denoted as 
\begin{eqnarray}
H^{120} = - J \sum_{i,\gamma=1...3} \hat{\pi}^{\gamma}_{i} \hat{\pi}^{\gamma}_{i + {\bm e}_{\gamma}},
\label{120p}
\end{eqnarray}
where $\hat{\pi}^{\gamma}_{i}$ are the three projections of $\tau$ along three equally spaced directions on a unit disk 
in the $xy$-plane:
\begin{eqnarray}
\begin{array}{l}
\hat{\pi}^1=\tau^{x}, \\
\hat{\pi}^2=-(\tau^{x} - \sqrt{3} \tau^{y})/2 \\ 
\hat{\pi}^3=-(\tau^{z} + \sqrt{3} \tau^{x})/2.
\end{array} 
\label{eq:pi120}
\end{eqnarray}
%
%
Hence the name 120$^\circ$ model. In the notation of $H_{compass}$ in Eq.~(\ref{eq:general_compass}) the $120^\circ$ Hamiltonian on a 3D cubic lattice, represented in Fig.~\ref{fig:120compass}, takes the form
\begin{eqnarray}
H^{120}_{3\square} &=& - J\sum_{i,\gamma} \hat{\pi}^\gamma_i \hat{\pi}^\gamma_{i+\bm{e}_\gamma} \nonumber \\
&&with
\left\{ 
\begin{array}{l}
  \hat{\pi}^\gamma=\tau^x  \cos{\theta_\gamma} + \tau^y  \sin{\theta_\gamma}\\
  \{\bm{e}_\gamma \}=\{ \bm{e}_x,  \bm{e}_y, \bm{e}_z\} \\
    \{\theta_\gamma\}=\{0,2\pi/3,4\pi/3\}.
\end{array} 
\right.
\label{eq:c120}
\end{eqnarray}
Similar to the $90^\circ$ compass model, the bare $120^\circ$ model can be extended to include anisotropy of the coupling constants $J_{\gamma}$ along the different crystalline directions and external fields~\cite{Rynbach10}.  
On a honeycomb lattice the $120^\circ$ Hamiltonian~\cite{Nasu08,Wu08,Zhao08} can be thought of as a breed of  $H^{120}_{3\square}$ and $H^{Kitaev}_{\varhexagon}$:
\begin{eqnarray}
H^{120}_{3\varhexagon} &=& - J\sum_{i,\gamma} \pi^\gamma_i \pi^\gamma_{i+\bm{e}_\gamma} \nonumber \\
&&with
\left\{ 
\begin{array}{l}
  \pi^\gamma=\tau^x  \cos{\theta_\gamma} + \tau^y  \sin{\theta_\gamma}\\
  \bm{e}_\gamma=\bm{e}_x\cos{\theta_\gamma}  +  \bm{e}_y \sin{\theta_\gamma} \\
    \{\theta_\gamma\}=\{0,2\pi/3,4\pi/3\}.
\end{array} 
\right.
\label{eq:c120_honeycomb}
\end{eqnarray}
It is worth highlighting the differences and similarity between the models of Eqs. (\ref{eq:c120}, \ref{eq:c120_honeycomb}) on the cubic and honeycomb lattices respectively. Although the pseudo-spin operators that appear in these two equations have an identical form, they correspond to different physical links. In the cubic lattice, bonds of the type $ \hat{\pi}^\gamma_i \hat{\pi}^\gamma_{i+\bm{e}_\gamma}$ are associated with links along the Cartesian $\gamma$ directions; on the honeycomb lattice, bonds of the type $ \pi^\gamma_i \pi^\gamma_{i+\bm{e}_\gamma}$ correspond to links along the three possible orientations of nearest neighbor links in the two dimensional honeycomb lattice.  

In {\it 120$^\circ$ compass models} the interactions involve only two of the components of $\bf{\tau}$ (so that $n=2$) as opposed to three component ``Heisenberg" character of the three dimensional $90^\circ$ compass system, having $n=3$.  In that sense 120$^\circ$ models are similar XY models. On bipartite lattices, the ferromagnetic (with $J > 0$) and antiferromagnetic ($J < 0$) variants of the 120$^\circ$ compass model are equivalent to one another up to the standard canonical transformation involving every second site of the bipartite lattice. This can be made explicit by defining the operator
\begin{eqnarray}
\label{ABAB}
U = \prod_{i=odd} \tau^{z}_{i},
\end{eqnarray}
with the product taken over all sites $i$ that belong to, e.g., the {\it odd} sublattice for which the sum of the components of the lattice site along the three Cartesian directions, $i_{x}+i_{y}+i_{z}$, is an odd integer. The unitary mapping $U^{\dagger} H_{120} U$ then effects a change of sign of the interaction constant $J$ (i.e., $J \to  -J $). The ferro and antiferro square lattice $90^\circ$ compass model $(H^{90^\circ}_{\square})$ are related to one another in the same way as, similarly, in this case $n=2$. It should be noted that this mapping does not hold for the 3D rendition of the $90^\circ$ model: in this case the interactions also involve $\tau^z$ and consequently $H^{90^\circ}_{3\square}$ has different low temperature statistical mechanical properties for $J>0$ and $J<0$.


The 120$^{\circ}$ models have also appeared in various physical contexts on non bipartite lattices. On the triangular lattice \cite{Mostovoy2002,Wu08,Zhao08}, the model is given by 
\begin{eqnarray}
H^{120}_{3\vartriangle}  &=& - J\sum_{i,\gamma,\eta} \pi^\gamma_i \pi^\gamma_{i+ \eta \bm{e}_\gamma} \nonumber \\
&&with
\left\{ 
\begin{array}{l}
  \pi^\gamma=\tau^x  \cos{\theta_\gamma} + \tau^y  \sin{\theta_\gamma}\\
  \bm{e}_\gamma=\bm{e}_x\cos \frac{\theta_\gamma}{2}  +  \bm{e}_y \sin \frac{\theta_\gamma}{2} \\
    \{\theta_\gamma\}=\{0,2\pi/3,4\pi/3\} \\
    \eta = \pm 1.
\end{array} 
\right.
\label{eq:c120_triangle}
\end{eqnarray}
The triangular model is very similar to the honeycomb lattice model of Eq. (\ref{eq:c120_honeycomb}). 
The notable difference is that in the triangular lattice there are additional links: In the triangular lattice, each site has six nearest neighbor whereas on the honeycomb lattice, each site has three nearest neighbors. In the Hamiltonian of Eq. (\ref{eq:c120_triangle}), nearest neighbor interactions of the $\pi^{1} \pi^{1}$ type appear for nearest neighbor interactions along the rays parallel to the ${\bm{e}}_{x}$ direction (i.e., appear, for a given site to its two neighbors at angles of zero or 180$^{\circ}$ relative to the ${\bm{e}}_{1}$ crystalline directions). Similarly, interactions of the $\pi^{2,3} \pi^{2,3}$ type appear for rays parallel to the other two crystalline directions.

\subsection{Hybrid Compass Models}
\label{hybrid}
 An interesting and relevant extension of the bare compass models is one in which both usual SU(2) symmetric Heisenberg-type exchange terms $\bm{\tau_i} \cdot \bm{\tau_j}$ appear in unison with the directional bonds of the bare  $90^\circ$ or $120^\circ$ compass model, resulting in {\it compass-Heisenberg} Hamiltonians of the type
\begin{eqnarray}
\label{hybrid_compass}
H= - \sum_{i,\gamma} (J_{H}  \bm{\tau_i} \cdot \bm{\tau}_{i+\bm{e}_\gamma} + J_{K} \tau^\gamma_i \tau^\gamma_{i+\bm{e}_\gamma}), 
\end{eqnarray}
where $J_H$ denotes the coupling constant for the interactions of Heisenberg form and $J_{K}$ the coupling constant of the compass or Kitaev terms in the Hamiltonian. For instance the $120^\circ$ rendition of this Hamiltonian lattice has been considered on a honeycomb lattice, where it describes exchange interactions between the magnetic moments Ir$^{4+}$ ions in a family of layered iridates A$_2$IrO$_3$ (A= Li, Na) -- materials in which the relativistic spin-orbit coupling plays an important role \cite{Chaloupka10,Trousselet11}. The hybrid $90^\circ$ Heisenberg-compass model was introduced in the context of interacting $t_{2g}$ orbital degrees of freedom \cite{Brink04} and its 2D quantum incarnation was studied  by~\onlinecite{Trousselet10}.  Another physical context in which such a hybrid model appears is the modeling of the consequences of the presence of orbital degrees of freedom in LaTiO$_3$ on the magnetic interactions in this material~\cite{Khaliullin01}. We will review in detail these resulting Heisenberg-compass and Heisenberg-Kitaev models \cite{Chaloupka10,Reuther11} and its physical motivations in Sec.~\ref{sec:sporb} and Sec.~\ref{HKCS}. 

In a very similar manner hybrids of Ising and compass models be constructed. An {\it Ising-compass} Hamiltonian of the form $H_\square^{90^\circ}+H_\square^{Ising}$ has for instance been introduced and studied by \onlinecite{Brzezicki10}.

\section{Generalized \& Extended Compass Models}
\label{sec:generalized_compass}

Thus far, we focused solely only a single pseudospin at a given site. It  is also possible to consider situations in which more than one pseudospin appears at a site or with a coupling between pseudospins and usual spin degrees of freedom -- a situation equivalent to having two pseudospin degrees of freedom per site. Kugel-Khomskii (KK) models comprise a class of Hamiltonians that are characterized by having both spin and pseudospin (orbital) degrees of freedom on each site. These models are introduced in the next Section but their physical incarnations will be reviewed in detail in Sec.~\ref{sec:motivation}. The KK models are reviewed 
in Sec.~\ref{sec_KK} followed by a possible generalization that we briefly introduce and discuss which includes multiple pseudo-spin degrees of freedom. We will then discuss, in Sec.~\ref{sec:clas}, extensions of the quantum compass models introduced earlier to the classical arena, to higher dimensions and to large number of spin components $n$. In Sec.~\ref{sec:other_compass} we collect other compass model extensions.

\subsection{Kugel-Khomskii Spin-Orbital Models} 
\label{sec_KK}
The situation in which at a site both pseudospin and usual spin degrees of freedom are present naturally occurs in the realm of orbital physics. It arises when (electron) spins can occupy different orbital states of an ion -- the orbital degree of freedom or pseudospin. 
The spin and orbital degree of freedom couple to each other because the inter-site spin-spin interaction depends on the orbital states of the two spins involved. Hamiltonians that result from such a coupling of spin and orbital degrees of freedom are generally know as Kugel-Khomskii (KK) model Hamiltonians, after the authors that have first derived~\cite{Kugel72,Kugel73} and reviewed them~\cite{Kugel82} in a series of seminal papers. Later reviews include~\cite{Tokura00,Khaliullin05}

The physical motivation and incarnations of such KK spin-orbital models will be discussed in Sec.~\ref{sec:orbitals}. In Sec.~\ref{sec:KK} they will be derived for certain classes of materials from models of their microscopic electronic structure, in particular from the multi-orbital Hubbard model in which the electron-hopping integrals $t_{i,j}^{\alpha \beta}$ between orbital $\alpha$ on lattice site $i$ and $\beta$ on site $j$ and the Coulomb interactions between electrons in orbitals on the same site are the essential ingredients. A KK Hamiltonian then emerges as the low-energy effective model of a multi-orbital Hubbard system in the Mott insulating regime, when there is on average an integer number of electrons per site and Coulomb interactions are strong. In that case charge excitations are suppressed because of a large gap and the low energy dynamics is governed entirely by the spin and orbital degrees of freedom. In this Section we introduce the generic structure of KK models. Generally speaking the interaction between spin and orbital degrees of freedom on site $i$ and neighboring site $i+\bm{e}_\gamma$ is the product of usual spin-spin exchange interactions and compass-type orbital-orbital interactions on this particular bond. 
The generic structure of the KK models therefore is
\begin{equation}
H^{KK}=- J_{KK} \sum_{i,\gamma} 
H^{orbital}_{i,i+\bm{e}_\gamma} H^{spin}_{i,i+\bm{e}_\gamma}.  
\label{eq:KK}
\end{equation}
$H^{orbital}_{i,i+\bm{e}_\gamma} $ are operators that act on the pseudospin (orbital) degrees of freedom ${\bm \tau}_i$ and  ${\bm \tau}_{i+\bm{e}_\gamma}$ on sites $i$ and $i+\bm{e}_\gamma$ and $H^{spin}_{i,i+\bm{e}_\gamma}$ acts on the spins ${\bm S}_i$ and  ${\bm S}_{i+\bm{e}_\gamma}$ at these same sites. 

When the interaction between spin degrees is considered to be rotational invariant so that it only depends on the relative orientation of two spins, $H^{spin}_{i,i+\bm{e}_\gamma}$ takes the simple Heisenberg form ${\bm S}_i \cdot {\bm S}_{i+\bm{e}_\gamma}  + c_S$.  This is the usual rotationally invariant interaction between spins if orbital (pseudospin) degrees of freedom are not considered. $H^{orbital}_{i,i+\bm{e}_\gamma}$, in contrast, is a Hamiltonian of the compass type. KK Hamiltonians can thus be viewed as particular extensions of compass models, where the interaction strength on each bond is determined by the relative orientation of the spins on the two sites connected by the bond. 

Electrons in the open $3d$ shell of for instance transition metal ions can, depending on the local symmetry of the ion in the lattice and the number of electrons in the $3d$ shell an orbital degree of freedom. In case of orbital degrees of freedom of so-called $e_g$ symmetry two distinct orbital flavors are present  (corresponding to an electron in either a $3z^2-r^2$ or a $x^2-y^2$ orbital). On a 3D cubic lattice the purely orbital part of the superexchange Hamiltonian $H^{orbital}_{i,i+\bm{e}_\gamma}$ takes the $120^\circ$ compass form~\cite{Kugel82}:
\begin{eqnarray}
\label{orbHam}
H^{orbital}_{i,i+\bm{e}_\gamma}= \left(\frac{1}{2} + \hat{\pi}_i^{\gamma}\right)\left(\frac{1}{2} + \hat{\pi}_{i+\bm{e}_\gamma}^{\gamma}\right),
\end{eqnarray}
where $\hat{\pi}_i^{\gamma}$ are the orbital pseudospins  and, as in the earlier discussion of compass models, ~$\gamma$ is the direction of the bond~$\langle ij\rangle$ . The pseudospins $\hat{\pi}_{i}^{\gamma}$ are defined in terms of $\tau_{i}^{\gamma}$ cf. Eq.~(\ref{eq:pi120}) as the 120$^\circ$ type compass variables. If the spin degrees of freedom in the KK Hamiltonian Eq.~(\ref{eq:KK}) are considered as forming static and homogenous bonds, then on the lattice only the orbital exchange part of the Hamiltonian is active. The Hamiltonian $\sum_{i,\gamma}H^{orbital}_{i,i+\bm{e}_\gamma}$ then reduces to $H^{120}_{3\square}$, up to a constant, as for the 120$^\circ$ compass variables $\sum_{\gamma} \tau_i^{\gamma}=0$.

For transition metal $3d$ orbitals of $t_{2g}$ symmetry, there are three orbital flavors ($xy$, $yz$ and $zx$), a situation similar to $p$ orbitals (that have the three flavors $x$, $y$ and $z$).  As one is dealing with a three-component spinor, the most natural representation of three-flavor compass models is in terms of the generators of the SU(3) algebra, using the Gell-Mann matrices, which are the SU(3) analog of the Pauli matrices for SU(2). Such three-flavor compass models also arise in the context of ultra-cold atomic gases, where they describe the interactions between bosons or fermions with a $p$-like orbital degree of freedom~\cite{Chern11}, which will be further reviewed in Sec.~\ref{sec:motivation}. In descriptions of transition metal systems, which we will explore in more detail in section \ref{sec:orbitals}, with pseudo-spin (orbital) and spin degrees of freedom, usual spin exchange interactions are augmented by both pseudo-spin interactions and KK type terms describing pseudo-spin (i.e., orbital) dependent spin exchange interactions. 

In principle, even richer situations may arise when, aside from spins, one does not have a single additional pseudospin degree of freedom per site, as in the KK models, but two or more.  As far as we aware, such models have so far not been considered in the literature. The simplest variants involving two pseudospins at all sites give rise to compass type Hamiltonians of the form
\begin{eqnarray}
H  &=& \sum_{i,\gamma} [J_{\gamma} \tau^\gamma_i \tau^\gamma_{i+\bm{e}_\gamma}  +
J'_{\gamma} \tau^{\prime \gamma}_i \tau'^{\gamma}_{i+\bm{e}_\gamma} ] \nonumber
\\ &+&  \sum_{i, \gamma, \gamma'} [V_{\gamma \gamma'} 
 \tau^{\gamma}_{i} \tau'^{\gamma'}_{i} + W_{\gamma \gamma'} 
 \tau^\gamma_i \tau^\gamma_{i+\bm{e}_\gamma} 
 \tau'^\gamma_i \tau'^\gamma_{i+\bm{e}_\gamma'}]  \nonumber
 \\ &+& \cdots .
 \label{multi-pseudo}
\end{eqnarray}
Such interactions may, of course, be multiplied by a spin-spin interaction as in the Kugel-Khomskii Hamiltonian of Eq.~(\ref{eq:KK}). 

\subsection{Classical, Higher D and Large $n$ Generalizations}
\label{sec:clas}
A generalization to larger pseudospins is possible in all compass models \cite{Nussinov04, Biskup05, Mishra04} and proceeds by replacing the Pauli operators $\tau_{i}^{\gamma}$ by corresponding angular momentum matrix representations of size $(2T+1) \times (2T+1)$ with $T> 1/2$. The limit $T \to \infty$ then corresponds to a classical model. For the classical renditions of the $H^{90^\circ}_\square$ and $H^{120^\circ}_\square$ compass models  $\bm{T}$ is a two component ($n=2$) vector of unit length,
\begin{eqnarray}
\label{bnormal}
(T_{i}^{x})^{2} + (T_{i}^{y})^{2} =1,
\end{eqnarray}
on each lattice site $i$. this is simply because $\tau^z$ does not appear in the Hamiltonian. In a similar manner, for $n=3$ renditions of the compass model, as for instance in $H^{90^\circ}_{3\square}$, the vector ${\bm{T}}$ has unit norm and three components. 

\begin{figure}
\centering
\includegraphics[width=\columnwidth]{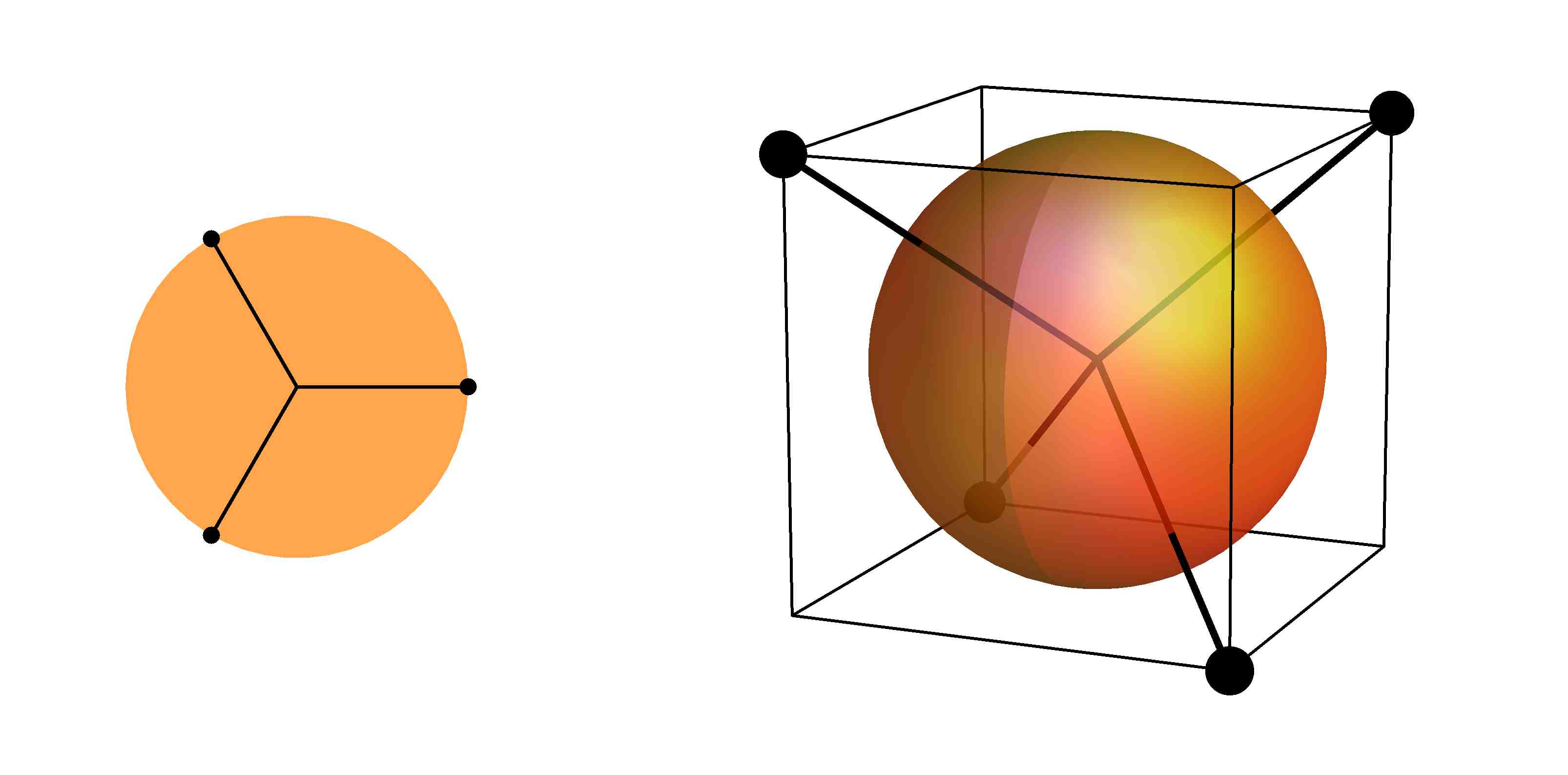}
\caption{Left: unit disk with three uniformly spaced vectors, the building blocks for the $120^\circ$ model with $n=2$, on for instance a $3D$ cubic or the $2D$ honeycomb lattice. Right: generalization to higher dimensions with four uniformly spaced vectors on the $n=3$ dimensional unit sphere, relevant to a $4D$ hyper-cubic lattice, or the $3D$ diamond lattice.
}
\label{fig:unit120}
\end{figure}

An obvious extension is to consider vectors ${\bm{T}}$ with a general number of components $n$. The $90^\circ$ compass models (Eq.~(\ref{c90comp})) generalize straightforwardly to any system having $n$ independent directions $\gamma$. The simplest variant of this type is a hyper-cubic lattice in $D=n$ dimensions wherein along each axis $\gamma$ (all at 90$^\circ$ relative to each other) the interaction is of the form 
\begin{eqnarray}
\label{highD90}
H_{\square}^{~\sf classical ~90^\circ} = - \sum_{i,\gamma} J_{\gamma} T_{i}^{\gamma} T_{i+\hat{e}_{\gamma}}^{\gamma}.
\end{eqnarray}
[More generally, we will set in general classical analogs, $T^{\gamma}_{i} \equiv {\bf T}_{i} \cdot {\bf e}_{\gamma}$.]
  When looked at through this prism, the one dimensional Ising model can be viewed as a classical one dimensional rendition of a compass model. 

In the classical arena, when ${\bm{\tau}}$ is replaced by vectors ${\bm{T}}$ of unit norm, there is also a natural generalization of the 120$^\circ$ compass model to hyper-cubic lattices in arbitrary spatial dimension $D$. To formulate this generalization, it is useful to introduce the unit sphere in $n$ dimensions. In the classical 120$^\circ$ compass model on the $D=3$ cubic lattice, the three two-component vectors $T^\gamma$ are uniformly partitioned on the unit disk (the $n=2$ unit sphere). These form $D$ equally spaced directions $\bm{e}_{\gamma}$ on the $n$ unit sphere. The angle $\theta$ between any pair of differing vectors is therefore same (and for $n=2$ equal to $2\pi/3$). The generic requirement of uniform angular spacing of $D$ vectors on a sphere in $n$  dimensions is possible only when $n=D-1$. The angle $\theta$ between the unit vectors is then given by
\begin{eqnarray}
\label{general_d_120}
\bm{e}_{\gamma} \cdot \bm{e}_{\gamma'} = \cos \theta = - \frac{1}{D-1}.
\end{eqnarray}
If $n=3$, for instance, the four equally spaced vectors can be used to describe the interactions on any lattice having 4 independent directions $\gamma$, for instance the $4D$ hyper-cubic one, or the $3D$ diamond lattice, see Fig.~\ref{fig:unit120}.

It is interesting to note that formally, in the limit of high spatial dimension of a hyper-cubic lattice rendition of the 120$^\circ$ model, the angle $\theta \to$ 90$^\circ$ and the  two most prominent types of compass models discussed above (the 90$^\circ$ and 120$^\circ$ compass models) become similar (albeit differing by one dimension of the $n$ dimensional unit sphere on which ${\bm{T}}$ is defined). 

From here one can return to the quantum arena. The quantum analogues of these $D$ dimensional classical compass models (including extensions of the 120$^\circ$ model on a $3D$ cubic lattice) can be attained by replacing ${\bm{T}}$ by corresponding quantum operators ${\bm{\tau}}$ that are the generators of spin angular momentum in $n$ dimensional space. 
These are then finite size representations of the quantum spin angular momentum generators in an $n$ dimensional space (e.g., the representations $T=1/2, 1, 3/2, ... )$ of SU(2) for a three component vector just discussed earlier (including the pertinent $T=1/2$ representation), representations of SU(2) $\times$ SU(2) for a four component ${\bm{\tau}}$, representations of Sp(2) and SU(4) for a five and six component ${\bm{\tau}}$, and so on). 

These dimensional extensions and definitions of the 90$^\circ$ and 120$^\circ$ models are not unique. The so-called ``one dimensional 90$^\circ$ compass model''  (sometimes also referred to as the one-dimensional
Kitaev model) was studied in multiple works, e.g., \cite{Brzezicki07,You08,sun-compass}. In its simplest initial rendition \cite{Brzezicki07}, this model is defined on a chain in which nearest neighbor interactions sequentially toggle between being of the $\tau^x_{2i} \tau^x_{2i+1}$ and $\tau^{y}_{2i+1} \tau^{y}_{2i+2}$ variants as one proceeds along the chain direction for even/odd numbered bonds. Many aspects of this model have been investigated such as
its quench dynamics \cite{Divakaran2009,mondal2008}. Such a system is, in fact, dual to the well-studied one-dimensional transverse field Ising model, e.g., 
 \cite{Brzezicki07,nussinov-bond,Eriksson09}. A two leg ladder rendition of Kitaev's honeycomb model (and, in particular, the quench dynamics in this system) was investigated in \cite{Sen2010}.
 A very interesting two-dimensional realization of the 120$^\circ$ model was further introduced and studied ~\cite{You08} wherein only two of the directions $\gamma$ are active in Eq.~(\ref{eq:c120}). 

Lastly, we comment on these models (in their classical or quantum realization) in the ``large $n$ limit'' wherein the number of Cartesian components of the pseudo-spins ${\bf T}$ becomes large. 
This limit, albeit seemingly academic, is special. The $n \to \infty$ limit has the virtue that it is exactly solvable, where it reduces to the ``spherical model'', \cite{Berlin1952,Stanley1968} and further 
amenable to perturbative corrections in ``$1/n$ expansions'' \cite{Ma1973}. We will return to discuss some aspects of the large $n$ limit in 
section \ref{sec:diso}.

\subsection{Other Extended Compass Models}
\label{sec:other_compass}

\subsubsection{Arbitrary angle}
Several additional extensions of the more standard models have been proposed and studied in various contexts. One of these includes a generalized angle that need not be 90$^\circ$ or 120$^\circ$ or another special value Ref. \cite{Cincio} considered a variant of Eq.~(\ref{120p}) on the square lattice in which, instead of Eq.~(\ref{eq:c120}),
one has
\begin{eqnarray}
\hat{\pi}_{i}^{x}=  \cos (\theta/2) \tau_{i}^{x} +  \sin (\theta/2) \tau_{i}^{y} \nonumber \\ 
\hat{\pi}_{i}^{y} =  \cos (\theta/2) \tau_{i}^{x} -  \sin (\theta/2) \tau_{i}^{y}
\label{generalized_ocm}
\end{eqnarray} 
with a tunable angle $\theta$.  

\subsubsection{Plaquette and Checkerboard (sub-)lattices}
\label{plaq&check}
\begin{figure}
\centering
\includegraphics[width=.6\columnwidth]{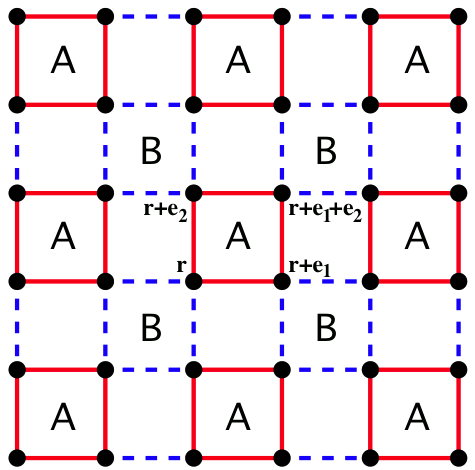}
\caption{The configuration underlying the definition of the plaquette orbital model. Here the $x$ components of the spins are coupled over the red (solid) edges and the $z$ components are coupled over the blue (dashed) edges~\cite{Biskup10}.}
\label{fig:Biskup10_1}
\end{figure}

Another variant that has been considered, initially introduced to better enable simulation~\cite{Wenzel09}, is one in which the angle $\theta$ is held fixed ($\theta =90^\circ$) but the distribution of various bonds is permuted over the lattice~\cite{Biskup10}. Specifically, the {\it plaquette orbital model} is defined on the square lattice via
\begin{eqnarray}
H_{POM}= - J_A \sum_{\langle i j \rangle \in A}  \tau^{x}_{i} \tau^{x}_{j} 
- J_B  \sum_{\langle i j \rangle \in B}  \tau^{y}_{i} \tau^{y}_{j} ,
\label{plaq_ocm}
\end{eqnarray}
where $A$ and $B$ denote two plaquette sublattices, see Fig.~\ref{fig:Biskup10_1}. Bonds are summed over according to whether the physical link $\langle i j \rangle$ resides in sublattice A or sublattice B. 
Although this system is quite distinct from the models introduced thus far, it does share some common features, including a {\it bond algebra} 
(the notion of bond-algebra  \cite{cobanera,ADP,Nussinov08b,nussinov-bond} will be introduced and applied to the Kitaev model in subsection~\ref{solnus})
which as the reader may verify in the Appendix (Section \ref{Aplaq})) is, locally, similar to that of the 90$^\circ$ compass model on the square lattice. 

\begin{figure}
\centering
\includegraphics[width=.6\columnwidth]{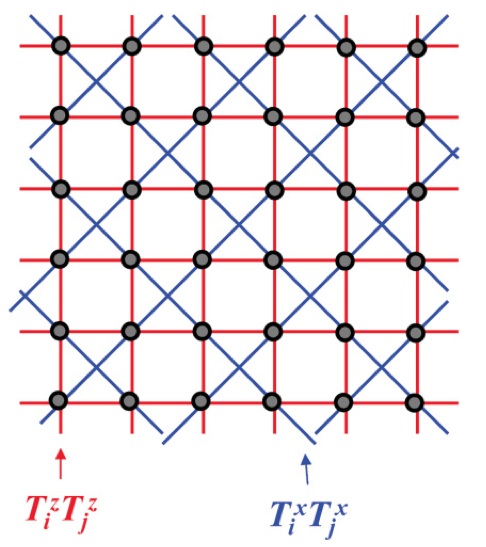}
\caption{A schematic representation for the orbital compass model on a checkerboard lattice.~\cite{Nasu12}.}
\label{fig:Nasu12_1}
\end{figure}

The checkerboard lattice (a two-dimensional variant of the three-dimensional pyrochlore lattice) is composed of corner sharing crossed plaquettes.  This lattice may be regarded as a
square lattice in which on every other square plaquettes, there are additional diagonal links, see Fig.~\ref{fig:Nasu12_1}. 
On this lattice, a compass model may be defined by the following Hamiltonian \cite{Nasu11b,Nasu12}
\begin{eqnarray}
H_{checkerboard}= - J_{x} \sum_{(ij)} \tau_{i}^{x} \tau_{j}^{x} - J_{z} \sum_{\langle i j \rangle} \tau^{z}_{i} \tau^{z}_{j}.
\label{checker}
\end{eqnarray}
In the first term of Eq. (\ref{checker}), the sum is over diagonal (or next nearest neighbor) pairs in crossed plaquettes. The second term in Eq. (\ref{checker}) contains a sum over all nearest neighbor (i.e., horizontal or vertical) pairs on the lattice.

\subsubsection{Longer-range and Ring Interactions} 
In a similar vein, compass models can be defined by pair interactions of varying range and orientation on other general lattices. For instance in the study of layered oxides, \onlinecite{Kargarian2012} introduced a hybrid compass model of Heisenberg-Kitaev type with nearest-neighbor and next-neighbor interactions on the honeycomb lattice, which we will return to in section \ref{sec:sporb}.

One should keep in mind that models in which different spin components couple for different spatial separations may be similar to compass models that we have considered in the previous sections, yet on enlarged lattices. A case in point is that of a one dimensional spin system with the Hamiltonian
\begin{eqnarray}
H_{chain} = - J_{x} \sum_{i} \tau_{i}^{x} \tau_{i+1}^{x} - J_{z} \sum_{i} \tau^{z}_{i} \tau^{z}_{i+2}.
\label{checker*}
\end{eqnarray}
Here, the interactions on the chain defined by the Hamiltonian of Eq. (\ref{checker*}) are topologically equivalent to  a system composed on two parallel chains that are horizontally displaced from one another by half a lattice constant. On one of these chains, we  label the sites by odd integers, i.e., $i=1,3,5, ... $ while the other chain hosts the even sites $i=2,4, ...$. On this lattice, the Hamiltonian of Eq. (\ref{checker*}) assumes a form similar to that of Eq. (\ref{checker}) when the $J_{x}$ interactions appear along diagonally connected sites between the two chains while $J_{z}$ coupling occurs between spins that lie on the same chain. Thus, the one dimensional system with interactions that vary with the range of the coupling between spins is equivalent to a compass model wherein the spin coupling is dependent on the orientation between neighboring  spin pairs. 

Compass models need not involve only pair interactions. A key feature of models that go beyond pair interactions is that the internal pseudospin components appearing in the interaction terms that depend on a external spatial direction can be extended to any number of interacting pseudospins. A very natural variant was considered in~\cite{Nasu11a} for ring exchange interactions involving four spins around basic square plaquette in a cubic lattice. Specifically, these interactions are defined via the Hamiltonian
\begin{eqnarray}
H_{ring}= K \sum_{[ijkl]_{\gamma}} 
(\tau^{\gamma +}_{i} \tau^{\gamma -}_{j} \tau^{\gamma +}_{k} \tau^{\gamma -}_{l} + h.c.).
\label{ring_exchange}
\end{eqnarray}
In ~Eq.(\ref{ring_exchange}), 
$\tau^{\pm \gamma}_{i} = \tau^{\gamma}_{i} \pm i \frac{\sqrt{3}}{2} \tau^{y}_{i}$
where, similar to the 120$^\circ$ model, 
$\tau_{i}^{\gamma} = \cos(2 \pi n_{\gamma}/3) \tau^{z}_{i} - \sin(2 \pi n_{\gamma}/3) \tau_{i}^{x}.$
In Eq.~(\ref{ring_exchange}), the subscript $[ijkl]_{\gamma}$ denotes sites $[ijkl]$
forming a four-site plaquette that is perpendicular to the cubic lattice direction $\gamma$.
In the definition of $\tau_{i}^{\gamma}$, 
$n_{\gamma}=1$ for a direction $\gamma$ parallel to the x-axis
(i.e., the plaquette $[ijkl]$ is orthogonal to the x direction).  
Similarly, $n_{\gamma} =2$ or 3 for an orientation $\gamma$ parallel to the cubic lattice
y- or z- axis. The physically motivated Hamiltonian of Eq.(\ref{ring_exchange}) 
with its definitions of  $\tau_{i}^{\gamma}$ corresponds to a ring-exchange of interactions of the 120$^\circ$ type. One may similarly consider extensions for other angles $\theta$.

\section{Compass Model Representations}
\label{sec:representations}

\subsection{Continuum Representation}
A standard approach in statistical mechanics is to construct effective continuum descriptions of discrete models. A continuum representation of a compass models can be attained by coarse-graining its discrete counterpart with pseudo-spins attached to each point on a lattice. Such coarse-grained continuum representations can offer much insight into the low-energy, long-wave-length behavior and properties of lattice models. We therefore briefly discuss the particular field-theoretic incarnation of compass type systems, both classical and quantum. 

\subsubsection{Classical Compass Models}
For a classical pseudospin $\bm{T}$ one defines 
$
T_i^{\gamma} = \bm{T}_{i} \cdot \bf{n}_{\gamma},
$
with the angles defining $\bf{n}_{\gamma} = (\cos \theta_{\gamma}, \sin \theta_{\gamma})$ given by
Eq.(\ref{eq:c120}) for the 120$^{o}$ model. Similarly, in the 90$^{o}$ compass model
in three dimensions, the three internal pseudospin 
polarization directions ${\bf{n}}$ are defined by ${\bf{n}} = {\bf{e}}_{x},
{\bf{e}}_{y}$ or ${\bf{e}}_{z}$.  In going
over from the discrete lattice model to its continuum representation
one uses 
\begin{eqnarray}
\label{triv_grad}
-T_{i}^{\gamma} T_{i+{\bf{e}}_{\gamma}}^{\gamma} &\to& \frac{a}{2} (T^{\gamma}_{i+{\bf{e}}_{\gamma}}
-  T_{i}^{\gamma})^{2} - \frac{a}{2} [(T_{i+{\bf{e}}_{\gamma}}^{\gamma})^{2} + (T_{i}^{\gamma})^{2}] \nonumber
\\ &\to& \frac{a}{2} (\partial_{\gamma} 
T ^{\gamma})^{2},
\end{eqnarray}
where $a$ is the lattice constant and the normalization of the pseudo-vector $\sum_{\gamma}   (T_{i}^{\gamma})^{2}$ has been invoked. Classical compass models will be reviewed in detail later. 
For now, we note that if $\bm{T}$ is a vector of unit norm then, in the 120$^{0}$ model in $D=3$ dimensions, regardless of the orientation of that vector on the unit disk,  $\sum_{\gamma}  (T_{i}^{\gamma})^{2} = 3/2$ identically. (For a rendition of the 120$^{o}$ model of the form of Eq.~(\ref{general_d_120}) in $D$ dimensions the general result is $D/(D-1)$.)  In a similar fashion, for the classical 90$^{o}$ model $\sum_{\gamma}   (T_{i}^{\gamma})^{2} = 1$. In all such instances, $\sum_{\gamma}  (T_{i}^{\gamma})^{2} $ identically amounts to an innocuous constant and as such may be discarded. 
 
In what follows, briefly the ``soft-spin''  approximation will be discussed, in which the ``hard-spin'' constraint ${\bf{T}}^{2} =1$ is replaced by a quartic term 
of order $\lambda$ that enforces it weakly. Such a term is of the form  $(\lambda/4!)({\bf{T}}^{2} -1)^{2}$ with small positive $\lambda$. The limit $\lambda \to \infty$ corresponds to the ``hard-spin'' situation in which the pseudospin is strictly normalized at every point.

With the definition of $T_i^{\gamma}$ and simple preliminaries, the continuum limit Ginzburg-Landau  type free energy in $D$ spatial dimensions is
\begin{eqnarray}
F =   \int d^{D}x  \Big[ \sum_{\gamma} \frac{(\partial_{\gamma} 
T ^{\gamma})^{2}} {2g} + \frac{r}{2} {\bf{T}}^{2} + \frac{\lambda}{4!} ({\bf{T}}^{2})^{2} \Big] ,
\label{continuum_lim} 
\end{eqnarray}
with $g$ an inverse coupling constant and $r$ a parameter that emulates the effect of temperature, $r=c(\cal{T}-\cal{T}_{0})$ with $c$ a positive constant and $\cal{T}_{0}$ the mean-field temperature. The partition function of the theory is then given by a functional integration over all pseudospin configurations at all lattice sites, $Z = \int DT e^{-F}$. What differentiates this form from standard field theories is that it {\em does not transform as a simple scalar} under rotations. Inspecting Eq.~(\ref{continuum_lim}), one sees that there is no implicit immediate summation over the repeated index $\gamma$ in the argument of the square. In Eq.~(\ref{continuum_lim}), the summation over
$\gamma$ is performed at the end after the squares of the various gradients have been taken.  Written long-hand for, e.g.,  the 90$^{o}$ compass model in two dimensions, the integrand is
\begin{eqnarray}
\left( \frac{\partial T^{x}}{\partial x} \right)^{2} + \left(\frac{\partial T^{y}}{\partial y}\right)^{2}.
\end{eqnarray} 
This is to be distinguished from the square of the divergence of $\bf{T}$
(in which the sum over $\gamma$ would be made prior to taking the 
square) which would
read
\begin{eqnarray}
\left(\frac{\partial T^{x}}{\partial x}\right)^{2} + 2 \frac{\partial T^{x}}{\partial x}\frac{\partial T^{y}}
{\partial y} + \left(\frac{\partial T^{y}}{\partial y}\right)^{2}.
\end{eqnarray}
This is also different from the square of the gradient of components $T^{\gamma}$ and their sums thereof for which, rather explicitly, one would have {\em for any single component} $\gamma =x$ or $y$, 
\begin{eqnarray}
(\nabla T^{\gamma})^{2}  = \left(\frac{\partial T^{\gamma}}{\partial x}\right)^{2}
+ \left(\frac{\partial T^{\gamma}}{\partial y}\right)^{2}.
\end{eqnarray}
In the present case, ${\bf{T}}$ indeed represents an internal degree of freedom that does not transform under a rotation of space. By comparison to standard field-theories,  Eq.~(\ref{continuum_lim}) manifestly breaks rotational invariance -- a feature that is inherited from the original lattice models that it emulates. In Sec.~\ref{sec:sym} the investigations of symmetries as well as of the classical compass models will be reviewed in detail. 

\subsubsection{Quantum Compass Models}

As with usual spin models, the quantum pseudospin systems differ from their classical counterparts by the addition of Berry phase terms. This phase, identical in form to that appearing in spin systems, can be written both in the real time and the imaginary time (Euclidean) formalisms. \cite{Eduardo_book, Subir_book} In the quantum arena, one considers the dynamics in  imaginary time $\tau$  where $0 \le u \le \beta$ with $\beta$ the inverse temperature. The pseudospin ${\bf T}(u)$ evolves on a sphere of radius $T$
with the boundary conditions that ${\bf T}(u =0) = {\bf T}(u = \beta)$. Thus, the pseudospin describes a closed trajectory  on a sphere of radius $T$.  The Berry phase for quantum spin systems (also known as the Wess-Zumino-Witten term (WZW))  is,
for each single pseudospin at site $j$, given by  $S^{WZW}_{j}=-iTA_{j}$ with $A$ the area the spherical cap circumscribed by the  closed pseudospin trajectory at that site. 
That is, there is a quantum mechanical 
(Aharonov-Bohm type) phase that
would be associated with a magnetic 
monopole of strength $T$ situated 
at the origin.  Denoting the orientation
on the unit sphere by ${\bf{n}}$, that monopole
may be described by a vector potential
${\cal{A}}$ is a function of ${\bf{n}}$  that solves the equation
$\epsilon^{abc} (\partial {\cal{A}}^{b}/\partial n^{c}) = T n^{a}$.
The partition function is given by for ferromagnetic
variants of the compass models is given by
\begin{eqnarray}
Z &=& \int Dn^{a}({\bf x}, u) \delta((n^{a})^{2} -1) \exp(-S), \nonumber
\\ S &=& iT  \int_{0}^{\beta} du  \int d^{D} x {\cal{A}}^{a} \frac{dn^{a}}{du} \nonumber
\\ &+&  T^{2} \int_{0}^{\beta} du \int d^{D}x  \sum_{\gamma} \frac{(\partial_{\gamma} 
n^{\gamma})^{2}}{2g} .
\label{ZSn}
\end{eqnarray}
As in the classical case, we note that here summation over $\gamma$ is
performed only after the squares have been taken. 
Similar to the ``soft-spin'' classical model, it is possible to construct 
approximations in which the delta function in Eq.~(\ref{ZSn}) is replaced by soft
quartic potentials of the form $\frac{\lambda}{4!}({\bf n}^{2} -1)^{2}$. 
In the classical case as well as for XY quantum systems 
(such as the 120$^{o}$ compass), the behavior of $J>0$ and $J<0$ systems is
identical.  As noted earlier, this is no longer 
true in quantum compass systems in which all three components
of the spin appear. Similar to the case of usual quantum spin systems, the role of the 
Berry phase terms is quite different for ferromagnetic and anti-ferromagnetic 
renditions of the three component compass models. Although the squared gradient
exchange involving ${\bf n}$ can be made similar when looking at
the staggered pseudospin on the lattice, the Berry phase term
will change upon such staggering and may lead to non-trivial
effects. 

\subsection{Momentum Space Representations}
The directional dependence of the interactions in compass models is, of course, manifest also in momentum space. Such a momentum space representation strongly hint that the 90$^{o}$ compass models may exhibit  a dimensional reduction~\cite{Batista05}. A general pseudospin model having $n$ components can be Fourier transformed and cast into the form
\begin{eqnarray} 
H= \frac{1}{2} \sum_{\bf k} {\bf T}^{\dagger}({\bf k}) \hat{V}({\bf k}) {\bf T}({\bf k}).
\label{Fourier_space}
\end{eqnarray}
In Eq.~(\ref{Fourier_space}),  ${\bf k}$ is the momentum space index, the row vector ${\bf T}^{\dagger}({\bf k}) = (T^{1}({\bf k}), T^{2}({\bf k}), ..., T^{n}({\bf k}))^{*}$ with $^*$ representing complex conjugation is the hermitian conjugate of ${\bf T}({\bf k})$ and ${\hat{V}}({\bf k})$ is a momentum space kernel- an $n \times n$ matrix whose elements depend on the $D$ components of the momenta ${\bf k}$. 

In usual isotropic spin exchange systems (i.e., those with isotropic interactions of the form ${\bf T}_{i} \cdot {\bf T}_{j}$ between (real-space) nearest neighbor lattice sites $i$ and $j$), the kernel ${\hat{V}}({\bf k})$ has a particularly simple form,
\begin{eqnarray}
\label{iso}
{\hat{V}}_{isotropic} = \left(-2 \sum_{l=1}^{D} \cos k_{l}\right)  \dblone_{n},
\end{eqnarray}
with $k_{l}$ the $l$th Cartesian component of ${\bf k}$ and $\dblone_{n}$ the $n \times n$ identity matrix.  There is a redundancy in the form of Eq.~(\ref{iso}) following from spin normalization. At each lattice site $i$ the sum $\sum_{\gamma} (T_{i}^{\gamma})^{2}$ is a constant so that $\sum_{i}  (T_{i}^{\gamma})^{2}$ is a constant
proportional to the total number of sites. From this follows that $\sum_{\bf{k}} {\bf T}^{\dagger}({\bf k}) {\bf{T}}({\bf k})$ is a constant. Consequently, any constant term (i.e., any constant (non-momentum dependent) multiple of the identity matrix) may be added to the right-hand side of Eq.~(\ref{iso}). Choosing this constant to be $2D$, in the 
 continuum limit, the right hand of Eq.~(\ref{iso}) disperses as ${\bf k}^{2}$ for small wave vectors ${\bf k}$. This is, of course, a manifestation of the usual squared gradient term that appears in standard field theories whose Fourier transform is given by ${\bf k}^{2}$.  Thus, in the standard case, the momentum space kernel ${\hat{V}}_{isotropic}$ has a single zero (or lowest energy state) with a dispersion that rises, for small ${\bf{k}}$ quadratically in all directions.

\subsubsection{Dimensional Reduction}
The form of the interactions is drastically different for compass models. As will be discussed in e.g., Sec.~\ref{sec:NBCB} in greater depth, this may lead to a flat momentum space dispersion in which lines of zeros of ${\hat{V}}({\bf k})$ appear much unlike the typical quadratic dispersion about low energy modes. The relation between the directional character of the interactions in external space (that of $D$ dimensions) and the internal space (the $n$ components of ${\bf T}$). {\bf sentences misses verb}
The $n \times n$ kernel ${\hat{V}}$ can be  written down for all
of the compass models introduced earlier by replacing any
appearance of $(J_{\gamma \gamma' l} T^{\gamma}_{i} T^{\gamma'}_{j}$)
in the Hamiltonian where the real space between nearest neighbor sites $i$ and $j$
are separated along the $l$-th lattice Cartesian direction
(on a hypercubic lattice) by a corresponding matrix element
of ${\hat{V}}$ that is given by
$\langle \gamma| {\hat V}| \gamma' \rangle = 2 J_{\gamma \gamma ' l} \cos k_{l}$.
By contrast to the usual isotropic spin exchange interactions,
the resulting ${\hat{V}}$ for compass models
is no longer an identity matrix in the internal 
$n$ dimensional space spanning the components
of ${\bf T}$. Rather, each component of ${\hat{V}}$ can have
a very different dependence on ${\bf k}$. For the 90$^{o}$ compass
models this allows expression of the Hamiltonian in the form of a one-dimensional system in disguise. 
One sets 
${\hat{V}}$ to be a diagonal matrix whose diagonal elements are
given by 
\begin{eqnarray}
\langle \gamma | {\hat{V}}_{90^{o}}| \gamma \rangle = -2J \cos k_{\gamma},
\label{90f}
\end{eqnarray}
the 90$^{o}$ compass model on an $n=D$-dimensional hyper-cubic lattice is recovered. The contrast between Eq.~(\ref{iso}) and Eq.~(\ref{90f}) is marked and directly captures the directional character of the interactions in the compass model.  As in the various compass models  (including, trivially, the 90$^{o}$ compass models),
$ \sum_{i} (T_{i}^{\gamma})^{2}$ is constant at every lattice site $i$, one may as before add to the right hand side of Eq.~(\ref{90f}) any constant times the identity matrix. One can then formally recast Eq.~(\ref{90f}) in a form very similar to a one dimensional variant of Eq.~(\ref{iso}) -- one which depends on only one momentum space ``coordinate'' but with that coordinate no longer being a $k$ but rather a matrix. Towards that end, one may define a diagonal matrix $\hat{K}$ whose diagonal matrix elements are given by $(k_{1}, ..., k_{n})$ and cast  Eq.~(\ref{90f}) as 
\begin{eqnarray}
{\hat{V}}_{90^{o}} = -2J \cos \hat{K}.
\label{90_as_1d}
\end{eqnarray}
In this form, Eq.~(\ref{90_as_1d}) looks like a one dimensional $(D=1)$ model by comparison to Eq.~(\ref{iso}). The only difference is that instead of having a  real scalar quantity $k$ in $1D$ one now formally has an $D \times D$ dimensional matrix (or a quaternion form for the $D=2$ dimensional 90$^\circ$ compass model) but otherwise it looks very much similar. 

Indeed, to lowest orders in various approximations ($1/n$, high temperature series expansions, etc.) the 90$^{o}$ compass models appears to be one dimensional. This is evident in the spin-wave spectrum: naively, to lowest orders in all of these approaches, there seems to be a decoupling of excitations along different directions. That is, in the continuum (small ${\bf k}$ limit), one may replace $2(1- \cos k_{\gamma})$ by $k_{\gamma}^{2}$ and the spectrum for excitations involving $T^{\gamma}$  is identical to that of a one dimensional system parallel to the Cartesian $\gamma$ direction.  This is a manifestation of the unusual gradient terms that appear in the continuum representation of the compass model -- Eqs.~(\ref{triv_grad},\ref{continuum_lim}). In reality, though, the compass models express the character expected from systems in $D$ dimensions (not one-dimensional systems) along their finite temperature phase transitions and universality classes. In the field theory representation of Eq.~(\ref{continuum_lim}), this occurs due to the quartic term that couples the different pseudospin polarization directions (e.g., $T^x$ and $T^y$) to one another.  However, an exact remnant of the dimensional reduction suggested by this form still persists in the form of symmetries \cite{Batista05}, see Sec.~\ref{sec:sym}.

\subsubsection{(In-)Commensurate Ground States}
In what follows below and in later sections, the eigenvalues of ${\bf V}({\bf k})$ for each ${\bf k}$ are denoted by $v_{\alpha}({\bf k})$ with $\alpha = 1,2, ..., n$ with $n$ the number of pseudo-spin components.  In rotationally symmetric, isotropic systems when $v_{\alpha}({\bf k})$ is independent of  the pseudo-spin index $\alpha$ and $ \pm {\bf q}^{*}$ are two wave-vectors that minimize $v$ then, it is easy to see that two-component spirals of the  form  ${\bf{T}}({\bf r}) = (\cos{\bm q^*}\cdot{\bm r}, \sin{\bm q^*}\cdot{\bm r})$ are classical 
ground states of the normalized pseudo-spins ${\bf T}$. Similar extensions appear for $n=3$
(and higher) component pseudo-spins. It has been proven that for general incommensurate wave-vectors ${\bf q}^{*}$,  {\em all} ground states must be spirals of this form \cite{Nussinov2001,Nussinov1999}. When the wave-vectors that minimize $v$ are related to one another by commensurability conditions \cite{Nussinov2001} then more complicated (e.g., stripe or checkerboard type) configurations can arise. 

In several compass type systems that are reviewed here (e.g., the 90$^\circ$ compass and Kitaev's honeycomb model),  the interaction kernel $v$ will still be diagonal in the original internal pseudo-spin component basis ($\alpha =1,2, \cdots,  n$) yet $v_{\alpha}({\bf k})$ will different functions for different $\alpha$. Depending on the model at hand, these functions for different components $\alpha$ may be related to one another by a point group rotation of ${\bf k}$ from one lattice direction to another.
We briefly remark on the case when the wave-vectors ${\bf q}^{*}$ that minimize, for each $\alpha$,
the kernel $v_{\alpha}({\bf k})$ are commensurate and allow the construction
of Ising type ground states  \cite{Nussinov2001} such as commensurate 
stripes or checkerboard states. In such a case it is possible to construct
$n$ component ground states by having Ising type states for each component $\alpha$.
That is, for each internal spin direction $\alpha = 1,2, ..., n$, we can construct states
with $(T_{\alpha}^{2} =a_{\alpha}^{2}$ with $a_{\alpha}$ 
(for an Ising system, only one component $n=1$ exists and 
$a_{\alpha} =1$, for Ising type states, we scale each Cartesian 
component by a uniform factor of $a_{\alpha}$ at all lattice site 
and require that $\sum_{\alpha=1}^{n} a_{\alpha}^{2} =1$ to ensure
global pseudo-spin normalization. As we will review in later sections,
the symmetries that compass type systems exhibit ensures that in many 
cases there is a multitude of ground states that extends beyond expectations 
in most other (pseudo)spin systems.

\subsection{Ising Model Representations}
It is well-known that using the Feynman mapping, one can relate zero temperature quantum system in $D$ spatial dimensions to classical systems in $(D+1)$ dimensions~\cite{Subir_book}. In the current context, one can express many of the quantum compass systems as classical Ising models in one higher dimension. The key idea of such Feynman maps is to work in a classical Ising basis ($\{\sigma^{z}_{i,u}\}$) at each point in space $i$ and imaginary time $u$ and to write the transfer matrix elements of the imaginary time evolution operator between the system and itself at two temporally separated times. The derivation will not be reviewed here, see e.g.,~\cite{Subir_book}.  

A simple variant of the Feynman mapping invokes duality considerations \cite{Nussinov05, cobanera, ADP} to another quantum system \cite{Xu04, Xu05} prior to the use of the standard transfer matrix technique. Here we merely quote the results. The two-dimensional 90$^{o}$ compass model of Eq.~(\ref{eq:general_compass}) in the absence
of an external field ($h=0$) maps onto a classical model in 2+1 dimensions with the action \cite{Nussinov05, ADP}
\begin{eqnarray}
\label{Ising_action}
S = &-K &\sum_{\Box \in (xu)~{\rm plane}} \sigma^{z}_{i,u} \sigma^{z}_{i+ {\bf{e}}_{x},u}
 \sigma^{z}_{i,u+ \Delta u}   \sigma^{z}_{i+ {\bf{e}}_{x},u + \Delta u} \nonumber \\  
 &&-J_{z}  \Delta u \sum_{i} \sigma^{z}_{i,u} \sigma^{z}_{i+{\bf{e}}_{z},u},
 \end{eqnarray}
 with $K$ and $J_{z} \Delta u$ constants that will be detailed later on.
 The Ising spins $\{\sigma^{z}_{i,u} \}$ are situated at lattice points in the 2+1
 dimensional lattice in space-time. A particular separation $\Delta u$ 
 along the imaginary time axis has to be specified in performing
 the mapping of the quantum system onto a classical lattice
 system in space-time. The coupling constants
 in Eq.~(\ref{Ising_action}) are directly related to those
 in Eq.~(\ref{eq:general_compass}). We
 aim to keep the form of Eq.~(\ref{Ising_action}) general
 and cast it in the form of a gauge type theory 
 (with spins at the vertices of the lattice
 instead of on links). The plaquette coupling
 $K$ is related to the coupling constant $J_{x}$
 of Eq.~(\ref{eq:general_compass}) via 
 \begin{eqnarray}
 \sinh 2 (J_{x} \Delta u) \sinh 2K =1.
 \end{eqnarray}
 The particular {\em anisotropic directional} character of the compass model
 rears its head in Eq.~(\ref{Ising_action}). Unlike canonical systems in 
 which the form of the interactions is the same in all plaquettes 
 regardless of their orientation, here four-spin interactions appear
 only for plaquette that lie parallel to the $(xu)$ plane- that is, the plane
 spanned by one of Cartesian spatial directions ($x$) and the imaginary
 time axis ($u$).  Similarly, exchange interactions (of strength $(J_{z} \Delta u)$)
 appears between pairs of spins that are separated along links parallel to the 
 spatial Cartesian $z$ direction.
 
 The zero temperature effective classical Ising action of Eq.~(\ref{Ising_action}) enables
 the study of the character of the zero temperature transition that occurs as $J_x/J_z$
 is varied. From the original compass model of Eq.~(\ref{2dpocm}), it is clear
 that when $|J_{z}|$ exceeds $|J_{x}|$ there is a preferential orientation
 of the spins along the $z$ axis (and, vice versa, when $|J_x|$ exceeds
 $|J_z$ an ordering along the $x$ axis is preferred).  The point $J_x = J_z$
 (a ``self-dual'' point for reasons which will be elaborated on
 later)  marks a transition which has been studied by
various other beautiful means and found to be first order \cite{Dorier05,Chen07a,Urus09}.

 \subsection{Dynamics -- Equation of Motion}
 \label{d-eom}
The anisotropic form of the interactions leads to 
equations of motion that formally appear similar to
those in magnetic systems but are highly anisotropic. 
In general spin and pseudospin systems, time evolution
(both classical (i.e., classical magnetic moments) and quantum) is governed by the equation
of motion
\begin{eqnarray}
\label{Larmor_eq}
\frac{\partial {\bf T}_{i}}{\partial t} = {\bf T}_{i} \times {\bf h}_{i},
\end{eqnarray}
where ${\bf h}_{i}$ is the local magnetic (pseudo-magnetic) field
at site $i$. For a stationary field ${\bf h}$, this leads to a
"Larmor precession''- the spin rotates at constant
rate about the field direction. This well known spin effect
has a simple incarnation for pseudospins  where
it may further implies a non-trivial time evolution of
electronic orbitals \cite{Nussinov08b} or any other degree
of freedom that the pseudospin represents. 

For uniform ferromagnetic variants of the compass models (with a single constant $J$), the equation of motion is
\begin{eqnarray}
\label{eom_x}
\frac{\partial {\bf T}_{i}}{\partial t} = J {\bf T}_{i} \times  \sum_{j}  ({\bf T}_{j} \cdot 
{\bf e}_{\gamma || \langle i j \rangle}) {\bf e}_{\gamma || \langle i j \rangle},
\end{eqnarray}
which directly follows from Eq.~(\ref{Larmor_eq}). 
In Eq.~(\ref{eom_x}), the sum is over sites $j$ that are nearest neighbors of $i$.
By the designation ${\bf e}_{\gamma || \langle i j \rangle}$,
we make explicit that the internal pseudospin direction
${\bf e}_{\gamma}$ is set by that  particular value of $\gamma$ that corresponds to the direction
from site $i$ to site $j$ on the lattice itself (i.e., by the direction of the lattice
link $\langle i j \rangle$). If the effective pseudo-magnetic
field at site $i$ is parallel to the pseudospin at that site,
$ \Big( \sum_{j}  ({\bf T}_{j} \cdot 
{\bf e}_{\gamma || \langle i j \rangle}) {\bf e}_{\gamma || \langle i j \rangle} \Big) || {\bf T}_{i}$
then semi-classicaly the pseudospin is stationary (i.e., $\partial {\bf T}_{i}/\partial t =0$).
Such a case arises, for instance, for any semi-classical uniform pseudospin configuration:
${\bf T}_{i}$ = constant vector for all $i$ which we denote below by ${\bf T}$. 
In such a case, for the 90$^{o}$ compass, $ \sum_{j}  ({\bf T}_{j} \cdot 
{\bf e}_{\gamma || \langle i j \rangle}) {\bf e}_{\gamma || \langle i j \rangle} = 2 {\bf T}$
whereas for the cubic lattice 120$^{o}$ compass, $ \sum_{j}  ({\bf T}_{j} \cdot 
{\bf e}_{\gamma || \langle i j \rangle}) {\bf e}_{\gamma || \langle i j \rangle}  = 3 {\bf T}$.

As, classically, ${\bf T} \times {\bf T} =0$, all uniform pseudospin states are stationary
states (which correspond to classical ground states at strictly zero temperature). 
Similarly, of course, a staggered uniform configuration in which ${\bf T}_{i}$ is 
equal to one constant value (${\bf T}$) on one sublattice and is equal to
$(-{\bf T})$ on the other sublattice, will also lead to a stationary
state (that of highest energy for $J>0$).
Such semi-classical uniform states are also ground states of usual spin ferromagnets.
The interesting twist here is that the effective field ${\bf h}_{i}$ is not given by $J\sum_{j} {\bf S}_{j}$
as for usual spin systems but rather by $ \sum_{j}  ({\bf T}_{j} \cdot 
{\bf e}_{\gamma || \langle i j \rangle}) {\bf e}_{\gamma || \langle i j \rangle} $. 

\section{Physical Motivations \&  Incarnations}
\label{sec:motivation}

In this section we review the different physical contexts that motivate compass models and how they can emerge as low-energy effective models of systems with strongly interacting electrons.
There are quite a few classes of materials where the microscopic interactions between electrons are typically described by an extended Hubbard model. Typically such materials contain transition-metal ions. Hubbard-type models incorporate both the hopping of electrons from lattice-site to lattice-site and the Coulomb interaction $U$ between electrons that meet on the same site, typically the transition-metal ion.
Particularly in the situation that electron-electron interactions are strong, effective low-energy models can be derived by expanding the Hubbard Hamiltonian in $1/U$ -- the inverse interaction strength. In such a low-energy model interactions are not anymore between electrons, but between the remaining {\it spin} and {\it orbital} degrees of freedom of the electrons.

Compass model Hamiltonians arise, in particular, when orbital degrees of freedom interact with each other, which we will survey in detail (Sec.~\ref{sec:orbitals}), but can also emerge in the description of chiral degrees of freedom in frustrated magnets (Sec.~\ref{sec:chiral}). 

In the situation that both orbital and spin degrees of freedom are present and their interactions are intertwined, so-called Kugel-Khomskii models arise. We will briefly review in Sec.~\ref{sec:KK} how such models are relevant for strongly correlated electron systems such as transition metal (TM) oxides, when the low-energy electronic behavior is dominated by the presence of a very strong electron-electron interactions. Within the standard Kugel-Khomskii models, the orbital degrees of freedom are represented via SU(2) type pseudo-spins.  

So-called $e_g$ and $t_{2g}$ orbital degrees of freedom that can emerge in transition metal compounds with electrons in partially filled TM $d$-shells, give rise to two-flavor compass models (for $e_g$) and to three-flavor compass models (for $t_{2g}$) which, as we will explain in this section, are conveniently cast in an SU(3) Gell-Mann matrix form. Precisely these type of compass models emerge in cold atom systems in optical 

\subsection{Orbital Degrees of Freedom}  
\label{sec:orbitals}

\begin{figure*}
\centering
\includegraphics[width=1.5\columnwidth]{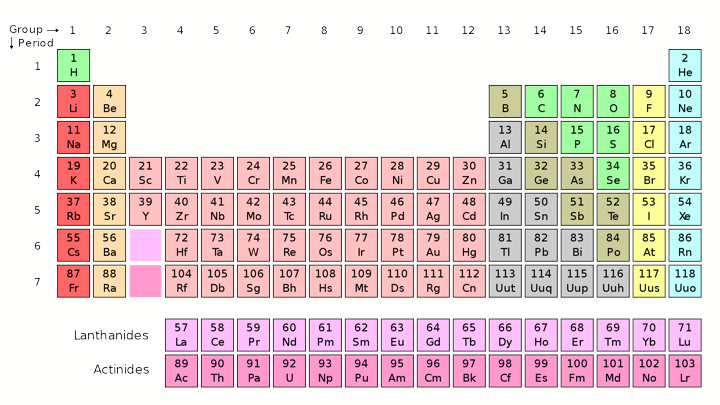}
\caption{In the periodic table the transition metal (TM) elements with ions that have partially filled $d$-shells are in the 4th row, the $3d$ elements Ti to Cu, below it the $3d$ elements in the 5th row, and the $5d$ elements in the 6th row. Electrons in such partially filled TM $d$-shell can have besides a spin, also an orbital degree of freedom.}
\label{fig:periodic_table}
\end{figure*}

Understanding the structure and interplay of orbital degrees of freedom has garnered much attention in various fields. Amongst many others, these include studies of the colossal-magnetoresistance manganites \cite{Tokura1999,Brink03} and pnictide superconductors \cite{Nakayama2009,Cvetkovic2009,Kuroki2008,Kruger2009}. 

Orbital degrees of freedom are already present in the electronic wavefunctions of the hydrogen atom. A brief discussion of the hydrogen atoms with just a single electron can thus serve as a first conceptual introduction to orbital physics (Sec.~\ref{sec:atomic}). These concepts translate to transition metal ions, where electrons in partially filled TM $d$-shells can have so-called $e_g$ and $t_{2g}$ orbital degrees of freedom~\cite{Griffith71,Fazekas99}.  These orbital states, which can be represented as spinors  (Sec.~\ref{sec:representation}), of ions on neighboring lattice sites can interact via electronic superexchange interactions (Sec.~\ref{sec:orbital-orbital}), which in the most general situation also depends on the spin orientation of the electrons. The relevant Hamiltonians that govern orbital-orbital interactions are derived, and we will briefly review how spin-spin interactions affect the interactions between orbitals in Kugel-Khomskii models (Sec.~\ref{sec:KK}). Reviews on this subject are Refs.~\cite{Kugel82,Tokura00}. Sec.~\ref{sec:JT} is devoted to effective orbital interactions that arise via lattice deformations and give rise to cooperative Jahn-Teller distortions. For $t_{2g}$ orbitals in addition relativistic spin-orbit coupling is relevant, especially for correlated materials build from heavy transition metal ions (those of the 6th row of the periodic table, see Fig.~\ref{fig:periodic_table}), for instance iridates, which will be discussed at the end of this section (Sec.~\ref{sec:orbitals}). The basic concepts relevant to strongly correlated electron systems can be found in the books~\cite{Goodenough63,Griffith71,Fazekas99,Khomskii10}.

We will first review orbital systems on cubic and other unfrustrated lattices. Caricatures of these systems lead to the most prominent realizations of compass models.
It is notable that on frustrated lattices, coupling with the orbital degrees of freedom may lead to rather unconventional states. These include, e.g., on spinel type geometries, 
spin-orbital molecules in AlV$_{2}$O$_{4}$ \cite{Horibe2006}, a viable cascade of transitions in ZnV$_{2}$O$_{4}$ \cite{Motome2004}. Resonating valence 
bond states were suggested to occur in the layered triangular compound LiNiO$_{2}$ \cite{Vernay04}.

\subsubsection{Atomic-like States in Correlated Solids}
\label{sec:atomic}
\begin{figure}
\centering
\includegraphics[width=.7\columnwidth]{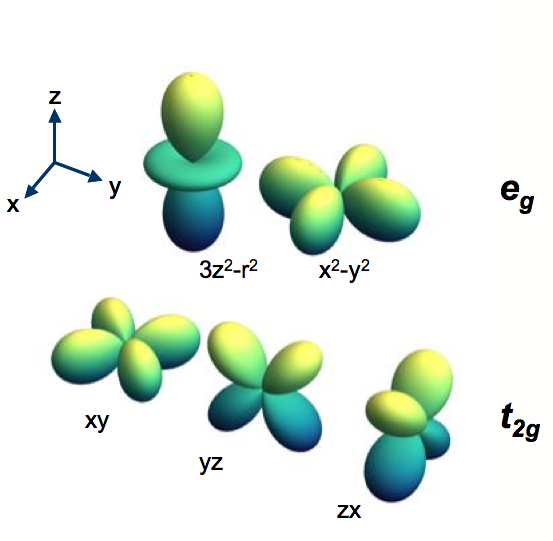}
\caption{The five orthogonal $d$-orbitals.
Crystal field effects lift the five-fold degeneracy of the $d$ atomic orbitals into an $e_{g}$ doublet (top) and a $t_{2g}$ triplet of states.}
\label{fig:orbitals}
\end{figure}

The well-know hydrogen wave-functions are the product of a radial part $R_{nl}$ and an angular part $Y_l^m$, with principle quantum number $n$ and angular quantum numbers $l$ and $m$:
\begin{eqnarray}
\psi_{nlm}(r,\theta,\phi)&=&  R_{nl}(2r/n) \cdot Y^m_l (\theta,\phi),
\end{eqnarray}
where the radial coordinate $r$ is measured in Bohr radii, $\theta$, $\phi$ are the angular coordinates and $n$ any positive integer, $l=0,...,n-1$ and $m=-l,...,l$. States with $l=0,1,2,3$ correspond to $s, p, d$ and $f$ states, respectively. The energy levels of hydrogen are $E_n=-13.6 eV/n^2$ when the small spin-orbit coupling is neglected. The energy therefore does not depend on the angular quantum numbers $l$ and $m$, implying that for any $l \geq 1$ the hydrogen energy levels are $2m+1$-fold degenerate -- this degeneracy of constitutes the orbital degeneracy and with it the orbital degree of freedom is associated. Thus hydrogen $p$ states are 3-fold degenerate, $d$-states 5-fold and $f$-states 7-fold. In explicit terms the angular wave-functions for the $d$ states, the spherical harmonics $Y_2^m$ are:
\begin{eqnarray}
Y_l^{-m}= (-1)^m (Y_l^m)^* \ and
\left\{ 
\begin{array}{l}
Y_2^0=  \sqrt{\frac{5}{16\pi}} \left( 3 \cos^2 \theta -1 \right)  \\
Y_2^1=   \sqrt{\frac{15}{8\pi}} \sin \theta \cos \theta e^{i \phi}  \\
Y_2^2=    \sqrt{\frac{15}{32\pi}} \sin^2 \theta e^{i 2\phi} \nonumber
\end{array} 
\right.
\end{eqnarray}
Introducing the radial coordinates $x=r\sin \theta \cos \phi $, $y=r\sin \theta \sin \phi $ and $z=r\cos \theta $  the angular basis-functions can be combined into real basis-states, for instance $(Y^{-2}_2+Y^2_2)/\sqrt{2} = \sqrt{\frac{15}{16\pi}}\sqrt{\frac{1}{r^2}} (x^2-y^2)$. Apart from an over-all normalization constant the resulting $d$ orbitals are
\begin{eqnarray}
e_g \ &orbitals& \
\left\{ 
\begin{array}{c | l}
Y^{-2}_2+Y^2_2  & x^2-y^2 \\
\sqrt{2}Y_2^0  & (3z^2-r^2)/\sqrt{3}  \nonumber
\end{array} 
\right.\\
t_{2g} \ &orbitals& \
\left\{ 
\begin{array}{c | l}
Y^{-2}_2-Y^2_2  & xy \\
Y^{-1}_2+Y^1_2  & yz \\
Y^{-1}_2-Y^1_2  & zx  \nonumber
\end{array} 
\right.
\end{eqnarray}
where a distinction between so-called $e_g$ and $t_{2g}$ orbitals is made, which is based on their different local symmetry properties, as will shortly become clear from crystal field considerations. These orbitals are pictured in Fig.~\ref{fig:orbitals}. In atoms and ions further down the periodic table (Fig.~\ref{fig:periodic_table}), this orbital degree of freedom can persist, depending on the number of electrons filling a particular electronic shell.

In solids $p$ wavefunctions of atoms tend to be rather delocalized, forming wide bands. When such wide bands form and a material is consequently a metal or semiconductor, the different atomic states mix and local orbital degeneracies are completely lifted.  However, $d$ and $f$ states tend to retain to certain extent their atomic character and especially the $3d$ and $4f$ states are particularly localized --  $4d$, $5d$ and $5f$ wave-functions are again more extended than $3d$ and $4f$, respectively. In the periodic table (Fig.~\ref{fig:periodic_table}) ions with open $d$-shells are in the group of transition metals and open $f$-shells are found in the lanthanides and actinides.

The localized nature of $3d$ and $4f$ states has as a consequence that in the solid the interactions between electrons in an open $3d$/$4f$ shell are much like in the atom~\cite{Griffith71}. For instance Hund's first rule -- stating that when possible the electrons form high-spin states and maximize their total spin -- keeps it relevance for these ions and for the $3d$'s leads to an energy lowering of  $J_H\sim0.8$ eV for a pair of electrons having parallel spins. Another large energy scale is the Coulomb interaction $U$ between electrons in the same localized shell. In a solid $U$ is substantially screened from its atomic value and its precise value therefore depends critically on the details of the screening processes -- it for instance reduces the Cu d-d Coulomb interactions in copper-oxides from an atomic value of 16 eV to a solid state value of about 5 eV~\cite{Brink95}. But in many cases it is still the dominant energy scale compared to the bandwidth $W$ of the $3d$ electrons~\cite{Imada98}. If $U$ is strong enough, roughly when $U>W$, this causes a collective localization of the electrons and the system becomes a Mott insulator~\cite{Mott90,Fazekas99,Khomskii10}. 

In a Mott insulator, that is driven by strong Coulomb interactions, electrons in an open $d$-shell can partially retain their orbital degree of freedom. The full 5-fold degeneracy of the hydrogen-like $d$ states is broken down by the fact that in a solid a positively charged TM ion is surrounded by other ions, which manifestly breaks the rotational invariance that is present in a free hydrogen atom and on the basis of which the atomic wave-functions were derived in the first place. How precisely the 5-fold degeneracy is broken depends on the point group symmetry of the lattice~\cite{Ballhausen62,Fazekas99}. 

The simplest -- and rather common -- case is the one of cubic symmetry, in which a TM ion is in the center of a cube, with ligand ions at the center of each of its six faces. The negatively charged ligand ions produce an electrical field at the center of the cube. Expanding this field in its multipoles, the first non-vanishing contribution is quadrupolar. This quadrupole field splits the $d$ states into the two $e_g$'s and the three $t_{2g}$'s, where the $t_{2g}$'s are lower in energy because the lobes of their electronic wavefunctions point away from the negatively ligand ions~\cite{Ballhausen62,Fazekas99}, see Fig. \ref{fig:orbitals}. Also, the electronic hybridization of these two classes of states with the ligand states is different, which further adds to the energy splitting between the $e_g$'s and $t_{2g}$'s.  But for a cubic ligand field (also referred to as crystal field) a two-fold orbital degeneracy remains if their is an electron (or a hole) in the $e_g$ orbitals and a three-fold degeneracy for an electron/hole in the $t_{2g}$ orbitals. 


The two $e_{g}$ states and the three $t_{2g}$ states relate, respectively, 
to two- and three-dimensional vector spaces (or two- and three-component pseudo-vectors ${\bf T}$). This, combined with the real space anisotropic directional character of the orbitals
leads to Hamiltonians similar to compass models that we introduced in earlier sections.

 A further lowering of the lattice point-group symmetry, from for instance cubic to tetragonal, will cause a further splitting of degeneracies.
The existence of degenerate orbital freedom raises the specter of cooperative effects- i.e., orbital ordering. Indeed, in many of the materials
in which they occur, orbital orders appear at high temperatures- often at temperatures far higher than magnetic orders.


\subsubsection{Representations of Orbital States}
\label{sec:representation}

\begin{figure}
\centering
\includegraphics[width=\columnwidth]{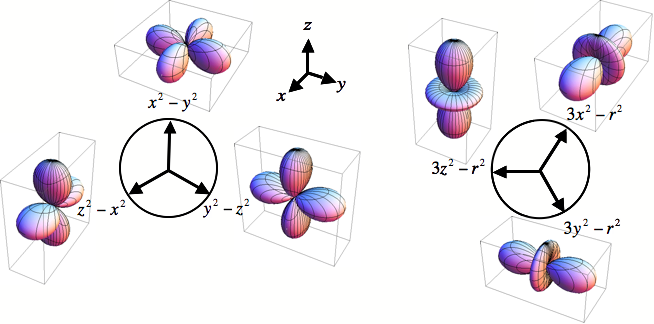}
\caption{Result of the rotations of  the $e_g$ orbital spinor by an angle  $\phi/2=2 \pi / 3 $.}
\label{fig:orbital_rotation}
\end{figure}

For the $e_g$ doublet the orbital pseudospin can be represented by a spinor, where $1 \choose 0 $ corresponds to an electron in the $x^2-y^2$ orbital and $0 \choose 1$ to the electron in the $(3z^2-r^2)/\sqrt{3}$ orbital. It is instructive to consider the rotations of this spinor, which are generated by the Pauli matrices $\sigma_{1}$, $\sigma_{2}$ and $\sigma_{3}$, the generators of the SU(2) algebra; the identity matrix is $\sigma_0$. Rotation by an angle $\phi$ around the $2$-axis is denoted by the operator $\hat{R_2}(\phi)$, where 
\begin{eqnarray}
\hat{R_2}(\phi)=e^{i \sigma_2 \phi/2} = \sigma_0 \cos \phi/2 + i \sigma_2 \sin \phi/2. 
\label{eq:rot_2}
\end{eqnarray}
It is easily checked that for $\phi/2=\pm 2 \pi / 3 $, rotation of the spinor corresponding to $x^2-y^2$ leads to $\hat{R_2^\pm} {1 \choose 0} = -{1 \over 2} {1 \choose \mp \sqrt{3}} = -{1 \over 2}(x^2-y^2 \mp (3z^2-r^2))=y^2-z^2, z^2-x^2$ and similarly $3z^2-r^2 \rightarrow 3x^2-r^2, 3y^2-r^2$. Rotations of the orbital wavefunction by $\phi/2=2 \pi / 3 $, thus cause the successive cyclic permutations $xyz \rightarrow yzx \rightarrow zxy \rightarrow xyz$ in the wavefunctions, as is depicted in Fig.~\ref{fig:orbital_rotation}.

Next we consider how the pseudospin operator ${\bm \tau}$ transforms under these rotations~\cite{Brink99b}. As ${\bm \tau}=\frac{1}{2} \sum_{\alpha \beta}c^\dagger_\alpha {\bm \sigma}_{\alpha \beta}c_\beta$, where the sum if over the two different orbital states for each $\alpha$ and $\beta$, after the rotation it is ${\bm \tau}=\frac{1}{2} \sum_{\alpha \beta}c^\dagger_\alpha  \hat{R}_2^\mp {\bm \sigma}_{\alpha \beta} \hat{R}^\pm_2 c_\beta$. For the vector component $\tau^3$ this implies for instance that successive rotations by an angle $\phi/2=\pm 2 \pi / 3 $ transform it as $\tau^3 \rightarrow -\frac{1}{2} (\tau^3+\sqrt{3} \tau^1) \rightarrow -\frac{1}{2} (\tau^3-\sqrt{3} \tau^1)\rightarrow \tau^3$.

The same procedure can be applied to the three $t_{2g}$ states, with can be represented by three-component spinors 
$ xy= \bigl(  \begin{smallmatrix}  1  \\  0  \\  0 \end{smallmatrix}\bigr)$, $yz= \bigl(  \begin{smallmatrix}  0  \\  1  \\  0 \end{smallmatrix}\bigr)$  and $ zx= \bigl(  \begin{smallmatrix}  0  \\  0  \\  1 \end{smallmatrix}\bigr) $. The operators acting on the three-flavor spinors form a SU(3) algebra, which is generated by the eight Gell-Mann matrices $\lambda^{1...8}$, see Appendix~\ref{sec:appendix}. This implies that pseudospin operator for $t_{2g}$ orbitals ${\bm \tau}=\frac{1}{2} \sum_{\alpha \beta}c^\dagger_\alpha {\bm \lambda}_{\alpha \beta}c_\beta$ is an eight-component vector. The operator 
$
\hat{R}^+= \bigl( \begin{smallmatrix}  0 & 0 & 1 \\  1 & 0 &0  \\  0 & 1 & 0 \end{smallmatrix} \bigr)
$
brings about the cyclic permutations $xyz \rightarrow yzx \rightarrow zxy \rightarrow xyz$ in the $t_{2g}$ wavefunctions and $\hat{R}^-=(\hat{R}^+)^T$.  $\hat{R}^{\pm}$ applied to the Gell-Man matrices transforms the $t_{2g}$ pseudospin operators accordingly.

\subsubsection{Orbital-Orbital Interactions}
\label{sec:orbital-orbital}

\begin{figure}
\centering
\includegraphics[width=.8\columnwidth]{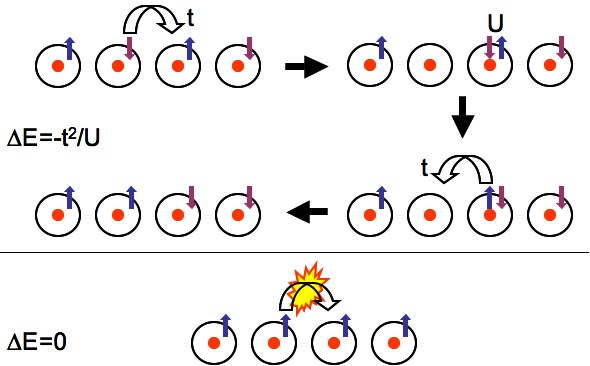}
\caption{Superexchange between spin $1/2$ electrons, resulting into the effective antiferromagnetic Heisenberg Hamiltonian  $H= J \sum_{i,j} \left( {\bf S}_i \cdot {\bf S}_j - \frac{1}{4} \right)$, with $J=4t^2/U$  }
\label{fig:superexchange}
\end{figure}

Even if in a Mott insulator electrons are localized in their atomic-like orbitals, they are not completely confined and can hop between neighboring sites. For electrons in non-degenerate $s$-like orbitals, this lead to the magnetic superexchange interactions between the spins of different electrons, see Fig.~\ref{fig:superexchange}.
The competition between the strong Coulomb interaction that electrons experience when they are in the same orbital, which tends to localize electrons, and the hopping, which tends to delocalize them is captured by the isotropic Hubbard Hamiltonian~\cite{Hubbard63}
\begin{eqnarray}
H^{iso}_{Hub}= \sum_{\substack{\langle ij \rangle, \alpha= \uparrow,\downarrow} } t ( c^\dagger_{i  \alpha} c^{\phantom\dagger}_{j \alpha} + h.c.) +U\sum_{i} n_{i  \uparrow} n_{i  \downarrow},
\label{eq:Hub}
\end{eqnarray}
where $ c^\dagger_{i  \alpha}$ creates and electron with spin $ \alpha= \uparrow,\downarrow$ on site $i$ and  $c^{\phantom\dagger}_{j \alpha}$ annihilates it on neighboring site $j$, $t$ is the hopping amplitude and the Hubbard $U$ the energy penalty when two electrons meet on the same site and thus are in the same $s$-like orbital~\cite{Fazekas99,Khomskii10}. 

It is convenient to introduce here for later purposes the two by two hopping {\it matrix} $t^{\gamma}_{\alpha \beta}$, where with spin $ \alpha= \uparrow,\downarrow$ and $ \beta= \uparrow,\downarrow$, which determines how an electron changes its spin from $\alpha$ to $\beta$ when it hops from site $i$ to $j$ on the bond $\langle ij \rangle $ in the direction $\gamma$.  Using this notation the first term in the Hubbard Hamiltonian $H_{Hub}$ is
\begin{eqnarray}
\sum_{\substack{\langle ij \rangle, \alpha} } t c^\dagger_{i  \alpha} c^{\phantom\dagger}_{j \alpha}  =
\sum_{\substack{i, \gamma, \alpha,  \beta} } t^{\gamma}_{\alpha, \beta} c^\dagger_{i  \alpha} c^{\phantom\dagger}_{i+ {\bm e}_\gamma \beta},
\label{eq:Hub_hop1}
\end{eqnarray}
so that
\begin{eqnarray}
H_{Hub}= \sum_{\substack{i, \gamma, \sigma,  \sigma\prime} } (t^{\gamma}_{\alpha \beta} c^\dagger_{i  \alpha} c^{\phantom\dagger}_{i+ {\bm e}_\gamma \beta} + h.c.)  +U \sum_{\substack{i}} n_{i  \uparrow} n_{i  \downarrow},
\label{eq:Hub}
\end{eqnarray}
where for the isotropic Hubbard Hamiltonian $H^{iso}_{Hub}$ of Eq. (\ref{eq:Hub}), since hopping does not depend on the direction $\gamma$ of the bond and spin is conserved during the hopping process, we simply have
\begin{eqnarray}
 t^{\gamma}_{\alpha \beta} = t \left(  \begin{matrix}  1 & 0 \\  0 & 1 \end{matrix}\right),
\label{eq:Hub_hop2}
\end{eqnarray}
for all $\gamma$. Compass and Kitaev models are related to Hubbard models with more involved, bond direction depend, forms of $t^{\gamma}_{\alpha \beta}$.

\begin{figure}
\centering
\includegraphics[width=\columnwidth]{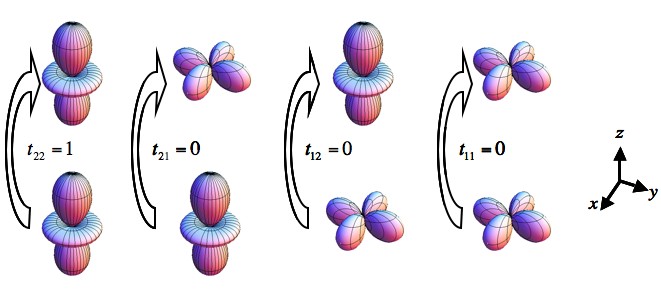}
\caption{Hopping amplitudes between $e_g$ orbitals along the $\hat{z}$ axis: the hopping matrix is $t^{\hat{z}}_{\alpha \beta}= t\delta_{\alpha,2} \delta_{\beta,2}$. Three matrix elements vanish because of the symmetry of the $x^2-y^2$ orbitals, with a wavefunction on adjacent lobes that has opposite sign.}
\label{fig:hopping_eg}
\end{figure}

For $U \gg t$ and half filling (i.e. the number of electrons equal to the number of sites in the system) the resulting Heisenberg-type interaction between spins is $H= J \sum_{i,j} \left( {\bf S}_i \cdot {\bf S}_j - \frac{1}{4} \right)$, which is antiferomagnetic: $J=4t^2/U$. The high symmetry of the Heisenberg Hamiltonian -- the interaction ${\bf S}_i \cdot {\bf S}_j$ is rotationally invariant -- is rooted in the fact that the hopping amplitude $t$ is equal for spin up and spin down electrons and thus does not depend on spin at all. 
This is again reflected by the hopping matrix of an electron on site $i$ and spin $\alpha$ to site $j$ and spin $\beta$ being diagonal: $t_{\alpha \beta}= t \bigl(  \begin{smallmatrix}  1 & 0 \\  0 & 1 \end{smallmatrix}\bigr)$.
For orbital degrees of freedom the situation is very different, because hopping amplitudes strongly depend on the type of orbitals involved and thus on the orbital pseudospin. This anisotropy is rather extreme as it not only depends on the local symmetry of the two orbitals involved, but also on their relative position in the lattice: for instance the hopping amplitude between two $3z^2-r^2$ orbitals is very different when the two sites are positioned above each other, along the $z$-axis, or next to each other, e.g. on the $x$-axis, see Fig.~\ref{fig:hopping_eg}.

\vspace{1em}
\underline{ \it $e_g$ orbital-only Hamiltonians}\\
For the $e_g$ orbitals the hopping matrix between sites $i$ and $j$ along the $\hat{z}$ direction is $t_{\alpha \beta}^{\hat{z}}= t\delta_{\alpha,2} \delta_{\beta,2}=t \bigl(  \begin{smallmatrix}  0 & 0 \\  0 & 1 \end{smallmatrix}\bigr)$ in the basis $x^2-y^2$, $3z^2-r^2$. This fully species the hopping between orbitals on a cubic lattice, as the hopping along $\hat{x}$ and $\hat{y}$ are dictated by symmetry. The corresponding hopping matrices can be determined with the help of the rotations introduced in the previous subsection, Sec.~\ref{sec:representation}. The hopping matrix  $t_{\alpha \beta}^{\hat{x}}$, is obtained by first the full coordinate system is rotated by $\pi/2$ around the $y$-axis, so that $t_{\alpha \beta}^{\hat{z}} \rightarrow t^{\hat{x}}_{\alpha \beta}= t \delta_{\alpha,2} \delta_{\beta,2}$, now with basis states $z^2-y^2$, $3x^2-r^2$. A subsequent rotation of the orbital spinors by $\phi/2=-2\pi/3$ around the $2$-axis brings the matrix back in the original  $x^2-y^2$, $3z^2-r^2$ basis and transforms $t_{\alpha \beta}^{\hat{x}}  \rightarrow R_2^{+} t_{\alpha \beta}^{\hat{x}} R_2^{-}$. After the rotations one finds $t_{\alpha \beta}^{\hat{x}}=\frac{t}{4} \bigl(  \begin{smallmatrix}  3 & \sqrt{3} \\  \sqrt{3} & 1 \end{smallmatrix}\bigr)$ and similarly first rotating around  the $\hat{y}$-axis and transforming $t_{\alpha \beta}^{\hat{y}}  \rightarrow R_2^{-} t_{\alpha \beta}^{\hat{x}} R_2^{+}$, leads to $t_{\alpha \beta}^{\hat{y}}=\frac{t}{4} \bigl(  \begin{smallmatrix}  3 & -\sqrt{3} \\  -\sqrt{3} & 1 \end{smallmatrix}\bigr)$, a well-known result~\cite{Kugel82,Brink99a,Ederer07} that is in accordance with microscopic tightbinding considerations~\cite{Harrison04}.

Orbital-orbital interactions are generated by superexchange processes between electrons in $e_g$ orbitals. When the electron spin is disregarded, the most basic form of the orbital-orbital interaction Hamiltonian is obtained. 
Superexchange with spin-full electrons leads to Kugel-Khomskii Hamiltonians which will be derived and discussed in the following section. 
For spin-less fermions the exchange interactions along the $\hat{z}$ axis take a particularly simple form. If the electron on site $i$ is in an $x^2-y^2$ orbital, corresponding to $\tau_i^3=-\frac{1}{2}$, and the one on site $j$ in a $3z^2-r^2$ orbital ($\tau_j^3=\frac{1}{2}$)  a virtual hopping process is possible, giving rise to an energy gain of $-t^2/U$ in second order perturbation theory, where $U$ is the energy penalty of having to spinless fermions on the same site (which are by definition in different orbitals). The only other configuration with non-zero energy gain is the one with $i$ and $j$ interchanged. The Hamiltonian on the bond $ij$ is therefore $H^{\hat{z}}_{ij} = -\frac{t^2}{U}\left[ (\frac{1}{2} - \tau^3_i) (\frac{1}{2} + \tau^3_j)+(\frac{1}{2} - \tau^3_j) (\frac{1}{2} + \tau^3_i)\right] = \frac{J}{2} ( \tau^3_i  \tau^3_j - \frac{1}{4}) $. With the same rotations as above, but now acting on the operator $\tau^z$, the Hamiltonian on the bonds in the other two directions can be determined: along the $\hat{x}$ and $\hat{y}$ axis, respectively
\begin{eqnarray}  
\tau^3 \rightarrow R_2^{+}\tau^3 R_2^{-} \ {\rm along} \ \hat{x} \nonumber \\
 \tau^3 \rightarrow R_2^{-}\tau^3 R_2^{+}  \ {\rm along} \ \hat{y}
 \label{eq:transform}
\end{eqnarray}
so that
$H^{\hat{x}}_{ij} = \frac{J}{8} [ (\tau^3_i+\sqrt{3} \tau^1_i) (\tau^3_j+\sqrt{3} \tau^1_j) - 1]=\frac{J}{4} \pi^x_i \pi^x_j$, where the last step defines $\pi^\gamma$, (see Eq.~(\ref{eq:eg120})) similarly as in Eq.~(\ref{eq:pi120}),
and along $\hat{y}$ one obtains
$H^{\hat{y}}_{ij} = \frac{J}{8} [ (\tau^3_i-\sqrt{3} \tau^1_i) (\tau^3_j-\sqrt{3} \tau^1_j) - 1]=\frac{J}{4} \pi^y_i \pi^y_j$.
The orbital-only Hamiltonian for $e_g$ orbital pseudospins therefore is exactly the 120$^{\circ}$ compass model of Eqs.(\ref{120p}, \ref{eq:pi120})~\cite{Brink99b}
\begin{eqnarray}
H^{e_g}_{3\square} &=& \frac{J}{2}\sum_{i,\gamma} \left( \pi^\gamma_i \pi^\gamma_{i+\bm{e}_\gamma} -\frac{1}{4} \right) \nonumber \\
&&with
\left\{ 
\begin{array}{l}
  \pi^\gamma=\tau^3  \cos{\theta_\gamma} + \tau^1  \sin{\theta_\gamma}\\
  \{\bm{e}_\gamma \}=\{ \bm{e}_x,  \bm{e}_y, \bm{e}_z\} \\
    \{\theta_\gamma\}=\{0,2\pi/3,4\pi/3\}.
\end{array} 
\right.
\label{eq:eg120}
\end{eqnarray}
with $J=4t^2/U$, which is the 120$^\circ$ quantum compass model on a cubic lattice, Eq.~(\ref{eq:eg120}), with "antiferro" orbital-orbital interactions, driving a tendency towards the formation of staggered orbital ordering patterns.

\begin{figure}
\centering
\includegraphics[width=.8\columnwidth]{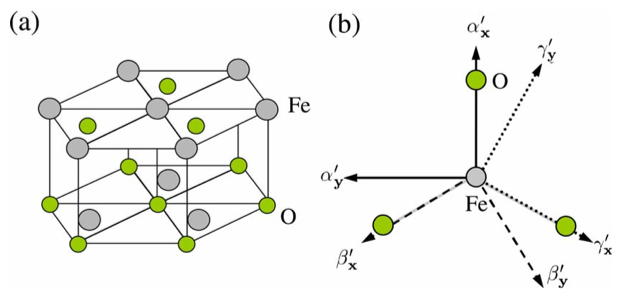} \\
\includegraphics[width=.6\columnwidth]{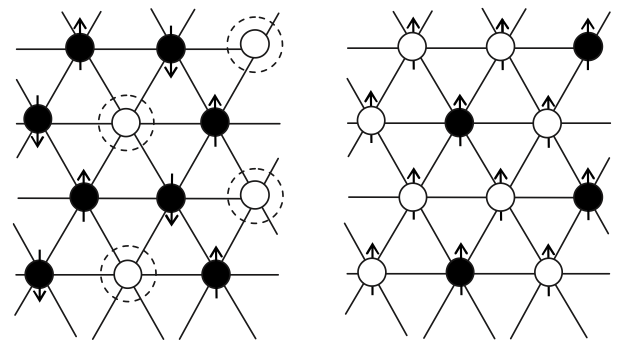}
\caption{$H_{\varhexagon}^{120^\circ}$ models orbital-orbital interactions in RFe$_2$O$_4$   (a) A pair of triangular planes  and (b) three Fe-O bond directions in a triangular lattice in RFe$_2$O$_4$.~\cite{Nasu08}. Below: Schematic of the charge and spin structures in 2Fe$^{2+}$-Fe$^{3+}$ plane (right) and in Fe$^{2+}$-2Fe$^{3+}$ plane (left) for RFe$_2$O$_4$. Filled and open circles represent Fe$^{3+}$ and Fe$^{2+}$, respectively. At sites surrounded by dotted circles, spin directions are not uniquely determined due to frustration. 
}
\label{fig:Nasu08_4}
\end{figure}

The 120$^{\circ}$ compass model on the honeycomb lattice $H_{\varhexagon}^{120^\circ}$, Eq.~(\ref{eq:c120_honeycomb}), was motivated by \cite{Nasu08} in a study of the layered iron oxides $R$Fe$_{2}$O$_{4}$ ($R$=Lu, Y, Yb), see Fig.~\ref{fig:Nasu08_4}.  These oxides are multiferroic systems in which both the magnetic and electric response are dominated by Fe $3d$ electrons.  The nominal valence of the Fe ions is 2.5 $^{+}$ and thus an equal number of Fe$^{2+}$ and Fe$^{+3}$ are present. One of the $e_{g}$ levels in the Fe$^{2+}$ ions is doubly occupied where all of the five 3d orbitals in the Fe$^{3+}$ ions are singly occupied.  The system assumes the form of a stack of pairs of triangular lattice planes along the $c$ axis of the form Fe$^{2+}$-2Fe$^{3+}$ and 2Fe$^{2+}$-Fe$^{3+}$. In the 2Fe$^{2+}$-Fe$^{3+}$ member of this pair, Fe$^{2+}$ ions (with a doubly degenerate $e_{g}$ orbital degree of freedom) form a honeycomb lattice. Superexchange with the Fe$^{3+}$ ions leads to directly the Hamiltonian of Eq. (\ref{eq:c120_honeycomb}).

\begin{figure}
\centering
\includegraphics[width=.7\columnwidth]{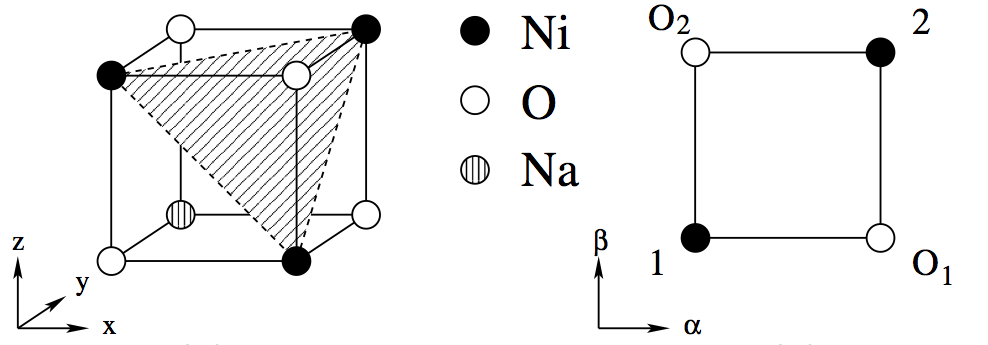}
\caption{Left: the crystal structure of NaNiO$_2$. Right: a plaquette in the $\alpha \beta$ plane $(\alpha, \beta=x,y, z)$ formed by two nearest-neighbor Ni ions, 1 and 2, and two oxygens, O$_1$ and O$_2$ \cite{Mostovoy2002}}
\label{fig:NaNiO2}
\end{figure}

The 120$^\circ$ model has been proposed to account for the physics of materials such as NaNiO$_{2}$ in which the transition metal ions [with doubly degenerate $e_{g}$ orbitals occupied by a single electron
or hole] lie on weakly coupled triangular layers \cite{Mostovoy2002}. In NaNiO$_{2}$, Na and Ni ions occupy alternate [111] planes as seen in Fig. (\ref{fig:NaNiO2}).

In NaNiO$_{2}$, consecutive low spin $Ni^{3+}$ triangular layers are weakly coupled to
each other. Within each such layer the dominant interactions between
Ni ions involve exchange paths via intermediate oxygens. 
The bonds between neighboring Ni and oxygen ions
form a 90$^{\circ}$ angle. Direct calculations lead to
the triangular lattice 120$^{\circ}$ Hamiltonian
of Eq. (\ref{eq:c120_triangle}). In section \ref{charge_transfer},
we will further review charge transfer via
intermediate ligand (e.g., oxygen) sites and how they may lead to
orbital interactions. Augmenting the orbital only interactions of the 120$^{\circ}$ compass type, 
an additional orbital dependent ferromagnetic spin exchange is
found \cite{Mostovoy2002}. Although the strength of this coupled spin and orbital Hamiltonian varies with
the orbital state, the (ferromagnetic) sign of the spin exchange is 
independent of the orbital configuration; even if the orbitals are in a disordered state,
ferromagnetic spin coupling will be present. The spin exchange coupling
is far weaker than associated with the orbital interactions (governed
by Eq. (\ref{eq:c120_triangle}). The experimentally far lower value
of the spin ordering temperature 
is indeed consistent with this analysis \cite{Mostovoy2002} .  
The dominant interactions are those of the orbital-orbital type. 

\vspace{1em}
\underline{ \it Compass and Kitaev Hamiltonians}\\
Compass and Kitaev Hamiltonians are the low-energy effective description of certain two-flavor Hubbard Hamiltonians of the type $H_{Hub}$ given by Eq.~(\ref{eq:Hub}). When the two flavors are spin up and down, the hopping matrix corresponds to the one of the simple isotropic Hubbard model $H^{iso}_{Hub}$, see Eqs.(\ref{eq:Hub_hop1},\ref{eq:Hub_hop2}) and the low-energy effective spin Hamiltonian is the spin 1/2 Heisenberg model. Instead
the hopping matrix  $t^{\hat{z}}_{\alpha \beta}= \bigl(  \begin{smallmatrix}  0 & 0 \\  0 & 1 \end{smallmatrix}\bigr)$, as for $e_g$ orbitals along $\hat{z}$, gives rise to a Ising type of  interaction $\tau^3_i \tau^3_j$ between pseudospins on the bond $\langle ij \rangle$ parallel to $\hat{z}$.  Such a hopping matrix is realized in the original Hubbard model (Eq.~\ref{eq:Hub}), if only spin $\downarrow$ electrons would be permitted to hop between the sites $i$ and $j$.  

When the hopping matrix has a different form along different bonds a compass model can arises. The 90$^\circ$ compass model, for instance, has a Ising-type interaction $\tau^3_i \tau^3_j$ along $\hat{z}$, corresponding to $t^{\hat{z}}_{\alpha \beta}= \bigl(  \begin{smallmatrix}  0 & 0 \\  0 & 1 \end{smallmatrix}\bigr)$, but on the bond along $\hat{x}$ $\tau^1_i \tau^1_j$ has to be active, which implies a rotation of (pseudo)spin with angle $\phi=\pi/2$ around the 2-axis, where in the rotated basis the hopping matrix again takes the shape $\bigl(  \begin{smallmatrix}  0 & 0 \\  0 & 1 \end{smallmatrix}\bigr)$. This requires an  specific form of the original, unrotated hopping matrix $t_{\alpha \beta}$ along $\hat{x}$. It is easy to check by performing these rotations that for $t^{\hat{x}}_{\alpha \beta}=\frac{1}{2} \bigl(  \begin{smallmatrix}  1 & -1 \\  -1 & 1 \end{smallmatrix}\bigr)$ and $t^{\hat{y}}_{\alpha \beta}=\frac{1}{2} \bigl(  \begin{smallmatrix}  1 & -i \\ i & 1 \end{smallmatrix}\bigr)$ the cubic 90$^\circ$ compass model $H_{3\square}^{90^\circ}$ (Eq.~(\ref{eq:90compass})) arises. 

It thus follows that for hopping matrices in the Hubbard Hamiltonian (Eq.~(\ref{eq:Hub})) that have the form
\begin{eqnarray}
 t^{\hat x}_{\alpha, \beta} = \frac{1-\sigma^x}{2}, \ \
 t^{\hat y}_{\alpha, \beta} = \frac{1-\sigma^y}{2}\  {\rm and} \ 
 t^{\hat z}_{\alpha, \beta} = \frac{1-\sigma^z}{2}
 \label{eq:hop_compass}
\end{eqnarray}
on a cubic lattice in the large $U$ limit and at half-filling, the low energy effective Hamiltonian is the 90$^\circ$ compass model $H_{3\square}^{90^\circ}$ (Eq.~(\ref{eq:90compass})). A hopping matrix of this type can be realized physically for electrons in the 5d states of iridium ions, where a strong relativistic spin-orbit coupling locks to spin to the orbital degree of freedom~\cite{Jackeli09}. Controlling the (pseudo)spin dependence of the hopping amplitudes on different bonds thus suffices to generate any type of compass Hamiltonian as the effective low-energy (pseudo)spin model of the Hubbard Hamiltonian. 

\begin{figure}
\centering
\includegraphics[width=.9\columnwidth]{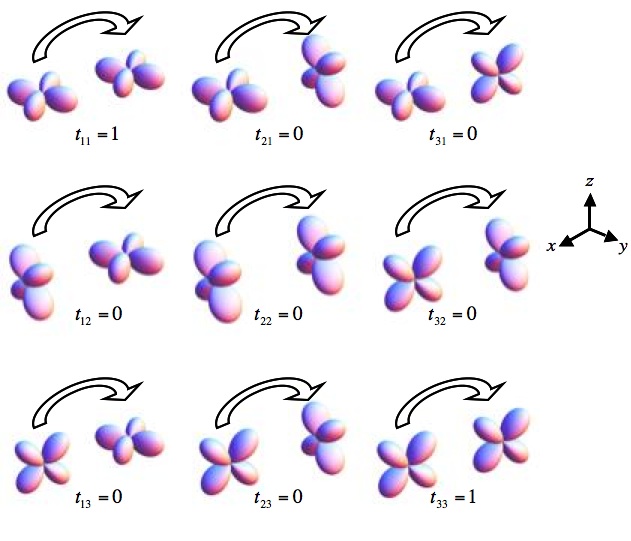}
\caption{Hopping amplitudes between $t_{2g}$ orbitals along the $\hat{x}$ axis, assuming that the hopping is via an ligand intermediate $p$ state (not shown here), for instance of an oxygen atom between two TM ions, see Figs.~\ref{fig:Jackeli09_2} and \ref{fig:Batista05}.}
\label{fig:hopping_t2g}
\end{figure}

\vspace{1em}
\underline{ \it $t_{2g}$ orbital-only Hamiltonian}\\
The three flavors of $t_{2g}$ orbitals $xy$, $yz$, $zx$ are most naturally represented by a three-component spinor so that the hopping $t_{\alpha\beta}$ is a 3x3 matrix. The structure of the hopping matrix is rather simple (Fig.~\ref{fig:hopping_t2g}), as between site $i$ and $j$ electrons can only hop between orbitals of the same symmetry so that orbital-flavor is conserved in the hopping process, which renders $t_{\alpha\beta}$ diagonal. Moreover, along the $\hat{x}$ axis the hopping between $yz$ orbitals vanishes. This determines the hopping matrices in all three directions, which  can be constructed via rotations, similar as for the $e_g$'s, (see Appendix, Sec.~\ref{sec:appendix}):
\begin{eqnarray}
t^{\hat{x}}=   \begin{pmatrix}  1 & 0 & 0 \\  0 & 0 &0  \\  0 & 0 & 1 \end{pmatrix};  \
t^{\hat{y}}=   \begin{pmatrix}  1 & 0 & 0 \\  0 & 1 &0  \\  0 & 0 & 0 \end{pmatrix};  \
t^{\hat{z}}=   \begin{pmatrix}  0 & 0 & 0 \\  0 & 1 &0  \\  0 & 0 & 1 \end{pmatrix}.
\end{eqnarray}

As along, for instance, the $\hat{y}$ the hopping matrix is diagonal for the two orbitals involved, the exchange interaction for two (spin-less) fermions in these two active orbitals on site $i$ and $j$ is of Heisenberg type. In terms of Gell-Mann matrices it is $\frac{J}{4}(\lambda_{1,i}\lambda_{1,j}+\lambda_{2,i}\lambda_{2,j}+\lambda_{3,i}\lambda_{3,j}-1)$, which is SU(2) invariant. Because both fermions need not be in the two active orbitals, an additional diagonal term $\rho_{1,i}\rho_{1,j}$ is present, where $\rho_1= \frac{1}{3}(\lambda_0-\sqrt{3} \lambda_8$). As $\rho_1$ commutes with $\lambda_{1...3}$, it does not break the SU(2) invariance. Defining the vector ${\bm \mu}^1= (\lambda_1,\lambda_2,\lambda_3,\rho_1)$ along the $\hat{x}$ direction $H^{\hat{x}}_{ij}=\frac{J}{4}({\bm \mu}^1_i \cdot {\bm \mu}^1_j - 1)$.
Rotation of the coordinate system and subsequently of the orbital basis produce the interactions along the other two directions, ${\bm \mu}^2_i \cdot {\bm \mu}^2_j - 1$ along $\hat{y}$ and ${\bm \mu}^3_i \cdot {\bm \mu}^3_j - 1$ along $\hat{z}$, so that
\begin{eqnarray}
H^{t_{2g}}_{3\square} &=& \frac{J}{4} \sum_{i,\gamma} \left( {\bm \mu}^\gamma_i {\bm \mu}^\gamma_{i+\bm{e}_\gamma} -1 \right) 
\label{eq:t2g}
\end{eqnarray}
with $\{\bm{e}_\gamma \}=\{ \bm{e}_x,  \bm{e}_y, \bm{e}_z\}$.  Along each of the bonds one of the SU(2) subgroups corresponding to the elements of ${\bm \mu}^\gamma$ is active and the Hamiltonian is rotational invariant in terms of that subgroup. This aspect emphasizes the compass character of the ensuing Hamiltonian. The situation is complicated by the fact that all three ${\bm \mu}^\gamma$ belong to the same SU(3) algebra, so that the elements of ${\bm \mu}^\gamma$ and ${\bm \mu}^{\gamma'}$ in general do not commute.

As the 3-flavor exchange Hamiltonian is represented Gell-Mann matrices, it is natural to refer to it as a {\it Gell-Mann matrix model}. This approach allows for a representation of the interactions between $t_{2g}$ orbitals that goes beyond the current well studied orbital Hamiltonians in which SU(2) representations are used. In the context of ultracold gas systems such a number of this type of models has been proposed. Of course the Gell-Mann representation is not unique. Gell-Mann matrices can for example be expressed in polynomials of the three $L=1$ angular momentum matrices $L_x$, $L_y$ and $L_z$, which thus also can be used to represent $H^{t_{2g}}_{3\square}$~\cite{Kugel82}.

\subsubsection{Spin-spin and orbital-orbital interactions}
\label{sec:KK}
Going beyond the case of spin-less fermions, requires considering the local Coulomb and exchange interactions between electrons in various orbital configurations, via a multi-orbital Hubbard Hamiltonian. This opens an entire field, of which reviews can be found in~\cite{Kugel82,Tokura00,khalS2005}. Here we restrict ourselves to indicating how compass models are decorated with spin-spin interactions, with a particular focus the $120^\circ$ compass model for $e_g$ electrons.

The considerations concerning the hopping amplitudes of $e_g$ electrons directly enter into the kinetic part of the $e_g$-orbital Hamiltonian
\begin{eqnarray}
H^{multi}_{Hub}&=& \sum_{\substack{i,\gamma\\ \alpha, \beta , \sigma} } t^\gamma_{\alpha\beta} ( c^\dagger_{i,  \alpha \sigma} c^{\phantom\dagger}_{i+{\bf e}_{\gamma}, \beta \sigma} + h.c.) +H_C  \\
{\rm with} &&
\left\{ 
\begin{array}{l}  
  t^\gamma_{\alpha \beta}=  \frac{t}{2} \begin{pmatrix} 1 -\cos 2 \theta_\gamma  &   \sin 2 \theta_\gamma  \\   \sin 2 \theta_\gamma & 1+\cos 2 \theta_\gamma \end{pmatrix}   \\
    \{\theta_\gamma\}=\{0,2\pi/3,4\pi/3\},
\end{array} \nonumber
\right.
\label{eq:eg_Hub}
\end{eqnarray}
where the on-site electron-electron interaction terms are~\cite{Dworin70,Kugel82,Han98} 
\begin{eqnarray}
H_C=&&
(U+J_H)\sum_{i,\alpha} n_{i \alpha \uparrow} n_{i \alpha \downarrow} 
+(U-J_H)\sum_{\substack{i, \sigma \\ \alpha<\beta}} n_{i \alpha \sigma} n_{i \beta \sigma} \nonumber \\
&&+U \sum_{\substack{i, \sigma \\ \alpha \neq \beta}} n_{i \alpha \uparrow} n_{i \beta \downarrow }
+J_H \sum_{\substack{i \\ \alpha \neq \beta}} c^\dagger_{i \alpha \downarrow} c^\dagger_{i \beta \uparrow} c^{\phantom\dagger}_{i \alpha \uparrow}  c^{\phantom\dagger}_{i \beta \downarrow}. \ \ \
\end{eqnarray}
Here not only the Hubbard $U$, but also Hund's rule $J_H$ enters, and in such a form that $H_C$ does not break the local rotational symmetry in the spin-orbital basis. Normally the regime $U \gg J_H$ is considered, which is the most physical.

A second order perturbation expansion in $t/U$ directly lead to exchange interactions between spin and $e_g$ orbital degrees of freedom, resulting in an effective low-energy Kugel-Khomskii (KK) Hamiltonian~\cite{Kugel72,Kugel73,Kugel82}. The KK Hamiltonian can also be derived from symmetry arguments. In doing so, first the case $J_H=0$ is considered. With regard to the orbital-only $e_g$ Hamiltonian, in the spin-full case in addition spin-superexchange is possible along the $\hat{z}$ direction if both electrons are $3z^2-r^2$ orbitals ($\tau_j^3=\frac{1}{2}$), so that $J({\bm S}_i \cdot {\bm S}_j -\frac{1}{4})(\frac{1}{2} - \tau^3_i) (\frac{1}{2} - \tau^3_j)$ has to be added to $\frac{J}{2}(\tau^3_i \tau^3_j -\frac{1}{4})$ from $H^{e_g}_{3\square}$ in $\hat{z}$ direction, see Eq.~(\ref{eq:eg120}), so that
\begin{eqnarray}
H^{\hat{z}}_{ij}&=& J\left({\bm S}_i \cdot {\bm S}_j +\frac{1}{4}\right) \left(\frac{1}{2} - \tau^3_i \right)  \left(\frac{1}{2} - \tau^3_j\right) \nonumber \\
&+&J/4 \left( \tau^3_i+\tau^3_j - 1\right)
\end{eqnarray}
The Hamiltonian along the other two axis is generated by the rotations of the orbital basis specified in Eqs.~(\ref{eq:rot_2}, \ref{eq:transform}). This Kugel-Khomskii Hamiltonian is, up to a constant, of the form (cf. Eq.~(\ref{eq:KK}))
\begin{eqnarray}
H^{KK}_U&=& J \sum_{i,\gamma} H^{U, orb}_{i,i+\bm{e}_\gamma} H^{U, spin}_{i,i+\bm{e}_\gamma} \\
{\rm with} &&
\left\{ 
\begin{array}{l}  
 H^{U, spin}_{i,i+\bm{e}_\gamma} = {\bm S}_i \cdot {\bm S}_{i+{\bm e}_\gamma} +\frac{1}{4} \\  
    H^{U, orb}_{i,i+\bm{e}_\gamma} = \left(\frac{1}{2} - \pi^\gamma_i \right)  \left(\frac{1}{2} - \pi^\gamma_{i+{\bm e}_\gamma} \right), \\
\end{array} \nonumber
\right.
\end{eqnarray}
where the operators $\pi^\gamma$ are defined in Eq.~(\ref{eq:pi120}). Interestingly, the energy of the classical antiferromagnetic N\'eel state, where ${\bm S}_i \cdot {\bm S}_j =-1/4$ is identically zero independent of any orbital configuration and therefore macroscopically degenerate, which opens the possibility to stabilize spin-orbital liquid states~\cite{Feiner97,Oles00} or drive the formation of quasi one-dimensional spin states that are stabilized by quantum fluctuations~\cite{Khaliullin97}. However, the presence of a finite $J_H$ will lift this degeneracy of the N\'eel ordered spin state. In leading order in $\eta=J_H/U$, this generates the spin-orbital Hamiltonian
\begin{eqnarray}
H^{KK}_{J_H}&=& \eta J \sum_{i,\gamma} H^{J_H, orb}_{i,i+\bm{e}_\gamma} H^{J_H, spin}_{i,i+\bm{e}_\gamma} \\
{\rm with} &&
\left\{ 
\begin{array}{l}  
 H^{J_H, spin}_{i,i+\bm{e}_\gamma} = {\bm S}_i \cdot {\bm S}_{i+{\bm e}_\gamma} +\frac{3}{4} \\  
 H^{J_H, orb}_{i,i+\bm{e}_\gamma} = \pi^\gamma_i \pi^\gamma_{i+{\bm e}_\gamma} -\frac{1}{4}  \\
\end{array} \nonumber
\right.
\label{eq:H_KK_JH}
\end{eqnarray}
and the full Kugel-Khomskii~\cite{Kugel82} model for electrons in $e_g$ orbitals on a cubic lattice given by
\begin{eqnarray}
H^{KK} = H^{KK}_U + H^{KK}_{J_H}.
\end{eqnarray}
It is interesting to note that when on two neighboring sites different orbitals are occupied, i.e. $\langle \pi^\gamma_i \pi^\gamma_{i+{\bm e}_\gamma }\rangle < 0$ the resulting spin-spin interaction according to Eq.~(\ref{eq:H_KK_JH}) is ferromagnetic. If instead different orbital are occupied and $\langle \pi^\gamma_i \pi^\gamma_{i+{\bm e}_\gamma} \rangle > 1/4$, the magnetic exchange is antiferromagnetic.  This correlation between orbital occupation and magnetic exchange interactions reflect the well-known Goodenough-Kanamori-Anderson rules for superexchange~\cite{Goodenough63,Kanamori59,Anderson59}.

Similar models describe magnetic systems with $e_g$ orbital degrees of freedom on different lattices, for instance the checkerboard one~\cite{Nasu2012a} and with different types of bonds between the ions, for instance $90^o$ ones~\cite{Mostovoy2002} and have been extended to systems with $t_{2g}$ orbital degrees of freedom~\cite{Kugel82,Khaliullin05,Khaliullin01}.

\subsubsection{Compass Hubbard Models}
\label{sec:EHCM}
Compass type hopping amplitudes leads to more complex variants of the standard Hubbard model \cite{Hubbard63} and lead to further impetus in the study of compass systems. In this subsection, we describe an extended compass Hubbard model (ECHM)  on the square lattice that contains both standard kinetic hopping terms  (as in the Hubbard model) as well as pairing terms. As we will elaborate on in section \ref{sec:SECHM},  this system has the virtue of being exactly reducible to well studied quantum gauge systems at a point of symmetry. At this point, this symmetric extended compass Hubbard model (SECHM) is given by
\begin{eqnarray}
\label{SECHM}
H_{\sf SECHM} = &-& \sum_{i,\gamma=x,y}
 t_{i,i+ {\bf{e}}_{\gamma}}
 \Big[ (c_{i \sigma_{\gamma}}^{\dagger}+c_{i \sigma_{\gamma}} ) \nonumber
\\ &&
( c_{i + {\bf{e}}_{\gamma}, \sigma_{\gamma}}^{\dagger}-c_{i + {\bf{e}}_{\gamma}, \sigma_{\gamma}}) \Big] \nonumber
\\ &+& \sum_{i} U_{i} n_{i \uparrow} n_{i \downarrow} - \sum_{i} U_{i} n_{i}.
\end{eqnarray}
Here both the Coulomb penalty $U_{i}$ as well the hopping amplitudes ($t$) linking sites $i$ and $i+{\hat{e}}_{\gamma}$ are allowed to vary spatially with the site $i$ and direction $\gamma$.
The operators $c_{i \sigma_{\gamma}}$ (and $c_{i \sigma_{\gamma}}^{\dagger}$)  denote the annihilation (creation) of an electron
of spin polarization $\sigma_{\gamma}$ at site $i$. The shorthand $\sigma_{\gamma}$ (with $\gamma = x,y$) is defined via $\sigma_{x} = \uparrow$ and $\sigma_{y} = \downarrow$. 
The dependence of a hopping amplitude for an electron of spin polarization $\sigma$ on the lattice direction $\gamma$ along which the electron may hop embodies a compass type feature. 
In sections (\ref{sec:orbital-orbital}, \ref{sec:KK}) we will review how such hopping amplitudes precisely appear for the pseudo-spin orbital degrees of freedom. 
The number operators $n_{i \sigma}$ with the spin polarization $\sigma = \uparrow, \downarrow$ are, as usual, given by $n_{i \sigma} = c_{i \sigma_{\gamma}}^{\dagger} c_{i \sigma_{\gamma}}$.
The total number operator at site $i$ is $n_{i} = n_{i, \uparrow} + n_{i, \downarrow}$. The Hamiltonian of Eq. (\ref{SECHM}) is symmetric inasmuch as the pairing and hopping terms are of equal magnitudes. Somewhat similarly to $H_{\sf SECHM}$, equal strength pairing and hopping terms appear soluble antiferromagnetic spin chains \cite{LSM1} and related fermionic representations of the two-dimensional Ising model \cite{Schultz1964}). An extended compass Hubbard model arises away from the particular point of symmetry in Eq. (\ref{SECHM}); such a system allows for differing ratios of the pairing and hopping terms as well as a general chemical potential term $\sum_{i} \mu_{i} n_{i}$ where $\mu_{i} \neq U_{i}$. Further extensions to other lattices are possible as well. 

\subsubsection{Lattice Mediated Interactions}
\label{sec:JT}

\begin{figure}
\centering
\includegraphics[width=.9\columnwidth]{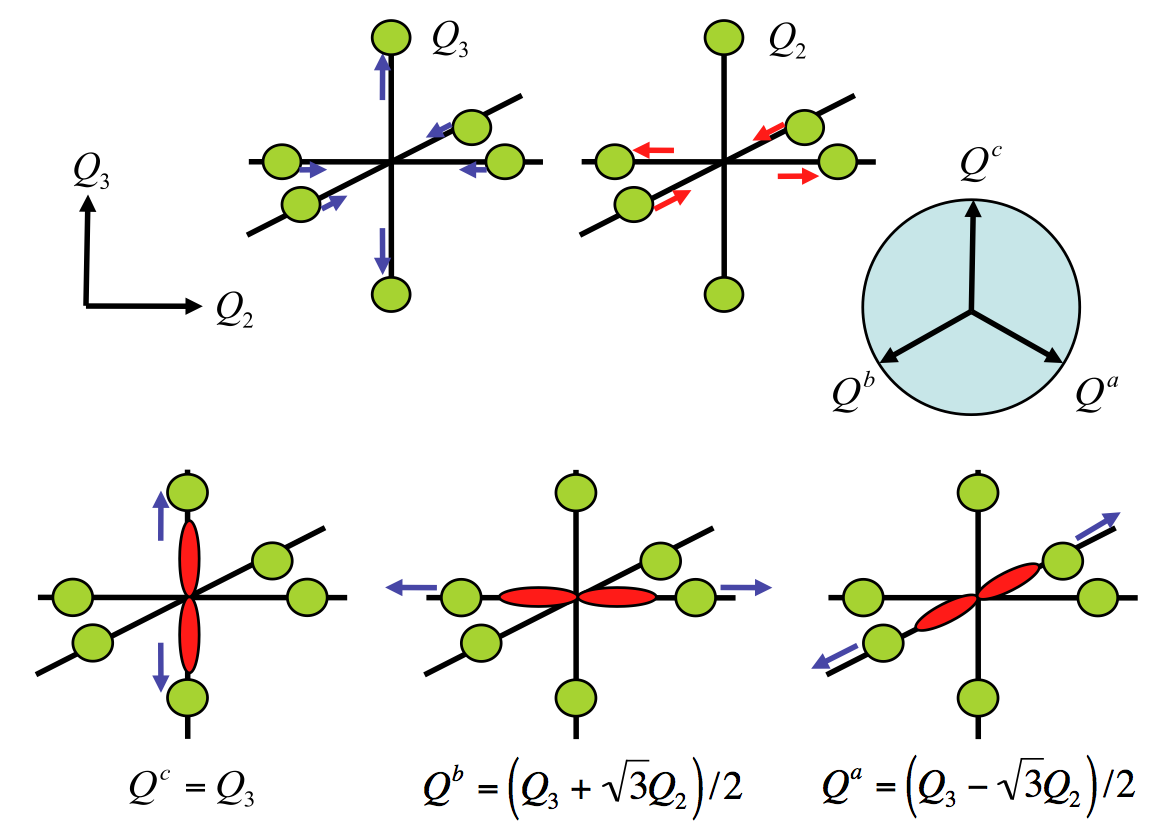}
\caption{Jahn-Teller distortions of $e_g$ symmetry, $Q_{2}$ and $Q_{3}$, of  a transition metal - oxygen octahedron. The orbital degree of freedom $(T_i^z,T_i^x)$ is locked to the distortion $(Q_3,Q_2)$ }
\label{fig:Q2_Q3}
\end{figure}

A convenient mathematical way to describe the pseudospin $\frac12$ of an orbital doublet on the classical level is to introduce the vector ${\bm T}_i=(T_i^z,T_i^x)$  to describe the orbital occupation, so that e.g. the state $|T^z=\frac12\rangle$ corresponds to the occupied $|3z^2-r^2\rangle$ orbital, and $|T^z=-\frac12\rangle$ to $|x^2-y^2\rangle$ one. With these  vectors at hand the effective, classical Hamiltonian for $e_g$ electrons on a cubic lattice interacting via Jahn-Teller distortions can be obtained. If the elongated orbital  $3z^2-r^2$ is occupied on site $i$, the octahedron elongates with a  $Q_3$ distortion, see Fig.~\ref{fig:Q2_Q3}.

Denoting the crystalographic axes of the solid by $a$, $b$ and $c$ and consider how the Jahn-Teller distortions of  neighboring octahedra interact. If the orbital  $3z^2-r^2$ is occupied on site $i$, and the octahedron elongates with a  $Q_3$ distortion the octahedron connected to it along the $c$ axis is automatically compressed: a distortion $-Q_3$ is induced on  the neighboring site along the $c$ axis. Thus along this direction the  interaction between the distortions at nearest neighbors $\{i,j\}$ is  $Q_{3,i} Q_{3,j}$. One can, however, rotate the orbitals in any direction: by choosing $\theta=2 \pi/3$ one obtains an orbital that is elongated along  the $a$ axis: the $3x^2-r^2$ orbital. As discussed earlier,  an $3x^2-r^2$ orbital corresponds to the linear  combination $\frac{1}{2}(-|3z^2-r^2\rangle + \sqrt{3}|x^2-y^2\rangle)$.  The distortion that goes along with it is  $\frac{1}{2} (-Q_3 + \sqrt{3} \ Q_2)$. Therefore it is this linear combination of distortions that determines the  interaction along the $a$ axis. Along the $b$ axis the situation is  analogous with $\theta = -2 \pi/3$. One arrives at the Hamiltonian for  $e_g$ orbitals on a cubic lattice with corner sharing octahedra \cite{Kanamori60,Brink04}
\begin{eqnarray}
H_{120} = \sum_{i,\gamma} 
Q_{i}^{\gamma}Q_{i+{\bm e}_\gamma}^{\gamma},
\label{120Ham_Q2_Q3}
\end{eqnarray}
where $\gamma = a,b,c$ and $Q^{a} =\frac12(Q_{3} - \sqrt{3} Q_{2})$,  $Q^{b} =\frac12(Q_{3} + \sqrt{3} Q_{2})$, $Q^{c}=Q_3$, see Fig. (\ref{fig:Q2_Q3}). 

This model is that of the 120$^{\circ}$ model of Eqs.(\ref{120p}, \ref{eq:pi120}). Note that unlike the realization of Eq. (\ref{eq:eg120}), the 120$^{\circ}$ model of Eq. (\ref{120Ham_Q2_Q3}) derived from Jahn-Teller distortions is essentially classical: the zero point quantum oscillations of the heavy oxygen ions that mediate the orbital-orbital interactions (or equivalently, the interactions between Jahn-Teller centers) are negligible.

\subsubsection{Charge Transfer Effects Through Ligand Sites}
\label{charge_transfer}
So far, the electronic hopping have been implied to occur directly from a $d$-orbital on one site to a  $d$-orbital on a neighboring site. In many oxides however such hoppings from $d$ to $d$ state occur via an oxygen $p$ orbital of an oxygen ion that is bridging two transition metal ions. This is particularly relevant for oxides that are charge transfer isolators \cite{Zaanen1985}. In these materials the charge transfer through ligand sites is dominant when the energy for an electron transfer $\Delta$ between the ligand and the transition metal ion is smaller than the energy penalty $U$
for direct charge transfer between two transition metal ions.

However, it can be easily shown that the effective hoping integrals between $e_g$ and $t_{2g}$ states
do not change their symmetry if hopping is occurring via an oxygen ligand bridging the two transition-metal sites and the emerging Kugel-Khomskii and compass models for the orbital and/or spin degrees of freedom in the strong coupling limit of large $U$, remain unaltered. This situation changes fundamentally when the TM-oxygen-TM bond is not 180 degrees, which is in particular the case for edge-sharing octahedra, where this bond is (close to) 90 degrees~\cite{Mostovoy2002}.

The effective orbital-only and orbital-dependent spin exchange Hamiltonians that result when charge transfer though ligand sites is the dominant conduit for charge excitations between transition metal ions leads to compass type Hamiltonians which are different from which we discussed thus far \cite{Mostovoy2004}. Most notably, an orbital only Hamiltonian appears
which does not result from Jahn-Teller distortions which may account for the far higher orbital ordering transition temperature as opposed to spin ordering in these materials.
Unlike the KK model (in which spin and orbital degrees of freedom are correlated), the orbital only Hamiltonian which remains in the limit of $U \to \infty$ is asymmetric between hole and electron excitations. 
When pairs of transition metal ions with a single {\it hole} (h)  on the doubly degenerate e$_{g}$ orbitals [e.g., Cu$^{2+}$ ions that have an outer-shell structure of $t^{6}_{2g} e^{3}_{2g}$] interact
with one another through ligand sites (the ligand holes are assumed to be dispersion-less), \cite{Mostovoy2004} in the limit $U \to \infty$ (leaving the charge transfer $\Delta$ as the only remaining finite energy scale),
the effective resultant charge transfer orbital-only Hamiltonian assumes the form
\begin{eqnarray}
\label{hct}
H^{(h)}_{CT} =  \frac{2t^{2}}{\Delta^{3}} \sum_{i,\gamma} (\frac{1}{2} + \pi^{\gamma}_{i})(\frac{1}{2} + \pi^{\gamma}_{i+ {\bm{e}}_{\gamma}}).
\end{eqnarray}
In Eq. (\ref{hct}), $t$ is the hopping amplitude between the transition metal ion and the ligand site. The operators
$\pi^{\gamma}$ are of the same form as in Eqs. (\ref{eq:pi120},\ref{eq:eg120}). 
Similarly, for transition metal ions that have one {\it electron} (e) in the doubly degenerate $e_{g}$ states, the effective interaction that remains in the large $U$ limit is
of the form
\begin{eqnarray}
\label{ect}
H^{(e)}_{CT} =  \frac{2t^{2}}{\Delta^{3}} \sum_{i,\gamma} (\frac{3}{2} - \pi^{\gamma}_{i})(\frac{3}{2} - \pi^{\gamma}_{i+ {\bm{e}}_{\gamma}}).
\end{eqnarray}
A single electron in the $e_{g}$ for which Eq. (\ref{ect}) my be relevant can be that of ions such as Mn$^{3+}$ Cr$^{2+}$ (both of which have an ($t_{2g}^{3} e_{g}^{1}$) structure) as
well as the low spin Ni$^{3+}$ ($t_{2g}^{6} e_{g}^{1}$) which we earlier encountered when reviewing NaNiO$_{2}$ \cite{Mostovoy2002}. 
The Hamiltonians of Eqs. (\ref{hct},\ref{ect}) capture the effect of common ligand sites which are shared by the transition metal ions. 
For finite values of $U$, a compass type coupled spin and orbital Hamiltonian different from the Kugel-Khomskii Hamiltonian further appears.
The fact that this spin and orbital Hamiltonian is of a smaller size, for large $U$, than the orbital only Hamiltonians of Eqs. (\ref{hct}, \ref{ect}) 
may, as hinted at above, rationalize the higher values of the orbital ordering temperatures as compared to the spin ordering temperatures
that are commonly observed. Similarly for finite $U$, there are additional corrections to the orbital only
Hamiltonians of Eqs. (\ref{hct}, \ref{ect})). The energetics associated with these Hamiltonians favors orbital and spin states which differ
from those that would be chosen by the Jahn-Teller or Kugel-Khomskii Hamiltonians alone. An additional marked feature of the orbital
only interactions that result is, as is clearly seen in Eqs. (\ref{hct}, \ref{ect}), the appearance of linear terms in the pseudospins.
Such terms are not present in the Jahn-Teller Hamiltonian.
These linear terms effectively act as external effective fields that couple to the pseudospins and may help 
account for empirically observed orbital structure which is not favored by Jahn-Teller nor Hubbard (and thus also Kugel-Khomskii) type
Hamiltonians \cite{Mostovoy2004}. 

\subsubsection{Strong Relativistic Spin-Orbit Coupling}
\label{sec:sporb}

As noted in Section \ref{hybrid}, hybrid models  interpolating between a Heisenberg model and compass models have been introduced  in various physical contexts (\cite{Brink04}, \cite{Chaloupka10}, \cite{Chern10}). Such models are relevant to describe superexchange interactions in transition metal systems with large {\em spin-orbit coupling}, which appears 4d and 5d transition metal ions (see Fig.~\ref{fig:periodic_table}) such as Rh, Ru, Os, and Ir. This spin-orbit coupling mixes the orbital and spin degrees of freedom of an ion, into an effective moment that carries both orbital and spin character. In a Mott insulator the moments on different ions couple to each other via superexchange process.
  
\begin{figure}
\centering
\includegraphics[width=\columnwidth]{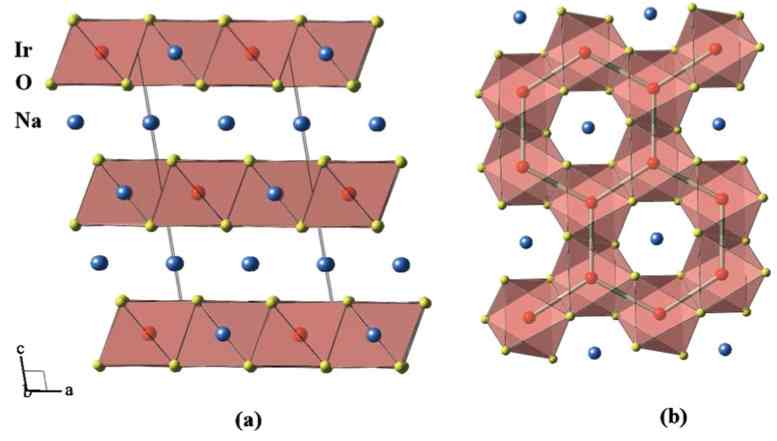}
\caption{The crystallographic structure of Na$_2$IrO$_3$. The Na, Ir, and O atoms are shown as blue (black), red (dark gray), and yellow (light gray) spheres, respectively. (a) The view perpendicular to the $c$ axis showing the layered structure with layers containing only Na atoms alternating slabs of NaIr$_2$O$_6$ stacked along the $c$ axis. The IrO$_6$ octahedra are shown in pink with the (red) Ir atoms sitting in the middle. (b) One of the NaIr$_2$O$_6$ slabs viewed down the c axis to highlight the honeycomb lattice of Ir atoms within the layer. The Na atoms occupy voids between the  IrO$_6$ octahedra~\cite{Singh10}.}
\label{fig:Singh10_1}
\end{figure}

In what follows, we will discuss the viable physical realization of the Heisenberg-Kitaev model. The properties and phase diagram of this type of Heisenberg-Kitaev model are reviewed in Sec.~\ref{HKCS}. 
We review a specific type first of these interactions as first introduced  by \cite{Jackeli09} in their study of Mott insulators and, specifically, several iridates. More recently, some other iridates (including the compounds A$_{2}$IrO$_{3}$ with A an alkaline metal such as Na or Li or a lanthanide that will focus on in this section) have witnessed a flurry of activity due to interest in their possible potential as topological insulators,  e.g., \cite{Shitade,Yang10,Xiangang10,Balents11}.  Other notable families include iridates with a pyrochlore A$_{2}$Ir$_{2}$O$_{7}$ or a hyper-kagome  structure A$_{4}$Ir$_{3}$O$_{8}$. Strong spin-orbit couplings were seen  to lead to unique Mott insulating states \cite{KimBJ2008}.  The couplings between the spin and orbital degrees of freedom on such lattices can lead to compass type interactions.  We first review the physical considerations underlying the specific ``honeycomb iridates''  that form the focus of present interest. In compounds of the A$_{2}$IrO$_{3}$ type, iridium (Ir$^{4+}$) ions at the centers of IrO$_{6}$ octahedra form a layered honeycomb lattice (see Fig.~\ref{fig:Singh10_1}).

Large {\em spin-orbit coupling} appears 4d and 5d transition metal ions such as Rh, Ru, Os, and Ir. The large mass of the iridium ion ($Z=77$) leads to a spin-orbit coupling constant as large as $\lambda \sim 380$ meV in Ir$^{4+}$ ions \cite{Schirmer}.  Direct measurements of the insulating iridates Sr$_{2}$IrO$_{4}$ \cite{Kim2012a}, Sr$_{3}$Ir$_{2}$O$_{7}$ \cite{Kim2012b}, as well as Na$_{2}$IrO$_{3}$ directly attest to the spin-orbit couplings
that they exhibit. Strong crystal field effects of octahedral  oxygen cage split the 5d$^{5}$ configuration of  the typical Ir$^{4+}$ valency and lead to the occupation of the five electrons (or single hole) in the three low energy t$_{2g}$ states. The considerations below apply to magnetic Ir$^{4+}$ and other (e.g., Rh$^{4+}$) transition metal ions with a single hole in the t$_{2g}$ triplet. 

The hole in the t$_{2g}$ orbitals has an effective orbital angular momentum $L =1$ \cite{Abragam}. Specifically, $|L_{z} =0 \rangle \equiv |xy \rangle$ and $| L_{z} = \pm 1 \rangle  \equiv - \frac{1}{\sqrt{2}}(i |x z \rangle \pm |y z \rangle)$. The t$_{2g}$ sector is further splintered as follows. The local, ionic low energy Hamiltonian for the single hole
 \begin{eqnarray}
 \label{jk_ham}
 H = \lambda {\bm L} \cdot {\bm S} + \Delta L_{z}^{2}
 \end{eqnarray}
contains both spin-orbit effects and a tetragonal splitting $\Delta$, which might be finite or zero depending on the coordination of the ligand ions. For an octahedral IrO$_{6}$ oxygen cage that is elongated along the $c$ axis direction the tetragonal splitting $\Delta >0$.  

Spin-orbit coupling splits up the six basis states $|L_{z},S_{z}\rangle$ spanning the  azimuthal angular momentum ($L_{z}$) and spin component $S_{z}$  (up or down) of the hole, into doublet and quartet. The low energy sector of the Hamiltonian of Eq. (\ref{jk_ham}) is spanned by the following two states for the hole \cite{Jackeli09} 
 \begin{eqnarray}
 | \Uparrow \rangle &=& \sin (\theta/2) |L_{z} = 0, \uparrow \rangle - 
 \cos (\theta/2) |L_{z} = 1,  \downarrow \rangle \ \ \ \  \nonumber
 \\ |\Downarrow \rangle &=& \sin (\theta/2) |L_{z}=0, \downarrow \rangle - \cos(\theta/2) |L_{z}=-1,\downarrow \rangle \ \ \ \
 \label{low_so}
 \end{eqnarray}
with $\tan \theta = 2 \sqrt{2 \lambda}/(\lambda - 2 \Delta)$. The two states in Eq. (\ref{low_so}) enable the definition of a  SU(2) pseudo-spin operator such that $| \Uparrow \rangle$ and $|\Downarrow \rangle$ correspond, respectively, to its up and down eigenstates ($|\tau_{z} = 1/2 \rangle = | \Uparrow \rangle$ and $|\tau_{z} = -1/2 \rangle = |\Downarrow \rangle$). These two states in Eq. (\ref{low_so}) are related to each other by time reversal (and thus form a Kramers doublet).  As a consequence of spin-orbit effects, the hole lies within the sector formed by these two states. 

\begin{figure}
\centering
\includegraphics[width=.8\columnwidth]{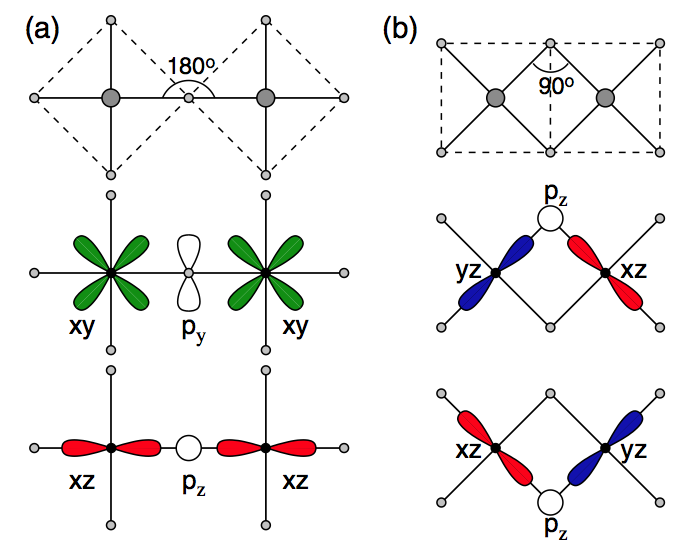}
\caption{Two possible geometries of a TM-O-TM bond with corresponding orbitals active along these bonds. The large (small) dots stand for the transition metal (oxygen) ions. (a) A 180$^\circ$-bond formed by corner-shared octahedra (see also Fig.~\ref{fig:hopping_t2g}), and (b) a 90$^\circ$-bond formed by edge-shared octahedra.
~\cite{Jackeli09}.}
\label{fig:Jackeli09_2}
\end{figure}

The Ir moments in Na$_2$IrO$_3$ become antiferromagnetically ordered at a Neel temperature $T_{N} \simeq 15 K$ \cite{Liu11}- a temperature which is notably lower than the paramagnetic Curie temperature (-125 K \cite{Sing11}).
Such a reduced ordering temperature is to be expected in frustrated systems and is natural within the Heisenberg-Kitaev'' (HK) model ~\cite{Jackeli09,Chaloupka10} which we will study in greater depth in Section \ref{HKCS}.
We now review the processes which may lead to precisely such a Hamiltonian in these systems. 

{\underline{Hopping through intermediate oxygen ligands:}}

Similar to our discussion in subsection \ref{charge_transfer}, we now consider the effect of intermediate ligand (i.e., oxygen) sites. In the case of cubic symmetry ($\Delta =0, \sin \theta = 1/ \sqrt{3}$), there are, generally, two dominant exchange paths for an exchange between two Ir ions at lattice sites $i$ and $j$ via an intermediate oxygen ion \cite{Jackeli09}, see Fig.~\ref{fig:Jackeli09_2}:

(i)  180$^{\circ}$ paths. These are schematically depicted in the lefthand panel of  Fig.~\ref{fig:Jackeli09_2}.
In this case, the hopping is diagonal is the orbital index. 
For instance, both a bond along the $x$ direction, as seen in panel (a) 
of  Fig.~\ref{fig:Jackeli09_2}, an $|x y \rangle$ state on the left TM ion may be 
coupled to an $|x y \rangle$ state on the right TM ion.  Similarly, along such a link
parallel to the cubic lattice $x$ direction, an $|x z \rangle$
state may be coupled to another $| x z \rangle$ state. 
The same applies with trivial alterations for bonds parallel to the $y$ or
$z$ axis cubic lattice directions. 
Within the two state subspace of Eq. (\ref{low_so}), 
the effective exchange interaction between the effective spin-orbital moments is given by
\begin{eqnarray}
H_{ij} = {\cal J}_{1} {\bm{\tau}}_{i} \cdot {\bm{\tau}}_{j}   + {\cal J}_{2} ({\bm{\tau}}_{i} \cdot \bm{e}_{ij}) ({\bm{\tau}}_{j} \cdot \bm{e}_{ij}),
\label{180path}
\end{eqnarray}
where the coupling constants $J_{1,2}$ are dependent on hopping amplitudes, the (Hubbard type) Coulomb repulsion $U$ between two $t_{2g}$ electrons on the same ion and Hund's coupling $J_{H}$ \cite{Jackeli09}.  
As throughout in this review, in Eq. (\ref{180path}), $\bf{e}_{ij}$ denotes a unit vector along the direction from point $i$ to point $j$.  
In the pertinent limit of strong spin-orbit coupling, the interaction of Eq. (\ref{180path}) is dominated by the isotropic
Heisenberg type interaction with the anisotropy set by Hund's coupling. In Na$_{2}$IrO$_{4}$ and other iridates, 
the dominant exchange paths are {\it not those of the 180$^{\circ}$ type} via an intermediate oxygen lying on the line between neighboring
Ir ions. Rather, the dominant mechanisms via intermediate oxygen orbitals are those involving oxygen ions that lie at 90$^{\circ}$ off relative 
to the line between the TM ions (see panel (b) of Fig.~\ref{fig:Jackeli09_2}). We turn to these next. 

(ii) 90$^{\circ}$ bonds.
In such geometries, several mechanisms are possible. In these, unlike the 180$^{\circ}$ bonds reviewed above, the dominant couplings between the $d$ electron orbital states are non-diagonal in the orbital index: different orbital states are coupled to each other.
We first comment on the case of (intra-orbital sector) coupling between two $t_{2g}$ (orthogonal) states and then discuss an important inter-orbital process that couples an $e_{g}$ orbital to a $t_{2g}$ state.

(a)  There are two dominant intra-orbital $t_{2g}$ couplings: \newline

\begin{figure}
\centering
\includegraphics[width=.8\columnwidth]{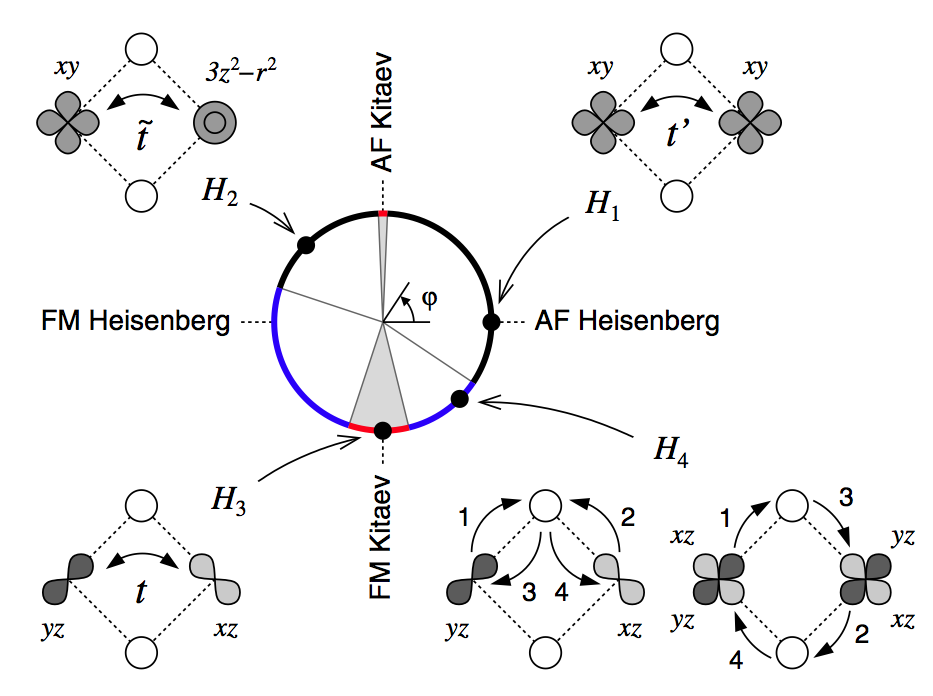}
\caption{Schematics of four different exchange processes~\cite{Chaloupka13}.}
\label{fig:Chaloupka13_2}
\end{figure}

$\bullet$  $t_{2g}$-$t_{2g}$ coupling resulting from hopping between two TM ions via intermediate oxygen ions (See also process $H_{3}$ of Fig. \ref{fig:Chaloupka13_2}.)
as well as the righthand panel of Fig.~\ref{fig:Jackeli09_2}).  
For links $(ij)$ with a TM-O-TM angle  90$^{\circ}$, the superexchange between theTM ions along the cubic $\gamma$ axis, is mediated via two ligands that are perpendicular to that axis, the effective exchange reads
\begin{eqnarray}
H^{superexchage}_{ij} = - {\cal J}^{ex} \tau_{i}^{\gamma} \tau_{j}^{\gamma}.
\label{90path}
\end{eqnarray}  
The exchange coupling ${\cal J}^{ex} \simeq \frac{8}{3} \frac{t^{2} J_{H}}{U^{2}}$ where $t$ is the indirect hopping amplitude between nearest neighboring $t_{2g}$ orbitals via intermediate oxygen ions.  \newline

$\bullet$ $t_{2g}$-$t_{2g}$ coupling via $pd$ charge transfer excitations (of energy $\Delta_{pd}$) 

such as those schematically illustrated in (process $H_{4}$ of Fig. \ref{fig:Chaloupka13_2}) \newline
-- In the first of such processes (that of the lefthand side of  process $H_{4}$ of Fig. \ref{fig:Chaloupka13_2}), two TM holes may hop to the same oxygen site wherein there is a Coulomb repulsion of strength $U_{p}$).
The energy associated with such a process is  
\begin{eqnarray}
H^{transfer}_{ij} \simeq   -  \frac{8}{9(\Delta_{pd}+ \frac{U_{p}}{2})} \tau_{i}^{\gamma} \tau_{j}^{\gamma}.
\label{90path4a}
\end{eqnarray}
 \newline 

-- In a second process, two holes may go (either clockwise or counter-clockwise within a loop) through the sites of
a plaquette (as in the righthand process of $H_{4}$ of Fig. \ref{fig:Chaloupka13_2}). This leads to a contribution 
\begin{eqnarray}
H^{plaquette} \simeq \frac{8}{9 \Delta_{pd}}   \tau_{i}^{\gamma} \tau_{j}^{\gamma}.
\label{90path4b}
\end{eqnarray}   \newline

(b)  Inter-orbital $t_{2g}$- $e_{g}$ hopping $\tilde{t}$ via an intermediate oxygen (as in the process of $H_{2}$ of Fig. \ref{fig:Chaloupka13_2}). 
This is the dominant  \cite{Chaloupka13} mechanism in a 90$^{\circ}$ bonding
geometry. In the iridates, this contribution is borne by the sizable $t_{p d \sigma}$ overlap between the 
oxygen $2p$ and $e_{g}$ orbital of the TM ion.  Along a cubic lattice direction $\gamma$, \cite{Khaliullin05} 
\begin{eqnarray}
\label{interorbit}
H^{inter-orbital}_{ij} = {\cal{J}}^{i-o} [2 \tau_{i}^{\gamma} \tau_{j}^{\gamma} - {\bm{\tau}}_{i} \cdot {\bm{\tau}}_{j}].
\end{eqnarray} 
Within the Mott-insulating iridates, ${\cal{J}}^{i-o} \simeq \frac{4}{9}({\tilde{t}}/{\tilde{U}})^{2} {\tilde{J}}_{H}$
where \cite{Chaloupka13} $\tilde{J}_{H}$ is Hund's coupling between $t_{2g}$ and $e_{g}$ orbitals and
$\tilde{U}$ is the Coulomb penalty associated with $t_{2g}$-$e_{g}$ hopping.  Typically, in the iridates, ${\tilde{t}}/t /t \sim 2$ rendering these
interactions very notable. The origin of the form of ${\cal{J}}^{i-o}$ is transparent as we briefly review. The factor of $(\tilde{t}/\tilde{U})^{2}$ provides the 
probability that a $t_{2g}$ spin is transferred to a nearest neighbor $e_{g}$ orbital; on arrival
at that orbital of the nearest neighbor ion, the transferred spin has to obey Hund's rule with the ``host'' $t_{2g}$ spin \cite{Khaliullin05,Chaloupka13}. 

{\underline{Direct hopping between $t_{2g}$ orbitals:}}

Supplanting all of these exchange paths above through an intermediate oxygen (whether of the 90$^{circ}$ or 180$^{\circ}$ type), there are also direct exchange interactions  (as in the process $H_{1}$ of Fig. \ref{fig:Chaloupka13_2}).  These lead to a Heisenberg type standard spin exchange result 
\begin{eqnarray}
H^{direct} = \frac{4}{9} \frac{(t')^{2}}{U} [{\bm{\tau}}_{i} \cdot {\bm{\tau}}_{j}],
\label{direct_t2g_}
\end{eqnarray}
where $t'$ denotes the direct hopping amplitude between
the $t_{2g}$ electrons. 

\begin{figure}
\centering
\includegraphics[width=.8\columnwidth]{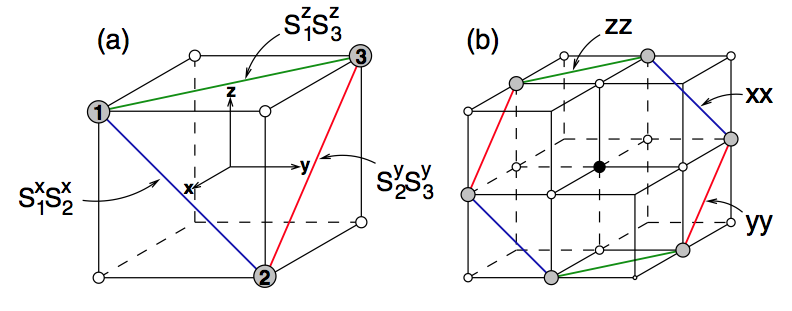}
\caption{Examples of the structural units formed by $90^\circ$ TM-O-TM bonds and corresponding spin-coupling patterns. Gray circles stand for magnetic ions, and small open circles denote oxygen sites. (a) Triangular unit cell of ABO$_2$-type layered compounds, periodic sequence of this unit forms a triangular lattice of magnetic ions. The model on this structure is a realization of a quantum compass model on a triangular lattice: e.g., on a bond $1-2$, laying perpendicular to x-axis, the interaction is $\tau^x_1 \tau^x_2$. (b) Hexagonal unit cell of A$_2$BO$_3$-type layered compound,  in which magnetic ions (B-sites) form a honeycomb lattice. (Black dot: nonmagnetic A-site). On an $xx$-bond, the interaction is $\tau^x_i \tau^x_j$, etc. For this structure, the model (3) is identical to the Kitaev model~\cite{Jackeli09}.
}
\label{fig:Jackeli09_3}
\end{figure}

In an ABO$_{2}$ system with A (and B) being non-magnetic alkali (TM magnetic) ions, a layered triangular geometry with 90$^{\circ}$ type exchange processes
between the magnetic B ions appear. [See also our earlier discussion of NaNiO$_{2}$ in subsection \ref{sec:orbital-orbital}.] 
Further dilution of this system, as in A$_{2}$BO$_{3}$ systems 
leads to a layered honeycomb type structure of the magnetic B ions, see Fig.~\ref{fig:Jackeli09_3}. On such a geometry, the 90$^{\circ}$ processes lead to Kitaev type interactions augmenting rotationally symmetric Heisenberg type exchange processes. These considerations apply to layered honeycomb iridates (B = Ir and A = Na or Li). 

Putting all of the pieces together, exchange processes lead to an effective Hamiltonian in the projected subspace of Eq.(\ref{low_so}) that is of the 
Heisenberg-compass form of Section~\ref{hybrid}. In A$_{2}$BO$_{3}$ systems this compass model is of the Kitaev type. This Heisenberg-Kitaev (HK) model, in the form that it is typically studied~\cite{Chaloupka10}, is
\begin{eqnarray}
H_{HK} &=& -  J_{1} \sum_{\langle ij \rangle_{\gamma}}  \tau_{i}^{\gamma} \tau_{j}^{\gamma} 
+ J_{2} \sum_{\langle i j \rangle} {\bm \tau}_{i} \cdot {\bm \tau}_{j} \nonumber \ \ \ \ \\ 
& \equiv & C \Big[ -2 \alpha \sum_{\langle ij \rangle_{\gamma}} \tau_{i}^{\gamma} \tau_{j}^{\gamma} 
+ (1- \alpha) \sum_{\langle i j \rangle} {\bm \tau}_{i} \cdot {\bm \tau}_{j} \Big] \nonumber  \ \ \ \ \\
& \equiv &  A \Big[  2  \sin \varphi  \sum_{\langle ij \rangle_{\gamma}}
 \tau_{i}^{\gamma} \tau_{j}^{\gamma} 
+ \cos \varphi \sum_{\langle i j \rangle} {\bm \tau}_{i} \cdot {\bm \tau}_{j} \Big].
\label{HKI}
\end{eqnarray} 
Eq. (\ref{HKI}) constitutes a particular realization of Eq. (\ref{hybrid_compass}). 
The parameterizations of the coupling constants $J_{1}$ and $J_2$ in terms of $(C, \alpha)$ or $(A, \varphi)$ are prevalent in the literature and are natural
in disparate contexts. We will return to these in subsection \ref{HKCS}. In early works within the field, the focus has been 
on systems with ferromagnetic $(J_{1} >0$) Kitaev couplings. It was later realized \cite{Chaloupka13} that antiferromagnetic 
Kitaev type interactions can be notable due to the process of Eq. (\ref{interorbit}); these contributions lead, consistent various experimental
measurements, to zig-zag type magnetic ordering \cite{Singh10,Liu11,Sing11,Choi2012}. 

Effective related models with longer ranger  interactions were advanced by \onlinecite{Kargarian2012}, introducing a Hubbard model with next nearest neighbor hopping on a honeycomb lattice -- the pertinent geometry of the layers. The Hamiltonian is given by
\begin{eqnarray}
\label{LRCHM}
H = &&- t \sum_{\langle i j \rangle, \alpha} c_{i \alpha}^{\dagger} c_{j \alpha} + \sum_{\langle \langle i j \rangle \rangle_{\alpha}} \sum_{\alpha \beta} t_{\alpha \beta}^{\gamma} c_{i \alpha}^{\dagger} c_{j \beta} \nonumber
\\ && +  U \sum_{i} n_{i \uparrow} n_{i \downarrow}.
\end{eqnarray}
In Eq. (\ref{LRCHM}), $\alpha,\beta = \uparrow, \downarrow$ denote spin-orbital coupled pseudo-spin states, $U$ is the standard Hubbard Coulomb penalty term, $t$ is the hopping amplitude between nearest neighbor sites on the lattice, and the compass-type next nearest neighbor hopping $t^{\gamma} = - t' + i t'' \sigma^{\gamma}$ where $\sigma^{\gamma}$ are Pauli operators; the flavors $\gamma= x,y,$ or $z$ of the Pauli operators are given by the relative orientation of the next nearest sites $i$ and $j$. On the honeycomb lattice, the vector connecting next neighbor sites can be along one of three directions. 
In the strong coupling limit, $U \gg t$, Eq. (\ref{LRCHM}) reduces to a longer range counterpart of the Heisenberg- Kitaev models discussed in section \ref{hybrid}. See Eq. (\ref{hybrid_compass}) in particular.
This is given by \cite{Kargarian2012} 
\begin{eqnarray}
\label{LRHK}
H &=& J_{1} \sum_{\langle i j \rangle} {\bf{\tau}}_{i} \cdot  {\bf{\tau}}_{j} - J_{2}  \sum_{\langle \langle i j \rangle \rangle} {\bf{\tau}}_{i} \cdot  {\bf{\tau}}_{j}  \nonumber
\\  &&+J_{3} \sum_{\langle \langle i j \rangle \rangle} \tau_{i}^{\gamma} \tau_{j}^{\gamma}.
\end{eqnarray}
where $J_{1} = 4t^{2}/U, J_{2} = 4[(t')^{2}-(t'')^{2}]/U$ and $J_{3} = 8(t'')^{2}/U$. 

\subsection{Vacancy Centers and Trapped Ions}
\label{VCsection}

As noted earlier, the Heisenberg-Kitaev model of Eq. (\ref{HKI}) represents a specific type of the compass-Heisenberg models discussed in Section \ref{hybrid}.
Hamiltonians of the form of Eq. (\ref{hybrid_compass}) may also describe vacancy center systems \cite{Trousselet12}.Such systems are afforded by, e.g., arrays of nitrogen vacancy (NV) centers 
in diamond \cite{Gaebel2006,Neumann2010}. When hyperfine and other effects may be neglected, vacancy arrays may, in some instances, be modeled by quantum pseudo-spins that are predominantly coupled
via effective dipolar interactions that contain both a couplings of the Heisenberg exchange variety along with compass type interactions,
\begin{eqnarray}
\label{effdip}
H_{dipolar} = \sum_{\langle i j \rangle_{\gamma}}  \frac{A}{r_{ij}^{3}} ({\bf{T}}_{i} \cdot {\bf{T}}_{j} - 3 T_{i}^{\gamma} T_{j}^{\gamma}),
\end{eqnarray}
where the pseudo-spin ${\bf T}_{i}$ now represents the dipole moment at lattice site $i$ and $A$ is a system dependent constant. In such lattices formed by vacancies,
the dipole interactions have a compass type, anisotropic, component (the second term in Eq. (\ref{effdip})) which depends on the 
projection of the effective dipoles along the external lattice directions $\gamma$ (the latter directions are defined by the directions of lattice vectors linking interacting nearest neighbor lattice sites). 
The sum of the two (anisotropic and isotropic) terms in Eq. (\ref{effdip}), capturing interactions between these effective dipoles on the lattice, leads to a compass-Heisenberg type system \cite{Trousselet12}. 
Dipole type interactions also arise in trapped ion systems \cite{Schmeid11}. Already on the classical level, the direction dependence of the dipolar interactions leads to notable differences
in thermodynamic properties and correlations \cite{Chakrabarty2011} from dipolar interactions that include the isotropic component alone. 

\subsection{Cold Atom Systems}  
\label{sec:cold_atoms}

In recent years, the ability to manipulate cold atom (and molecule) systems in standing wave laser beams has enabled the generation of systems with tunable interactions. In essence, laser beams enable to generate confining potentials and a {\it crystal of light} in which the lattice sites serve as energy minima for the location of dilute atoms or molecules. 

Gaining understanding of electronic and magnetic effects is in a solid typically complicated by, for example, the presence of impurities, and the long-range nature of Coulomb interactions and in general the rather limited possibility to change parameters and interactions. Ultracold atoms in optical lattices provide a great advantage in allowing to probe model Hamiltonians that capture the essential many-body physics of strongly correlated electron systems in a controllable and clean experimental setting~\cite{Jaksch05,Bloch08}. Relevant parameters can be independently controlled, thus allowing quantitative comparisons of the experiment and theory. 

In particular the Hubbard Hamiltonian for both bosonic~\cite{Jaksch98,Greiner02,Stoferle04} and fermionic particles~\cite{Schneider08} on optical lattices has been realized, also in the Mott insulating regime. This has opened the road to prepare other effective spinor models with ultracold atoms on the lattice, such as the ones of compass and Kitaev type, which we review in this section.

Proposals for the creation of compass-type models in the ultra-cold gas setting can be classified into three categories. The first one is to use an ensemble of ultra-cold bosonic or fermionic atoms with two relevant internal states and engineer the hopping amplitudes by additional laser fields~\cite{Duan03}.  The second category is to use atoms that are in $p$-like states, the orbital degeneracy of which constitutes the pseudo-spin degree of freedom, which can be created either by excitation out of $s$-like states or by filling a site with more than one fermionic atom~\cite{Isacsson05,Kohl05,Browaeys05,Muller07,Anderlini07,Kuklov06,Liu06,Wu06,Wu08,Wu08b,Wu08c,Zhao08}. Finally, by manipulating ultra cold dipolar molecules anisotropic spin interactions can be generated~\cite{Micheli06,Weimer13}. 



\subsubsection{Engineering Tunneling Amplitudes}
In an ensemble of ultra-cold bosonic or fermionic atoms with two relevant internal states, a T=1/2 pseudospin, confined in an optical lattice, the pseudospin dependent tunneling between neighboring atoms in the lattice can be controlled. As reviewed in Sec.~\ref{sec:orbital-orbital}, full control of these hopping amplitudes is in the Mott insulating regime of the Hubbard model enough to construct any compass-type Hamiltonian.  In both Bose and Fermi systems, the anisotropy of the exchange in particular tunneling directions can be engineered by applying blue-detuned standing-wave laser beams along those directions ~\cite{Duan03,Kuklov03}.

\subsubsection{Bosonic Gases with Orbital Degree of Freedom}
\label{sec:bosegas}
In the ground state, the atoms in an optical lattice are centered about their local 
minima provided by the confining potential of the laser beams which in
the vicinity of its minima is harmonic. The atomic states
in the lowest Bloch band are, essentially, the ground 
of the harmonic oscillator (more precisely, the product of single harmonic
oscillator centered about each of the minima of the periodic confining potential
generated by the laser beams) and those within the first excited
Bloch band correspond to the first excited states of a harmonic
oscillator.

Several approaches are available for transferring cold atoms to the first excited $p$-orbital band, for instance by applying an appropriate vibrational pulse with frequency on resonance with the $s$-$p$ state transition~\cite{Liu06}. A theory for the interactions in a dilute system of bosons in which the two lowest Bloch bands of a three dimensional optical lattice are considered was developed by \cite{Isacsson05}.  

The central point in all of this is that in the cold atomic gas there are three such excited state corresponding to an "excitation" along each of the three Cartesian directions (which for a single atom about its local confining potential minimum, which for symmetric confining potentials along all three directions, are of the form $ x e^{-(r/a)^2}$, $y  e^{-(r/a)^2}$ and $z e^{-(r/a)^2}$, with $r^2=x^2+y^2+z^2$ and $a$ the  harmonic confining potential length scale. Henceforth these excitations are labelled as $p=X$, $Y$, $Z$.  

The $p$-states are rather confined along all Cartesian directions apart from one and in that sense resemble atomic $p$-orbital. In the presence of Hubbard-type local interactions between the bosons the resulting system is thus of a compass type, where the pseudo-spins emerge from bosonic degrees of freedom.  The strength of the confining potential along the three Cartesian directions
can be tuned by the optical lattice. In the symmetric case, the resulting
effective Hubbard type model taking into account on-site 
interactions of strength $U$ between the atoms is the
form \cite{Isacsson05}

\begin{eqnarray}
H_{IG}&=& \sum_{i,p}  \Big( E_{i}(i) n_{i}^{(p)} + \frac{U_{pp}}{2} n_{i}^{p}(n_{i}^{p}-1) 
\Big) \nonumber\\ 
&+& \sum_{i,p \neq p'} U_{pp'} \Big(n_{i}^{p} n_{i}^{p'}  + \frac{1}{2}
(p_{i}^{\dagger} p_{i}^{\dagger} p_{i}^{\prime} p_{i}^{\prime} + h.c.) \Big) \nonumber
\\ &-& t \sum_{\langle i,i' \rangle_{p},p}  (p_{i}^{\dagger} p_{i'} + h.c.). 
\label{ig3}
\end{eqnarray}
The operators $p_{i}^{\dagger}$ and $p_{i}$ correspond to the creation and annihilation operators for an excited boson of 
flavor $p=X,Y,Z$ at site $i$. 
The constants $U_{pp'}$, $U_{pp}$, and $E_{i}$ are determined by the parameters
describing the confining optical potential. 
In a similar vein, if the confining potential along, say, the $z$ direction is much larger than along
the $x$ and $y$ directions, the system is effectively two dimensional ($p=X,Y$ in Eq. \ref{ig3} 
above). 
Physically, the Hamiltonian then describes two boson species (of type $X$ and $Y$) each of which may propagate only along one direction. The interaction terms enable two bosons of type $X$ to fuse and generate two bosons of type $Y$ (and vice versa). 

There is a formal connection between a system of {\it hard core bosons} where the on-site repulsion $U \to \infty$ and no two bosons can occupy the same site and the pseudo-spin variants of the compass models. Towards this end, one can employ the Matsuda-Matsuda transformation \cite{Matsuda-Matsuda} relating a two flavor system of hard core bosons (e.g., bosons of type $X$ and $Y$) and the two states of a pseudo-spin $T=1/2$ particle. 

\subsubsection{Fermionic Gases with Orbital Degree of Freedom}

Fermionic realizations of compass type systems have also been considered in optical lattices \cite{Wu08, Zhao08}.  A situation with a strong confining potential along e.g. the spatial $z$ direction will again lead to a {\em two dimensional} system. \cite{Wu08} focused on atomic orbitals and considered a situation in which there are two fermions per site with one of the fermions in an inert $s$ shell and the other occupying the $p$ bands (which in  the case of strong optical confinement along the vertical ($z$) direction
is restricted to the {\em one of the two $p-$states}  (i.e., $p_{x}$ and $p_{y}$ orbitals). Hopping within
the $p$ band states can be of either of the $\sigma$ bonding ($t_{||}$)-
wherein there is a head on overlap of one electronic lobe of one site with
another (parallel) single electronic $p$ lobe on a neighbor site- the wave-functions
are parallel to the spatial direction linking the two sites
or of the $\pi$ bonding ($t_{\perp}$) type where
the $p$ wave-functions on two neighboring sites 
are orthogonal to the axis that links these two sites. Due
to the far smaller overlaps involved in $\pi$ bonding, the
$\sigma$ bonding is typically far stronger ($t_{||}/t_{\perp} \gg 1$).
In what follows $\pi$ effects will be neglected.
{\em The directional character of the $\sigma$ bonding underlies
the compass type interactions in this system.}
Orbitals in the $p_{x}$ state have a high tunneling amplitude
only the $x$ direction and similarly orbitals in the $p_{y}$ state
have a high tunneling and lead to consequent effective interactions
only along the $y$ direction.
Scattering in the $p$ wave channel as well as enhancements by magnetic effects 
and proximity to the Feshbach resonance can lead to a substantial Hubbard like interaction
\begin{eqnarray}
H_{Hubbard} = U \sum_{i} n_{i,x} n_{i,y}.
\label{Hubbard_ferm_opt}
\end{eqnarray}
In Eq. \ref{Hubbard_ferm_opt}, with $p_{i,x}^{\dagger}$ and $p_{i,x}$
denoting the creation and annihilation operators for an electron in
the $p_{x}$ orbital at site $i$, the operators 
$n_{i,x} = p_{i,x}^{\dagger} p_{i,x}$ and $n_{i,y}=p_{i,y}^{\dagger} 
p_{i,y}$ are the number operators
for states of the $p_{x}$ and $p_{y}$ type respectively on the lattice site $i$. 
We may define $T=1/2$ pseudo-spin operators to be \cite{Wu08, Zhao08}
\begin{eqnarray}
\tau_{1} &=& \frac{1}{2}(n_{x} - n_{y}), \nonumber
\\ \tau_{2} &=& \frac{1}{2}(p_{x}^{\dagger} p_{y} + H.c.), \nonumber
\\ \tau_{3} &=& - \frac{i}{2}(p_{x}^{\dagger} p_{y} - H.c.).
\end{eqnarray} 
The $p_{x,y}$ states are eigenstates of $\tau_{1}$ with eigenvalues $\pm 1/2$ respectively. 
The compass type character emerges naturally.  The $\sigma$-bonding exchange between 
two sites separated along, say, the Cartesian $x$ lattice direction. In that case, for large $U$
where a perturbative expansion in $t_{||}/U$ about the degenerate ground
state of Eq. \ref{Hubbard_ferm_opt} (that of a single $p_{x}$ or $p_{y}$state per
site) is possible. Second order perturbation theory in the kinetic $t_{||}$ term 
gives rise to an effective Ising type exchange $H_{ex} = J_{||} \tau_{i,1} \tau_{i+ {\bf{e}}_{x},1}$
with $J_{II} = 2 t_{||}^{2}/U$ \cite{Wu08, Zhao08}. Let us now consider the case
of general quantization axis and separation between neighboring sites on the lattice.
Similar to compass models in other arenas (in particular in orbital physics
of the transition metal oxides), a simple but important feature of the underlying
quintessential physics is that 
{\em the Ising quantization axis will change with different orientations 
of the link connecting neighboring lattice sites}. For a lattice link 
of general direction ${\bf{e}}_{\theta} = \cos \theta {\bf{e}}_{x} + \sin \theta {\bf{e}}_{y}$,
it is possible to rotate the $p_{x,y}$ orbitals by $\theta$ to restore the situation above.
This change of basis effects $p_{x}^{\prime} =   p_{x} \cos \theta+ p_{y} \sin \theta$
and $p_{y}^{\prime} = p_{y}  \cos \theta - p_{x} \sin \theta$. These two states $p_{x,y}^{\prime}$
are eigenstates of the operator $\tau_{1}^{\prime} = (\tau_{1} \cos 2 \theta + \tau_{2} \sin 2 \theta)$.
The exchange interaction for general orientation of a link between nearest neighbor sites
is thus \cite{Wu08, Zhao08}
\begin{eqnarray}
H_{ex} (i,i+ {\bf{e}}_{\theta}) = J_{||} [{\bm{\tau}}_{i}  \cdot {\bf{e}}_{2 \theta}]
[{\bm{\tau}}_{i + {\bf{e}}_{\theta}}  \cdot {\bf{e}}_{2 \theta}].
\label{ex_theta}
\end{eqnarray}
As in other orbital systems, once the interaction along one 
link (Eq. \ref{ex_theta}) in known, the Hamiltonian for the entire lattice 
can be pieced together by summing
over all links in the lattice (taking into account their 
different spatial orientation ${\bf{e}}_{\theta}$). 

\subsubsection{Fermions in an Optical Lattice}
\label{f3cw}

In {\em three dimensions}, similar considerations recently led to the 
introduction of the {\it Gell-mann compass models}
of Chern and Wu \cite{Chern11} on the cubic and diamond 
(Eqs. (\ref{Gell-mann1},\ref{Gell-mann2})) and more general lattices as we now
review. As in the two-dimensional case, each site of the lattice hosts
two fermions with one electron filling the inert $s$-orbital. In three-dimensions, 
the remaining electron can be in {\em any one of the three} $p-$ orbitals ($p_{x}, p_y$ or $p_z$).
Replicating the arguments presented above for two-dimensions \cite{Chern11}, 
in the limit $U \gg t_{||} \gg t_{\perp}$,
Chern and Wu arrived at the following Hamiltonian \cite{Chern11}
\begin{eqnarray}
H_{CW}= - J \sum_{\langle ij \rangle} [P_{i}^{{\bm{e}}_{ij}} (1- P_{j}^{{\bm{e}}_{ij}})
+ [(1-P_{i}^{{\bm{e}}_{ij}}) P_{j}^{{\bm{e}}_{ij}}].
\label{CW3p}
\end{eqnarray}
In Eq. (\ref{CW3p}), ${\bm{e}}_{ij} = (e_{ij}^{x}, e_{ij}^{y}, e_{ij}^{z})$ 
is the bond direction (along which $t_{||}$ dominates
for the orbital $|{\bm{e}}_{ij} \rangle =e_{ij}^{x} |p_{x} \rangle +
e_{ij}^{y} |p_{y} \rangle + e_{ij}^{z} |p_{z} \rangle$
(over the transverse hopping $t_{\perp}$)). 
The projection operator $P^{{\bm{e}}_{ij}} = | {\bm{e}}_{ij} \rangle
\langle {\bm{e}}_{ij} |$. The Hamiltonian of Eq. (\ref{CW3p})
embodies the ability of an electron in state $|{\bm{e}}_{ij} \rangle$
on site $i$ to hop in a direction parallel to ${\bm{e}}_{ij} $
to site $j$ if that site is unoccupied in that state
(and vice versa).  As in the standard Hubbard model,
and the two-dimensional Hubbard type model
discussed above, this kinetic hopping leads,
for large $U$, to an effective exchange 
Hamiltonian in the presence of one 
relevant electronic degree of freedom 
per site. 

When applied to the cubic and diamond lattice,
this Hamiltonian reduces to the form provided
in Eqs. (\ref{Gell-mann1}, \ref{Gell-mann2}) \cite{Chern11}. 

Expressing, in the case of the cubic lattice model,
the projection operators along the three crystalline directions ($\gamma =x,y,z$)
as $P^{\gamma} = \frac{1}{3}(1 + 2 {\bm{\lambda}} \cdot {\bm{e}}_{\gamma}$
and inserting this form into Eq. (\ref{CW3p}) leads, up an innocuous additive
constant, to Eq. (\ref{Gell-mann1}). Similarly, in the case of the diamond lattice,
the projection operators may be written as
$P= \frac{1}{3} (1 + \sqrt{3} {\bm{\lambda}} \cdot  {\bm{n}}_{\gamma})$
which reduces Eq. (\ref{CW3p}) to Eq. (\ref{Gell-mann2}). 

\subsubsection{Spin interactions on a lattice}
Ref. \cite{Micheli06} discussed how to design general lattice spin systems by cold 
systems of polar molecules. 

In cold gases of polar molecules, the spin degree of freedom originates 
the spin of an electron outside a closed shell of a hetero-nuclear molecule in its 
rotational ground state. The complete energy of the system is given by the sum
of the translational kinetic and potential energies representing the confining
potential of the laser system 
and two contributions which are of paramount importance in this setup-
the individual rotational excitation energies of each molecule (that contains the nuclear 
angular momentum energy $B {\bf N}^{2}$ (with ${\bf N}$ the nuclear orbital angular momentum)
and spin-rotation coupling (${\bf S} \cdot {\bf N}$)), 
and the dipole-dipole interactions between two molecules
with the dipoles induced by the (nuclear) orbital angular momentum of each molecule. A key point is that large dipole-dipole interactions may be induced by 
a microwave field with frequency is near resonance with the transition $N=0 \to N=1$ 
transition. An effective second order Hamiltonian in the ground
state basis was obtained \cite{Micheli06} which when averaged over the 
inter-molecular relative distance between members of a pair of molecules 
leads to an effective spin only interaction. The final effective Hamiltonian 
enables rather general interactions.  The effective spin interactions
are borne by the dipolar interactions induced by the microwave field. 
The interactions depend on the orientation of inter-molecular 
separations relative to the microwave field direction. 
In this setup spin orientation dependent compass 
type interactions appear very naturally.

\subsubsection{Three-Flavor Compass Models}
\label{3fs}
%
Even though the current focus on compass type interactions within various pseudospin systems,
such interactions can arise in many other systems. One of their most natural incarnations is
within bosonic and fermionic type systems. Of these the multi-orbital Hubbard model, from which the Kugel-Khomskii models are derived, is the simplest example~\cite{Kugel82}.
As discussed in Section~\ref{sec:bosegas}, 
also in bosonic systems \cite{Isacsson05}, there may be two different types of particle species each of which may propagate along only one spatial direction. These
particle species may interact with one another via on-site terms wherein two particles
"collide" along one axis and then convert into two particles that may propagate along
an orthogonal direction. 
%
%

One can also similarly define models in which there are several fermionic
species- each of which have ``compass type'' hopping amplitudes and 
may, e.g., propagate only along one direction or more generally have anisotropic
hopping amplitudes that differ from one species to another. Different types of such systems have
been investigated \cite{Chern11} 
Here the concept is illustrated by specifically considering the incarnation
of such system recently introduced by Chern and Wu \cite{Chern11}.
It leads to compass type systems referred to as {\it Gell-mann matrix compass models}.
Unlike the $SU(2)$ isospins that
formed the focus of our discussion thus far,
the basic degree of freedom in these systems
are Gell-mann operators.

Specifically, on the cubic lattice, these take the form \cite{Chern11}
\begin{eqnarray}
H^{Gell-mann}_{3\Box} = \frac{8J}{9}  \sum_{a=x,y,z} \sum_{\langle i j  \rangle || \gamma} 
({\bm{\lambda}}_{i} \cdot {\bm{e}}_{\gamma}) ({\bm{\lambda}}_{j} \cdot {\bm{e}}_{\gamma}).
\label{Gell-mann1}
\end{eqnarray}
In Eq. (\ref{Gell-mann1}), ${\bm{\lambda}} = \frac{\sqrt{3}}{2}  (\lambda^{(3)}, \lambda^{(8)})$
where the standard Gell-mann matrices $\lambda^{(3)}$ and $\lambda^{(8)}$ are diagonal and given 
by $\lambda^{(3)} = $diag$(1,-1,0)$ and $\lambda^{(8)} =$ diag$(1,1,-2)/\sqrt{3}$. 
As in the earlier compass model that we introduced thus far, $\gamma$ denotes the direction
of the link between the nearest neighbor sites $i$ and $j$. Similar to the 120$^{\circ}$ model,
the three unit vectors in Eq. (\ref{Gell-mann1}) are equidistant on a disk,
${\bm{e}}_{x,y} = (\pm \sqrt{3}, 1)/2$ and ${\bm{e}}_{z} = (0,-1)$.

On the diamond lattice \cite{Chern11}, 
\begin{eqnarray}
H^{Gell-mann}_{3\diamond} =  \frac{2J}{3} \sum_{\gamma =0}^{3} \sum_{\langle ij \rangle || \gamma}
({\bm{\lambda}}_{i} \cdot {\bm{n}}_{\gamma}) ({\bm{\lambda}}_{j} \cdot {\bm{n}}_{\gamma}).
\label{Gell-mann2}
\end{eqnarray}
In this case, in Eq. (\ref{Gell-mann2}), the vector ${\vec{\lambda}} = (\lambda^{(6)}, \lambda^{(4)},
\lambda^{(1)})$. The Gell-mann matrices $\lambda^{(1)},\lambda^{(4)}$, and $\lambda^{(6)}$ are non-diagonal (and do not commute amongst themselves).
The index $\gamma=0,1,2,3$ denotes the four nearest neighbor directions
on the diamond lattice with correspondingly $\{{\bm{n}}_{\gamma}\}$ denoting 
the unit vectors from a given lattice site
to its nearest neighbors. (Specifically, 
when expressed in the Cartesian coordinate system,
${\bm{n}}_{0} = ({\bm{e}}_{x} + {\bm{e}}_{y} + {\bm{e}}_{z})/\sqrt{3}, {\bm{n}}_{1} = ({\bm{e}}_{x} - {\bm{e}}_{y} - {\bm{e}}_{z})/\sqrt{3}, {\bm{n}}_{2} = (-{\bm{e}}_{x} + {\bm{e}}_{y} - {\bm{e}}_{z})/\sqrt{3}$, and
${\bm{n}}_{3} = (-{\bm{e}}_{x} - {\bm{e}}_{y} + {\bm{e}}_{z}))/\sqrt{3}$. 
The motivation and properties of these models are reviewed in Sections \ref{f3cw}, \ref{santa-fe}. 

\subsection{Chiral Degrees of Freedom in Frustrated Magnets}  
\label{sec:chiral}

\label{sec:chiral_spin}

\begin{figure}
\centering
\includegraphics[width=.8\columnwidth]{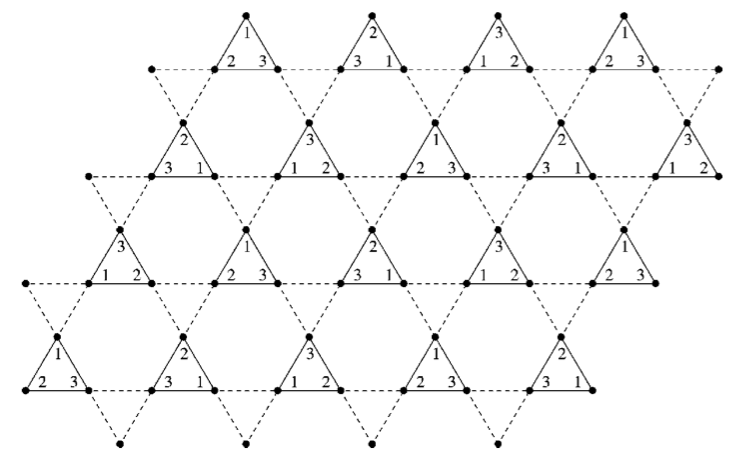}
\caption{The trimerized kagome lattice. The solid and dashed lines indicate the antiferromagnetic coupling $J$ and $J'$, respectively. The numbers 1, 2, and 3 indicate the site indexing inside the elementary triangles which defines the gauge. From \cite{Ferrero03}.}
\label{fig:Ferrero03_1}
\end{figure}

Compass models also appear in effective low energy description of quantum
magnets that have a chiral degree of freedom \cite{Budnik04,Capponi04,Ferrero03,Mila07}.
In these systems, the chirality plays the role of the pseudo-spin
with non-trivial directional dependence of the coupling. 
Similar to the pseudo-spin in orbital systems that enables us
to track the different degenerate orbital states (belonging,
e.g., to the different degenerate orbital sectors ($e_{g}
$ and $t_{2g}$ in transition metal ions), in frustrated 
magnets with a basic building block (e.g., triangle or other)
that leads to a multitude of ground states, the chirality tracks the 
extra degeneracy of ground states. In the quantum magnets
that we will detail below, there are within each building block
several degenerate ground states that are labeled by different
values of the chirality.  This degeneracy is lifted by interactions
between the different building blocks (e.g., interactions between
different triangular units in a kagome lattice) that rise to
effective interactions involving chiralities on different basic
units (triangles) which are precisely of the compass type. 
To date, two variants of the kagome lattice antiferromagnet
were investigated in their low energy sector. These are the
{\em trimerized kagome lattice antiferromagnet} \cite{Ferrero03} 
and the {\em uniform kagome antiferromagnet} \cite{Budnik04}. 
Both of these systems were investigated for a spin $S=1/2$ rendition
of the original antiferromagnet.   One way to describe the 
kagome lattice- which was made use of for both 
the trimerized and uniform systems- 
is, indeed, as {\em a triangular lattice of triangles}, see Fig.~\ref{fig:Ferrero03_1}.
%
%

The kagome lattice has a very low
coordination number. This along with the frustrated nature of
the antiferromagnetic interactions around individual triangular
loops lead the system to have a richness of low energy 
states and an extremely high degeneracy of classical
ground states. 

Below we will elaborate on
the effective low energy description and 
consequent origin of the compass type interactions
in both systems. 

\subsubsection{Non-uniform Trimerized Kagome Lattice Antiferromagnet}

In systems such as the spin $S=3/2$ antiferromagnet SrCr$_{8-x}$Ga$_{4+x}$O$_{19}$,
the existence of triangular layers between the kagome lattice planes generates two types
of effective bond strengths inside the kagome lattice plane. The resulting effective planar
system- the trimerized kagome lattice antiferromagnet- highlights the geometry of
the kagome lattice as a triangular lattice of triangles. Focusing on the 
upwards facing triangles, we see that they form a triangular lattice.  \cite{Ferrero03, Capponi04} 
considered a spin $S=1/2$ model in which the nearest neighbor couplings inside the triangles
($J$) were far larger than the nearest neighbor couplings between sites on different'
triangles ($J'$).  
In the limit $J'/J \ll 1$, the trimerized kagome lattice antiferromagnet
becomes a set of decoupled triangular units (with an antiferromagnetic
exchange constant of $J$ within each triangular unit).  The idea is then 
to employ perturbation theory in $J'/J$ about this limit of decoupled antiferromagnetic
triangular units.  

\begin{figure}
\centering
\includegraphics[width=.7\columnwidth]{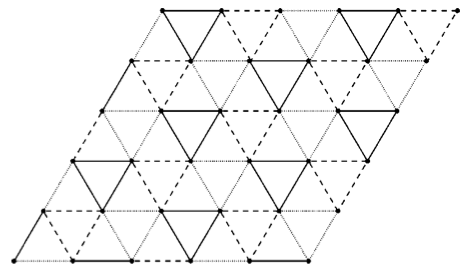}
\caption{Triangular lattice on which the effective Hamiltonian is defined. The unitary vector for the bond is indicated by solid lines (${\bf e}_\mu={\bf e}_1$), dashed lines  (${\bf e}_\mu={\bf e}_2$), and dotted lines  (${\bf e}_\mu={\bf e}_3$). From \cite{Ferrero03}.}
\label{fig:Ferrero03_2}
\end{figure}

Now, the problem of three spin $S=1/2$ on an antiferromagnetic ring
(i.e., a basic triangular unit of the kagome lattice) spans $2^{3} =8$ states.
In the total spin basis it can be decomposed into a Hilbert space sector 
that has a total  spin $S_{tot} =3/2$ (spanning
four states) as well as two sectors with total spin $S_{tot} =1/2$
(with each of these latter sectors, of course, spanning two states).  Formally, that is,
the direct product basis can decomposed in the total spin basis
as $1/2 \otimes 1/2 \otimes 1/2 = 3/2 \oplus 1/2 \oplus 1/2$.
In the antiferromagnetic problem, the tendency is to minimize 
the spin as much as possible. Indeed an immediate calculation
that we will perform now shows that at low energies we can confine our attention
to the four lower lying $S_{tot} =1/2$ ground states.  
Towards that end, we very explicitly note that for a three-site antiferromagnetic
problem on a triangle,
\begin{eqnarray}
J({\bf S}_{1} \cdot {\bf S}_{2} + {\bf S}_{1} \cdot {\bf S}_{3} + {\bf S}_{2} \cdot {\bf S}_{3})
=  \frac{J}{2} {\bf S}_{tot}^{2} - \frac{9J}{8}.
\end{eqnarray} 
with ${\bf S}_{tot} = {\bf S}_{1} + {\bf S}_{2} + {\bf S}_{3}$ and ${\bf S}_{tot}^{2} 
= S_{tot} (S_{tot}+1)$. In the ground state, we thus minimize the total spin $S_{tot}$.
For the three spins that we consider, the minimal value of $S_{tot}$ is 1/2.
Physically, these states in which the total spin is smaller than the maximal
one (i.e., $S_{tot}<3/2$) are superpositions of states in which two 
of the three spins combine to form a singlet.  
This is a particular
instance of a more general result that states that when the total
spin is smaller than the maximal possible in a plaquette, all plaquette states are superpositions
of states that contain (at least) one singlet connecting two sites \cite{Nussinov_Klein}.
The four ground states that are spanned by the two $S_{tot} =1/2$ sectors 
can be parameterized in terms of eigenvalue of a spin and a chirality
pseudo-spin each of  size $S=T=1/2$. 
These are defined via
\cite{Mila98, Capponi04}
\begin{eqnarray}
\sigma^{z}|\alpha R \rangle &=& \alpha |\alpha R \rangle,~ \sigma^{z}|\alpha L \rangle = \alpha |\alpha L\rangle\nonumber \\ 
\tau^{z}  |\alpha R  \rangle &=&   |\alpha R \rangle, ~ \tau^{z} |\alpha L \rangle = - |\alpha L \rangle.
\label{def_st_chiral} 
\end{eqnarray}
That is, $\alpha$ and R/L denote the eigenvalues of the two operators $S_{z}$ and $T_{z}$. 
Written in terms of the original degrees of freedom of the three spins on a triangular unit
($|\alpha_{1}, \alpha_{2}, \alpha_{3} \rangle$), with, e.g., $\alpha_1$ corresponding to
the ``top-most'' spin of the upward facing triangles, we have \cite{Mila98}
\begin{eqnarray}
\label{spin_chiral}
|\alpha R \rangle &=& \frac{1}{\sqrt{3}}(|-\alpha \alpha \alpha \rangle + \omega |\alpha -\alpha \alpha \rangle + \omega^{2} |\alpha \alpha -\alpha \rangle),\nonumber
\\  |\alpha L \rangle &=& \frac{1}{\sqrt{3}} (|-\alpha \alpha \alpha \rangle + \omega^{2} |\alpha -\alpha \alpha \rangle + \omega |\alpha \alpha -\alpha \rangle), \nonumber
\end{eqnarray}
with $\omega \equiv \exp(2 \pi i/3)$. When $J'=0$, the system exhibits an exponential in size 
ground state degeneracy. That is, the degeneracy is equal to $4^{N_{\triangle}}$ with $N_{\triangle}$
equal to the number of triangular units.  This degeneracy is lifted once $J'$ is no longer zero. 
For small $J'/J$, we can work in the ground state basis of the $J'=0$ problem 
and employ perturbation theory to write down an effective Hamiltonian in
that basis. The resulting
effective low energy Hamiltonian is of a compass type
(more precisely, of a form akin to the Kugel-Khomskii Hamiltonian
augmenting usual uniform spin exchange)
that is defined on a triangular lattice
in which each site represents a triangle of the original kagome
lattice. Unlike the definition of ${\bf e}_{\gamma}$ in the compass
models that we considered earlier, now ${\bf e}_{\gamma}$ does depend not
only on the orientation of the link connecting two sites. Rather, it differs from bond to
bond depending on its physical location on the lattice. A certain ``gauge''
for ${\bf e}_{\gamma}$ is to be chosen. Such a gauge is shown in Fig. \ref{fig:Ferrero03_2}.
%
%
%
Explicitly, the effective low energy Hamiltonian reads \cite{Ferrero03, Capponi04}
\begin{eqnarray}
H = \frac{J'}{9} \sum_{\langle i j \rangle} {\bm \sigma}_{i} \cdot {\bm \sigma}_{j}
(1- 4 {\bf e}_{ij} \cdot {\bm{\tau}}_{i})(1- 4 {\bf e}_{ij} \cdot {\bm {\tau}}_{j}).
\end{eqnarray}

\subsubsection{Uniform Kagome Antiferromagnet}
Several groups \cite{Budnik04, Capponi04}  employed the ``contractor renormalization method" (CORE)
to investigate kagome antiferromagnets. This method has been invoked to
find an effective low energy Hamiltonian for the uniform kagome antiferromagnet
wherein all exchange couplings are the same. In a spirit similar to that earlier,
the individual triangular units are examined and, to lowest order in CORE, an
effective low energy Hamiltonian is constructed that embodies interactions
between different triangular units. A notable difference with the earlier
approach is that perturbation theory was not invoked. Rather the system
is solved on larger size units and effective Hamiltonians involving
the more primitive basic units are constructed.  In \cite{Budnik04}, a related yet, 
by comparison to  \cite{Capponi04, Ferrero03}, 
different definition of the spin and chiral degrees of freedom is employed. 
Rather explicitly, with $s$ an $S_{z}$ eigenvalue of a spin operator ${\bf S}$
and $\Uparrow$ and $\Downarrow$ denoting states of eigenvalues $\pm 1/2$
of a pseudo-spin operator ${\bf T}$, \cite{Budnik04}
\begin{eqnarray}
|s, \Uparrow \rangle &=& \frac{|s \uparrow \downarrow \rangle - |s \downarrow \uparrow \rangle}{\sqrt{2}}, \nonumber \\ 
|s, \Downarrow \rangle &=& \frac{|s \uparrow \downarrow \rangle + |s \downarrow \uparrow \rangle}{\sqrt{6}} - \sqrt{\frac{2}{3}} |(-s) s s \rangle.
\end{eqnarray}
As in the perturbative
treatment, the resulting effective Hamiltonian \cite{Budnik04} contains effective interactions
similar to those of the Kugel-Khomskii model  augmenting standard 
spin exchange and pseudo-spin exchange. These are further 
augmented by direct compass type interactions
(i.e., pseudo-spin interactions uncoupled from spin) similar to 
those that arise from Jahn-Teller interactions in orbital systems
as well as non-trivial compass type coupled pseudo-spin spin interactions of the form 
\begin{eqnarray}
{\bf S}_{i} \cdot {\bf S}_{j} (J_{1} {\bf T}_{i} \cdot {\bf e}_{ij}
+ J_{2} {\bf T}_{j} \cdot {\bf e}_{ji})
\end{eqnarray} with $J_{1}$ and $J_{2}$ being fixed multiples
of the uniform exchange constant $J$ in the kagome lattice antiferromagnet.
The direct pseudo-spin interactions that couple the chiralities on
neighboring triangles favor the formation of aligning singlets parallel to
one another along particular directions.

\section{Symmetries of Compass Models} 
\label{sec:sym}

\subsection{Global, Topological, and Intermediate symmetries and invariances}
\label{global_inter}

In terms of symmetries, compass systems are particularly rich.
In what follows, we will discuss the invariances 
that these systems exhibit, but first recall the classification of orders and their relation to symmetry: 

({\bf i}) {\em Global symmetry}. In many
condensed matter systems (e.g. ferromagnets, liquids), there is an 
invariance of the basic interactions with 
respect to global symmetry operations 
(e.g. continuous rotations in the case of ferromagnets,
uniform translations and rotations in liquids) 
that are to be simultaneously performed on all of the 
constituents of the system. At sufficiently low temperatures
(or strong enough interactions), such symmetries might be {\it spontaneously}
broken. 

({\bf ii}){\em  Topological invariants and orders}. Topological orders have been the object of some fascination in more recent years \cite{Wen04}.
In the condensed matter community, part of the activity in analyzing 
these types of order is stimulated by the prospects of tolerant free 
quantum computation  --  an issue which we will return to in Sec.~\ref{sec:kit}. 
What lies at the crux of topological order is the observation
is that even if, in some cases, global symmetry breaking cannot 
occur, systems may nevertheless still exhibit a robust order 
of a non-local, topological, type.  

The most prominent examples
of topological order  --  long studied by high energy theorists  --  
are afforded by gauge theories \cite{Kogut,Wen04}. Gauge theories
display {\em local gauge symmetries} and indeed, in pure gauge theories -- theories 
that have only gauge bosons yet no matter sources -- the only
measurable quantities pertain to correlators defined 
on loops, the so-called {\it Wilson loops}.  Related products pertain to 
open contours in some cases when matter sources are present \cite{Kogut,Fradkin1979,Nussinov2005}.

({\bf ii}){\em  Intermediate symmetry}. The crucial point is that 
many compass systems display symmetries 
which, generally, lie midway between the above two extremes of global symmetries
and local gauge symmetries. 
To make this statement precise, one can rephrase it in a formal way
as it applies to general systems \cite{Batista05,holography}. Consider a theory with fields $\{\phi_{\bf i}\}$
that is characterized by a Hamiltonian $H$ (or action $\cal{S}$).

{\underline{Definition:}}  A {\it $d$-dimensional gauge-like symmetry} of a theory
is a group of symmetry transformations such that the minimal non-empty 
set of fields $\{\phi_{\bf i}\}$ changed by the group 
operations occupies a $d$-dimensional subset ($\cal{C}$) 
of the the full $D$-dimensional region on which the 
theory is defined. In the following we will refer to such symmetries as {\it $d$-dimensional symmetries}.

To exercise this notion it is useful to make contact with known cases. Clearly local gauge symmetries 
correspond to symmetries of dimension $d=0$. That is, gauge transformations
can be applied locally at any point in space -- a region of dimension $d=0$. 
At the opposite extreme, e.g.,  
in a nearest neighbor ferromagnet on a $D$-dimensional lattice, 
described by the Heisenberg Hamiltonian
$H = -J \sum_{\langle i j \rangle} {\bm{S}}_{i} \cdot {\bm{S}}_{j}$,
the system is invariant under a global rotation of
all spins. As the volume influenced by the 
symmetry operation occupies a $D$-dimensional region and in this case $d=D$. Sections~\ref{sst2gs}, \ref{sec:cubic120class}, \ref{120honey}, \ref{120triangle}, \ref{3compKK}
exemplify how symmetries of {\it intermediate} dimension $0<d<D$ arise in compass systems. 

In their simplest form, one which typically appears in compass models, $d-$ dimensional symmetries are of the form
\begin{eqnarray} 
\prod_{j \in P} g_{j}
\end{eqnarray}
where $g_{j}$ are group elements associated with a site $j$ and $P$  is a $d-$ dimensional spatial region. In many cases, depending on the boundary conditions of the system, 
$P$ correspond to entire open $d-$ dimensional planes (as in 90$^{\circ}$ compass models
that we will review in subsection \ref{sst2gs}; see, e.g., Fig. \ref{fig:Nussinov08}) or closed contours (when compass models are endowed with periodic 
boundary conditions). Defect creation operators (those that restore symmetries) and translations of defects are typically products of local group elements that do not span such an entire region $P$ but rather 
a fragment of it (see, e.g., the open finite string in Fig. \ref{fig:Nussinov08} with domain wall boundaries) generally leading to defects at the boundaries where the group element operations are applied \cite{PNAS}. 
We will pay particular attention to defect creation operators in the Kitaev compass model in Section \ref{sec:kit} when we will discuss anyons and how 
they can be moved. 

\subsection{Exact and Emergent Symmetries}
\label{sec:exact_emergent}

A Hamiltonian $H$, and by extension the system it describes, can have two principal kinds of symmetries: exact and emergent ones. These are defined as follows.

{\bf{(i)}}  {\it Exact symmetries}.
By this, one refers to the existence operators ${\hat{O}}$ that commute with the Hamiltonian
\begin{eqnarray}
\label{symm_defn}
[H,{\hat{O}}]=0. 
\end{eqnarray}
Such operators, indicted in this review by a hat $\hat{}$, reflect symmetries of the Hamiltonian. 

{\bf{(ii)}} {\it Emergent symmetries}.
In many compass (and numerous other) systems, there are operators $\tilde{O}$ that do not commute 
with the Hamiltonian,
\begin{eqnarray}
[H, \tilde{O}] \neq 0
\end{eqnarray}
i.e., do not satisfy Eq. (\ref{symm_defn}) and are therefore indicated throughout this review by a tilde $\tilde{}$. 
Yet these operators do become symmetries when projected to a particular sector -- a particular subset of states on which the Hamiltonian acts. That is, 
\begin{eqnarray}
[H,{\cal{P}} \tilde{O} {\cal{P}}]=0,
\end{eqnarray}
where ${\cal{P}}$ is the relevant projection operator that sector. 
In this case, if one defines $ {\cal{P}} \tilde{O} {\cal{P}}= {\hat{O}}$ then ${\hat{O}}$ will be an exact symmetry satisfying Eq. (\ref{symm_defn}).

The most prominent cases in condensed matter systems, compass models in particular, relate to symmetries that appear
in the {\it ground state sector} alone. In such instances, the symmetries
are sometimes said to {\it emerge} in the low energy sector of the theory. 

Although the formulation above is for quantum Hamiltonians, the same can, of course, be said for classical systems. There are numerous classical systems in which the application of a particular operation on an initial configuration will yield, in general, a new configuration with a differing energy. However, when such an operation is performed on a particular subset of configurations, such as the classical ground states, it will lead to other configurations that have precisely the same energy as the initial state. Similarly, certain quantum systems exhibit such
particular symmetries only in their large pseudo-spin (or classical) limit. In such cases,  symmetries may be said to emerge in the large pseudo-spin (or classical) limit.  As will be reviewed in sections~\ref{sec:intermediate90cube},
\ref{sec:cubic120class},
\ref{120honey},
 particularly in certain compass-type models, symmetries may emerge within a sector of the combined large pseudo-spin and/or low energy (or temperature) limit.

One should note that emergent low-energy symmetries are notably different from the far more standard situation of spontaneous symmetry breaking, wherein an invariance of the Hamiltonian (or action) is spontaneously broken in individual low energy states (which are related to one other by the symmetry operation at hand). In the condensed matter arena, the canonical example is rotationally symmetric ferromagnets in a spatial dimension larger than three, in which at sufficiently low temperature a finite magnetization points along a certain direction -- thus breaking the rotational symmetry. Another canonical example is the discrete ({\it up} $\leftrightarrow$ {\it down} or) time reversal symmetry is broken in Ising ferromagnets. Spontaneous symmetry breaking appears in systems that exhibit long-range order of some sort such as crystallization (breaking translational and rotational symmetries), superconductors (local gauge invariance and a Anderson-Higgs mechanism), superfluid Helium. Other examples include the Higgs mechanism of particle physics, chiral symmetry breaking in quantum chromodynamics, nucleon pairing in nuclei, electro-weak symmetry breaking at low energies, and related mass generation. 

In all of these textbook examples, the system is symmetric at high energies and exhibits low-energy states that do not have that symmetry. However, in low energy emergent symmetries, the situation is reversed: the system may become {\it more} symmetric in the low-energy sector. 
We will discuss explicit examples of exact and emergent symmetries in compass models in the following sections.

\subsection{Consequences of Intermediate Symmetry}
\label{inter_expt}

In this subsection, we review the consequences of intermediate symmetries. In later subsections, we will see how these the intermediate symmetries the below features 
appear in various compass models. Aside from the earlier results
reviewed below, in Section \ref{new_theorem_on_bands}, we will further report on a new consequence concerning the link between these symmetries and ``flat bands'' and illustrate 
how this relation appears throughout the compass models investigated. 

\subsubsection{Degeneracy of Spectrum}

We now briefly discuss how the presence of a $d$-dimensional intermediate symmetry, either classical or quantum, implies an exponential degeneracy of the energy spectrum that corresponds to the Hamiltonian. The application of intermediate symmetries on disparate $d$-dimensional planes leads to inequivalent states that all share the same energy. If a symmetry transformation $\tilde{O}_{P}$ has its support on a $d$-dimensional plane $P$, then one can concoct the composite symmetry operators
\begin{eqnarray}
\label{composite}
\tilde{O}_{composite} = \tilde{O}_{P_{1}} \tilde{O}_{P_{2}} .. \tilde{O}_{P_{R}}.
\end{eqnarray}
For a hypercubic lattice in $D$ dimensions which is of size $L \times L \times L ... \times L$, the number
of independent planes ($R$) in Eq. (\ref{composite}) scales as $R = {\cal{O}}(L^{d'})$ 
where
\begin{eqnarray}
d' = D-d.
\label{d'Dd}
\end{eqnarray}
If each individual $d$-dimensional
symmetry operation (exact or emergent) $U_{P_{i}}$ leads to a degeneracy 
factor of $m$ then the composite operation of Eq. (\ref{composite})
can lead to a degeneracy (of any state (for exact symmetries) or of the ground state
(for emergent symmetries)) whose logarithm is of magnitude 
\begin{eqnarray}
\label{domd}
\log_{m} {\sf degeneracy}= {\cal{O}}(L^{D-d}).
\end{eqnarray} 
That this is indeed the case is clearer for classical system with discrete symmetries 
than for quantum systems. Nevertheless, 
in the thermodynamic limit and/or on lattices whose boundaries are tilted 
the degeneracy factor of Eq. (\ref{domd}) associated with the 
intermediate $d$-dimensional symmetries becomes exact \cite{Nussinov2012a}. 
On hypercubic lattices, such as the square lattice of the planar 90$^{\circ}$ compass
model discussed in subsection \ref{sst2gs}, whose boundaries are the same along the $d'$ directions
orthogonal to the planes $P$, the application of the operators of 
Eq. (\ref{composite}) does not lead to independent states for
finite size systems. However, in the thermodynamic limit, 
the application of disparate operators of the form of Eq. (\ref{composite})
on a given initial state may lead to orthogonal states. 

\subsubsection{Dimensional Reduction}
\label{mishpat}

The existence of intermediate symmetries has important consequences: it implies a dimensional reduction. The corresponding dimensional reduction is only with respect to expectation values of local quantities: the free energies of these systems and the transitions that they exhibit are
generally those of systems in high dimensions \cite{Batista05,holography}.

\paragraph{Theorem on Dimensional Reduction}
More precisely, the expectation value of any such quantity $\langle f \rangle$
in the original system (of dimension $D$) is bounded from above by the
expectation value of the same quantity evaluated on a $d$ dimensional
region: 
\begin{eqnarray}
|\langle f \rangle| \le |\langle f \rangle|_{H_{d}}.
\label{boundfi}
\end{eqnarray} 
The expectation value $\langle f \rangle$ refers to that
done in the original system (or lattice) that resides in $D$ spatial
dimensions. The Hamiltonian $H_{d}$ on the righthand side is
defined on a $d$ dimensional subregion of the full lattice (system). 
The dimensionality $d \le D$. The Hamiltonian $H_{d}$ preserves the range of the interactions
of the original systems. It is formed by pulling out of the full Hamiltonian on the 
complete ($D$ dimensional) lattice, the parts of the Hamiltonian 
that appear within the $d$ dimensional sub-region ($\cal{C}$) on which 
the symmetry operates. Fields (spins) external to $\cal{C}$ act as
non-symmetry breaking external fields in $H_{d}$. The bound of Eq.~(\ref{boundfi}) becomes
most powerful for quantities that are not symmetry invariant as 
then the expectation values $\langle f \rangle_{H_{d}}$ need to vanish for low 
spatial dimensions $d$ (as no spontaneous symmetry breaking
can occur). This, together with Eq.~(\ref{boundfi}),
then implies that the expectation value of $\langle f \rangle$
on the full $D$ dimensional spatial lattice must vanish.
By ``non invariant", we mean that $f(\phi_{\bf i})$ vanishes
when summed over all arguments related to each other
a $d$ dimensional symmetry operation,  
$\sum_k f[{\bf g}_{{\bf i}k}(\phi_{\bf i})] = 0$. 
For continuous symmetries, non-invariance explicitly translates into
an integral over the group elements
$\int f[{\bf g}_{\bf i}(\phi_{\bf i})] d{\bf g} = 0$.

We will now summarize for completeness general
corollaries of such symmetry based analysis
for general systems. 

\paragraph{Corollaries}
By choosing $f$ to be the order parameter or
a two-particle correlator, one arrives at the following general 
corollaries \cite{Batista05,Nussinov06,holography}:

{\it Corollary I:\/}
Any local quantity that is not invariant under local symmetries ($d=0$)
or symmetries that act on one dimensional regions ($d=1$) 
has a vanishing expectation value  $\langle f \rangle_{H_{d}}$
any finite temperature. This follows as both zero and one dimensional
systems cannot exhibit symmetry breaking: in one and two
dimensional systems, the expectation value of any 
local quantities not invariant under global symmetries:
$\langle f \rangle =0$. 

\begin{figure}
\centering
\includegraphics[angle=0,width=\columnwidth]{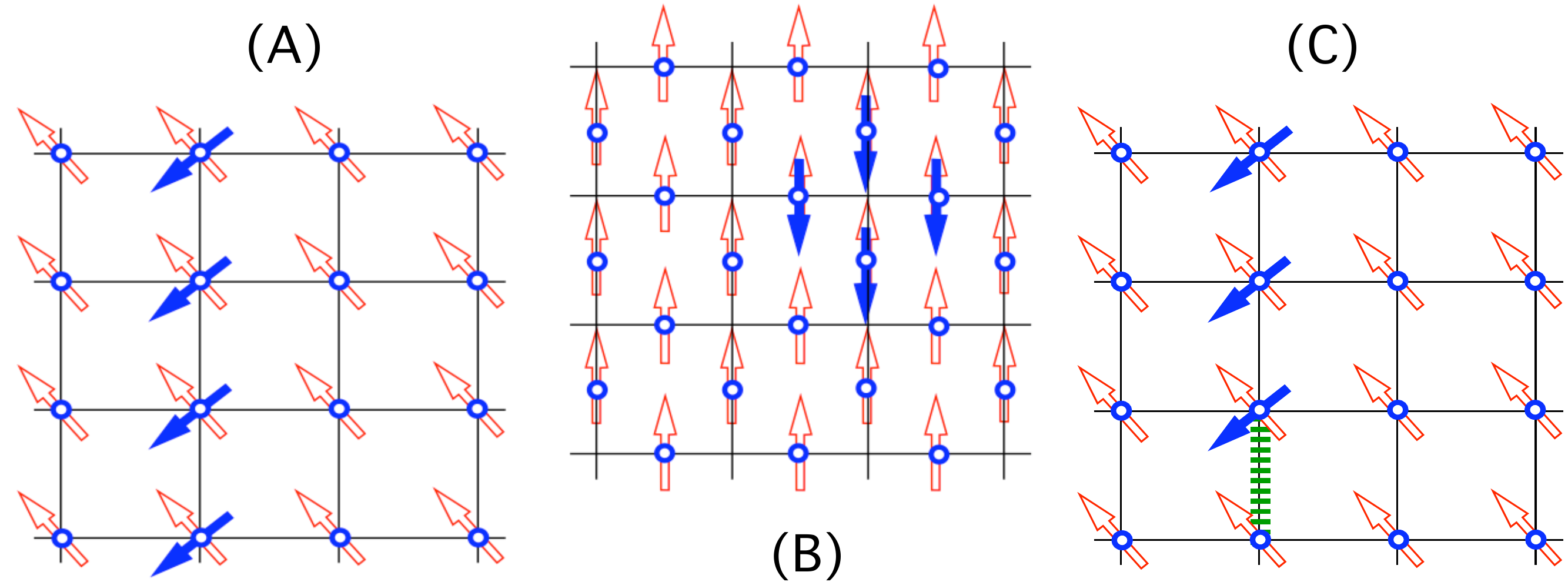}
\caption{(A) The $90^{\circ}$ square lattice compass model. The action of the $d=1$ symmetry operation of Eq.~(\ref{symorb}) when the "plane" $P$ is chosen to lie along the vertical axis. (B) A $d=0$ (local) gauge symmetry. Defects within a gauge theory cost a finite amount of energy. Local symmetries such as the one depicted above for an Ising lattice gauge theory cannot be broken. (C) A defect in a semi-classical ground state of the two dimensional orbital compass model. Defects such as this do not allow for a finite on-site magnetization. The energy penalty for this defect is finite (there is only one bad bond- the dashed line) whereas, precisely as in $d=1$ Ising systems, the entropy associated with such defects is monotonically increasing in system size \cite{PNAS}. Reproduced with permission.}
\label{fig:Nussinov08}
\end{figure}

Physically, entropy overwhelms energetic penalties
and forbids a symmetry breaking. Just 
as in zero and one dimensional systems, much
more entropy is gained by introducing defects
(e.g., domain walls in discrete systems),
the same energy-entropy calculus is replicated 
when these symmetries are
embedded in higher dimensions. 
An example with $d=1$ domain walls
in a two-dimensional systems is afforded
by the planar 90$^{\circ}$ compass model
[see Fig. \ref{fig:Nussinov08}]; even though the planar 
compass model is two-dimensional, the energy
cost of these domain walls is identical
to that in a $d=1$ system. 
The particular case of local ($d=0$) 
symmetry is that of Elitzur's 
theorem \cite{Elitzur75} so well 
known in gauge theories. We may
see it more generally
as a consequence of dimensional 
reduction.

A discussion of how, by virtue of this consequence, such symmetries may protect and lead to 
topological quantum orders in systems at both finite and zero temperature appears in \cite{AOP,PNAS}. 

{\it Corollary II:\/}
One can push the consequences further by recalling that
no symmetry breaking occurs for continuous
symmetries in two spatial dimensions.
Here again, free energy penalties are not
sufficiently strong to induce order.
When embedding continuous two dimensional symmetries
in higher dimensions, the energy entropy 
balance is the same and the same result
is attained $\langle f \rangle =0$ at all finite 
temperatures for any quantity $f$ that is not 
invariant under continuous $d \le 2$ symmetries. 

Further noting that order does not exist in continuous two dimensional systems
also at zero temperature in the presence of a gap between
ground and the next excited state, one similarly finds that for 
a $d  \le  2$ dimensional continuous symmetry
the expectation value of 
any local quantity not invariant under this symmetry, 
strictly vanishes at zero temperature. Though local
order cannot appear, multi-particle (including topological) 
order can exist. In standard gauge ($d=0$) theories,
the product of gauge degrees of freedom along a closed
loop (the Wilson loop) can attain a non-zero value as
it may be invariant under all $d=0$ symmetries. 
In more general theories with higher $d$ dimensional symmetries,
similar considerations may lead to loop (or ``brane'') type correlators
that involve multiple fields and are invariant under all low dimensional
symmetries. As we will discuss in Section \ref{sec:kit}, precisely such non-local correlation functions appear in Kitaev's honeycomb model
and many other systems with topological orders \cite{Chen08,PNAS,AOP,Pererz-Garcia2008}.

In section \ref{sec:diso}, we will review how 
when it is indeed allowed by symmetry,
symmetry breaking in the highly degenerate compass
models often transpires by a fluctuation driven 
mechanism (``order by disorder'') \cite{Villain1972,Shender82,Henley89}. In
this mechanism, entropic contributions to the free energy play a key role.

{\it Corollary III:} 
Not only can one make statements about 
the absence of symmetry breaking,
we can also adduce fractionalization
of non-symmetry invariant quantities in 
high dimensional system. That occurs if no 
(quasi-particle type) resonant terms appear 
in the lower dimensional spectral functions~\cite{Nussinov06}.

This corollary  allows for fractionalization in quantum systems, where $d= 1, 2$. It enables symmetry invariant quasi-particles excitations to {\em coexist} with non-symmetry invariant fractionalized excitations. Fractionalized excitations  may propagate in $(D- d)$ dimensional regions.  Examples afforded by several frustrated spin models where spinons may drift along lines on the square lattice \cite{Batista04} and in $D$ dimensional regions on the pyrochlore lattice \cite{Nussinov07}.


In what follows, we explicitly enumerate the symmetries that appear in various compass 
models. 
The {\em physical origin of dimensional reduction} in these systems can be seen examining intermediate symmetry restoring defects. 

\subsection{Symmetries of the 90$^{\circ}$ Compass Model} 
\label{sst2gs}

We now classify symmetries of the 90$^{\circ}$ compass model in various spatial dimensions, reviewing both quantum and classical versions. To highlight some aspects of the symmetries of this system, it is profitable to discuss the general anisotropic compass model, as given for $D=2$ in Eq. (\ref{2dpocm}) with general couplings $J_{x}$ and $J_{y}$ and in general spatial dimension $D$ given by Eq.(\ref{eq:general_compass}), without field:
\begin{eqnarray} 
H^{90^{\circ}}_{D\square} =  -\sum_{i,\gamma}  J_\gamma \tau^\gamma_i \tau^\gamma_{i+\bm{e}_\gamma}.
\label{eq:an_general_compass1/2}
\end{eqnarray} 
The equivalent classical Hamiltonian on a $D$-dimensional hyper cubic lattice is
\begin{eqnarray} 
H^{90^{\circ} {\rm , class}}_{D\square} =  -\sum_{i,\gamma}  J_\gamma T^\gamma_i T^\gamma_{i+\bm{e}_\gamma}.
\label{eq:an_general_compassT}
\end{eqnarray} 
In the quantum systems, $T^{\gamma}$ are generators of the representations of SU(2) of size $(2T+1)$.
For a pseudo-spin 1/2 system, $T^{\gamma} = \tau^{\gamma}/2$. 
In the classical arena, $T^{\gamma}$ are are the Cartesian 
components of normalized vector ${\bm{T}}$, as discussed in subsection \ref{sec:clas}. These classical and quantum Hamiltonian systems exhibit both exact and emergent symmetries.

\subsubsection{Exact discrete intermediate symmetries}
\label{exact90cds}
Exact symmetries of both the square lattice and cubic lattice 90$^{\circ}$ compass model in any pseudo-spin representation are given
by \cite{Batista05,Nussinov04,Biskup05,Ioffe05,Nussinov05,Dorier05}
\begin{eqnarray}
\hat{O}^{(\gamma)} = \prod_{i\in P_{\gamma}} e^{i \pi T^{\gamma}_i}\,
\label{symorb}
\end{eqnarray}
where,  as in Eq. (\ref{symorb-e}),  $P_{\gamma}$ is any line (in the case of the two-dimensional model) or plane (in the case of the cubic lattice model) which is orthogonal to the external ${\bf{e}_{\gamma}}$ axis of the lattice. 
A schematic for the $D=2$ dimensional case is provided in panel (a) of Fig. \ref{fig:Nussinov08}.

Albeit appearances, Eq. (\ref{symorb}) is, when written longhand, quite different from Eq. (\ref{symorb-e}). In Eq. (\ref{symorb-e}) describing the emergent symmetries of the 120$^{\circ}$ model, ${\bm{T}}$ is a two-component vector that is projected along three different equidistant non-orthogonal planar directions. That is, in Eq. (\ref{symorb-e}), the unit vectors ${\bf{e}}_{\gamma}$ in the argument of the exponential
correspond, with $\gamma =1,2,$ and $3$ to the equidistant non-orthogonal internal pseudo-spin directions  ${\bm{a}}, {\bm{b}},$ and ${\bm{c}}$ that lie in the two-dimensional plane defined of the 120$^{\circ}$ model.
By contrast, in Eq. (\ref{symorb}), ${\bm{T}}$ is a $D=2$ (square lattice model) or $D=3$ (cubic lattice) vector and $T^{\gamma}$ are projections along orthogonal directions. The two operators appearing in Eqs. (\ref{symorb-e}, \ref{symorb}) differ from one another: ${\bm{T}} \cdot {\bm{b}} \neq T_{2}$, etc. In Fig. (\ref{fig:Nussinov08}), we provide a classical schematic of the action of such an operator when it acts on a uniform state. As in the case of the 120$^{\circ}$ model on the cubic lattice, these operators lead to stratified states. 

The exact nature of the symmetries of Eq. (\ref{symorb}) is readily seen: the operators of Eq. (\ref{symorb})
commute with the general Hamiltonian of Eq. (\ref{eq:an_general_compassT}): $[O^{(\gamma)},H]=0$. 
Thus, rotations of individual planes about an orthogonal axis 
leave the system invariant.  Written generally, for a 90$^{\circ}$ compass model in $D$ dimensions, the planes $P_{\gamma}$ are
objects of spatial dimensionality $d=(D-1)$. In the $D=3$ dimensional system, the symmetries
of Eq.(\ref{symorb}) are of dimension $d=2$ as the planes $P_{\gamma}$ are
two-dimensional objects. On the square lattice, the symmetries are of dimension $d=1$ as $P_{\gamma}$ are lines. 
These symmetries hold for both the quantum system with 
arbitrary size pseudo-spin as well as the classical
system of Eq. (\ref{highD90}) in a high number of
dimensions $D$. A consequence of these symmetries
is an exponential in $L^{D-1}$ degeneracy of each eigenstate of the Hamiltonian (including
but not limited to ground states) is, in systems 
with ``tilted'' boundary conditions that emulate the
thermodynamic limit \cite{Nussinov2012a}.
In pseudo-spin one half realizations of this
system [Eq.(\ref{eq:an_general_compass1/2})],
on an $L \times L$ square lattice, a $2^{L}$ degeneracy was
numerically adduced for anisotropic systems ($J_{x} \neq J_{y}$) in the thermodynamic limit \cite{Dorier05}.

Now, here is an important point to which we wish to reiterate- that of the {\it physical origin of the dimensional reduction
in this system}. In a $D=2$ dimensional 90$^{\circ}$ compass model system, the energy cost 
for creating defects (domain walls) is identical to that in a $d=1$ dimensional system [see Fig. \ref{fig:Nussinov08}]. 
With the aid of the bound of Eq.~(\ref{boundfi}), 
we then see the finite temperature expectation value $ \langle \sigma^{z}_{i} \rangle =0$
within the $D=2$ orbital compass model. The physical engine behind the 
loss of on-site order of $\langle \sigma^{z}_{i} \rangle$ 
is the proliferation of solitons, see Fig.~\ref{fig:Nussinov08}.
Just as in $d=1$ dimensional systems, domain walls (solitons)
cost only a finite amount of energy while their entropy
increases with system size. A schematic is provided in panel (c) of Fig. \ref{fig:Nussinov08}.
The Hamiltonian $H_{d=1}$ defined on the 
vertical chain of Fig.~\ref{fig:Nussinov08} where these operations appear
is none other than a one dimensional Ising Hamiltonian augmented by transverse
fields generated by spins outside the vertical
chain. Any fixed values of the spins outside the $d=1$ dimensional chain lead
to transverse fields that act on the chain. These along the Ising exchange interactions
between neighboring spins along the chain lead in this case to the pertinent $H_{d=1}$ in Eq.~(\ref{boundfi}):
that of a transverse field Ising model Hamiltonian. By virtue of their location outside the region
where the symmetry of Eq.~(\ref{symorb})
operates, the spins $\sigma^{x}_{i \not \in P_{x}}$ 
do not break the discrete $d=1$ symmetry associated
with the plane $P_{x}$. These defects do not enable a
finite temperature symmetry breaking. 
 
\subsubsection{Exact discrete global symmetries}

When the couplings are not completely anisotropic (e.g., $J_{x} = J_{y} \neq J_{z}$ or $J_{x} = J_{y} = J_{z}$ on the cubic lattice or $J_{x} = J_{y}$ on the square lattice) there are additional discrete symmetries augmenting the $d=D-1$ Ising symmetries detailed above. For instance, when $J_{x} = J_{y} \neq J_{z}$ a global discrete rotation of all pseudo-spins on the lattice by an angle of 90$^{\circ}$ about the $T^{z}$ direction leaves the Hamiltonian of Eq. (\ref{eq:an_general_compassT}) invariant. Such a discrete rotation essentially permutes the $x$ and $y$ oriented bonds which are all of equal weight in the isotropic case when these are summed over the entire square lattice. The same, of course, also applies for the square lattice model when $J_{x} = J_{y}$.

Yet another possible representation of essentially the same symmetry as it is pertinent to the exchange of couplings in the compass model is that of a uniform global rotation by 180$^{\circ}$ about the $(1/\sqrt{2},1/\sqrt{2})$ direction of the pseudo-spins. Such a representation will return in Eq. (\ref{OR}) later on. Similarly, when $J_{x} = J_{y} = J_{z}$, a uniform global rotation by 120$^{\circ}$ of all pseudo-spins about the internal $(1/\sqrt{3}, 1/\sqrt{3},1/\sqrt{3})$ pseudo-spin direction is also a discrete symmetry; this latter symmetry is of the $Z_{3}$ type- if performed three times in a row, this will give back the identity operation. 

These additional discrete symmetries endow the system with a higher degeneracy.  For isotropic systems ($J_{x}=J_{y}$), numerically a $2^{L+1}$ fold degeneracy is seen in the pseudo-spin $T=1/2$ system \cite{Dorier05}; this additional doubling of the degeneracy is related to a global Ising operation of a rotation by 180$^{\circ}$ about a chosen pseudo-spin direction that leaves the system invariant.  These additional symmetries are global symmetries and thus of a dimension $d=D$ which is higher than that of the discrete lower dimensional that are present in both the anisotropic and isotropic systems ($d=(D-1)$). As a result, in, e.g., the isotropic $D=2$ dimensional 90$^{\circ}$ compass model may exhibit a finite temperature breaking of such a discrete global symmetry associated with such a discrete rotation. By contrast, the $d=1$ symmetries of the two-dimensional 90$^{\circ}$ compass model cannot be broken as will discussed in section \ref{mishpat}. 
 
We note that in the classical anisotropic rendition of this system the degeneracy is exactly the same- i.e., $2^{L}$, aside from continuous emergent symmetries that will be discussed in the next section. The classical isotropic case is somewhat richer. There, each uniform pseudo-spin state (each such state is a ground state as will be elaborated in section (\ref{new_theorem_on_bands}) and there is an additional degeneracy factor of $2^{2L}$ associated with the $2L$ independent classical $d=1$ Ising symmetries. 

\subsubsection{Emergent Intermediate Discrete Symmetries: Cubic 90$^{\circ}$ Model}
\label{sec:intermediate90cube}

We now turn to intermediate symmetries that appear in the large pseudo-spin (or classical) limit of the  90$^{\circ}$ compass model in three dimensions.
In its classical limit, the classical 90$^{\circ}$ compass model on the cubic
lattice has $d=1$ inversion (or reflection)
 symmetries along lines parallel to each of the
three Cartesian axes $x_{a}$. Along these
lines, we may set $\tau^{a}_{i} \to -\tau^{a}_{i}$
and not touch the other components. This corresponds to,
e.g, a reflection in the internal $xy$ pseudo-spin plane
when we invert $\tau^{z}$ and not alter the $x$ or $y$
components. 

We explicitly note that this transformation is not canonical and does not satisfy
the commutation relation and is thus disallowed
quantum mechanically; indeed, this appears only 
as an emergent symmetry in the classical limit of large pseudo-spin. 
Instead in the 90$^{\circ}$ compass model on the cubic lattice, quantum mechanically
we have the $d=2$ symmetries which we wrote earlier
(which of course trivially also hold for the classical system).
Thus, the quantum system is less symmetric than
its classical counterpart. 

By contrast to the cubic lattice case, for the square lattice 90$^{\circ}$ compass model, the intermediate $d=1$ symmetries of Eqs. \ref{symorb} are
are exact quantum (as well as classical) symmetries.

\subsubsection{Emergent Continuous Global Symmetries}
 
 In addition to its exact symmetries, the 90$^{\circ}$ model also exhibits emergent symmetries
 in its isotropic version.  As mentioned earlier, globally uniform pseudo-vector configurations are ground
 states of any classical isotropic ferromagnetic compass model. Thus, similar to the considerations
 presented for the 120$^{\circ}$ compass model, any global rotation of all pseudo-spins
 is an emergent symmetry of the 90$^{\circ}$ models. In the $D=2$ system, this corresponds
 to a global $U(1)$ rotation of all angles of the planar pseudo-spins. In the $D=3$
 cubic lattice system, any SO(3) rotation of the three-dimensional pseudo-spins  
 is an emergent symmetry. That a rotation does not change the energy of 
 any uniform configuration is clear in the 90$^{\circ}$ model.
 Imagine that all pseudo-spins in the planar 90$^{\circ}$ model
 are oriented at an angle $\theta$ relative to the $T^{x}$ axis.
 In such a case, the energy associated with the horizontal bonds, $T^{x}_{i} T^{x}_{i+{\bm{e}}_{x}}$
 will vary as $\cos^{2} \theta$ whereas that associated with the 
 vertical bonds varies as $\sin^{2} \theta$.  As $J_{x} = J_{y} =J$ in the isotropic system
 and as $\sin^{2}\theta+ \cos^{2} \theta =1$, any uniform pseudo-spin state will have
 the same energy (which is, in fact, the ground state energy as 
 we be discussed in section \ref{unifcm}) and global rotations will not alter this energy.

\subsection{Emergent Symmetries: Classical Cubic 120$^{\circ}$ Compass Model} 
\label{sec:cubic120class}

The 120$^{\circ}$ compass model on a 3D cubic lattice, Eqs.(\ref{120p}, \ref{eq:pi120}), exhibits non-trivial symmetries which emerge in the {\it ground state} sector in the large pseudo-spin $T$ (classical) limit \cite{Lieb73,Simon80} (see also section \ref{sec:clas}). In the classification of section \ref{sec:exact_emergent}, all of the symmetries which we detail below correspond to emergent symmetries.
Before explicitly describing these symmetries, we briefly recount how to define this classical system from the quantum one, which we briefly alluded to in subsection \ref{sec:JT}. 

The classical 120$^{\circ}$ compass model
may, following the discussion in subsection \ref{sec:clas},
be specified as follows.
At each site we assign a unit length two-component 
spin denoted by~${\bm{T}}$. Let $\bf{a}$, $\bf{b}$ and~$\bf{c}$ be evenly-spaced 
vectors on the unit circle that are separated from one another by~120 
degrees. To conform with the operators of Eq. (\ref{120p}),
one sets~$\bf{c}$ to point at~$0^\circ$ 
and~$\bf{a}$ and~$\bf{b}$ to be at~$\pm 120^\circ$, respectively. 
Next, one defines $T^{(c)}={\bm{T}} \cdot {\bm{c}}$, 
and similarly for $T^{(a,b)}$.
These projections onto the above unit vectors, $T^{(a,b,c)}$,
 are the classical counterpart of the pseudo-spin 1/2 operators of Eq. (\ref{120p}).
The classical 120$^{\circ}$ compass model Hamiltonian is then
given by
\begin{equation}
\label{H120}
H_{3\square}^{120 {\rm , class}}= -\sum_{i}\left(
 T_{i}^{(a)}T_{i+{\bf e}_x}^{(a)}
+T_{i}^{(b)}T_{i+{\bf e}_y}^{(b)}
+T_{i}^{(c)}T_{i+{\bf e}_z}^{(c)}\right),
\end{equation}
where the interaction strength $J$ is set to unity. The ferromagnetic and antiferromagnetic model are related by symmetry, so that for convenience the interaction strength is chosen as negative, so that low-temeprature ordering patterns of pseudospins tend to be uniform. This model exhibits two types of emergent symmetries in its ground state sector.

\subsubsection{Emergent Continuous Global Symmetries}

All uniform pseudo-spin configurations, i.e., ones with constant pseudo-spin ${\bm T}_{i} = {\bm T}$ or uniform angular orientation of the classical two component pseudo-spins in the XY plane, are ground states of  $H_{3\square}^{120  {\rm , class}}$ in Eq.(\ref{H120}) \cite{Nussinov04}. Therefore {\em any} configuration for which 
\begin{eqnarray}
T^\gamma_{i} = T^\gamma_{i+\bm{e}_\gamma}
\label{eq:TTGS}
\end{eqnarray}
on all sites $i$ is also a ground state configuration. Thus, when the system is restricted to this subspace of uniform configurations, any uniform rotation of all of the pseudo-spin angles $\theta_{i} \to \theta_{i} + \delta \theta$ does not change the energy. This global rotation operation -- formally a $U(1)$ symmetry -- emerges as a symmetry when the system is restricted to these ground states. It can be readily verified that this emergent symmetry is not an exact symmetry of the system. When a global rotation is applied to any initial pseudo-spin configuration that is {\it not} uniform, it will generally lead to a new state that has an energy from that of the initial configuration. 

Formally therefore the classical cubic lattice 120$^{\circ}$ compass model exhibits a {\it global (i.e., a dimension $d=D=3$) emergent U(1) symmetry} within the ground state sector. It turns out that on top of this there are additional non-uniform {\it stratified} classical ground states
for which this global rotation is not a symmetry, which will be discussed next.

\subsubsection{Emergent Discrete $d=2$ Symmetries}

\begin{figure}
\centering
\includegraphics[width=2.7cm]{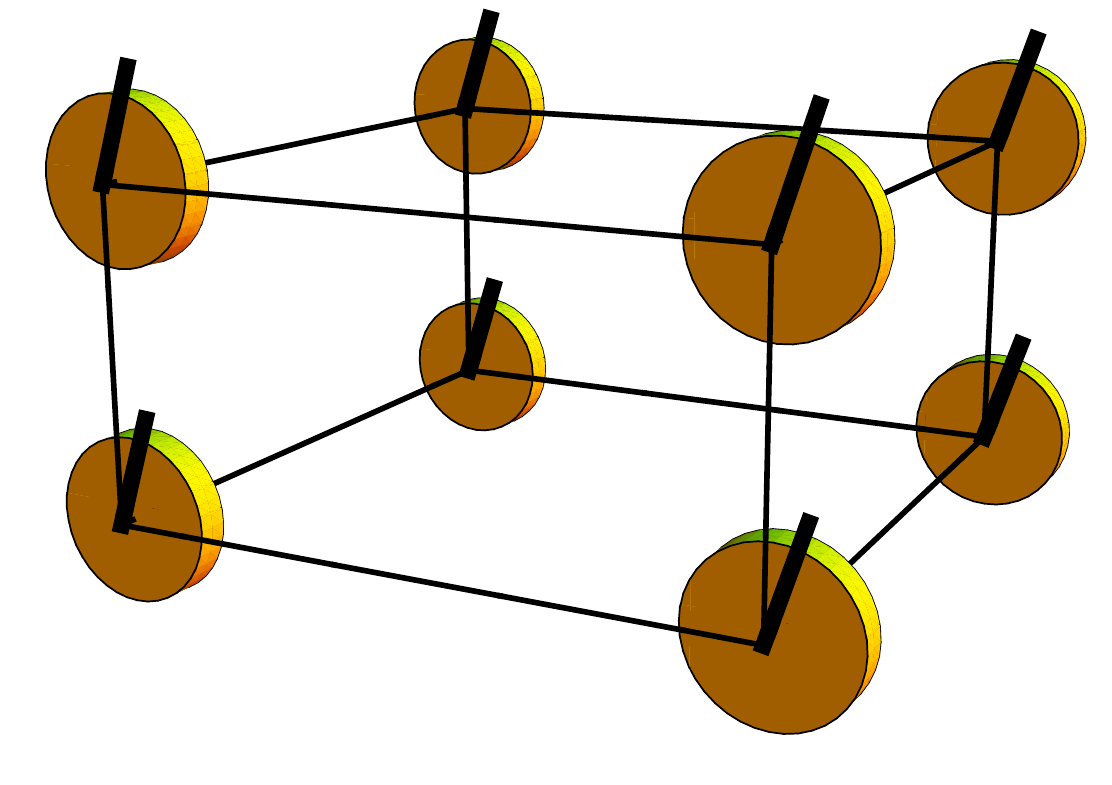}
\includegraphics[width=2.7cm]{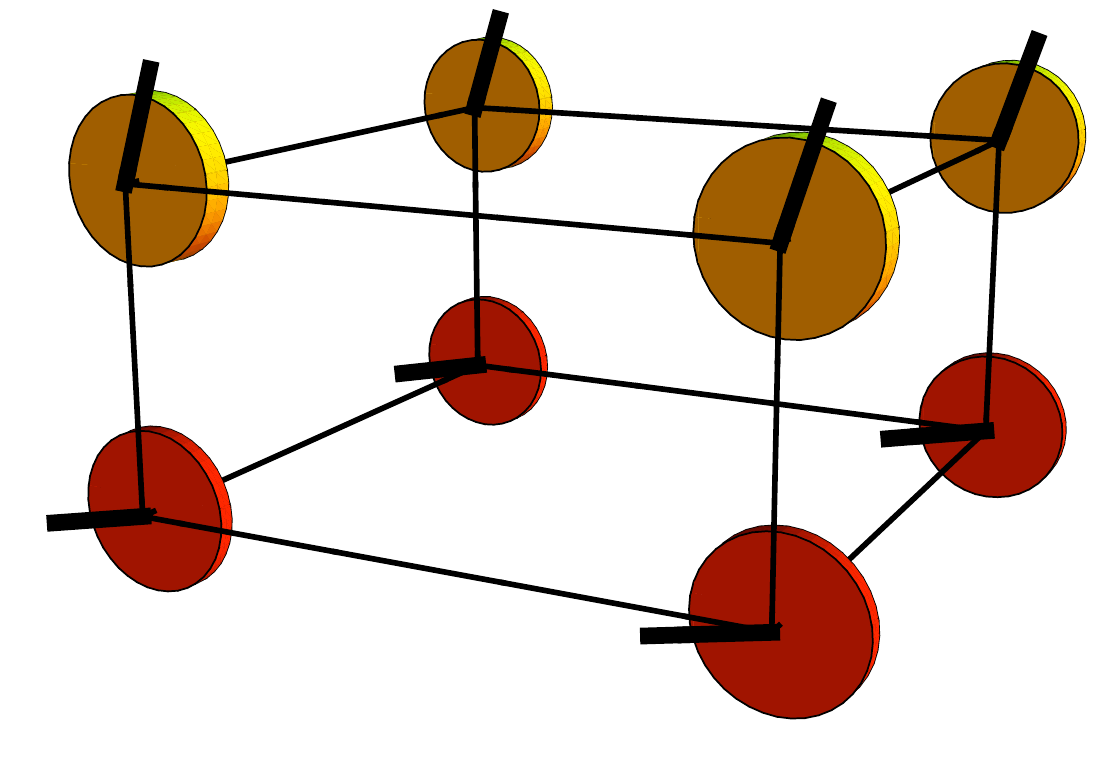}
\includegraphics[width=2.7cm]{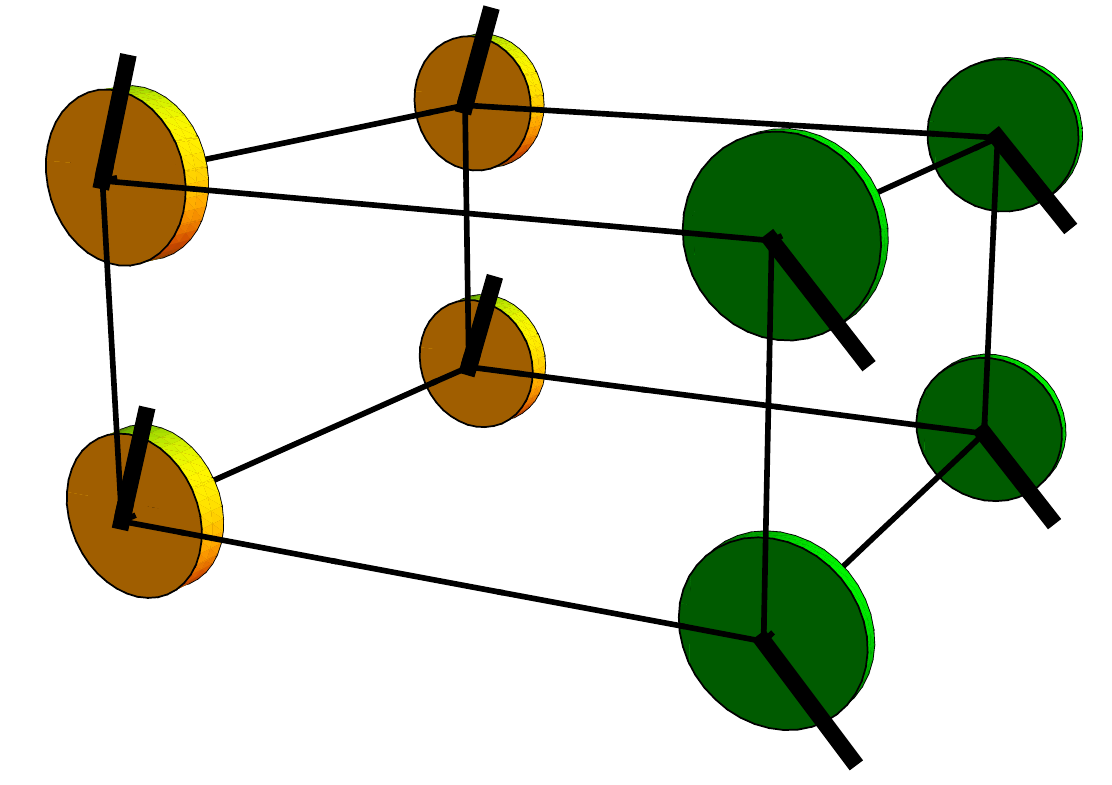}
\includegraphics[width=2.7cm]{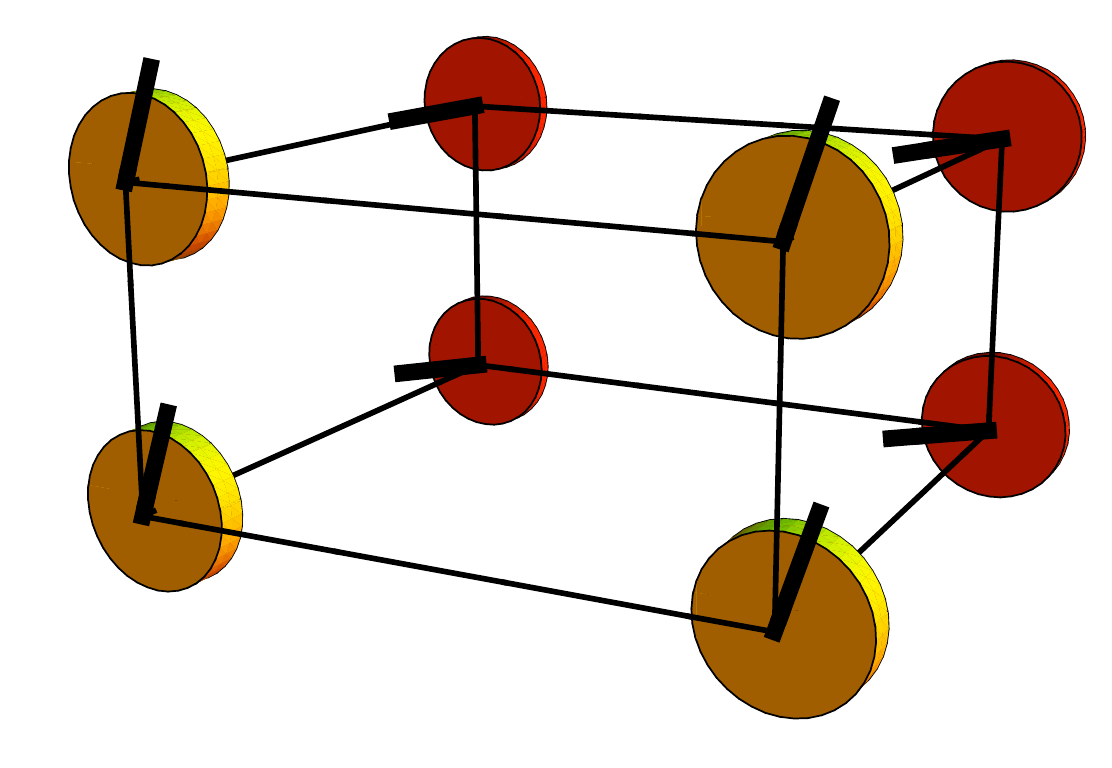}
\medskip
\caption{ The symmetries of Eq. (\ref{symorb-e}) applied a uniform ground state (top left).
}
\label{fig:120symmetries}
\end{figure}

The existence of a global rotational symmetry, as discussed in the previous section, is pervasive in physical systems -- although usually these are  exact symmetries. Much more peculiar to the 120$^{\circ}$ compass and related models is the existence of numerous low dimensional ($d<D$) symmetries. These symmetries relate to ground states that will be stabilized at finite (yet low) temperatures. An explanation of what these symmetries are
is given best done pictorially. In the top lefthand corner of Fig.~\ref{fig:120symmetries}, a general uniform configuration is shown -- a ground state of the classical system.
Starting with any such state, it is possible to {\it reflect} pseudo-spins in individual planes to generate myriad other configurations
which are also ground states of the classical 120$^{\circ}$ compass model. For instance, one may take any plane that is orthogonal
to the ${\bf{e}}_{x}$ direction and reflect all of the pseudo-spins in that plane about the ${\bf a}$ direction.
Under such an operation, $T_{i}^{(a}$ is unchanged but the pseudo-spin component along the direction that is orthogonal
to ${\bf{a}}$ flips its sign. This will lead to a state that has exactly the same energy as that of the uniform state. Similarly, 
one may reflect all pseudo-spins in planes orthogonal to the ${\bf{e}}_{y}$ or ${\bf{e}}_{z}$ directions by ${\bf b}$ or ${\bf{c}}$ respectively.  All of these three cases are depicted in Fig.~\ref{fig:120symmetries}. 

These reflections are Ising symmetry operations or, formally, $Z_{2}$ symmetries. Any reflection performed twice will lead to the original state and is thus an Ising type operation. Going beyond the $2 \times 2 \times 2$ cube shown in Fig.~\ref{fig:120symmetries}, one can consider a cubic lattice of dimension $L \times L \times L$ with $L \gg 1$. On such a lattice, these reflections which are emergent ($d=2$) $[Z_{2}]^{3L}$ gauge-like symmetry operators~\cite{Batista05,Nussinov04,Biskup05}. The power of $(3L)$ relates to the number of planes ($d=2$ dimensional objects) in which such reflections can be applied: there are $L$ such planes which orthogonal to one of the three cubic lattice directions.

Formally, these operations, rotations of all pseudo-spins by an angle of 180$^{\circ}$ about the internal $T^{\gamma}$ axis, can be written
as quantum operators in the limit of large pseudospin size (where they correspond to classical rotations). These operations are
\begin{eqnarray}
\tilde{O}^{(\gamma)} = \prod_{i\in P_{\gamma}} e^{i\pi{\bm{T}}_i \cdot {\bf{e}}_{\gamma}},
\label{symorb-e}
\end{eqnarray}
where $P_{\gamma}$ is any plane orthogonal to the corresponding cubic ${\bf{e}}_{\gamma}$ axis. It is important to re-iterate that these are not {\it bona fide} symmetries over the entire spectrum -- these are not {\it exact} symmetries of the Hamiltonian. That is, these operations are symmetries when restricted to classical ground states and {\it emerge} in those combined limits, i.e., the classical limits of (i) high pseudo-spin
and (ii) zero temperature. 

It is well-known that two-dimensional Ising symmetries can be broken at finite temperatures. Thus, the symmetries of Eq. (\ref{symorb-e}) of the classical 120$^{\circ}$ can be broken. And indeed they are, as will be discussed in section \ref{tf}.   
 
\subsection{Emergent Symmetries: Classical Honeycomb 120$^{\circ}$ Compass Model}
\label{120honey}

We will now review the  ground states and associated low energy emergent symmetries of the classical (or large pseudo-spin limit of the) 120$^{\circ}$ model on the honeycomb lattice \cite{Nasu08,Zhao08,Wu08},.
This model is given by Eq. (\ref{eq:c120_honeycomb}).
In what follows, we will invoke a decomposition of the honeycomb lattice into two interpenetrating triangular sublattices, referred to as sublattices A and B. Two neighboring sites of the honeycomb lattices thus belong to different sublattices.
 
The $120^{\circ}$ model on the honeycomb lattice shares a number of similarities with the 120$^{\circ}$ model on the cubic lattice discussed above and the key elements of the discussion will be the same. Nevertheless, in some respects, this system is even richer largely as a result of the larger number of emergent symmetries in the ground state sector. 



One may generally may seek to find all of the ground states of this system using Eq. (\ref{eq:TTGS}) -- a condition for finding {\it all} ground states of classical ferromagnetic compass model. It is instructive, within the framework of symmetries, to compare the consequences of this constraint as they apply to both the cubic lattice 120$^{\circ}$ model whose symmetries we enumerated above and the honeycomb lattice 120$^{\circ}$ model.

The coordination number of honeycomb lattice ($z=3$) is far smaller than that of the cubic lattice ($z=6$). Thus, the number of independent conditions of the type of Eq. (\ref{eq:TTGS}) will be halved.  As a result of this simple counting argument, we see that the ground state manifold might be far richer. This indeed turns out to be the case and emergent {\em local} ($d=0$) symmetries appear. 

We first review the ground states of this classical system and stratification procedures that are more similar in nature to those of the 120$^{\circ}$ model on the cubic lattice (i.e., involve the application of emergent intermediate and global symmetries on a uniform ground state) and then review additional local symmetry operations that appear in this case. 

\subsubsection{Ground States and Emergent Intermediate Symmetries}

\begin{figure}
\centering
\includegraphics[width=.8\columnwidth]{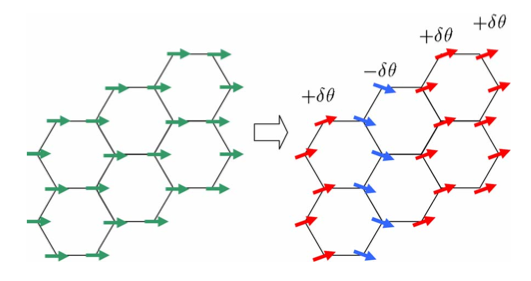}
\caption{Left: pseudospin configuration for $\theta^*$=0. Right: configuration obtained by $\pm \delta \theta$ rotations of pseudospins in each zigzag chain.~\cite{Nasu08}.}
\label{fig:Nasu08_8}
\end{figure}

In the classical limit, the pseudo-spins in Eq. (\ref{eq:c120_honeycomb}) become two-component (XY) type variables which may parameterized 
by (with some abuse of notation) a continuous angular variable $\theta_{i}$ at the different lattice sites $i$. Here, $\{\theta_{i}\}$ denote the orientation
of the classical pseudo-vectors ${\bf T}_{i}$ (the large pseudo-spin limit variant of ${\bf \tau}_{i}$ in Eq. (\ref{eq:c120_honeycomb})). 

As in the cubic lattice case reviewed in Sec.~\ref{sec:cubic120class} all uniform states (${\bf T}_{i} = {\bf T}$) are ground states and these may be stratified by the application
of low dimensional emergent symmetry operations. The $d=2$ emergent symmetries of Eq. (\ref{symorb-e}) and Fig.~\ref{fig:120symmetries}
have their counterparts in $d=1$ symmetries in the 120$^{\circ}$ model on the honeycomb lattice \cite{Nasu08}. As shown in Fig.~\ref{fig:Nasu08_8} it is possible, starting from a uniform state to generate other ground states by varying $\theta_{i} \to \theta_{i} + \delta \theta_{i}$. In this case, by considering (the $d=1$) zig-zag chains along one of the three crystalline directions \cite{Nasu08}, it is possible to generate other ground states by a reflection of all of the spins
in these chains as in Fig.~\ref{fig:Nasu08_8}.

\begin{figure}
\centering
\includegraphics[width=.8\columnwidth]{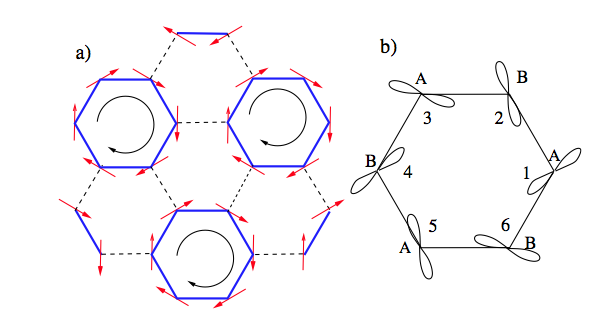}
\caption{The fully packed oriented loop configurations in which $\tau$-vectors lie in directions of $\phi = \pm 30^\circ,  \pm 90^\circ,  \pm 150^\circ$. (a) The closest packed loop configuration with all the loops in the same chirality. (b) The p-orbital configuration for one closed loop in (a). The azimuthal angles of the p-orbitals are $45^\circ, 105^\circ, 165^\circ, 225^\circ, 285^\circ, 345^\circ$~\cite{Wu08}. }
\label{fig:Wu08_2}
\end{figure}

\subsubsection{Emergent Local Symmetries}

\begin{figure}
\centering
\includegraphics[width=.7\columnwidth]{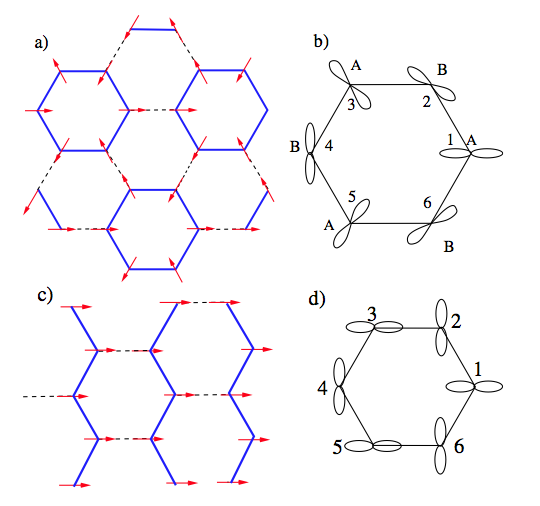}
\caption{The fully packed unoriented loop configurations in which $\tau$-vectors lie along the bond directions. (a), (c) are the $\tau$-vector configurations with the closest packed loops and the ferromagnetic state, respectively. (b), (d) are their corresponding p-orbital configurations~\cite{Wu08}.}
\label{fig:Wu08_3}
\end{figure}

Fig.~\ref{fig:Wu08_2} shows particular ground states found by \cite{Wu08} wherein the pseudo-spins ${\bf T}_{i}$ are oriented in the plane, at angles of $(\pm 30^{\circ}, \pm 90^{\circ}, \pm 150^{\circ})$  such that  they are tangential to the basic hexagonal plaquettes.  In Fig. \ref{fig:Wu08_2}, the explicitly shown clockwise (or anti-clockwise) {\it chirality}  [correspondingly, $C_{h} =1$ (or $C_{h}=-1$)] for each hexagon $h$ relates to the tangential direction of the pseudo-spins which can be flipped with no energy cost. Similar to our earlier considerations, chiral degrees of freedom in adhere to emergent discrete Ising like {\em gauge symmetries} (or $d=0$ symmetries in the classification of Section \ref{global_inter}).  These particular ground states lie within a larger space of classical states that are generated from  the chiral tangential patterns is shown in Figs.~\ref{fig:Wu08_2},\ref{fig:Wu08_3}. Panel (a) of Fig. \ref{fig:Wu08_3} corresponds to a staggered rotation by 90$^{\circ}$ of the chiral state
depicted in Fig. \ref{fig:Wu08_2}.  Generally, this larger set of ground states is generated by an application of a continuous $d=2$ symmetry on the ground states of Fig. \ref{fig:Wu08_2}. This set of classical configurations may be obtained as follows: Starting with any tangential state of the pseudo-spins an in Fig. \ref{fig:Wu08_2} about the various hexagons, one can apply a {\em global} staggered  ($U(1)$) rotation of all of the pseudo-spins in the plane  such that all of the spins that lie on sublattice A are rotated by an angle  of $ \delta \theta$ whereas all of the spins lying on sublattice B are rotated by an angle of $(-\delta \theta$). This leads to state such as those shown in Fig. \ref{fig:Wu08_3}. \cite{Wu08} provides a detailed analysis of these results. 

\subsection{Emergent Symmetries of the Triangular 120$^{\circ}$ Compass Model}
\label{120triangle}

\begin{figure}
\centering
\includegraphics[width=.8\columnwidth]{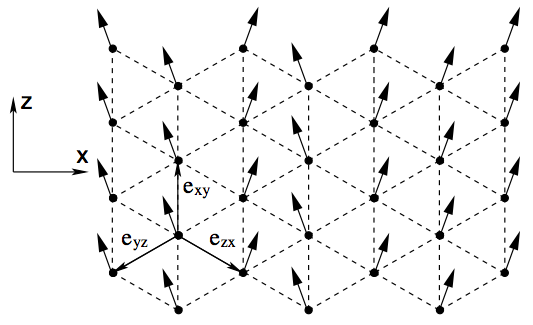}
\caption{The triangular lattice formed in the $[111]$ plane. Shown is a disordered mean-field ground state, in which the isospins form lines parallel to the unit vector ${\bf e}_{xy}$, such that $\langle T_j^z \rangle$ is the same on all lattice sites, while the sign of $\langle T_j^x \rangle$ varies arbitrarily from line to line.~\cite{Mostovoy2002}.}
\label{fig:Mostovoy02_2}
\end{figure}

In its ground state sector, the classical 120$^{\circ}$ model of Eq. (\ref{eq:c120_triangle}) exhibits $d=1$ dimensional emergent symmetries.
Similar to those discussed above, those relate to reflections of the pseudo-spins ($T_{i}^{\gamma} \to - T_{i}^{\gamma}$) for all sites $i$ that lie along a ``plane'' $P$ (a one-dimensional line in this case)
that is parallel to the direction ${\bm{e}}_{\gamma}$. This operation leads to stratified states once again. A schematic is shown in Fig.~\ref{fig:Mostovoy02_2}.

\subsection{Three component Kugel-Khomskii model}   
\label{3compKK}


In sections (\ref{sec_KK},\ref{sec:KK}), we discussed the Kugel-Khomskii (KK) model \cite{Kugel72,Kugel73,Kugel82}. In particular, we reviewed underlying physics of this Hamiltonian in subsection \ref{sec:KK}.
Its most prominent version is that for two component pseudo-spins wherein the KK Hamiltonian describes the two $e_{g}$ levels (represented by two-component pseudo-spins). 
We now return to the three-component variant of this model that is more pertinent to three $t_{2g}$ orbital states
We will label these as follows~\cite{Harris03}:
\begin{equation}
\label{abc}
|a\rangle\equiv |yz\rangle, \hskip .7cm
|b\rangle\equiv |xz\rangle, \hskip .7cm
|c\rangle\equiv |xy\rangle.
\end{equation}
To make the discussion self-contained, we write anew the KK Hamiltonian in its general form and focus on 
its three-component pseudo-spin version. The KK Hamiltonian is given by
\begin{equation}
\label{eq:Horb}
H = \sum_{\langle ij\rangle\parallel\gamma} H_{\rm orb}^{(\gamma)}(ij)
\left({\bm{S}}_i\cdot {\bm{S}}_j + \frac{1}{4}\right).
\end{equation}
Physically, ${\bm{S}}_{i}$ is the spin of the electron at site~$i$ and 
$H_{orb}^{(\gamma)}(ij)$ are operators that act on the orbital degrees of freedom.
For TM atoms arranged in a cubic lattice, wherein each TM atom is surrounded 
by an octahedral cage of oxygens, these operators are given by
\begin{eqnarray}
\label{eq:orbHam}
H_{\rm orb}^{(\gamma)}(ij) = J\left(4\hat{\pi}_i^{\gamma} 
\hat{\pi}_j^{\gamma}-2{\hat{\pi}}_i^{\gamma} 
- 2\hat{\pi}_j^{\gamma}+1\right),
\end{eqnarray}
where $\hat{\pi}_i^{\gamma}$ are pseudospin components,
and~$\gamma=a,b,c$ is the direction of the bond~$\langle ij\rangle$.
In the three-component realization that we wish to discuss now,
\begin{equation}
\hat{\pi}_i^{\gamma} =\frac{1}{2} \tau_i^{\gamma}.
\label{compass1}
\end{equation} 

\begin{figure}
\centering
\includegraphics[angle=0,width=.75\columnwidth]{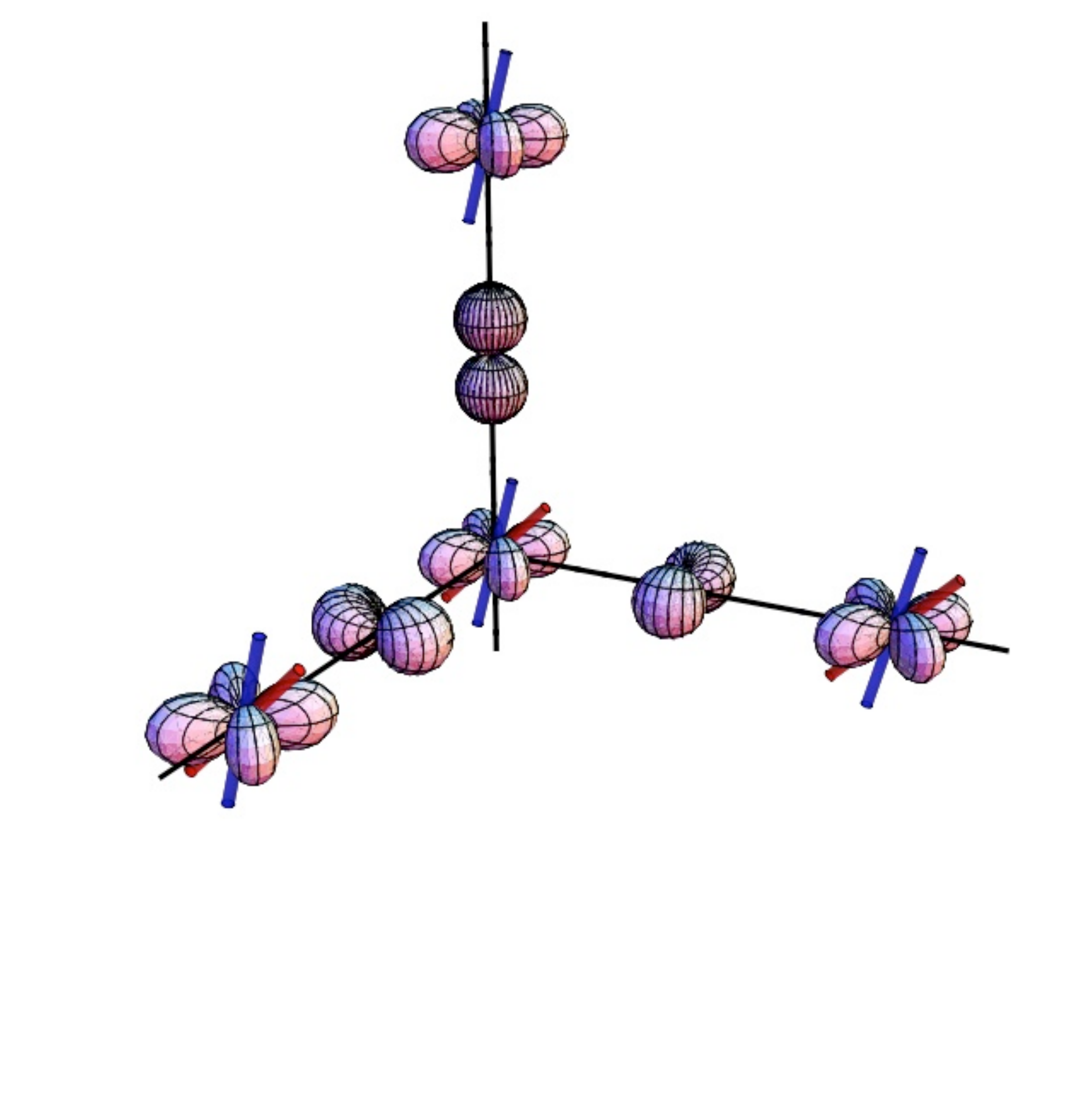}
\vspace*{-2.0cm}
\caption{ 
The anisotropic hopping amplitudes leading to the 
KK Hamiltonian after Ref. \cite{Batista05}. The spins are indicated by blue rods. Similar to Ref. \cite{Harris03}, 
the four-lobed states denote the $3d$ orbitals of a TM ion while the 
intermediate small $p$ orbitals are oxygen orbitals through which the 
superexchange process occurs. Due to orthogonality with intermediate oxygen $p$ states, in any orbital state 
$|\gamma \rangle$ (e.g. $| c \rangle\equiv | xy \rangle $ above), 
hopping is disallowed between sites separated along the cubic 
$\gamma$ ($c$ above) axis. The ensuing KK Hamiltonian has a $d=2$ 
$SU(2)$ symmetry that corresponds to a uniform rotation of all
spins whose orbital state is $|\gamma\rangle$ 
in any plane orthogonal to the cubic direction $\gamma$. Such a rotation in the $xy$ plane is indicated by the red spins in the figure.}
\label{fig:Batista05}
\end{figure}

The KK model in $t_{2g}$ systems exhibits a {\it continuous exact lower dimensional symmetry} as we now review.
In the $t_{2g}$ compounds, hopping is disallowed via intermediate 
oxygen $p$ orbitals between any two electronic
states of orbital flavor $|\gamma\rangle$ ($\gamma = a$, $b$, or $c$)
along the $\gamma$ axis of the cubic lattice 
(see Fig. \ref{fig:Batista05}). As a consequence,
as noted in \cite{Harris03}, 
a uniform rotation of all spins, whose electronic orbital state
is $|\gamma \rangle$, in 
any plane ($P$) orthogonal  
to the $\gamma$ axis 
$c^{\dagger}_{i \gamma \sigma} 
= \sum_{\eta} U^{(P)}_{\sigma, \eta}
d^{\dagger}_{i \gamma \eta}$
with $\sigma, \eta$ the spin 
directions, leaves Eq. (\ref{eq:Horb}) invariant.
The total spin of electrons
of orbital flavor $|\gamma \rangle$ 
in any plane orthogonal to the cubic $\gamma$ axis is conserved. 
Here, we have $d=2$ $SU(2)$ symmetries
\begin{eqnarray} 
\hat{O}_{P;\gamma} \equiv [\exp(i\bm{S}^{\gamma}_{P} \cdot 
{\bm{\theta}}^{\gamma}_{P})/\hbar], \hskip 1cm [H, \hat{O}_{P;\gamma}]=0,
\label{symt2g}
\end{eqnarray}
with $\bm{S}^{\gamma}_{P} = \sum_{i \in P} \bm{S}_{i}^{\gamma}$,
being the sum of all the spins $\bm{S}^{i, \gamma}$ in the orbital state
$\gamma$ in any plane $P$ orthogonal to the direction $\gamma$ 
(see Fig. \ref{fig:Batista05}).

We now, once again, turn to {\em the physical origin of dimensional reduction in this
system with continuous $d=2$ SU(2) symmetries}. The bound of Eq.~(\ref{boundfi}) prohibits,
at finite temperatures, local on-site order is provided by Eq.~(\ref{symt2g}) for the KK model.
Physically, this is so due to the proliferation and deleterious effect of
$d=2$ dimensional defects (i.e., spin waves) in SU(2) continuous pseudo-spin
systems. The energy/entropy balance associated with
these defects in the three-dimensional KK system 
is identical to that in a two-dimensional three-component
Heisenberg spin system.

\section{Intermediate Symmetries \&  Flat Bands in Classical Spin-wave Dispersion}
\label{new_theorem_on_bands}

In this section, we introduce a new result that will be of utility in understanding some aspects of the order-by-disorder physics and the role of the large degeneracy of these systems
as it pertains to simple ${\bm k}$-space classical spin wave type analysis.  We outline a new result that sheds light on the relation between spectral structure, degeneracy, and intermediate symmetries in general classical ferromagnetic compass systems in $D-$ spatial dimensions. In a nutshell, one asks what the consequences are of the existence of real-space stratified ground states found in section \ref{sec:diso}  [schematically
illustrated in Figs. (\ref{fig:120symmetries},\ref{fig:Nasu08_8},\ref{fig:Wu08_2},\ref{fig:Mostovoy02_2},\ref{fig:Batista05})] on the momentum space spectrum of 
pseudo-spin excitations. One finds that the low $d-$dimensional symmetries (either exact or emergent) that leads to the stratified states in real space
their application on the canonical uniform (${\bm k}=0$) ferromagnetic state,
lead, in momentum-space, to a redistribution of weights in $(D-d)$ dimensional regions. As all of these
states share the same energy,  one finds that the existence of $d$-dimensional symmetries ensures that
there are $(D-d)$ dimensional volumes which are ``flat'' and share the same mode energy as the 
${\bm k}=0$ point. Although $d$ dimensional symmetries imply flat bands in classical systems, the converse is not true-
in classical systems with a finite number of pseudo-spin components, flat bands generally do not imply the existence of $d-$ dimensional symmetries. 
However, in the large $n$ limit, $(D-d)$ dimensional flat bands indeed imply the existence of $d-$ dimensional real space symmetries. 
Large $n$ analysis of these systems is identical to that of $d$ dimensional
systems (i.e., in all directions orthogonal to the flat zero-energy regions in ${\bm k}$ space).
That is, in the large $n$ system, an effective dimensional reduction occurs (from $D-$ dimensions to $d-$ dimensions). 
Thus, for systems with, e.g., $d=2$ symmetries (such as the cubic lattice 120$^{\circ}$ compass model), large $n$ analysis
and related approximate methods relying on simple classical ${\bm k}$-space spin wave analysis will, incorrectly, predict incorrectly
that the finite $n$ classical system does not order and that quantum fluctuations are mandatory to explain the observed ordering
in these systems. Similar considerations to all of these results concerning the interesting link between symmetries and 
band structure may apply, in general (i.e., not necessarily ferromagnetic) systems for both ground states and excited states. 

\subsection{Uniform States as Ground States of Classical Compass Models}
\label{unifcm}
In the absence of an external field, the classical ground states corresponding to the general isotropic compass model  
Hamiltonian of Eq. (\ref{eq:general_compass}) are fairly
trivial. In the anisotropic (non-uniform $J_{\gamma}$), 
the pseudo-spins tend to align along the direction $\gamma'$- the direction associated with
the highest exchange coupling $J_{\gamma'}$. 
We now first explicitly turn to the isotropic situation wherein $J_{\gamma} =J>0$ \cite{Nussinov04,Biskup05}.
As discussed in subsection \ref{sec:clas}, in their classical rendition, the 
the pseudo-spins are normalized at all lattice sites, ${\bm{T}}_{i}^{2} =1$.
In such a case, for the classical rendition all of the systems that we focus on in this review,
up to an irrelevant additive constant  $C$, 
the Hamiltonian may be written as a sum of squares
\begin{eqnarray}
H_{isotropic}^{compass}=   \frac{J}{2} \sum_{i} \left[ \sum_{\gamma} (T^\gamma_i - T^\gamma_{i+\bm{e}_\gamma})^{2}  - 2C \right].
\label{iso_comp1}
\end{eqnarray} 
A direct computation shows yields the value of
$C= \sum_{j} ({\bm T}_{i} \cdot {\bm{e}}_{ij})^{2})$, which is independent 
of the orientation of ${\bm T}_{i}$.
For all classical compass models on regular lattices with two-component (i.e., XY) type spins
whose orientation may be
specified by a single angle $\theta_{i}$ on the unit disk, the constant $C=z/2$
with $z$ being the coordination number of the lattice (the number of nearest neighbors
of any given site). Values of the constant $C$ in Eq. (\ref{iso_comp1}) can be readily computed for
compass models with a higher number of spin components. The classical $D-$ dimensional 
90$^{\circ}$ compass model of Eq. (\ref{highD90}), the additive constant $C$ in Eq. (\ref{iso_comp1}) is given by
$C=2$. Similarly, for the classical counterpart of the Kitaev model of Eq. (\ref{eq:HKT}), $C=1$.

As all terms in the sum of Eq. (\ref{iso_comp1}) are positive or zero, minima are achieved when
${\bf{T}}_{i} = {\bf{T}}$ for all $i$ with ${\bf{T}}$ an arbitrary orientation. 
Thus, any uniform state is a ground state and a continuous global rotation
may relate one such ground state to another. These rotations are not bona
fide symmetries of the Hamiltonian and may {\em emerge} as such only
in the restricted ground state subspace. Thus, the ferromagnetic compass
models exhibit a {\em continuous emergent symmetry} of their ground states.
Starting from any uniform state (a ground state of the classical
system), any uniform global rotation of all pseudo-spins 
will lead to another ground state.

Although perhaps obvious, we remark on the relation between ferromagnetic and antiferromagnetic compass
models. On bipartite lattices, the sign of the exchange couplings
can be reversed ($J_{\gamma} \to - J_{\gamma}$) for classical systems.
The same trivially holds true for quantum XY spins (such as those in the 120$^{\circ}$
model) for which a canonical transformation (rotation by 180 degrees about the z axis)
can be performed. 

\subsection{Stratification in Classical Compass Models}
The richness of the classical compass models stems from the many
possible ground states that they may possess (aside from the uniform state).
Such stratified ground states  were depicted in Figs. (\ref{fig:120symmetries},\ref{fig:Nasu08_8},\ref{fig:Wu08_2},\ref{fig:Mostovoy02_2},\ref{fig:Batista05}).
Equal energy states (classical or quantum) are generally related to each other via the symmetries discussed 
in Section \ref{sec:exact_emergent}. Emergent (and exact) symmetries of the classical ferromagnetic compass models link the uniform ferromagnetic states discussed in subsection \ref{unifcm} to a plethora of other 
classical ground states. As will be elaborated on in section \ref{sec:diso}, this
 proliferation of low energy states lead to high entropic contributions and the failure of the simplest 
analysis to predict finite temperature order. We now explicitly determine all classical ground states of ferromagnetic compass models
and link those to the earlier depicted ground states. 
As can be seen from Eq. (\ref{iso_comp1}), 
{\em any} configuration for which 
\begin{eqnarray}
T^\gamma_{i} = T^\gamma_{i+\bm{e}_\gamma}
\label{TTGS}
\end{eqnarray}
on all sites $i$ is also a ground state configuration. 
That is, in standard compass models, the projections of any two nearest neighbor
${\bm T}$ along the bond direction $\gamma$ must 
be the same. [As noted several times earlier and made explicit in the original compass model definitions in subsection \ref{sec:clas},
the components in Eq. (\ref{TTGS}) are defined by $T^{\gamma} \equiv {\bf{T}} \cdot {\bf{e}}_{\gamma}$; in this scalar product, 
the corresponding internal pseudo-spin unit vectors ${\bf{e}}_{\gamma}$ are chosen differently for different compass systems.] 
 In Kitaev's model, the direction specified by $\gamma$ is dictated by the 
lattice link direction but it is not equal to it. At any rate,  generally, 
the number of conditions that Eq. (\ref{TTGS}) 
leads to is equal to the number of links on the lattice- 
$(Nz/2)$. Eq. (\ref{TTGS}) states that only the 
$\gamma$ component of the pseudo-spin ${\bf T}$ is important
as we examine the system along the $\gamma$ lattice direction.
It may therefore generally allow for numerous other configurations apart
from the uniform ferromagnetic states in which one transforms the pseudo-spins in planes 
orthogonal to the $\gamma$ direction in such a way as not alter the projection $T^{\gamma}$
of ${\bf T}$ on the $\gamma$ axis. This allows for the multitude of ground states
discussed in section(\ref{sec:exact_emergent}) that are related to the uniform ground
states via intermediate low dimensional operation (generally an emergent symmetry
of the ground state sector). 

\subsection{Flat bands: Momentum Space Consequences of Real Space Stratified Ground States}
 
A new prevalent aspect that has not been discussed before in the literature concerns a general relation between the classical ground states of the compass
models and the classical spin wave dispersions.  This new relation will be introduced shortly. Towards this end, it will be profitable to
examine the matrix ${\hat{V}}({\bf k})$ of Eq. (\ref{Fourier_space}) in its internal
pseudo-spin eigenbasis and 
write the classical compass Hamiltonians as 
\begin{eqnarray}
H = \frac{1}{2} \sum_{\alpha} \sum_{\bf k} v_{\alpha}({\bf k}) |t_{\alpha}({\bf k})|^{2}.
\label{long_ebasis}
\end{eqnarray}
In Eq. (\ref{long_ebasis}), the internal pseudo-spin space index $\alpha$ labels the eigenvalues $v_{\alpha}(\bf{k})$
of the matrix ${\hat{V}}({\bf k})$ and $t_{\alpha}({\bf k})$ are the 
internal pseudo-spin components of the vectors ${\bf T}({\bf k})$ when
expressed in this basis. 

These {\em emergent} symmetries within the ground state
sector lead to an enormous degeneracy of the classical
ground states. One can relate this to the eigenvalues of the matrix
${\hat{V}} ({\bf k})$ of Eq. (\ref{Fourier_space}).  Before doing so for 
the compass (and general systems),
we reflect on the situation in canonical nearest neighbor 
classical ferromagnets. In standard, isotropic, ferromagnetic systems, $v_{\alpha}({\bf k})$ attains its global 
minimum when ${\bf k} =0$. Thus, in standard ferromagnets,
only the uniform (${\bm k} =0$) states are ground states. 
Any other non-uniform state necessarily has non-vanishing
Fourier space amplitudes $t_{\alpha} ({\bf k}) \neq 0$ also for modes ${\bm k} \neq 0$
each of which costs some energy relative to the
lowest energy ${\bm k} =0$ state.  
By contrast, the multitude of non-uniform
ground states generated by the 
stratification operations of Fig. ~\ref{fig:120symmetries} 
prove that $v_{\alpha}({\bf k})$ no longer
attains its minimum at a single 
point in ${\bf k}$ space but rather
at many such points. Applying the general
stratification (or stacking) operations of, e.g., Fig. (\ref{fig:120symmetries})
on the uniform ${\bm k}= 0$ state 
(one for which the Fourier amplitudes $T_{\bm k \neq 0} =0$ leads to
new configurations for which the  
Fourier amplitudes $T_{\bm k} \neq 0$
where ${\bm k}$ lies along the $k_{z}$ axis. 

According to Eq. (\ref{long_ebasis}),
this suggests that the lowest values of $\min_{\alpha} \{v_{\alpha}({\bm k})\}$ define
lines along the $k_{x}$, $k_{y}$, or $k_{z}$ axis. This
can indeed be verified by a direct computation. 

More generally, if, one sets $\min_{\alpha, {\bm k}} v_{\alpha}({\bm k}) =0$
and the ground state energy happens to have a zero value
according to Eq. (\ref{long_ebasis}) 
In general, of course, when one applies
a general operation $U$ to get a new 
ground state, with $t_{\alpha'}({\bm k'}) \neq 0$
then for all of these values of $\alpha'$ 
and ${\bm k'}$ with a non-zero Fourier amplitude $t_{\alpha'}({\bm k'})$, 
one must have that  $v_{\alpha'}({\bm k'}) =0$. 
The fact that the uniform ground states at ${\bf k}=0$
are invariant under global rotation (i.e., a change of 
basis of the internal indices $\alpha'$ for all components
$\alpha'$  for which $t_{\alpha'}({\bm k'}=0)$) asserts
that states having components $\alpha'$ such that $\min_{\alpha} \{v_{\alpha}({\bm k}=0)\}
= v_{\alpha'}({\bm k}=0)$ can,
indeed, be materialized. This follows as 
whatever $\alpha'$ happens to be, for ${\bm k}=0$, the eigenvector ${\bm t} = (0 ..0 1 0 ...0)^{T}$
corresponding to it  will relate to some particular uniform real space vector ${\bf{\tilde{T}}}$
in the original basis. On the other hand, any uniform state is a ground state and thus
such a configuration with a vector ${\bf{\tilde{T}}}$ can be materialized.
That is, the lower bound on the energy stemming from the lowest energy eigenvector(s) 
of $\hat{V}$ of Eq. (\ref{Fourier_space}) can be saturated.

Thus, emergent symmetries mandate the appearance
of lines of nodes in the dispersion. [The
same, of course, also trivially holds
for exact symmetries of the Hamiltonian.]
The converse is of course not true:
the existence of flat regions of the dispersion
(those with $v_{\alpha'}({\bm k'}) =0$)
do not mandate that symmetries appear
in the ground state sector as although any
linear combination involving only $t_{\alpha'}({\bm k'}) $ it might not be possible to
construct real space states out of these amplitudes for which ${\bm T}_{i}^{2} =1$
at all sites $i$. 
 
The discussion above {\it relates the degeneracies brought about by 
(exact or emergent) intermediate
symmetries with the dispersion of $v_{\alpha}({\bf{k}})$ 
about its minimum}.  This general link between intermediate symmetries 
and (``flat") spin-wave type dispersion applies to many of the 
other compass models in this review. 

In general, if in a general compass model, a $d$ dimensional operation relates the different
ground states (such as the $d=2$ reflections of Fig. (\ref{fig:120symmetries}) and Eq. (\ref{symorb-e})
then the lowest bands $v_{\alpha}({\bm k})$ are zero (or,
more generally attain their lowest values) within $d'=(D-d)$ dimensional  
regions in $k-$space. This follows from the application,
on a uniform ferromagnetic state of the symmetry operators of 
the form of
Eq. (\ref{composite}). Different symmetries (either emergent ($\tilde{O}_{P}$)
or exact ($\hat{O}_{P}$)) can be 
chosen in the string product of Eq. (\ref{composite}) that when
acting on the uniform ferromagnetic state lead to  
disparate configurations that must all share the same energy. 
Thus, putting all of the pieces together, we have established a new
theorem:  \newline

{\it When a system of the general form of Eq. (\ref{long_ebasis}) exhibits a ferromagnetic state then
the existence of $d$-dimensional symmetries (exact or emergent) implies that 
$ v_{\alpha'}({\bf k})$ has a flat dispersion in a $(D-d))$-dimensional manifold that
connects to the ferromagnetic point of ${\bf k}=0$. }
\newline
 
As explained above, for classical pseudo-spins ${\bf T}_{i}$ with a finite number ($n$) of components, that have to be normalized
 at each lattice site $i$, 
the converse is not guaranteed to be true: if one has flat lowest energy bands then we are not guaranteed that we can generate
real space configurations with normalized pseudo-spins ${\bf T}_{i}$ whose sole Fourier amplitudes
are associated with wave-vectors ${\bf k}$ that belong to these flat bands.
 In the large $n$ limit  of the classical models (or, equivalently, in the corresponding spherical models) \cite{Berlin1952,Stanley1968,Nussinov2001}
 the local normalization conditions becomes relaxed and linear superpositions of Fourier modes
 on the flat band lead to allowed states that share the same energy. That is,
 in the large $n$ limit (and, generally, only in that limit), if there is a band $v_{\alpha'}({\bm k})$ that assumes a constant
 value $v_{\alpha'}({\bm k}) = {\sf const.}$ for wave-vectors ${\bm k}$ that belong to a manifold ${\cal{M}}$ of dimension $d'=D-d$
 then the system exhibits a $d$-dimensional symmetry: any transformation that acts as a unitary transformation on the modes
 ${\bm k} \in {\cal{M}}$ will not alter the energy of states whose sole non-vanishing Fourier amplitudes $t_{\alpha'}({\bm k})$ belong to this
 manifold. For related aspects, see \cite{holography,Batista05}. As the spectrum $v_{\alpha'}({\bm k})$ is pinned at its 
 minimum value along $d'=(D-d)$ dimensional regions in ${\bm k}$ space, large $n$ computations will, up to constant factors associated
 with the volume of these regions, reproduce 
 results associated with the non-vanishing dispersion in the remaining $(D-d') =d$ dimensional regions. 
 Thus, in the large $n$ limit, the behavior of compass model ferromagnets in $D$ spatial dimensions is identical to that
 of the ferromagnets in the large $n$ limit in $d$ dimensions. As the large $n$ ferromagnet does not exhibit long range order 
 in $d=2$ dimensions (and indeed any pseudo-vector system with $n \ge 2$ components), the large $n$ analysis of the classical
 cubic lattice 120$^{\circ}$ model will predict that it does not order at finite temperatures- an erroneous conclusion.
 As it turns out, simple large $n$ and other related approximations are not valid for the analysis of the classical 120$^{\circ}$ model
 and careful calculations are required for the free energy of the $n=2$ component classical system \cite{Biskup05, Nussinov04}.
 We will return to this point in section \ref{tf}.

In principle, the theorem can be replicated for any other commensurate real space ground state structure
for which the only non-vanishing Fourier components $t_{\alpha}({\bm k}) \neq 0$ are those
that minimize the kernel $ v_{\alpha}({\bf k})$ in Eq. (\ref{long_ebasis}).
In the above, we illustrated that the ferromagnetic compass model has, amongst many other
states, the uniform (${\bm k}=0$) state as a ground-state. corresponding to well defined ${\bf k}$ space points.
There are other commensurate structures (e.g., Neel states, 2 x 2 checkerboard states, etc.) that correspond to 
a particular set of wave-vectors \cite{Nussinov2001}. We proceed by discussing the particular realizations of this this theorem in compass models.

\subsubsection{Spin-waves of Cubic Lattice 120$^{\circ}$ Compass Model}

In the case of, e.g., the 120$^{\circ}$ model in $D=3$ dimensions this co-dimension is
$d'=1$ and the zeros of the modes lie along lines (which happen to be 
the Cartesian coordinate axes in momentum space).  We briefly remark that when local
symmetry operations are present (i.e., when $d=0$) as they are on some
of the more frustrated compass models that we will review later on then 
there will be flat bands where the corresponding $v_{\alpha'} ({\bm k}) =0$
for all ${\bm k}$ in the full $(d'=D)-$ dimensional $k-$ space
for some value(s) of the (band) index $\alpha'$.  

Although it is, of course, of less physical significance, the analysis for the highest energy
state is essentially identical to that for the ground states.  When the sign in Eq. (\ref{TTGS})
may be flipped (as on bipartite lattices), the resulting staggered configuration is that of highest energy possible.
Replicating all of the arguments made above {\em mutatis mutandis} 
it is seen that if the operations $U$ do not change the energy of these
states then the manifold of highest energy modes is of the dimensionality 
$d'$ of Eq. (\ref{d'Dd}).

\subsubsection{Honeycomb Lattice 120$^{\circ}$ Compass Model}

 We now discuss the system of Eq.  (\ref{eq:c120_honeycomb})
on the honeycomb lattice. 

As was noted from Eq. (\ref{long_ebasis}), the existence emergent $d-$ dimensional
symmetries of ground states that include the ferromagnetic state mandates (Eq. (\ref{d'Dd}))
that a $d'$ dimensional sub-volume of $k-$ space correspond to zero models ($v_{\alpha'}({\bm k'}) =0$
for one of more bands $\alpha'$). 

Given the appearance of the discrete chiral $d=0$ symmetries above \cite{Wu08}, 
given the earlier derivation above, one sees that $d'=D$
and thus flat bands may exist corresponding to the highest
and lowest possible energy states. Indeed, flat bands exist
in the spin wave dispersion about a state that has these
symmetries \cite{Wu08}. A more general
diagonalization of the $4 \times 4$ matrix $\hat{V}(\vec{k})$ of
Eq. (\ref{Fourier_space}) indeed illustrates that
there are two-flat bands with (in our convention)
values of $v_{1,2}({\bm k}) = 0, \frac{3J}{2}$ that correspond
to the lowest and highest energies attainable. There
are also two dispersing modes.  [This matrix
is of dimension four as a result of
two factors or two. Translation
invariance appears only for the honeycomb
lattice once it is considered as
a triangular lattice  (belonging to either the A or B
sublattices) with a basis
of  two sites. The second factor of two stems from
the number of components of each of the classical pseudo-spin
at each of these sites.

\section{Order by Disorder in Compass Models}
\label{sec:diso}

In sections (\ref{sec:sym}, \ref{new_theorem_on_bands}), we  illustrated how classical (and also quantum) compass systems might exhibit numerous ground states.  Aside from emergent global symmetries of the classical ferromagnetic compass model, both the classical and quantum models in $D$ spatial dimensions
exhibit a degeneracy which scales exponentially in $L^{D-d}$ where $d$
is the dimension of the intermediate symmetries  (see Eq. (\ref{domd})). As we will now review, 
this large degeneracy is generally lifted by by fluctuationsÑa process colloquially referred to as {\em order-by-disorder} \cite{Villain1980,Shender82,Henley89,Moessner00}. 
Although several states may appear to be equally valid candidate 
ground state, fluctuations can stabilize those states which have
the largest number of low energy fluctuations about them. 
These differences can be explicitly captured in values of the free energies for fluctuations 
about the contending states. Classically, fluctuations are driven by thermal effects and lead to entropic
contributions to the free energy. Quantum tunneling processes may fortify such ordering tendencies (``quantum order by disorder''  \cite{Henley89,Rastelli1987,Chubukov1992}),
especially so at zero temperature and stabilize a particular set of linear combinations of classically degenerate states. 
We note that albeit being very different, somewhat related physics concerning forces deriving from
the weight of zero-point ``fluctuations'' appears in the well-known Casimir effect of quantum electro-dynamics \cite{Casimir1948,Casimir1948a}. 
In the classical arena, similar effects appear- sea farers have long known about the tendency of closely separated ships to pull inwards towards each other as a result of hydrodynamic fluctuations. 
Other notions related to those in order by disorder physics concern entropy driven effects that lead to particular conformations appear in the funnel model for protein folding \cite{Bryngelson1995}. 

\subsection{Classical and Quantum Order out of Disorder}
\label{dcq}

Colloquially, quantum and classical systems may be anticipated to exhibit the same qualitative ``order out of disorder'' physics.
Although this is often the case, there is no fundamental reason 
for this to be so (and, indeed, the two effects may lead to very different results 
in some instances). Different sets of states can be stabilized by these fluctuations. 
An understanding of the quintessential physics may be obtained by
considering small (harmonic) fluctuations about classical ground states. 
To harmonic order, within the quantum arena, the fluctuations will be governed by a Bose distribution
(with frequencies $\omega_{i}$ that denote the energies of the various independent harmonic modes)
whereas the classical fluctuations obey a Boltzmann distribution with the same set of harmonic 
modes. The two may, obviously, be radically different 
at low temperatures especially insofar as they apply to zero mode fluctuations about the ground states.
In the appendix 

In Appendix \ref{appendixB}, section (\ref{tf}) [and in
our discussion the large $n$ structure factor of the 120$^{\circ}$ model] we will
aim to make this intuition more precise. In a nutshell, in 
many situations, quantum systems
may order more readily than their classical
counterparts. This may, in some
of the compass models that we consider.
be viewed as a consequence of ``order out of disorder''
effects at play which can be more pronounced in quantum
systems. We next examine order out of disorder effects in specific compass models.

\subsection{Cubic lattice 120$^\circ$ compass model}
\label{sec:NBCB}
When entropic contributions are omitted,  the spin-wave spectrum of the standard classical cubic lattice120$^\circ$ compass model is gapless\cite{Brink99}. This suggests that,  on the classical level, these orbital systems exhibit finite temperature disorder. Indeed the commonly held lore for some time was that quantum fluctuations (tunneling between the different contending classical ground states) are mandatory in order to lift the orbital degeneracy and account for the experimentally detected orbital orders. Most of the work on ``quantum order out of disorder'' focused on $1/S$ corrections (with $S$ the spin size) to the classical spin-wave spectrum.  

\subsubsection{Thermal fluctuations}
\label{tf}

The difficulties encountered in the simplest analysis of the classical model stem from the $d=2$ symmetries that it exhibits 
[see section \ref{sec:exact_emergent}] as was exemplified in Fig. (\ref{fig:120symmetries}). As we discussed in subsection \ref{new_theorem_on_bands},
these symmetries lead to flat $d'=(D-d)$ dimensional regions in ${\bm k}$ space along which the dispersion $v_{\alpha'}({\bm k})$
attains its minimum. In the case of the cubic lattice 120$^{\circ}$ model, there are lines ($d'=1$) along the Cartesian axis along which the dispersion is non-increasing.
In simple Gaussian calculations (such as that of the large $n$ or spherical models) \cite{Biskup05} this leads to a canonical
divergent fluctuations that inhibit low temperature order. The divergences are identical to those associated 
with canonical $D-d'=d$ ferromagnetic systems (or, in cubic lattice 120$^{\circ}$ compass systems, those associated with two-dimensional
continuous spin ferromagnetic systems). In various guises, this dispersion led to early difficulties 
in the analysis of this system and to the inclusion of quantum 
or thermal effects to lift this degeneracy.
To make this lucid, we briefly note that the structure
factor ${\cal{S}} ({\bf k})$ within spin wave theory (and classical large $n$ analysis \cite{Biskup05})
behaves, at low temperatures, as 
\begin{eqnarray}
{\cal{S}} ({\bf k}) \propto \frac{E_{x} + E_{y} + E_{z}}{E_{x} E_{y} + E_{x} E_{z} + E_{y} E_{z}},
\label{structure_factor}
\end{eqnarray}
with the shorthand $E_\gamma({\bm{k}}) \equiv 2-2\cos k_\gamma$.
As can be seen by inspection,  the structure factor of Eq. (\ref{structure_factor}) 
{\em diverges along lines} in ${\bf k}$ space (corresponding to momenta 
along the lattice directions $k_{x}$, $k_{y}$, or $k_{z}$).
As briefly alluded to in section \ref{dcq} [and elaborated on in the appendix],
in the simplest, large $n$ spin-wave type approaches, this divergence
of the classical system (as opposed to the convergence of the corresponding integral for its quantum
large $n$ counterpart as well as standard $1/S$ calculations)
leads to the false conclusion that there is no finite temperature
ordering in this system.  This divergence is removed by the proper inclusion of fluctuations about the ground states
of the $n=2$ component classical pseudo-spin system- an item which we turn to next.

Let us now, in particular, briefly review finite temperature effects
on the classical 120$^\circ$-model of Eq. (\ref{H120}) \cite{Nussinov04}.
The important thing to note is that the {\em free energy minima} 
(not the energy minima) determine the low energy states at finite temperatures.
The classical spins $\{\vec{S}_i\}$ are parameterized by the angles $\{\theta_i\}$ with the $a$ axis.
We may consider the finite temperature fluctuations about the
uniform ground states where each $\theta_i=\theta^\star$.
At low temperatures, the deviations $\vartheta_i=\theta_i-\theta^\star$ are small, 
and the quadratic [spin-wave (SW)] Hamiltonian corresponding
to Eq. (\ref{H120}) becomes \cite{Nussinov04, Biskup05}
\begin{equation}
H_{\rm SW}=\frac{1}{2}\;J \sum_{i,\gamma}
q_\gamma(\theta^\star)\;(\vartheta_{i}-\vartheta_{i+\vec{e}_\gamma})^2,
\end{equation}
where $\gamma=a,b,c$ while
$q_c(\theta^\star)=\sin^2(\theta^*)$, 
$q_a(\theta^\star)=\sin^2(\theta^\star+ 2 \pi/3)$ and $q_b(\theta^*)=\sin^2(\theta^\star-2 \pi/3)$.
On a cubic lattice with
periodic boundary conditions with
~$\theta^*$ the average of~$\theta_i$ on the 
lattice, at an inverse temperature $\beta = 1/(k_{B} T)$, 
the partition function \cite{Nussinov04, Biskup05}
\begin{equation}
\label{part}
Z(\theta^\star)=\int 
\delta\Bigl(\sum_i\vartheta_i=0\Bigr)\,
e^{-\beta H_{\rm SW}}
\prod_i\frac{d \vartheta_i}{\sqrt{2\pi}} \ .
\end{equation}
A Gaussian integration leads to 
\begin{equation}
\label{Gauss}
\log Z(\theta^\star)=-\frac{1}{2}
\sum_{k \ne\boldsymbol 0}\log\Bigl\{\sum_\gamma \beta 
J q_\gamma(\theta^\star)\,E_\gamma({\bm{k}})\Bigr\},
\end{equation}
where ${\bf{k}}=(k_x,k_y,k_z)$ is a reciprocal lattice vector.

The spin-wave free energy 
${\cal F}(\theta^*)$ of Eq.~(\ref{Gauss}) has
minima at 
\begin{eqnarray}
\theta_{n}^* = n \pi/3
\label{tn*}
\end{eqnarray}
with integer $n$~\cite{Biskup05,Nussinov04}.

The application of the $d=2$ 
stratification operations of Eq.~(\ref{symorb-e})
on each of these uniform configurations, see Fig.~\ref{fig:120symmetries}, leads to interface 
with an effective surface tension that leads to
a free energy energy penalty additive in 
the number of operations. The detailed
derivation is provided in ~\cite{Nussinov04, Biskup05}.
Below, we will provide physical intuition 
concerning the preference of uniform angles of the form of Eq. (\ref{tn*}) over
all others (i.e., why the minima of the free energy ${\cal F}(\theta^*)$ indeed
has its minima at the points $\theta_{n}^{*}$.

This analysis will build, once again, on the $d$ dimensional
emergent (i.e., ground state) symmetries of the problem.
Let us first start with the system when, for all lattice sites $i$, the angle
$\theta_{i} = \theta_n^*$ of Eq. (\ref{tn*}) with a particular value of $n$.
For concreteness, let us set $\theta_{i} = 0$ at all $i$. 
Let us next ask what occurs when we twist the 
angle between sequential planes (i.e., apply
the operation of Eq. (\ref{symorb-e}) leading
to a configuration such as 
\begin{eqnarray}
\theta_{i} = \delta \theta (-1)^{i_{z}}
\label{stag_eq}
\end{eqnarray}
(all other related ones in which the angle is uniform within each plane
orthogonal to the z axis) with $i_{z}$ the z coordinate of the lattice point $i$
and $\delta \theta$ being arbitrary. 
In this situation, as we emphasized
earlier, the energy of Eq. (\ref{H120}) 
does not change. This is the origin
of the large degeneracy that we have
been alluding to all along. 
Next, let us now consider the case when 
the system is uniformly oriented along an angle
that differs from the angles of Eq. (\ref{tn*}), i.e.,
$\theta_{i} = \theta^* \neq \theta_n^*$.
Now, if we perform a twist 
between any two consecutive planes separated,
e.g., $(\theta^* +  \delta \theta)$ on one plane of fixed $i_{z}$
and a uniform angle of $(\theta^* - \delta \theta)$
on a neighboring plane separated by one lattice
constant along the $z$ axis then as a simple
calculation shows the energy of Eq. (\ref{H120})
will be elevated. This simple picture
can be fleshed out in the full blown 
detailed calculation for the free energy of
the system about a chosen set of angles \cite{Nussinov04,Biskup05}.
Thus, the stratification (or stacking) ground state symmetry operation
of Eq. (\ref{symorb-e}) leads to the preference of the uniform 
states of Eq. (\ref{tn*}) over all others when thermal fluctuations
are included. Thus, while for all values of $\theta^{*}$,
a uniform spatial twist will lead to no energy
cost, a staggered twist in which consecutive
planes are rotated by $(\pm \delta \theta)$
costs no energy only for uniform 
states of Eq. (\ref{tn*}). 

Along similar lines of reasoning, 
if we consider the staggered state 
in which consecutive planes transverse
to the $z$ axis have the angles of Eq. (\ref{stag_eq})
then an additional staggered twist ($\pm \delta \varphi)$ of  the opposite
parity, i.e., one for which 
$\theta_{i_{z}} = \delta \theta (-1)^{i_{z}}  + \delta \varphi (-1)^{i_{z}+1}$,
will elevate the energy for general small $\delta \theta$ and $\delta \varphi$
(while, of course, the energy of a uniform state of, e.g., $\delta \theta =0$,
will not). This is, once again, the origin of the lower free energy
for a uniform state vis a vis a stratified one- there
are more low energy fluctuations about
the uniform states of Eq. (\ref{tn*})
then their stratified counterparts
with this increase being proportional
to the number of stratified interfaces
for which a twist was applied.  
  
 ``Blocking'' the lattice and 
employing reflection positivity
bounds \cite{Nussinov04,Biskup05}, it can indeed  be proven that the results
of the spin wave analysis are correct: the {\em free energy} has strict
minima for six uniform orientations \cite{Nussinov04,Biskup05}:
${\bm{T}}_{i}=\pm S{\bm{e}}_a,$ ${\bm{T}}_{i}=\pm S{\bm{e}}_b$, 
${\bm{T}}_{i}=\pm S{\bm{e}}_c$.
Thus, out of the large number of classical ground states,
only six are chosen. Orbital order already appears
within the classical ($S \to \infty$)
limit \cite{Nussinov04,Biskup05} and is not exclusively reliant on  
subtle quantum zero point fluctuations (captured by $1/S$ calculations
\cite{Tanaka05,Kubo02}) 
for its stabilization. Indeed, orbital order is detected 
up to relatively high temperatures
$({\cal{O}}(100 K)$) \cite{Tokura00,Murakami98}. Numerical 
work \cite{Dorier05} and an analysis with ``tilted'' boundary conditions \cite{Nussinov2012a} shows that quantum fluctuations do
{\em not lift the orbital degeneracy} in the simplest $S=1/2$ systems   -- 
the planar orbital compass model of Eq. (\ref{2dpocm}). A 2D pseudo-spin $T=1/2$ 
analogue of the cubic lattice120$^{\circ}$ compass model of Eqs. 
(\ref{120p}), a model of far less symmetry 
(and frustration) than the square lattice 90$^{\circ}$
compass model, has been shown to have a $S=0$ order \cite{Biskup05}. 
A mean-field analysis of the $T=1/2$ orbital compass
model on the square lattice \cite{Chen07a} suggests
that, at zero temperature, the symmetric point $J_{x} = J_{z}$
may mark a first order quantum transition, similar to the 1D case \cite{Brzezicki07}. 

\subsubsection{Quantum Order out of Disorder}

\begin{figure}
\begin{center}$
\begin{array}{cc}
\includegraphics[width=.6\columnwidth]{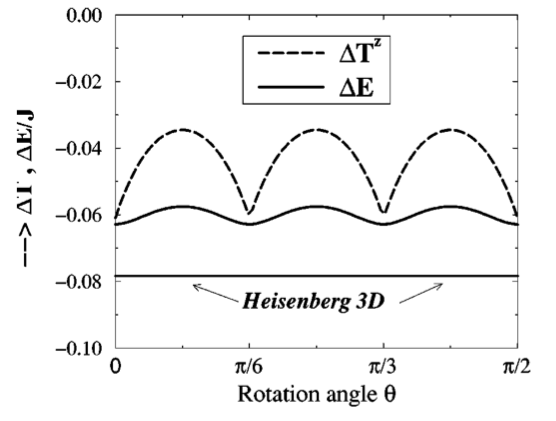} &
\vspace{0.5cm} \includegraphics[width=.39\columnwidth]{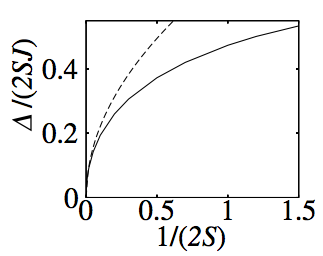} 
\end{array}$
\end{center}
\caption{Left: Quantum corrections for the cubic lattice 120$^{\circ}$ model system as functions of rotation angle $\theta$ for the renormalized order parameter $\Delta T^z$ (full lines) and the ground-state energy $\Delta E/J$ (dashed lines) \cite{Brink99}.
Right: the gap $\Delta$ as a function of 1/(2S) (solid curve) and the square root behavior at small 1/(2S) given by $\Delta^2/(2SJ)^2 = 0.49 / (2S)$, for pseudospin $S$  (dashed curve) \cite{Kubo02}.
}
\label{fig:Brink99_4}
\end{figure}

In certain geometrically frustrated systems, one encounters quantum order from disorder phenomena, that is, quantum fluctuations lifting the degeneracy of the ground states obtained within a mean field approach. Examples are the Heisenberg antiferromagnet on the triangular and pyrochlore lattice~\cite{Chubokov91,Tsunetsugu01}. The $120^\circ$ quantum compass model also exhibits this phenomenon, where quantum fluctuations not only select the ordered state, but also stabilize the selected state against thermal fluctuations which would destroy the ordering at finite temperatures.

If the ground state of the $120^\circ$ quantum compass model is considered to be ordered, the evaluation of the quantum corrections to the ground-state energy reveals pronounced minima for specific $\theta^*$, as illustrated in Fig.~\ref{fig:Brink99_4}. The quantum corrections to the energy in a $1/T$ expansion (also denoted as $1/S$ expansion in order to make a clear connection with the equivalent approach spin models.) and order parameter being finite, is consistent with the presumed presence of order \cite{Brink99,Kubo02}.

Thus globally rotating the pseudospins does not affect the energy of the classical ground state, which is therefore rotational invariant, but quantum corrections to the ground state energy restore the discrete symmetry of the Hamiltonian. When the quantum fluctuations are evaluated to lowest order the excitation spectra are found to be gapless and purely 2D, but higher order corrections cause the opening of an excitation gap of around 0.49 $J$~\cite{Kubo02}, which concurs with the quantum Monte Carlo simulations on this model and ints extensions~\cite{Rynbach10} will be reviewed in Sec.~\ref{sec:quantum120}.

\subsection{90$^{\circ}$ compass models}

We now focus on the planar and three-dimensional realizations of the 90$^{\circ}$ models in both the classical and quantum cases.

\subsubsection{Quantum Planar 90$^{\circ}$ Compass Models}

We first examine both the quantum 90$^{\circ}$ planar compass model. 
By the theorem reviewed in section \ref{mishpat} and, in particular, corollary I therein, at all positive temperatures, 
the average local ``magnetization'' $\langle {\bf \tau}_{i} \rangle =0$. In the quantum arena, this is so as the system admits the inversion symmetries of 
Eqs. (\ref{compass1}, \ref{symorb}), and thus, as reviewed in section \ref{exact90cds} and displayed in Fig. \ref{fig:Nussinov08}, 
insofar as the breaking of the Ising symmetries of Eqs. (\ref{symorb},\ref{compass1}), the system behaves as though it were one dimensional. 
As these Ising symmetries cannot be broken in $d=1$ dimensional symmetry, the finite temperature average $\langle {\bf \tau}_{i} \rangle =0$.
By contrast, bi-linears such as $\langle \tau^{x}_{i} \tau^{x}_{i+ {\bf e}_{x}} - \tau^{z}_{i} \tau^{z}_{i+ {\bf e}_{z}} \rangle$
are invariant under all of these $d=1$ symmetries and can attain non-zero values at finite positive temperatures \cite{Batista05,Nussinov05}.
Thus, nematic type order parameters may be constructed as linear combinations of these 
bi-linears. In particular, in a general anisotropic compass model [such as that of Eq.(\ref{eq:general_compass}) sans an applied field]
which we rewrite here (yet again) for clarity,
\begin{eqnarray} 
{\cal H}_{compass}=  -\sum_{i,\gamma}  J_\gamma \tau^\gamma_i \tau^\gamma_{i+\bm{e}_\gamma},
\label{eq:an_general_compass}
\end{eqnarray} 
the difference between the energy associated with bonds along the two lattice directions,
\begin{eqnarray}
\label{bond_diff_eq}
\langle J_{x} \tau^{x}_{i} \tau^{x}_{i+{\bm{e}}_{x} }- J_{y}  \tau^{y}_{i} \tau^{y}_{i+{\bm{e}}_{y}} \rangle
\end{eqnarray}
 may be used as
an order parameter \cite{Wenzel08}. 
In dimensions $D>2$, there are no $d=1$ symmetries of the quantum model (the symmetries of Eq. (\ref{compass1}, \ref{symorb})
are generally $d=(D-1)$ dimensional). As Ising symmetries can be broken in more than one-dimension, the local $\langle {\bf \tau}_{i} \rangle$ 
may be finite at low temperatures.

\subsubsection{Classical 90$^{\circ}$ Compass Models 
}
 
In the classical version of the 90$^{\circ}$ compass model in arbitrary spatial dimension, the considerations are identical. We elaborate on these below. 
As alluded to earlier (section \ref{sec:clas}), in considering the classical compass models, the Pauli operators ${\bf \tau}$ are replaced by a normalized
classical XY pseudo-spin ${\bf T}$ subject to Eq. (\ref{bnormal}), and the model becomes once again
of the form of Eq. (\ref{iso_comp1}). In the planar system, the lattice directions ${\bf e}_{\gamma }= {\bf e}_{1}, {\bf e}_{2}$. 
Along any line $\ell$ parallel to the lattice ${\bf e}_{\gamma}$ direction, the
classical planar system is trivially invariant under the global reflection (an identical 
Ising symmetry as that in the quantum case) about the $T_{\gamma}$ axis: $T_{i}^{\gamma' \neq \gamma} \to -T_{i}^{\gamma}, ~~ T_{i}^{\gamma} \to T_{i}^{\gamma}$ for all sites $i$ that lie such a line $\ell$. 
As such Ising symmetries cannot be broken in one dimension (for both the quantum and classical systems), they also cannot be broken, at finite temperatures, in the planar 
compass model and the local magnetization $\langle {\bf T}_{i} \rangle =0$. Similar to the quantum models, it is possible to construct nematic type two-site bilinears
such as that of Eq. (\ref{bond_diff_eq}) \cite{Wenzel08}. It is, in fact, also possible to construct single site quantities which are identical to those of the standard order parameters for classical nematic liquid crystals \cite{Nussinov05} 
which would be most appropriate for isotropic planar compass models (with $J_{\gamma} = J$ for all $\gamma$). In the planar case, a simple generalization of Eq. (\ref{bond_diff_eq}) is given by 
$Q= \langle J_{x} (T_{i}^{x})^{2} - J_{y}(T_{i}^{y})^{2} \rangle$. It is noteworthy that a quantity such as $Q$ is meaningful for all pseudo-spin representations of the planar compass model
with a pseudo-spin of size $S>1/2$. In the pseudo-spin 1/2 case, $Q$ is trivially zero.

\subsection{120$^{\circ}$ Honeycomb Model}
\label{120honey'}

We now discuss the system of Eq.  (\ref{eq:c120_honeycomb})
on the honeycomb lattice. \\

{\em Thermal fluctuations.} 

An order by disorder analysis for the classical version of the  
Hamiltonian of Eq. (\ref{eq:c120_honeycomb}) proceeds \cite{Nasu08,Wu08} along 
similar lines as of that in the section above for the cubic 
lattice 120$^{\circ}$ model \cite{Nussinov04,Biskup05}. By considering 
thermal fluctuations about a uniform state, it is seen that 
orientations with the values of Eq. (\ref{tn*} are preferred \cite{Nasu08}.
The underlying physics for the preference of these states 
(and the larger multitude of low energy states made
possible by stacking operations) is similar to our discussion
for the cubic lattice \cite{Nasu08}.

Work to date has not investigated thermal fluctuations about a 
non-uniform state such as that of panel (a) of Fig. \ref{fig:Wu08_3}
that resides in the sector of ground states that, as we reviewed above, are related by 
a local chiral emergent symmetry operation to each other. \\

{\em Quantum fluctuations.} 

The effect of quantum fluctuations (as seen in $1/S$ calculations) was investigated  
\cite{Wu08,Zhao08,Nasu08}.  The analysis is similar to that in the case of 120$^{\circ}$ model
on the cubic lattice. All investigations concluded that similar to the thermal fluctuation
analysis on this system \cite{Nasu08} and similar to the 120$^{\circ}$ system
on the cubic lattice, the preferred ground states are those of Eq. (\ref{tn*}).

A detailed calculation for the free energy due to thermal fluctuations (as well as the physical
considerations underlying the ``order by disorder'' mechanism as it favored by the application
of these symmetry operations in the ground state sector)  similar to that of that
of the cubic lattice 120$^{\circ}$ model discussed above 
shows that the low energy states are, once again, one of the six
uniform states of Eq. (\ref{tn*}).

\cite{Wu08} further considered fluctuations about
the non-uniform chiral state of Fig. \ref{fig:Wu08_3}
 with emergent chiral gauge symmetries  and found that
these had a lower free energy than those
resulting from fluctuations about 
the uniform states. The low free energy of these states is in accord
with the multitude of low energy fluctuations about them \cite{Wu08}. 
\cite{Wu08,Zhao08} both similarly also investigated the triangular and Kagome
lattice version of this system. Earlier work \cite{Mostovoy2002} introduced and examined the 
triangular ferromagnetic 120$^\circ$ model of Eq. (\ref{eq:c120_triangle})  to find that quantum fluctuations lift the degeneracy to
favor the six uniform pseudo-spin states.

\subsection{Effect of Dilution}

We conclude this section with a brief summary of some of the recent results on diluted (or ``doped") orbital compass-like systems \cite{Tanaka05,Tanaka07,Ishihara07,Tanaka09}. It was found  the critical doping fraction ($x=1/2$) necessary to remove order is smaller than the requisite doping needed to eradicate order in typical diluted magnets (e.g. KCu$_{1-x}$Zn$_x$F$_3$)\cite{Stinhcombe83,Breed70}; in typical magnetic systems, the
decrease in the ordering temperature and its saturation are governed by the percolation threshold (where the ordering temperature vanishes as the critical dopant concentration of $x_{c}$=0.69 for the simple cubic lattice). The faster degradation of orbital order with doping vis a vis simple percolation physics can be attributed to the directional character of the orbital exchange interactions. Similar effects have been found in related systems, 
as, e.g., in Ref. \cite{Honecker07}.

The concept of an {\em orbital order driven quantum critical point} was introduced \cite{Nussinov08b} by an  exact solution of diluted 2D and 3D orbital compass models. The solution relies on an exact gauge type symmetry
which results from dilution and the use of a bond algebra mapping \cite{Nussinov08b,nussinov-bond,cobanera,ADP,clock,holography}
wherein the system
is mapped onto decoupled one dimensional transverse field Ising chains \cite{Nussinov08b}
that exhibit quantum criticality at their isotropic point. The symmetries associated with the dilution increase the degeneracy of the system. Similar to charge and spin driven quantum critical fluctuations, orbital fluctuations may also drive the system to quantum criticality. The system may be driven to criticality by a combination of doping and uniaxial pressure/strain~\cite{Nussinov08b}. More recently, Ref.~\cite{Gang08} considered such a quantum critical point for spin-orbital singlets. An over-damped collective mode leading to non-Fermi liquid type response functions may emerge in systems that exhibit orbital ordering driven quantum critical points \cite{KaWaiLo2012}.
it can be shown that spin-glass type behavior can arise in doped orbital systems with random exchange constants. Here, the orbitals take on the role of spins in the usual spin-glass systems. 

In Section \ref{sec:diso}, we illustrated how low temperature orders in compass systems may be triggered by thermal and/or quantum fluctuations. We now remark 
on the opposite limit-- that of high temperatures. As illustrated in \cite{holography,Chakrabarty11} the high temperature limit of compass (and other) systems
as evinced by general correlation functions and thermodynamics 
coincides with that of the large $n$ (or spherical model) solution.  In the large $n$ limit, all thermodynamic quantities are directly given by integrals of
simple functions involving eigenvalues of the kernel $\hat{V}({\bf k})$ of Eq. (\ref{Fourier_space}). A brief review of some aspects of this limit is
provided in Section \ref{appendixB}. Flat bands, such as those discussed in 
Section \ref{new_theorem_on_bands}, in which these eigenvalues $v_{\alpha}({\bf k})$ depend on a reduced number of Cartesian components
of ${\bf{k}}$ lead, in the large $n$ or high temperature limit, to exact dimensional reductions (to a system whose dimensionality is given 
by the number of components of ${\bf{k}}$ on which $v_{\alpha}({\bf k})$ depends. Bolstered by their unique high temperature limit in which compass
models may effectively exhibit a reduced dimensionality, all large $n$ renditions
of the compass models that we considered are disordered. In Section \ref{sec:compass_phase_diagram}, we next discuss
the precise character of the transitions in a multitude of compass models between their low and high temperature phases.

\subsection{High Temperature Correlations \&  Dimensional Reduction}
\label{sec:high_t_section}

In the previous Section, it was illustrated how low temperature orders in compass systems may be triggered by thermal and/or quantum fluctuations. We now remark 
on the opposite limit-- that of high temperatures. As illustrated in \cite{holography,Chakrabarty11} the high temperature limit of compass (and other) systems
as evinced by general correlation functions and thermodynamics 
coincides with that of the large $n$ (or spherical model) solution.  In the large $n$ limit, all thermodynamic quantities are directly given by integrals of
simple functions involving eigenvalues of the kernel $\hat{V}({\bf k})$ of Eq. (\ref{Fourier_space}). A brief review of some aspects of this limit is
provided in Section \ref{appendixB}. Flat bands, such as those discussed in 
Section \ref{new_theorem_on_bands}, in which these eigenvalues $v_{\alpha}({\bf k})$ depend on a reduced number of Cartesian components
of ${\bf{k}}$ lead, in the large $n$ or high temperature limit, to exact dimensional reductions (to a system whose dimensionality is given 
by the number of components of ${\bf{k}}$ on which $v_{\alpha}({\bf k})$ depends. Bolstered by their unique high temperature limit in which compass
models may effectively exhibit a reduced dimensionality, all large $n$ renditions
of the compass models that we considered are disordered. In Section \ref{sec:compass_phase_diagram}, we next discuss
the precise character of the transitions in a multitude of compass models between their low and high temperature phases.  

\section{Phases \& Phase Transitions in Compass Models} 
\label{sec:compass_phase_diagram}

Transitions correspond to singularities in the free energy. When possible, transitions are most easily ascertained when an order parameter is found whose value differs from zero in a symmetry broken phase. This is not the case  for gauge theories that exhibit finite temperature transitions but do not have a simple corresponding order parameter \cite{Kogut, Fredenhagen86,Bricmont1983} as they display local ($d=0$) symmetries which according to our earlier discussion cannot, by Elitzur's theorem, be broken at any finite temperatures \cite{Elitzur75} due to an effective dimensional reduction \cite{Batista05,holography}. Via this extension of Elitzur's theorem concerning generalized dimensional reduction, topological order (see Section \ref{global_inter}) can be established in numerous
systems including, in particular, numerous compass models \cite{PNAS,AOP}.  In systems with topological orders (see Section \ref{global_inter}), analogs \cite{Cobanera2013,Gregor2011}
of the quantities discerning phases in gauge theories \cite{Kogut, Fredenhagen86,Bricmont1983} may be considered.
As reviewed in sections (\ref{sec:sym},\ref{sec:diso}), at low temperatures, most compass models exhibit broken symmetry states in which discrete symmetries of the compass Hamiltonians are broken.
While there are notable exceptions, such as Kitaev's model of Eq. (\ref{eq:HKT}) which, as we will review in section (\ref{sec:kit}) 
[and section (\ref{kith}) in particular], may (for some range of couplings) exhibit no ordered phases (or ``spin-liquid'' type states) down to zero temperature, 
the majority of the compass models exhibit low temperature broken symmetries. 
While symmetry arguments are powerful and while, as discussed in section (\ref{sec:diso}), it may be possible to rigorously prove the existence
of a phase transition, it is of great interest to get more insight on the qualitative and 
quantitative character of the transitions that these systems display by performing 
direct numerical and analytical analysis of various sorts. Both numerically
and analytically, this task is daunting as these systems are highly frustrated.
Moreover, numerically, many variants of the compass models currently suffer
from the ``minus sign'' problem.  

Many results have been attained in particular for the simpler compass models.
However, many more, including models pertinent to orbital ordering, 
are currently unknown. 

Below we review the results known to date on nearly all compass models. We reserve reviewing the Kitaev, and the related Kitaev-Heisenberg, and compass Heisenberg models to
sections (\ref{HKCS}, \ref{compass_heisen},\ref{sec:kit}).

We start with a summary of results on the classical
models and then turn the attention to the quantum systems. 

\subsection{90$^{\circ}$ Compass Models}
\subsubsection{Classical Square Lattice}
For ease, we rewrite anew the classical planar 90$^{\circ}$ model. [The general dimensional extension of this system was given in Eq. (\ref{highD90}).]
This planar system is typically defined on a square lattice and has its Hamiltonian given by
\begin{eqnarray}
\label{90ccc+}
H_{\square}^{~\sf classical ~90^\circ}=  - J_{x} \sum_{\langle ij \rangle_H} T^x_i T^x_j - J_{y} \sum_{\langle ij \rangle_V} T^y_i T^y_j,
\end{eqnarray}
with $\langle i j \rangle_{H}$ and $\langle i j \rangle_V$ denoting nearest neighbor links along the horizontal and vertical directions respectively.
Eq. (\ref{90ccc+}) is simply the classical counterpart of the quantum model of Eq. (\ref{2dpocm}). 
 
In the 90$^{\circ}$ compass model, unlike the 120$^{\circ}$ compass model, attention is required in order
to examine contending order parameters.
The sole symmetry of high dimension which can be broken
in the 90$^{\circ}$ compass model on the square lattice
is an Ising type reflection symmetry of the symmetric
compass model (with equal exchange constants along
the x and y directions, $J_{x}=J_{y} (=J)$) that involves
a global ($d=2$ dimensional) reflection of
all pseudo-spins in the plane. 
Formally, such a symmetry is given by
\begin{eqnarray}
O_{\sf Reflection} = \prod_{r} 
e^{i \pi \frac{\sqrt{2}}{4} (\sigma^{x}_{r} + \sigma^{y}_{r})}.  
\label{OR}
\end{eqnarray}
This global Ising reflection symmetry is related to a (self-)duality [$J_{x} \leftrightarrow J_{y}$] between the 
couplings. Along the self-dual line, $J_{x}=J_{y}$, the duality between the x and y bonds becomes
a symmetry [as in general self-dual systems \cite{cobanera,ADP}]. As such a $d=2$ dimensional
Ising type symmetry can be broken at finite temperature, this reflection
symmetry can (and indeed is) broken at finite temperatures.
However, the order parameter cannot be of the usual single
site type.  By the symmetry arguments that we outlined in section \ref{sec:sym},
it is clear that while spontaneous symmetry breaking of the pseudo-spin 
on a single site ($i$) is prohibited ($\langle {\bm T}_{i} \rangle =0$)
in the planar 90$^{\circ}$ compass model,
any quantity that is invariant under  all $d=1$ dimensional symmetries 
might serve as an order parameter. This implies that one should
consider quantities involving more than one on-site operator.

Indeed, $d=1$ symmetry invariant, low temperature nematic type order
is stabilized in this system by thermal fluctuations \cite{Nussinov04};
the physical considerations are similar to those presented earlier for the 120$^{\circ}$ compass
model in subsection \ref{tf}. An elegant study of the classical 
two dimensional 90$^{\circ}$ compass model was pursued in \cite{Mishra04}.   
Similar to the entropic stabilization in the 120$^{\circ}$ model, \cite{Nussinov04, Biskup05} 
a (pseudo)spin wave type dispersion about state with a particular uniform orientation $\theta^*$ 
of all of the classical pseudospins ${\bm{T}}_{i}$ may 
be computed. For the 90$^{\circ}$ square lattice compass model of Eq. (\ref{90ccc+}), the dispersion about $\theta^*=0$ 
is given by 
\begin{eqnarray}
m+  \gamma_{x} (1- \cos k_{x}) + \gamma_{y} (1- \cos k_{y}),
\end{eqnarray}
with $m$ and $\gamma_{x,y}$ denoting a self-consistent (pseudo)spin gap and moduli along the $x$ and $y$
axis respectively. At low temperatures, these scale as \cite{Mishra04}
\begin{eqnarray}
\gamma_{x} &=&0, ~ ~  \gamma_{y}= 1 -{\cal{O}}(T^{2/3}), \nonumber \\
m(T) &=& \frac{1}{2} T^{2/3}+ {\cal{O}}(T).
\end{eqnarray}
To emulate the ordering transition in a qualitative way, \cite{Mishra04} studied
the ``four-state Potts compass model'' given by
\begin{eqnarray}
H= -J \sum_{i} (n_{i \mu} n_{i+{\bf{e}}_{x} \mu} \mu_{i} \mu_{i + {\bf{e}}_{x}} 
+ n_{i  \nu} n_{i + {\bf{e}}_{y} \nu}  \nu_{i} \nu_{i+{\bf{e}}_{y}}), \nonumber
\end{eqnarray} 
where at each lattice site $i$ there are occupation numbers $n_{i \nu} =0,1$ and $n_{i \mu} =0,1$
for which $n_{i \mu} + n_{i \nu} =1$ and $\mu, \nu$ are classical Ising variables
($\mu = \pm 1$, $\nu = \pm 1$). This Hamiltonian captures the quintessential directionality
of the bonds in the compass model. By tracing over the Ising variables
$\mu$ and $\nu$ at all sites, this four state Potts compass can be mapped
onto the two dimensional Ising model from which it can be deduced that the Potts compass
model has a critical temperature of  \cite{Mishra04} $T_c= 0.4048 J.$

Ordering at lower temperatures corresponds to a dominance of
horizontal bonds over vertical ones or vice versa. That is, for temperatures below the 
critical temperature 
\cite{Mishra04}
\begin{eqnarray}
\langle n_{i , \mu}  \rangle - \langle n_{i, \nu} \rangle \neq 0.
\end{eqnarray}
In effect, this reflects an order of the nematic type present in the classical 
90$^\circ$ compass at low temperatures 
in which the four fold rotational symmetry of the square lattice is lifted. 
A natural nematic type 
order is given by \cite{Mishra04}
\begin{eqnarray}
q= \langle (T_{i}^{x})^{2} - (T_{i}^{y})^{2} \rangle.
\label{qm}
\end{eqnarray}
Using Monte Carlo calculations, it was found \cite{Mishra04} that this quantity $q$
becomes non-zero for temperatures lower than an estimated transition temperature 
of $T_{c} = (0.147 \pm 0.001) J$. Tour de force calculations further improved this estimate \cite{Wenzel08, Wenzel10} 
to a value for the classical  90$^{\circ}$  compass model of 
$
T_{c}= 0.14612 J.
$

In the 90$^{\circ}$ compass models (whether classical or quantum), related nematic type
order is also characterized by the energy difference between the 
vertical and horizontal bonds,
\begin{eqnarray}
\langle Q_{i} \rangle  \equiv \langle T_{i}^{x} T^{x}_{i+ {\bf{e}}_{x}} - T_{i}^{y} T^{y}_{i+{\bf{e}}_{y}} \rangle.
\label{energy_diff_Q}
\end{eqnarray}
The virtue of this form by comparison to that of Eq. (\ref{qm}) is that can be extended to 
quantum pseudo-spins $T=1/2$. Near a general critical point (including the one at hand for the 90$^{\circ}$ compass model in the vicinity of its
critical temperature), the connected correlation function canonically behaves as 
\begin{eqnarray}
\label{QQknow}
\langle Q_{i} Q_{j} \rangle - \langle Q_{i} \rangle \langle Q_{j} \rangle \simeq  \frac{e^{-r_{ij}/\xi}}{|r_{ij}|^{p}},
\end{eqnarray}
with $Q_{i}$ the corresponding local order parameter that attains a non-zero average value ($\langle Q_{i} \rangle$) in the ordered phase. 
In Eq. (\ref{QQknow}), $r_{ij}$ is the distance between sites $i$ and $j$, and $\xi$ is the correlation
length, $A$ is an amplitude, and $p$ a power.  Typically, a susceptibility $\chi = \langle Q^{2} \rangle - \langle Q \rangle^{2}$ (with $Q = \sum_{i=1}^{N} Q_{i}/N$)
diverges at the critical point.  The classical 90$^{\circ}$ compass model was indeed found to fit this form with $Q_{i}$ chosen to be the local nematic type order parameter of 
Eq. (\ref{energy_diff_Q}). As was discussed in section \ref{sec:sym}, any generally non-zero quantity (as such, involving any number of  bonds 
\cite{cobanera, nussinov-bond}) that is invariant under all low dimensional
gauge like symmetries can serve as an order parameter. That is, general composites
of such bonds can serve as order parameters. \cite{Batista05} A similar very interesting
measure was introduced in \cite{hidden-dimer} for the quantum 90$^{o}$ compass model.

Although order sets in at a temperature 
far lower than that of the two dimensional Ising model and its equivalent four state Potts clock model 
the transition was numerically found to be in the two-dimensional Ising universality 
class \cite{Mishra04, Wenzel08, Wenzel10}. 
The standard critical exponents
that describe the divergence of the correlation length ($\nu$)
and susceptibility $(\gamma$) as the temperature
approaches the critical temperature $T_{c}$,
\begin{eqnarray}
\xi \sim |T-T_{c}|^{-\nu}, ~~~
\chi \sim |T-T_{c}|^{-\gamma}.
\label{stan_exp}
\end{eqnarray}
For the two dimensional 
Ising model and all systems that belong to its universality class
are given by $\nu_{2D~ Ising} =1$ and $\gamma _{2D~Ising}= 1.75$.  These two
exponents were numerically measured in \cite{Wenzel10}. From any two exponents,
the values of all other exponents follow by scaling relations (in 
this case the values of all other critical exponents are identical to
those of the two-dimensional Ising model). Earlier work \cite{Mishra04} 
found Binder cumulants similar to those
in the two dimensional Ising model as a specific heat collapse
which is also similar to that of the two-dimensional Ising
model. This two-dimensional Ising type transition
is consistent with the transition in the Potts clock model
on the square lattice.

A technical issue that reflects the unusual nature of the system 
(its high degree of symmetry and proliferation of degenerate and 
nearly degenerate states) is that 
finite size effects are of far greater dominance 
here than in usual systems. \cite{Wenzel08, Wenzel10}
The most successful boundary conditions found to 
date to numerically study these systems are 
the so called ``screw periodic boundary conditions'' \cite{Wenzel10}
in which there is periodicity along a line that wraps around the system with a 
general non-zero pitch.

\subsubsection{Quantum Square  Lattice}
The planar $T=1/2$ planar 90$^{\circ}$ compass model of 
Eq. (\ref{2dpocm}) was investigated by
multiple groups using a variety of tools. The results to date belong
to two inter-related subclasses:  (i) The character of the
finite temperature transition between a low temperature
ordered state and the disordered high temperature phase
in the symmetric  ($J_{x}= J_{y} (=J)$) 90$^{\circ}$ compass model for which the 
global $d=2$ Ising type reflection symmetry can be broken 
and (ii) studies of the zero temperature transition 
in the extended anisotropic 90$^{\circ}$ compass model
of Eq. \ref{eq:general_compass} in the absence 
of an external field ($h=0$) at the
point $J_{x} = J_{y}$. As in the classical system, In the anisotropic 90$^{\circ}$ quantum compass model,  $J_{x} \neq J_{y}$, the global
reflection symmetry is not present. the sole symmetries
that remain  in the anisotropic model relate to the $d=1$ Ising type symmetries
of Eq. \ref{symorb}.

\paragraph{Finite temperature transitions}
\label{ft9pcm}
A few direct studies were carried out \cite{Wenzel08, Wenzel10}
on the finite temperature breaking of the ($d=2$ Ising type) reflection symmetries
in the symmetric ($J_{x}= J_{y}$) 90$^{\circ}$ compass model. 
The calculations of \cite{Wenzel08, Wenzel10} employed an order
parameter akin to Eq. (\ref{energy_diff_Q}) and a its related susceptibility to 
find that the two-dimensional 
quantum pseudospin $T=1/2$,~ 90$^{\circ}$ compass system also belongs to the universality class
of the classical two dimensional Ising model. While the exponents characterizing the 
transition are identical to those in the classical two dimensional Ising model
and thus also of the classical two dimensional 90$^{\circ}$ compass model,
the critical temperature is significantly reduced once again. The reduction
in the critical temperature is, however, far more severe in the quantum
case than in the classical rendition of the 90$^{\circ}$ compass model.
Specifically, within numerical
accuracy \onlinecite{Wenzel10} find for the quantum 90$^{o}$ compass model 
$
T_{c} = 0.0585 J.
$
Different numerical fitting schemes (e.g., allowing the critical (correlation length) exponent 
$\nu$ to differ from its value of $\nu =1$ and using it as an adjustable parameter)
lead to only an incremental shift in the value of the ascertained critical temperature (i.e., a shift only in the last decimal place). 
The factor of approximately 0.4 difference between the quantum $T=1/2$
compass model critical temperature value and the classical value shows that, at
least, in this simple compass models, quantum fluctuations inhibit 
order rather than fortify it contrary to what was thought some time 
ago to be universally true for compass models (and certain other highly frustrated
spin systems).

A slightly less accurate (by comparison to the numerical
values above) yet quite insightful and intensive high temperature series expansion 
\cite{oit11} to order $\beta^{24}$ in the inverse temperature $\beta=1/(k_{B} T)$
led to a similar value for $T_{c}$ ($T_{c} = 0.0625 J$. This was achieved by determining when 
the inverse susceptibility $\chi^{-1}$, evaluated with Pade approximants, extrapolated to zero.
By fitting the determined susceptibility from the high temperature 
series expansion with the standard form of Eq. (\ref{stan_exp}) 
while setting $T_{c}$ to the numerical value,
the critical exponent $\gamma$ was found to be 1.3 (of the same order of the
two-dimensional Ising value of $\gamma = 1.75$ yet still a bit removed from it) \cite{oit11}.

\paragraph{Zero Temperature Transitions}

Before focusing on transitions between ground states,  
we regress to a very simple discussion concerning the unimportance 
of the sign of the couplings $J_{x}$ and $J_{y}$ within the quantum (and classical) 90$^{o}$ model on the square lattice. This is so, as in other two component pseudo-spin systems,  
it is possible
to invert the sign of the individual couplings $J_{x}$ or $J_{y}$ (or both simultaneously 
as in Eq. \ref{ABAB}) by simple canonical transformations. In order to, e.g., set $J_{x} \to -J_{x}$
we may rotate all of the pseudo-spins that lie on odd numbered columns (wherein $i_{x}$- the x component
of the site $i$- is an odd integer) by 180$^{\circ}$ about the $\tau^{y}$ axis.  The simple
transformation 
\begin{eqnarray}
U = \prod_{i_{x} = odd} \exp(i \pi \tau_{i}^{y}/2)
\end{eqnarray}
implements this transformation.  One may, of course, similarly
rotate by 180$^{\circ}$ all pseudo-spins on odd numbered rows (odd $i_{y}$) to effect
$J_{y} \ to -J_{y}$. The combined effect of both transformations is encapsulated in 
the sublattice rotation of 
Eq. (\ref{ABAB}) as a result of which all of the exchange couplings have their sign flipped.
In the below we will at times refer to the system for positive $J_{x}, J_{y}$
and sometimes for general real $J_{x}$ and $J_{y}$.  Using the above
transformations, the results for positive $J_{x}$ and $J_{y}$ imply
identical conclusions for all $J_{x}$ and $J_{y}$ once their
modulus ($|J_{x,y}|$) is considered. 

The very existence of a finite temperature two-dimensional Ising type 
critical point within the symmetric 90$^{\circ}$ planar 
compass model ($J_{x}=J_{y}$)- both in the classical (proven by entropy stabilization
with detailed numerical results and further analysis) and quantum renditions
(thus far supported by numerical results alone)-  allows for, but does not prove, 
that for temperatures $T<T_{c}$ there may be a line of first order
transitions along the temperature axis when $J_{x} = J_{y}$. Across this line the system 
may switch from preferring ordering along the $x$ direction (when $|J_{x}|>|J_{y}|$) 
to ordering of the pseudo-spin parallel
to the $y$ direction (when $|J_{x}|<|J_{y}|$). The situation is reminiscent of, amongst other
systems, the ferromagnetic two dimensional Ising model in a magnetic field $h$,
\begin{eqnarray}
H = - J \sum_{\langle i j \rangle} \sigma_{i} \sigma_{j} - h \sum_{i} \sigma_{i}.
\end{eqnarray}
At $T=T_{c}$, the system is critical with the two-dimensional Ising model
critical exponents for small $|T-T_{c}|$ for $h=0$. For all temperatures 
$T<T_{c}$, there is a line of first order transitions along the 
temperature axis when $h=0$ where the system switches from preferably 
order with positive magnetization $\langle \sigma_{i} \rangle>0$ (when $h>0$) to negative 
magnetization $\langle \sigma_{i} \rangle<0$ (when $h<0$). Across the$h=0$ line for
$T <T_{c}$, there is a discontinuous jump in the value of $\langle \sigma_{i} \rangle$
between its values at $h=0^{+}$ and $h=0^{-}$
marking the first order transition. 

Similarly, establishing the existence of a first order phase transition in the $T=0$ system
as a function of $(|J_{x}|-|J_{y}|)$ when $|J_{x}|=|J_{y}|$ would suggest 
(but not prove) the existence of a finite temperature critical point $T_{c}>0$
at which the line of phase transitions terminates and above which ($T>T_{c}$),
the system exhibits no order of any kind. At arbitrarily high temperatures $T \gg J_{x},J_{y}$
the system is, of course, disordered. 

A natural question then concerns the direction investigation of the $T=0$ transition at 
$J_{x}=J_{y}$. We note that one approach for analyzing the character of the transition
at the point $J_{x}=J_{y}$ in the quantum system
would be to analyze the 2+1 dimensional corresponding
classical Ising model of Eq. \ref{Ising_action}. A first order transition
would suggest the possibility of a finite temperature critical point $T_{c}>0$
as seen by numerical studies. 

Many other approaches to investigate the zero temperature
transition have been put forth. The upshot of these studies 
is that the zero temperature transition at the both $J_{x}=J_{y}$
is indeed first order. As in the classical system, $J_x \leftrightarrow J_y$
is a ``self-dual'' transformation of the quantum system \cite{Nussinov05, Nussinov06, cobanera, ADP}
and the transition in question pertains to the system at its self-dual point. 

As any other zero temperature transition, 
the zero temperature transition at $J_{x}= J_{y}$ in the 90$^{\circ}$ compass model
corresponds to ``level crossing'' at which the low energy state(s) change from being
of one type for $J_{x}>J_{y}$ to another type for $J_{x}<J_{y}$. At the point $J_{x}= J_{y}$,
their energy levels cross. In order to understand the level crossing, one needs to understand
the structure of the low energy levels in general. 

In Section \ref{sec:exact_emergent}, we earlier reviewed the non-commutativity of the symmetries of 
Eq. \ref{symorb} as applied to the two dimensional 90$^{\circ}$ model
(where the planes $P_{\gamma}$ are one dimensional lines
orthogonal to the $\gamma$ axis) on all lattices as well as time reversal symmetry 
as applied to odd sized lattices both imply (at least) two-fold degeneracy 
of the ground state sector. (As it turns out, the two considerations are not independent.
Time reversal symmetry can be directly expressed in 
terms of the symmetries of Eq. \ref{symorb} \cite{PNAS}.)
 This implied two-fold degeneracy appears also
in the anisotropic case of $J_{x} \neq J_{y}$. 
The ground states can be characterized in terms of the set of eigenvalues $\{ \lambda_{1}, \lambda_{2}, ..., \lambda_{L}$ of, say, the $L$ symmetries of 
Eq. \ref{symorb} corresponding to vertical planes $P$ \cite{Ioffe05, Dorier05}. 
All of these symmetries commute with one another (while, 
as just highlighted below, anti-commuting with all
of the symmetries of Eq. \ref{symorb} corresponding to horizontal planes $P$).
The application of any horizontal plane symmetry will generate another ground states
with {\em all} of the eigenvalues flipped, 
$\lambda_{i} \to -\lambda_{i}$. 

The large number of symmetries ($(2L)$ for an $L \times L$
lattice) of the form of Eq. \ref{symorb} allows for (and, in fact, mandates \cite{Nussinov2012a}) a degeneracy which is exponential in the perimeter.
Crisp numerical results illustrate \cite{Dorier05} that in the square lattice 90$^{\circ}$ compass model, each level is $2^{L}$-fold  degenerate for $J_{x} \neq J_{y}$ and is $2^{L+1}$-fold degenerate when $J_{x} = J_{y}$).
This degeneracy rears its head in the thermodynamic limit $L \to \infty$. For finite $L$,
these states split to form a narrow band. There is a gap of size ${\cal{O}}(e^{-L/L_{0}})$, 
with a fixed length scale $L_{0}$, that separates the ground states from the next excited 
state \cite{Ioffe05, Dorier05}. In the thermodynamic limit,  these sets of $2^{L}$ degenerate 
states further merge at the point $J_{x}=J_{y}$ to form bands of $2^{L+1}$ degenerate states. Numerical and other analysis illustrates that the level crossing at $J_{x}=J_{y}$ is related to
a first-order (or discontinuous) transition  
of the lowest energy state as a function of ($J_{x}-J_{y}$) \cite{Urus09, Chen07a,Dorier05}.The
two sets of states for positive and negative values $(J_{x}- J_{y})$ are related to one another by the global Ising type reflection symmetry
of the 90$^{o}$ compass model which exchanges $J_{x} \leftrightarrow J_{y}$.
Particular forms for this global symmetry were written down in \cite{Nussinov05, AOP, Urus09}.
In essence, these correspond, e.g., to rotations in the internal pseudo-spin space 
about the $T^{z}$ axis by an angle of 90$^{\circ}$
or by 180$^{\circ}$ about the 45$^{\circ}$ line in the $(T^{x}, T^{y})$ plane compounded by an overall
external reflection of the lattice sites about the 45$^{o}$ line on the square lattice or a rotation
by 90$^{\circ}$ about the lattice $z$ axis that is orthogonal to the square lattice plane. 
The first order transition at $J_x = J_y$ found by various groups represents the crossing of two bands that are related by this global symmetry.
Similarly, 
although by the considerations outlined in earlier sections, 
$\langle T^{x,y}_{i} \rangle =0$ at any positive temperature,
within the ground state, ${\bf T}_{i}$ can attain a non-zero
expectation value. It is seen that the ``magnetization components" $\langle T^{x,y} \rangle$
exhibit a discontinuous jump at the point $J_{x} = J_{y}$ \cite{Dorier05, Urus09}.
[For $J_{x} > J_{y}$, the expectation value $\langle T^{x} \rangle$ is strictly positive;
this expectation value jumps discontinuously to zero when $J_{x} =J_{y}$
(and remains zero for all $J_{x} < J_{y}$). Similar results
are found when exchanging $J_{x} \leftrightarrow J_{y}$ and
$\langle T^{x} \rangle \leftrightarrow \langle T^{y} \rangle$.]
The free energy is similarly found to exhibit a discontinuity
in first derivative relative to $(J_{x}-J_{y})$ at the point
$J_{x}=J_{y}$ \cite{Urus09}.

It is also interesting to note that for when $J_{x}>J_{y}>0$,
the ground states $|\psi \rangle$ were found to be an eigenstate of the 
$T^{x}$ related symmetry operators of Eq. \ref{symorb}
with an eigenvalue of $(+1)$. That is, for the pseudo-spin $T=1/2$
analyzed, \cite{Urus09}
\begin{eqnarray}
\prod_{i_y, ~\mbox{fixed}~ i_{x}} \tau^{x}_{i} | \psi \rangle = + |\psi \rangle.
\end{eqnarray}
Similarly, for $J_{y}>J_{x}$ the same occurs with $x$ and $y$ interchanged,
\begin{eqnarray}
\prod_{i_x, ~\mbox{fixed}~ i_{y}} \tau^{y}_{i} |\psi \rangle = + |\psi \rangle.
\end{eqnarray}
A symmetry analysis starting from the decoupled chain limit is provided in \cite{Ioffe05, Dorier05}. 

An analytic mean-field type approximation was invoked by \cite{Chen07a} to the 
fermionic representation of the 90$^{o}$ compass model. In general, fermionization
cannot be done a useful way in dimensions larger than one. That is, on general lattice
a fermionization procedure (known as the Jordan-Wigner transformation) 
wherein pseudo-spins (or spins) are replaced by
spinless fermions gives rise, in spatial dimensions larger than one, to a system with arbitrarily long range interactions.
In the case of the 90$^{o}$ compass model, however, the special form of the interactions
and consequent symmetries of Eq.(\ref{symorb}) enable a reduction to a fermionic system in two-dimensions with local terms. The resulting fermionic Hamiltonian \cite{Chen07a} contains
both hopping and pairing terms along single (e.g., horizontal) chains. The chains
interact with one another along a transverse direction (e.g., vertical) 
via a nearest neighbor type density-density attractions ($J_{y}>0$)
or repulsion ($J_{y}<0$). The fermionic Hamiltonian reads 
\begin{eqnarray}
H =- \sum_{i} \Big[ J_{y} n_{i} n_{i + {\bm{e}}_{y}}  - J_{y} n_{i} \nonumber
\\ + \frac{J_{x}}{4} (c_{i}- c_{i}^{\dagger})
(c_{i+{\bm{e}}_{x}} + c^{\dagger}_{i+{\bm{e}}_{x}}) \Big].
\label{fermi-compass}
\end{eqnarray} 
The fermionic Hamiltonian of Eq. \ref{fermi-compass} was analyzed by a self-consistent
mean field type analysis and the analysis of these results to perturbations
beyond mean field  \cite{Chen07a}. This very interesting work suggests that a first order is indeed
present at $J_x = J_y$.  The self-consistent mean-field type calculation suggests
that the average values of $\langle T^{x,z} \rangle$
exhibit a discontinuous jump. 
This analytical result is in accord with the numerical approaches of
\cite{Dorier05, Urus09}. We pause to re-iterate and remark that while fermionization 
giving rise to local interactions is generally impossible in canonical systems,
in compass type systems fermionization is possible. A similar
occurrence will be encountered in Kitaev's honeycomb model where in fact the fermionization will enable us
to solve the problem exactly in different topological charge sectors whose content
will be explained later on. The possibility of fermionization in these systems in rooted
in the simple ``bond algebra'' that the interactions along different bonds satisfy further giving rise
to symmetries (giving rise to local conserved topological charges in Kitaev's model) 
\cite{nussinov-bond, cobanera, ADP}. 
As will be discussed later in connection to Kitaev's model, a direct Jordan-Wigner transformation
is not necessary in order to cast these and more general pseudo-spin (or spin) systems into a system
with local interactions that contains spinless fermions. 

The quantum 90$^{\circ}$ compass models that we have thus far focused on, were of
pseudo-spin $T=1/2$. For integer pseudo-spin $T=1,2,...$, all of the symmetries
of Eq. \ref{symorb} commute with one another. Unlike the case of all half odd integer pseudo-spins
where the anticommutator $\{\exp(i \pi T_{x}), \exp(i \pi T_{y})\} =0$ for integer $T$, the commutator
$[\exp(i \pi T_{x}), \exp(i \pi T_{y})]=0$. Thus, for integer pseudo-spin $T$, 
the two types of symmetry operators of Eq. \ref{symorb} 
with the two different possible orientations for the planes (in this case lines) $P_{\gamma}$
corresponding to vertical columns and horizontal rows commute with one another. 
As noted by \cite{Dorier05}, in this case the pseudo-spin 
$T=1/2$ argument concerning a minimal two-fold degeneracy
as a result of the incompatibility of the symmetry operators of Eq. (\ref{symorb}) 
no longer holds and a non-degenerate ground state can arise.
Indeed, numerical calculations on small finite size systems \cite{Dorier05} found the ground
state to be non-degenerate. In a similar fashion, time reversal no longer implies a two fold
degeneracy for integer pseudospin $T$ as it does for all half odd integer pseudospin values \cite{PNAS, AOP}. 
As in the considerations discussed in subsection \ref{inter_expt}, 
the $d=1$ symmetries of this system imply a degeneracy, for ``tilted'' boundary conditions, which is exponential in the system perimeter \cite{Nussinov2012a}.
Such boundary conditions may emulate the square lattice in the thermodynamic limit. 

We close this subsection by remarking that a solution of a one-dimensional (1D) variant of the quantum planar 90$^{\circ}$ compass model \cite{Brzezicki07} 
further illustrates how the energy spectrum collapses at the {\it quantum 
phase transition\/} between two possible kinds of order, with either 
$\sigma^z$-like or $\sigma^x$-like short-range correlations, and is thus 
highly degenerate, similar to the 2D case where, as alluded to above, the degeneracy 
scales exponentially in the perimeter size (i.e. as ${\cal{O}}(2^{L})$).

\subsubsection{Classical Cubic Lattice}

For the classical three dimensional 90$^{\circ}$ compass model, the existence of $d=1$ symmetry invariant nematic order
can be established, via entropic stabilization calculations along the same lines 
as for the classical 120$^{\circ}$ model \cite{Nussinov04}.  Clear signatures of nematic order were seen in
Monte Carlo simulations \cite{Wenzel2011a}. A particular three dimensional
extension of Eq. (\ref{energy_diff_Q}) was considered by Wenzel and Lauchli,
\begin{eqnarray}
Q_{WL} = N^{-1} \langle 
(\sum_{i} (T_{i}^{x} T^{x}_{i+ {\bf{e}}_{x}} -  T_{i}^{y} T^{y}_{i+{\bf{e}}_{y}}))^{2} \nonumber
\\ +  (\sum_{i}( T_{i}^{y} T^{y}_{i+{\bf{e}}_{y}} -
T_{i}^{z} T^{z}_{i+{\bf{e}}_{z}}))^{2} \nonumber
\\ + (\sum_{i}( T_{i}^{x} T^{x}_{i+ {\bf{e}}_{x}} - T_{i}^{z} T^{z}_{i+{\bf{e}}_{z}}))^{2} \rangle,
\label{qwl}
\end{eqnarray} 
with (as throughout) $N$ denoting the total number of sites in the lattice.
A discontinuous transition appeared at an ordering transition temperature $T_{o} \simeq 0.098 J$. That is, the 
nematic-type order parameter of Eq. (\ref{qwl}) was finite just below $T_{o}$
and exhibits a discontinuous jump at $T_{o}$. As noted by \cite{Wenzel2011a},
when present, the detection of a first order transition via the vanishing of $\chi^{-1}$, as we review
next for the quantum model, may lead to a null results. 

\subsubsection{Quantum Cubic  Lattice}

Using the same high temperature series methods \cite{oit11} discussed in subsection \ref{ft9pcm},
the authors of \cite{oit11} further examined also the pseudo-spin $T=1/2$ 
three dimensional 90$^{\circ}$ compass model.
The susceptibility, evaluated with the free energy 
associated with the inclusion of an external field 
coupled to a standard three dimensional version of the nematic
order parameter of Eq. (\ref{energy_diff_Q}), 
\begin{eqnarray}
Q_{3}=  \langle 
2 \tau_{i}^{x} \tau^{x}_{i+ {\bf{e}}_{x}} - \tau_{i}^{y} \tau^{y}_{i+{\bf{e}}_{y}} -
\tau_{i}^{z} \tau^{z}_{i+{\bf{e}}_{z}} \rangle,
\end{eqnarray}
did not, to order ${\cal{O}}(\beta^{20})$ with $\beta$ the inverse temperature, indicate the existence 
of a real zero of $\chi^{-1}$. This suggested that no finite critical transition temperature 
exists). The absence of divergence of $\chi$ does not rule out the existence
of a first order transition similar to that found in the classical model \cite{Wenzel2011a}.  

\subsection{Classical 120$^{\circ}$ Model}

Transitions in the 120$^{\circ}$ compass model on the cubic lattice were numerically examined by various groups. In the most recent study to date, \cite{Wenzel2011a,Wenzel2011b} examined the standard XY type order parameter 
\begin{eqnarray}
m = N^{-1} \sqrt{ (\sum_{i=1}^{N} T_{i}^{x})^{2} + 
(\sum_{i=1}^{N} T_{i}^{y})^{2}},
\label{mwl}
\end{eqnarray}
and the susceptibility $\chi = N( \langle m^{2} \rangle - \langle m \rangle^{2})$ as a function of temperature.
In accordance with earlier estimates  \cite{Rynbach10,Tanaka05},
the transition temperature between the ordered and disordered state was to determined to be 
\cite{Wenzel2011a} 
\begin{eqnarray}
T_{c; 120^{\circ}~ classical} \simeq 0.6775 J,
\label{tc120mo}
\end{eqnarray}
This value is, essentially, the same as that reported earlier by \cite{Rynbach10}. 
As the classical 120$^{\circ}$ model concerns XY type pseudospins in $D=3$ dimensions,
a natural expectation may be that the transition may be
in the same universality class as 3D XY systems- that 
turned out to not be the case. In fact, the collection of 
exponents found seem to suggest that the 120$^{\circ}$ compass
model lies in a new universality class. These results beg further analysis. 
Specifically, by examining the scaling of the $m$ and $\chi$ with system size, \cite{Wenzel2011a,Wenzel2011b} found that the critical exponents associated with the
transition at the critical temperature of Eq.(\ref{tc120mo}) are
\begin{eqnarray}
\nu_{120^{\circ}} = 0.668(6), ~~~ \eta_{120^{\circ}} = 0.15.
\label{crit_wen}
\end{eqnarray}
The ``anomalous'' exponent $\eta$ governs the algebraic decay of the correlation function at the critical point. That is, 
the two-point correlation function at the critical point scales as 
\begin{eqnarray}
\label{TTknow}
\langle {\bm{T}}_{{\bm{i}}}  \cdot {\bm{T}}_{{\bm{j}}} 
\rangle \sim \frac{1}{|{\bm{r}}_{{\bm{ij}}}|^{D-2+\eta}},
\end{eqnarray}
with, as in earlier expressions, $|{\bm{r}}_{{\bm{ij}}}|$ denoting the distance between point
${\bm{i}}$ and point ${\bm{j}}$, and $D$ being the spatial dimensionality
of the lattice.  To make a connection with the canonical form of the correlation function 
of Eq. (\ref{QQknow}), which is valid for general parameters, at the critical point $\xi$ diverges and 
an algebraic decay of correlations remains. For the bare fields ${\bm{T}}_{i}$,  at the critical temperature, 
the form of Eq. (\ref{TTknow}) appears. These reported exponents 
{\em do not fall into any of the typical universality classes}. In particular, 
although $\nu$ of Eq. (\ref{crit_wen}) is not that different from 
its value in a 3D XY type system (wherein $\nu_{3D~XY} = 0.671$),
the value of the anomalous exponent is {\em significantly larger}
($\nu_{120^{\circ}} \gg \nu_{3D~XY}  \simeq 0.038$) \cite{Wenzel2011a,Wenzel2011b}. 
Combined with the hyper-scaling relations, these critical
exponents are consistent with the numerically seen small specific heat exponent $\alpha$ ($C_{v} \sim |T-T_{c}|^{-\alpha}$) \cite{Wenzel2011a,Wenzel2011b}

A similar large discrepancy between the exponents
of the 120$^{\circ}$ model and those of known 
universality classes appears in the value 
of an exponent $``a_{6}''$ that will be introduced
next for a related discrete version of the 120$^{\circ}$ model.

\subsection{Discrete Classical 120$^{\circ}$ Compass Model}
A clock model version of the 120$^{\circ}$ compass
model was further introduced  \cite{Wenzel2011a,Wenzel2011b}. In this variant, the classical Hamiltonian of 
Eq. (\ref{H120}) is used when with the clasical
pseudo-vectors ${\bm{T}}_{i}$ at any site $i$
can only point along six discrete directions.
These directions correspond to the 
angles of Eq. (\ref{tn*}) along which the system
may be oriented at low temperatures \cite{Biskup05,Nussinov04}.
One of the virtues of this system is that it is easier to simulate
and enables numerical investigations of larger
size systems. 

The quantity $Q_{WL}$ of Eq. (\ref{qwl}) as well 
as the ``magnetization'' $m$ of Eq. (\ref{mwl}), 
attain non-zero values below a critical
temperature $T_{c~discrete~ 120^{\circ}} \simeq 0.67505 J$.
This value is numerically close yet slightly larger than the transition for the continuous
classical 120$^{\circ}$ model (Eq. (\ref{tc120mo})). As noted by \cite{Wenzel2011a,Wenzel2011b},
if this deviation in the values of the critical temperatures between the discrete version and 
the original continuous 120$^{\circ}$ model is indeed precise,
it may well be that the entropic stabilization of the 120$^{\circ}$ model driven by
continuous pseudospin fluctuations \cite{Biskup05, Nussinov04}
can be somewhat larger than in its discrete counterpart where
fluctuations are more inhibited.

The critical exponents, as attained numerically, for the discrete 120$^{\circ}$ compass 
model are almost identical to those of the continuous 120$^{\circ}$ model
(given by Eq. (\ref{crit_wen})). 

An analysis similar to that of \cite{JLou2007},  for $T<T_{c}$, examined the 
the distribution of the orientations,
as seen in the average 
\begin{eqnarray}
 {\bm{m}}  =  N^{-1} \sum_{i}  
 \sum_{i} {\bm{m}}_{{\bm{i}}} 
\end{eqnarray}
for individual systems 
of sufficiently small size,
($L < \Lambda_{6}$).
Similar to \cite{JLou2007}, it was found
that  when examined over an ensemble 
of such systems the probability $P({\bm{m}})$ of attaining a particular ${\bm{m}}$
was invariant under continuous (i.e., U(1)) rotations. Conversely, for 
larger systems, this continuous rotational symmetry was lifted.
That is, for systems of size $L>\Lambda_{6}$, the probability distribution $P({\bm{m}})$ 
exhibited only the discrete global six-fold global symmetry
of the system with clear peaks along the six angles 
along which each individual ${\bm{T}}_{i}$ may point.
This system size length scale $\Lambda$ 
at which this change onsets
scales with the correlation length $\xi$
as
\begin{eqnarray}
\Lambda_{6} \sim \xi^{a_{6}}.
\end{eqnarray}
This exponent was found to be $a_{6; ~ discrete ~120^{\circ}} \simeq 1.3$
which is {\em far removed} than that for the corresponding value ($a_{6;~six~state~clock} = 2.2$ 
\cite{JLou2007}) for
XY models perturbed by a tern of the type $(-h \sum_{i} \cos 6 \theta_{i})$.
[Such an external field term renders XY systems to be of the discrete (clock) type.]
The lack of breaking of continuous rotational symmetry as evinced in the distribution $P({\bm{m}})$
for sufficiently small systems thus enables a 
new exponent which, similar to the standard critical anomalous exponent $\eta$
of Eq. (\ref{crit_wen}), 
differs from that in known examples thus far.

\subsection{Extended 120$^{\circ}$ Model}
An extended 120$^{\circ}$ model was recently studied \cite{Rynbach10}. 
The model is defined  by the Hamiltonian 
\begin{eqnarray}
H_{120}^{extended} &= & - \sum_{i, \alpha=x,y} \frac{1}{4} \Big[
J_z T^{z}_{i} T^{z}_{i+{\bm{e}}_{\alpha}} + 3 J_{x} T^{x}_{i} T^{x}_{i+{\bm{e}}_{\alpha}} \nonumber
\\ &&\pm \sqrt{3} J_{mix} (T^{z}_{i} T^{x}_{i+{\bm{e}}_{\alpha}} + T^{x}_{i} T^{z}_{i + {\bm{e}}_{\alpha}}) 
\Big] \nonumber 
\\ &&- J_{z} \sum_{i}
T^{z}_{i} T^{z}_{i+{\bm{e}}_{z}} .
\end{eqnarray}
This model was studied in both its classical and quantum incarnations.
The symmetric point $J_{x}=J_{z}=J_{mix} (=J)$ corresponds to the 120$^{\circ}$ model
of Eq. (\ref{120p}). 
Below, we survey these results.

\subsubsection{Classical Extended 120$^{\circ}$ Model}
A free energy analysis similar to that in subsection{\ref{tf})
found that the six uniform states discussed earlier, at angles $\theta^{*} = 0,60^{\circ}, 120^{\circ}, 180^{\circ}, 240^{\circ}, 320^{\circ}$ relative to the $T^{x}$ axis,
as the entropically stabilized low energy states for the extended 120$^{\circ}$ model
over a region of parameter space where $ 0.8 \le J_{mix}/J_{z} <1$. This region, however, lies at the interface between two other phases \cite{Rynbach10}. For $J_{mix} >J_{z}$, low temperature states are energetically 
selected (and not entropically selected as discussed earlier for the 120$^{\circ}$ model) 
to be states in which there is the preferred angle 
that alternates in a staggered fashion. Pseudospins in a single xz or yz plane may 
have a value of $\theta^*$ while those on the next parallel plane may assume a 
value of $\theta^*$ + 180$^{\circ}$ and so on.
This value of $\theta^{*}$ varies continuously from 30$^{\circ}$ for $J_{mix}/J_{z} \to 1^{+}$
to value of $\theta^{*} = 45^{\circ}$ for asymptotically large $J_{mix}/J_{z}$.  
The transition between the regime with $J_{mix}/J_{z}  \le 1$ (where order is stabilized
by entropy) to that where $J_{mix}/J_{z} >1$ (where order is energetically stabilized) is
a first transition at zero temperature in which level crossing occurs.  For $J_{mix}/J_{z} \le 0.8$, entropic stabilization favors configurations for which the 
angle $\theta^*$ is uniform throughout the system and assumes a value 
that is an integer multiple of 90$^{\circ}$. Throughout the entire region $0  \le J_{mix}/J \le 1$,
the $d=2$ emergent symmetries of Eq. \ref{symorb-e} found earlier for the classical
system remain in tact. 

Similar to the 90$^{\circ}$ compass model, the extended 120$^{o}$ 
model exhibits finite temperature critical points concurrent 
with the first order transitions at zero temperature at the point 
of symmetry (the original 120$^{\circ}$ for which $J_{mix} = J_{z}$). 
In the extended 120$^{\circ}$ model, these critical points fuse to
form a continuous line as $J_{mix}/J_{z}$ is varied (while $J_{x}=J_{z}$). 
The critical nature is seen by specific heat divergence and the finite temperature 
expectation values of the pseudo-spins. \cite{Rynbach10} Reported that at the symmetric point, the 120$^{\circ}$ 
model exhibits a critical transition at a temperature of $T_{c} = (0.677 \pm 0.003) J$-  a value which is
very close to that of the later study of  \cite{Wenzel2011a} [See Eq.(\ref{tc120mo})].

\subsubsection{Quantum Extended 120$^{\circ}$ Model}
\label{sec:quantum120}
One of the major virtues of the extended model,  along the line $J_{mix}=0$, is that it is free of the ``sign problem''
that plagues quantum Monte Carlo simulations. Along the line 
$J_x=J_z=J$ (and $J_{mix}=0$), the system was found to undergo a continuous transition at 
a temperature $T_{c}= (0.41 \pm 0.1)J$ into an ordered state in which all pseudo-spins
point in up or down along the $T^{z}$ direction (the $\pm T^{z}$  directions). 
At zero temperature, as the ratio $J_{x}/J_{z}$ is varied, a first order 
transition corresponding to level crossing at $J_{x}=J_{z}$ appears. For $J_{x}/J_{z} < 1$,
the ground state is of the $\pm T^{z}$ form. Conversely, for $J_{x}/J_{z}>1$, the ground states
are of the $\pm T^{x}$ type. This situation is reminiscent of the first order transition found 
in the 90$^{\circ}$ compass model on the square lattice. In both cases, elementary 
excitations corresponding to a pseudo-spin flip (either of the $\pm T^{z}$ or $\pm T^{x}$ type)
are gapped. The gap is reduced at the point of symmetry ($J_{x}=J_{z}=J$) 
of this truncated model with $J_{mix}=0$ where it attains a value equal to 
$\Delta \approx (0.34 \pm 0.04) J$.

The main interest lies in the symmetric 120$^{\circ}$ and its environs. Towards that end,
\cite{Rynbach10} computed perturbatively the effect of a finite $J_{mix}/J_{z}$ to find
a very interesting suggestive result. These calculations suggest that the gap
closes in the vicinity of the symmetry point ($J_{x}=J_{z}=J_{mix}$). If this is
indeed the case then the states found in the ``un-mixed''  truncated model ($J_{mix}=0$)
are adiabatically connected to those near and at the original symmetric 120$^{o}$
model. On either side of the symmetry point, the ground states
are of the $\pm T^{x}$ type and $\pm T^{z}$ as discussed above.
These states must somehow evolve and merge into the states at the point
of symmetry. This suggests a greater degree of degeneracy within the ground state
sector of the symmetric 120$^{\circ}$ model.  Amongst other possibilities this raises the specter (compounded by 1/S calculations) of six states akin to those found classically \cite{Biskup05, Nussinov04} in the symmetric 120$^{o}$ model or a possibility of having 12 states with pseudospins all uniformly oriented at an angle  
$\theta^{*} = 0,30^{\circ}, ..., 330^{\circ}$ relative to the $\tau^{x}$ direction.

\subsection{Honeycomb Lattice 120$^{\circ}$ Compass Models}
\label{120_h_pd}

In subsections \ref{120honey} and \ref{120honey'}, we reviewed key physical aspects of
the 120$^{\circ}$ honeycomb model of Eq. (\ref{eq:c120_honeycomb}).
This included an analysis of the ground state sector, its emergent symmetries, and 
the order out of disorder free energy calculations. We now turn to further other more quantitative aspects.

\subsubsection{Classical model}

Following \cite{Nasu08,Wu08,Zhao08}, we reviewed, 
in Section \ref{120honey},  the presence of a continuous global ($d=2$) and chiral discrete $d=1$
emergent symmetries of the 120$^{\circ}$ compass models on the honeycomb lattice. The low temperature orders are unconventional. That is, the numerically observed usual pair correlations $\langle {\bm T}_{i} \cdot {\bm T}_{j} \rangle$ were found to be short ranged (and $\langle {\bm T}_{i} \rangle$ vanishes) as the system size increased \cite{Nasu08}.
Numerically, a continuous (or weakly first order) low temperature ordering transition circa $T_{c}= 0.0064 J$ is marked by an order parameter $q$ defined as \cite{Nasu08}
\begin{eqnarray}
q_{i}  = \cos 3 \theta_{i} , ~~~ q =  N^{-1} \sum_{i=1}^{N} q_{i}.
\end{eqnarray}
(Note that this quantity constitutes an analogue to the nematic type
order parameters in the two- and three- dimensional 90$^\circ$ models.)
The pair correlations $\langle q_{i} q_{j} \rangle$ exhibit a correlation
length of size $\xi$ that scales in accordance with Eq.(\ref{stan_exp})
with an exponent $\nu = 0.72 \pm 0.04$. Similarly, the  transition
at $T_{o}$ is evident as a peak in the specific heat.  
Within the ground states $|q| =1$ in accord with the 
order out of disorder analysis that, as reviewed
in Section \ref{120honey}  (similar to that
of the 120$^{\circ}$ model on the cubic lattice)
led the angles of Eq. (\ref{tn*}).

\subsubsection{Quantum Model}
\label{120qhn}

\begin{figure}
\centering
\includegraphics[width=.6\columnwidth]{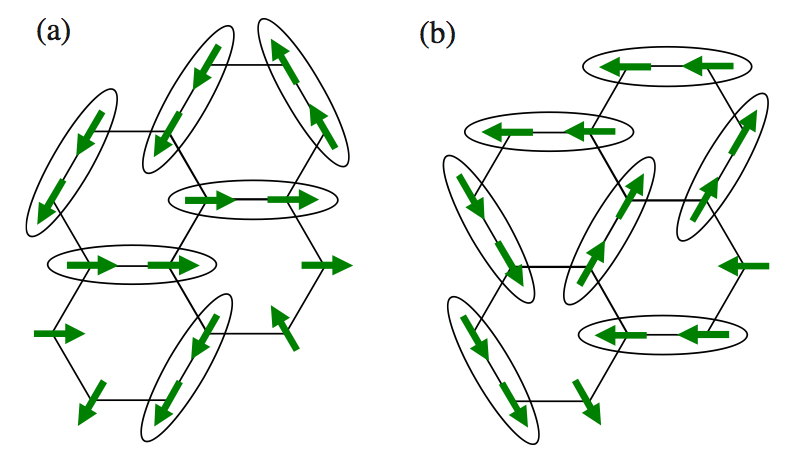}
\caption{Some of the pseudo spin configurations where the honeycomb lattice is covered by NN bonds with the minimum bond energy. One of the q=1 states in (a) and one of the q=-1 in (b). In NN bonds surrounded by ellipses, the bond energy is the lowest~\cite{Nasu08}.}
\label{fig:18Nasu08}
\end{figure}

The numerical value of the spectral gap between the ground state and the next
excited state was found to progressively diminish as the system size was increased \cite{Nasu08}.
Currently, it is not clear if this reflects the existence of gapless modes or point
to a degeneracy of the system. Generally, in 
many spin (and pseudo-spin) systems, similar results appear 
in simpler systems that harbor bona fide SU(2) symmetries
where the Lieb-Schultz-Mattis theorem and more recent
extensions exist \cite{LSM1,LSM2}. It was furthermore
found that the ground states might be approximated by 
an ansatz wavefunction of
the type \cite{Nasu08}
\begin{eqnarray}
| \Psi^{(\pm)} \rangle = {\cal{N}} \sum_{l}  {\cal{A}}_{l}  \{ | \psi_{l}^{(\uparrow)} \rangle
\pm |\psi_{l}^{(\downarrow)} \rangle \}.
\label{var_hex_120}
\end{eqnarray}
In Eq. (\ref{var_hex_120}), ${\cal{N}}$ is a normalization constant, $\{{\cal{A}}_{l}\}$ are variational parameters
and the states $| \psi_{l}^{(\uparrow,\downarrow)} \rangle$ schematically represented
in  Fig. \ref{fig:18Nasu08} are somewhat akin to the classical configurations discussed in panel (a) of Fig. \ref{fig:Wu08_3}.

Explicitly, 
\begin{eqnarray}
| \psi^{(\uparrow)}_{l} \rangle = \prod_{\langle i j \rangle_{l}} U(\phi_{\gamma}) | \uparrow \uparrow ... \uparrow \rangle.
\label{def_hex_l}
\end{eqnarray}
In the above, $l$ denotes a set of links $\langle i j \rangle$ for which the fully polarized state 
$|\uparrow ... \uparrow \rangle$ will be rotated so that the pseudo-spins will be parallel to
the links in the set $l$. In Eq. (\ref{def_hex_l}) we will, specifically, set  
for a single pair of sites $i$ and $j$ on the link $\langle ij \rangle$\cite{Nasu08}
\begin{eqnarray}
U(\phi_{\gamma})_{\langle i j \rangle} = \exp[- i \phi_{\gamma} (T_{i}^{y} + T_{j}^{y})] 
\end{eqnarray}
where $\gamma$ is set by the spatial direction of the link between $i$ and $j$:
$(\phi_{1},\phi_{2},\phi_{3}) = (0, 2 \pi/3, 4 \pi/3)$. 
Thus, the states of Eq. (\ref{var_hex_120}) correspond to a linear
superposition of ``dimer states'' , e.g.,\cite{NN1,RK,RK1,Nussinov07}.
In this case, the dimer states $|\psi^{\uparrow (\downarrow)}_{l} \rangle$ 
correspond to states wherein the pseudospins are parallel (or antiparallel) to the spatial direction. 
Kinetic tunneling between different dimer
states can lower the energy of such states, see Fig. \ref{fig:19Nasu08}.

Thus, in the space spanned by 
the dimer states $\{|\psi^{(\uparrow/\downarrow)}_{l} \rangle \}$ certain admixtures of
these states (with certain sets of the amplitudes $\{
{\cal{A}}_{l}
\}$ in Eq. (\ref{var_hex_120})
can be selected by quantum fluctuations.  

\begin{figure}
\centering
\includegraphics[width=.8\columnwidth]{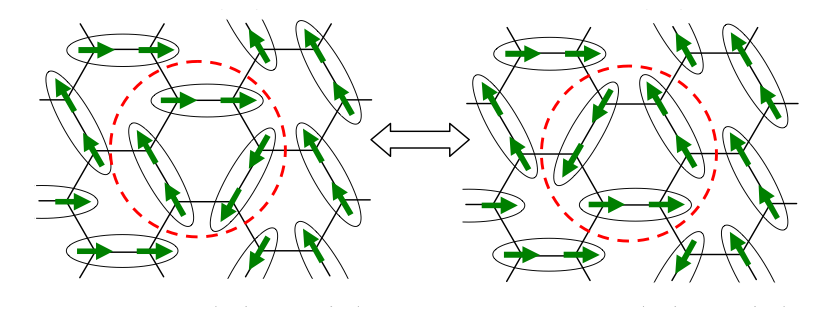}
\caption{One example for the two pseudospin configurations where a resonance state is possible due to the off-diagonal matrix element~\cite{Nasu08}.}
\label{fig:19Nasu08}
\end{figure}

\subsection{Checkerboard Lattice Compass Models} 

The most prominent compass models have been inspired by orbital or other interactions
on cubic or other geometrically unfrustrated lattices. We have briefly touched on some aspects
of geometric frustration in different arenas in sections (\ref{sec:chiral_spin},\ref{3fs}) and elsewhere.
We now explicitly turn to compass models on the checkerboard lattice. In subsection \ref{plaq&check}
[and in particular in Eq. (\ref{checker})], we briefly introduced the checkerboard on
the checkerboard lattice \cite{Nasu2011},\cite{Nasu2012}. The checkerboard lattice,
a two dimensional rendition of the pyrcholore lattice, 
is a prototypical frustrated lattice. The system of Eq. (\ref{checker}) was motivated by examining,
within second order perturbation theory (assuming the kinetic term is small relative to the Coulomb penalty),  
a spinless Hubbard model on this lattice. This model exhibits $d=1$ symmetries in the form  \cite{Nasu2012}
of 
\begin{eqnarray}
O_{l} = \prod_{i \in l} \tau^{z}_{i},
\end{eqnarray}
where $l$ denote diagonals that run across the system either in the $\langle 11 \rangle$ or $\langle 1 \overline{1} \rangle$ directions.   
By the generalization of Elitzur's theorem \cite{Batista05,holography}, these symmetries cannot be broken at finite temperatures. 
Some limits of the problem are obvious. When $|J_{x}| \gg |J_{z}|$, as each site lies on only one of the two diagonal directions ($\langle 11 \rangle$ or $\langle 1 {\overline{1}} \rangle$), 
the Hamiltonian of Eq. (\ref{checker}) reduces to that of decoupled diagonal chains with Ising $\tau_{i} \tau_{j}$ interactions between
nearest neighbors. In the other extreme limit, that of $|J_{z}| \gg |J_{x}|$, interactions along the diagonals become negligible and the system becomes a two dimensional Ising model 
on the square lattice with nearest neighbor $\tau_{i}^{z} \tau_{j}^{z}$ interactions. In tandem with these limits, it was reported \cite{Nasu2011,Nasu2012} that at low temperatures,
for $|J_{x}| \lesssim 2 |J_{z}|$, uniform or Neel (dependent on the sign of $J_{z}$) Ising order appears. By contrast, when $2 |J_{z}| \lesssim |J_{x}|$, the decoupled chain like character 
leads, on an $L \times L$ lattice, to a $2^{2L}$ degeneracy similar to that found for the square lattice 90$^{\circ}$ compass model.  In the antiferromagetic variant of
this system, at zero temperature, a first order transition between the two low phases was found at $J_{x} \simeq  2.7 J_{z}$. Several approaches \cite{Nasu2011,Nasu2012} 
suggest that there is a finite temperature tricritical point in the vicinity of $J_{x} = 2J_{z}$.

\subsection{Arbitrary Angle Compass Models}

We now discuss the arbitrary angle square lattice compass models \cite{Cincio} of Eq. (\ref{generalized_ocm}). 
The symmetry of the ground states of these systems changes character at an angle $\theta_{c}$
which is very close to the right angle value of the 90$^{\circ}$ compass model. The
second order transition at $\theta=\theta_{c}$ is associated with 
the doubling of the ground state degeneracy. Specifically, 
for $\theta<\theta_{c}$, the system of Eq. (\ref{generalized_ocm}) has two degenerate
ferromagnetic ground states 
with a spontaneous magnetization
that is parallel to anti-parallel to
the $\tau^x$ (or $\pi^{x} + \pi^y$) direction. Conversely when $\theta>\theta_{c}$, 
there are four degenerate ground states with pseudo-spins along the $\pm \pi^{x}$ or $\pm \pi^{y}$ directions. 
For the pseudo-spin 1/2 realization of Eq. (\ref{generalized_ocm}), it was numerically seen
that $\theta_{c} \simeq 84.8^{\circ}$. As the pseudo-spin value increases and
the system becomes more classical, $\theta_{c}$ monotonically increases
and veers to 90$^{\circ}$ in the classical limit. Thus, the four-fold degenerate phase
is promoted by quantum fluctuations.

\subsection{XXZ Honeycomb Compass Model}
\label{sec:xxz}
In section \ref{xxz:sec} [in particular, in Eq. (\ref{xxzhc})]  the XXZ honeycomb compass model \cite{Nussinov2012} was introduced
(see also Fig. \ref{fig:xxzhc}).
This model can be mapped onto a 
quantum Ising gauge theory on a square lattice \cite{Nussinov2012}
\begin{eqnarray}
\label{QIG}
H_{QIG,XXZ}= -  \sum_{\bf x ~bonds} J_{x}^{\ell} \sigma^{x}_{\ell} -  \sum_{\bf y ~bonds} J_{x}^{\ell} \sigma^{x}_{\ell}  \nonumber
\\ -  \sum_{\bf z ~bonds} J_{z}^{\ell'} \prod_{\ell \in P_{\ell'}} \sigma_{\ell}^{z}.
\end{eqnarray}
A few explanations are in order concerning this Hamiltonian. The links $\ell$ and the associated coupling constants $J^{\ell}$ refer to the 
links of the original honeycomb lattice; these links can be oriented along either the $x$, $y$,or $z$ directions of the honeycomb lattice.
In Eq. (\ref{QIG}), the Pauli operators $\sigma_{\ell}^{x,z}$ are located at the centers $\ell$ of the square lattice which is
formed by shrinking all of the vertical (or z-) links of the honeycomb lattice to individual point. After such an operation, the resulting (topologically square) lattice
is comprised of x- and y- type links. As seen in Eq. (\ref{QIG}), there is a field $h=J_{x}$ that couples to the 
Pauli x operator on each such link. This is augmented by a plaquette term (the last term in Eq. (\ref{QIG})). 
The plaquette $P_{\ell}'$ is formed by the centers of the four links (two x-type links and two y-type links) 
that are nearest neighbors to the center of a vertical z-type link $\ell$. The product $\prod_{\ell \in P_{\ell'}} \sigma_{\ell}^{z}$ 
denotes the product of all four $\sigma^{z}$ operators at the centers of links of the square plaquette 
that surrounds an original vertical link $\ell$ that has been shrunk to a point. The sum over the original
vertical links (z-bonds) becomes, in Eq. (\ref{QIG}), a sum over all plaquettes of a square lattice formed 
the shrinking of all vertical links. The link center-points of this square lattice
coincide with those formed by the center-points of the x- and y-type bonds of the original honeycomb lattice.
When all of the coupling constants $J_{x,z}^{\ell}$ are isotropic, the system is that of the canonical uniform 
standard transverse field Ising gauge theory which, as is well known, maps onto the 3D Ising gauge
theory. The 3D Ising gauge theory is dual to the standard 3D Ising model on the cubic lattice \cite{Wegner,Kogut}.
Thus, the uniform XXZ honeycomb compass model is dual to the 3D Ising model and exhibits a finite 
temperature phase transition with the standard 3D Ising exponents \cite{Nussinov2012}.  
As is evident in Eq. (\ref{QIG}), not all coupling constants $J_{x,z}^{\ell}$ need to be of the same strength. 
As the disordered transverse field Ising gauge theory can exhibit a spin-glass type transition, the XXZ honeycomb model
may also correspond to a spin glass when it is non-uniform \cite{Nussinov2012}.
Additional information concerning the quantum Ising gauge theory appears in 
section \ref{sec:SECHM}.

\subsection{Plaquette Orbital Model}

The authors of \cite{Biskup10} studied the classical realization of the
``plaquette orbital model"  \cite{Wenzel09} and certain quantum variants.
Below, these results are reviewed. 

\subsubsection{Exact Symmetries}
Examining the Hamiltonian of Eq. (\ref{plaq_ocm}) and Figure \ref{fig:Biskup10_1},
we note that the inversion of the four pseudo-spins $\tau^{x}_{i} \to - \tau^{x}_{i}$ on an A plaquette
(while leaving $\tau^{y}_{i}$ unchanged) constitutes a local symmetry.   
A similar effect occurs with $x$ and $y$ interchanged on any of the B-type plaquettes.
These local (i.e., gauge) symmetries are recast in terms
of the following 4-site symmetry operators of the $T=1/2$ quantum 
Hamiltonian of Eq. (\ref{plaq_ocm})
($[U_{\Box_{
{\cal{A}}
}}, H]=
[U_{\Box_{
{\cal{B}}
}} , H]=0$),
\begin{eqnarray}
U_{\Box_{
{\cal{A}}
}} = \prod_{i \in \Box_{
{\cal{A}}
}} \tau^{y}_{i}, ~ ~ U_{\Box_{
{\cal{B}}
}} = \prod_{j \in {\Box_{
{\cal{B}}
}}}
\tau^{x}_{j}.
\label{uabsym}
\end{eqnarray}
In Eq. (\ref{uabsym}),  ${\cal{A}}$ denotes any plaquette of the A type and,
similarly, ${\cal{B}}$ denotes any plaquette of the B type. 
By Elitzur's theorem, at any finite temperature ($T>0$), all expectation
values must be invariant under the symmetries of Eq.(\ref{uabsym}). 

\subsubsection{Classical Ground States \& Emergent Symmetries}
By rewriting, similar to the analysis for the classical 120 and 90 degree models,
\cite{Nussinov04, Biskup05}, the Hamiltonian of Eq.(\ref{plaq_ocm}) as a sum of squares and using 
uniform states as classical  ``variational states'', \cite{Biskup10}
demonstrated that all classical ground states of Eq. (\ref{plaq_ocm}) are
uniform states up to the application of the classical version of the local symmetries of 
Eqs.(\ref{uabsym}).  In particular, for $J_{A} > J_{B}$, a state which is fully polarized along
the $x$ axis constitutes a ground state; this state can be further mutated by the local inversion 
gauge transformations. Similar to the situation in the classical 120 degree and compass models
a continuous symmetry emerges in the classical ground state sector. When $J_{A} = J_{B}$,
any constant uniform state of the pseudo-spins ${\bm{T}}_{\bm r}$ is a ground state of the classical system. As these classical vectors can point anywhere on the unit disk, a continuous rotational
symmetry appears. 

\subsubsection{Finite Temperature Order out of Disorder}

Similar to the situation in the classical 120 degree and 90 degree compass models,
a finite temperature order out of disorder mechanism lifts the ground state degeneracy  
\cite{Nussinov04,Biskup05,Mishra04}, and leads, at low positive temperatures
($0<T<T_{0}$) to a nematic type order in the plaquette compass model 
wherein most of the configurations have
a majority of the pseudo-spins aligned along either the $(\pm {\bm e}_{x})$ 
or the $(\pm {\bm e}_{z})$ directions \cite{Biskup10}.  
Due to the (classical version of the) local symmetries of 
Eqs. (\ref{uabsym}), both sign of the orientation ($\pm$) are equally likely.   

Following \cite{Biskup07},  low temperature order was also proven to 
hold in the quantum model when the magnitude of
the pseudo-spin is sufficiently large ($|{\bm T}| > c \beta^{2}$ with $c$ a positive constant and $\beta$
the inverse temperature) \cite{Biskup10}. 
The technical reason for requiring a sufficiently large pseudo-spin
is that within the proof of \cite{Biskup10,Biskup07}, 
thermal fluctuations were assumed to dominate
of quantum fluctuations.

\subsection{Gell-mann Matrix Compass Models}
\label{santa-fe}

The two Gell-mann matrix compass models
of Eqs. (\ref{Gell-mann1}, \ref{Gell-mann2}),
derived from Eq. (\ref{CW3p}), 
have very interesting and distinct behaviors \cite{Chern11}.

\subsubsection{Cubic Lattice Gell-mann Matrix Compass Model}

As the two Gell-mann matrices $\lambda^{(3)}$ and $\lambda^{(8)}$ are diagonal 
and commute with one another, the quantum model of Eq. (\ref{Gell-mann1})
is essentially classical \cite{Chern11}. 

{{$T=0$:}} \newline
The ground state energy per site $E/N= -2J/3$
is consistent with 2/3 of the bonds being minimized
and the remaining 1/3 being frustrated. The two-point correlation
function $\langle {\bm{\lambda}}_{i} \cdot {\bm{\lambda}}_{j} \rangle$
exhibits rapidly decaying oscillations and is essentially vanishing
for distances $|{\bm{r}}_{ij}| \ge 5$ lattice constants \cite{Chern11}.

{{$T>0$:}} \newline
Monte Carlo simulations were performed. An integration from
the specific heat curve
indicates that there is a large residual entropy at zero temperature.  
Although not explicit estimate was given in \cite{Chern11}
for viable transitions,  judging from the data shown the sharp
specific heat peak occurs at a temperature $T \sim 0.7 J$. 

\subsubsection{Diamond Lattice Gell-mann Matrix Compass Model}

For a single pair of nearest neighbor sites on the lattice 
along the ${\bm{n}}_{0}$ direction ($i$ and $j$), the minimum
of the corresponding term in Eq. (\ref{Gell-mann2})
is achieved when the corresponding orbital states
are $3^{-1/2}(|p_{x} \rangle + |p_{y} \rangle+ |p_{z} \rangle)$
and $2^{-1/2}(|p_{x}  \rangle- |p_{y} \rangle)$.  
Similar to the case of the Gell-mann model on the cubic lattice
and, more generally, compass models, the system is frustrated and not all interactions can be simultaneously minimized. As shown in
 \cite{Chern11}, the ground states are of the form
 $| \psi \rangle =  \prod_{i} |\lambda_{i} $ with 
 for any site $i$, the local state $|{\bm{\lambda_{i}}} \rangle = | \pm {\bm{e}}_{x} \rangle, 
 | \pm {\bm{e}}_{y} \rangle$ or $ | \pm {\bm{e}}_{z} \rangle$ {\em such
 that for all nearest neighbor pairs} $\langle ij \rangle$,
 \begin{eqnarray}
 ({\bm{\lambda}}_{i} \cdot {\bm{e}}_{ij})({\bm{\lambda}}_{j} \cdot {\bm{e}}_{ij}) = - \frac{1}{3}.
 \label{condition_diamond}
 \end{eqnarray} 
 
 When expressed in terms of the original orbital degrees of freedom, the local states are,
 explicitly, $| \pm {\bm{e}}_{x} \rangle = 2^{-1/2} (|p_{y} \rangle  \pm |p_{z} \rangle)$
 and cyclic permutations thereof (i.e.,
 $| \pm {\bm{e}}_{y} \rangle = 2^{-1/2} (|p_{z} \rangle  \pm |p_{x} \rangle)$ and
 $| \pm {\bm{e}}_{z} \rangle = 2^{-1/2} (|p_{x} \rangle  \pm |p_{y} \rangle)$).
 
\begin{figure}
\centering
\includegraphics[width=.8\columnwidth]{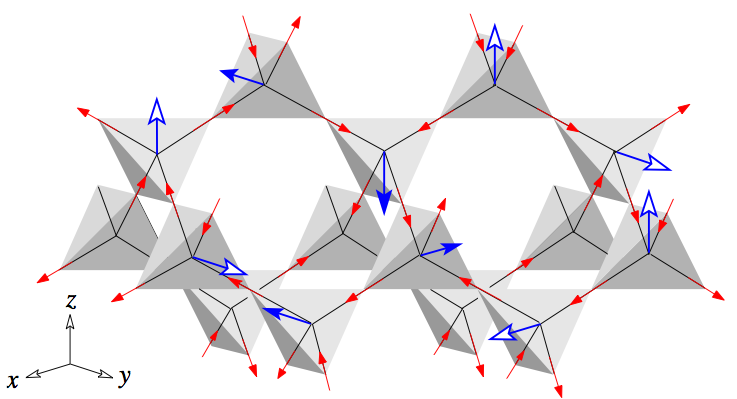}
\caption{A configuration of the pseudo-vectors on the diamond lattice and its mapping to the spin-ice state on the dual pyrochlore lattice. The pseudo-vector only assumes six different values $\langle \mu i\rangle = \pm \hat x, \pm \hat y, \pm \hat z$ in the ground states, corresponding to ($p_y \pm p_z $), ($p_z \pm p_x$), and ($p_x \pm p_y$) orbitals, respectively. These six orbital configurations are mapped to the six 2-in-2-out ice state on a tetrahedron.~\cite{Chern11}.}
\label{fig:Chern11_3}
\end{figure}

 
 As shown in Fig. (\ref{fig:Chern11_3}), 
 the states $| \pm {e}_{x,y,z} \rangle$ at any site $i$ can 
 be represented by corresponding red arrows on the pyrochlore
 lattice formed by the centers of all nearest neighbor links 
 $\langle ij \rangle$. Specifically, these arrows are given
 by 
 \begin{eqnarray}
 {\bm{R}}_{\langle ij \rangle} = \sigma_{i}^{\gamma} {\bm{e}}_{ij} (= - \sigma_{j}^{\gamma} {\bm{e}}_{ij}),
 \end{eqnarray}
 where, with ${\bm{e}}_{ij}$ denoting a unit vector from site $i$ to site $j$, the Ising type 
 variables $\sigma_{i}^{\gamma} = \pm 1$ are given by 
 $\sigma_{i}^{\gamma} = \sqrt{3} ({\bm{\lambda}}_{i} \cdot {\bm{e}}_{ij})$.
Following \cite{Chern11}, we next focus on the basic tetrahedrons of pyrochlore lattice (that have
 the vertices of the original diamond lattice at their centers).
 As a result of the condition of Eq. (\ref{condition_diamond}),
 there are two incoming and two outgoing arrows ${\bm{R}}$ 
 towards the center of each tetrahedron. This is the so-called
 ``ice condition'' which appears in many other systems 
 and leads to an extensive degeneracy \cite{Lieb1967, Nagle1966} 
 which according to the Pauling estimate
 would be $S \approx Nk_{B} \ln (3/2) \simeq 0.405 N k_{B}$ \cite{Chern11}.
 (Note that according to the more accurate estimate
 of Nagle \cite{Nagle1966} this would be $S \approx 0.4102 N k_{B}$).
 Similar to the Gell-mann matrix model on the cubic lattice,
 two-point correlations within the ground state are decaying.
 In general, the correlations associated with extensively degenerate
 ice states are dipolar type power law
 correlations $\langle {\bm{\lambda}}_{i} \cdot {\bm{\lambda}}_{j} \rangle \sim |{\bm{r}}_{ij}|^{-3}$. \cite{Villain1972,Stillinger1973,Youngblood1981,Ioffe1989,Huse2003,Henley2005,Nussinov07}.
Such correlations were indeed numerically verified by Chern and Wu in their system 
\cite{Chern11}.  The ice condition and its breaking are known 
to lead to effective fractional charges and related effects as found in different contexts \cite{Shannon2002,Nussinov07,Castel2008,Powell11,chang12}.
In particular, when the temperature $T>0$, thermal excitations out of the 
ground state ice condition manifold can lead to
deconfined fractional charges (with dipolar correlations).
It would be interesting to see what is the corresponding physics
in this orbital system. 

\subsection{Symmetric Extended Compass Hubbard Models}
\label{sec:SECHM}

In section \ref{sec:EHCM}, and in Eq. (\ref{SECHM}) in particular, 
a compass type Hubbard model was introduced that, aside from lattice hopping terms,
further included electronic pair creation and annihilation terms. 
Both of these terms (kinetic and pairing) were of
the compass type. In Eq. (\ref{SECHM}), the spatial indices of the electronic
creation and annihilation operators involved were determined by 
the spin polarization. The particular, symmetric,
variant written of the extended compass Hubbard model,
that of Eq. (\ref{SECHM}), in which the pairing and hopping amplitudes are 
of equal strength, is amenable to an exact result.
It can be demonstrated \cite{Nussinov2012}
that the square lattice system of Eq. (\ref{SECHM}) 
is dual to the quantum Ising
gauge theory (QIG) on the dual lattice.
This dual lattice (which is also a square lattice) is formed by regarding each site $i$
of the original square lattice as the center of a minimal
square (or plaquette) of the dual lattice.  The QIG theory was
already written down as its associated couplings pertain to the XXZ honeycomb compass
in Eq. (\ref{QIG}). We now do so anew 
for the symmetric extended compass Hubbard model. 
The Hamiltonian of the quantum Ising gauge theory which
is dual to the theory of Eq. (\ref{SECHM}) is given by
\begin{eqnarray}
H_{QIG,SEHCM} = - 2 \sum_{l} t_{l} \sigma^{x}_{l} - \sum_{P} U_{i} \prod_{l \in P_{i}} \sigma^{z}_{l}.
\label{QIGe}
\end{eqnarray}
The index $l$ in Eq. (\ref{QIGe}) denotes a link of the square lattice. 
In reference to the symmetric extended compass Hubbard model of Eq. (\ref{SECHM}), $t_{l}$ is the hopping amplitude between two sites 
in the original electronic system. In the spin Hamiltonian of Eq. (\ref{QIGe}), a Pauli operator is placed at the center of
each link $l$ of the square lattice. The first term in Eq. (\ref{QIGe}), thus physically corresponds, at each such link $l$, 
to a magnetic field term along the $x$ direction which is of strength $t_{l}$. The second term in Eq. (\ref{QIGe}) 
is the standard ``plaquette'' term of classical gauge theories. $P_{i}$ denotes any elementary plaquette of the square lattice
on which Eq. (\ref{QIGe}) is defined
(corresponding to a site $i$ on the original square lattice model of Eq. (\ref{SECHM})).
$\prod_{l \in P_{i}} \sigma^{z}_{l}$ is the product of the four $\sigma^{z}$ operators on
the links $l$ of such a minimal square plaquette $P_{i}$ of the lattice. In the absence of the first 
(magnetic field) term in Eq. (\ref{QIGe}), the Hamiltonian is that of the classical square lattice Ising gauge theory \cite{Kogut}
(which is trivially dual to an Ising chain). The field $t_{l}$ along the transverse $x$ direction leads to quantum 
fluctuations between different classical spin states. These fluctuations are the origin of the name ``quantum Ising gauge theory''.
As is well known, the square lattice quantum Ising gauge theory can be mapped onto the 3D classical Ising gauge theory 
(the theory given solely by square plaquettes terms on the cubic lattice). The 3D Ising gauge theory is, in turn, dual 
to the standard Ising model on the cubic lattice. 
Thus, similar to the discussion in section \ref{sec:xxz}, by the equivalence between the theories of Eqs.(\ref{SECHM}, \ref{QIGe})
one can adduce much information. These considerations and make specific remarks about  (i) the spatially uniform and (ii) disordered realizations of this theory.

{\it The spatially uniform system} 
When all of the pairing/hopping amplitudes $t_{l}$ and the Hubbard energy terms $U_{i}$ in Eq. (\ref{SECHM}) are spatially
uniform and equal to fixed values $t$ and $U$, the system is equivalent to and exhibits canonical 3D Ising behavior.
At zero temperature, a 3D Ising critical transition appears at a ratio of
\begin{eqnarray}
\Big( \frac{t}{U} \Big)_{crit} = 0.14556.
\end{eqnarray}

{\it Disordered systems} 
The mapping \cite{Nussinov2012} between the symmetric extended compass Hubbard model of Eq. (\ref{SECHM}) and the 
quantum Ising gauge theory of Eq. (\ref{QIGe}) is general applies to any set of couplings $\{t_{l}, U_{i}\}$. As is well
known, sufficiently disordered Ising models (in which couplings are non-uniform) may display a spin glass type behavior. 
Thus, by the correspondence between Eqs. (\ref{SECHM}, \ref{QIGe}), the electronic system given by random symmetric extended compass Hubbard model
may display spin glass behavior. 

\subsection{Heisenberg-Kitaev Models \& Honeycomb Iridates}
\label{HKCS}

In section \ref{sec:sporb}, we reviewed how spin-orbit effects may lead to the Heisenberg-Kitaev model of Eq. (\ref{HKI}).
We will now review various results obtained on the model. 

\subsubsection{Phase Diagram of Heisenberg-Kitaev Model}

The overall multiplicative 
constant $C$ in Eq. (\ref{HKI}) can be absorbed by scaling the temperature $T \to T/C$. 
As a function of $\alpha$, the Heisenberg-Kitaev model has a wealth
of phases. Recent experiments \cite{Singh10,Liu11,Sing11,Choi2012} 
seem to agree with theoretical
analysis \cite{Chaloupka10,Reuther11,Jackeli09,Jiang11,Chaloupka13}.  
These notably include the viable presence of a zigzag phase and 
its associated magnetic susceptibility and spin-wave spectra.
We remark that some phases similar to those that are found in the Heisenberg-Kitaev model appear in certain 
frustrated U(1) symmetric XY models on the honeycomb lattice \cite{varney2011}.
Below we first examine specific properties of the Heisenberg-Kitaev model.

 {\it Exact self-duality of the Heisenberg-Kitaev model:}\\
Although not explicitly stated earlier in these terms, the Heisenberg-Kitaev model of Eq. (\ref{HKI}) exhibits an elegant and exact {\it self-duality}-
i.e., a self-duality which exactly applies to the entire energy spectrum of the theory (not only that which emerges at low energies (or other sectors) as in many,
far weaker, emergent self-dualities). This exact self-duality is made explicit via a unitary transformation (a sublattice dependent canonical 180$^{\circ}$ rotation)
that maps the Hamiltonian of Eq. (\ref{HKI}) back to itself yet with rescaled couplings \cite{Chaloupka10,khalS2005,Chaloupka13}.
Before explicitly reviewing this transformation, one rewrites Eq. (\ref{HKI}).
The first (Heisenberg type) term 
in Eq.(\ref{HKI}) contains also a 
(Kitaev type) contribution
of the form of $\tau^{\gamma}_{i} \tau^{\gamma}_{j}$ but with a sign that is opposite
to the first. As throughout this review, we denote by ${\bf{e}}_{x,y,z}$
the three unit vector directions on the honeycomb lattice. Specifically,
we define these to point vertically up (${\bf{e}}_z$)
or downwards while veering to the right or left (${\bf{e}}_{x,y}$).
With $i \in A$ denoting lattice sites that belong to the even
sublattice of the honeycomb lattice (i.e., all lattice sites that lie
at the bottom of all vertical links of the lattice), 
the Hamiltonian of Eq. (\ref{HKI})
reads
\begin{eqnarray}
H_{HK} =   C \sum_{i \in A} 
 \sum_{\gamma = x,y,z} \Big[ (1-3 \alpha)  \tau^{\gamma}_{i} \tau^{\gamma}_{i+ {\bf{e}}_{\gamma}}  \nonumber 
 \\ + (1-\alpha) \sum_{\gamma'  \neq \gamma}  \tau^{\gamma'}_{i} 
\tau^{\gamma'}_{i+ {\bf{e}}_{\gamma'}} \Big].
\label{HKII}
\end{eqnarray}
In the second line of Eq. (\ref{HKII}), the sum is performed over all honeycomb lattice directions $\gamma' = x,y$ or $z$ that
differ from a given $\gamma$. A spatially dependent rotation \cite{khalS2005,Chaloupka10} can simplify this problem. Specifically, the system is partitioned into 
four sublattices [see Fig. \ref{fig:Chaloupka10_1b}] and then a site dependent rotation operator is applied, 
\begin{eqnarray}
\label{uical}
{\bm{\tau}}_{i} \to {\bm{\tilde{\tau}}}_{i} = U^{\dagger}_{i} {\bm{\tau}} U_{i}.
\end{eqnarray}
The site dependent unitary operator $U_{i}$ may,
depending to which sublattice the site $i$ belongs to (see Fig. (\ref{fig:Chaloupka10_1b}), rotate on three of the sublattices 
the iso-spin vector ${\bm{\tau}}_{i}$ by 180 degrees about one of three associated ($x$, $y$, or $z$) iso-spin directions while acting as the identity operator [$ {\bm{\tau}}_{i} = {\bm{\tilde{\tau}}}_{i}$]
for sites $i$ that belong to the fourth sublattice. As in other general cases, this self-duality implies that couplngs related to one another by 
the self-duality lead to identical thermodynamics and dynamics. 

\begin{figure}
\centering
\includegraphics[width=.3\columnwidth]{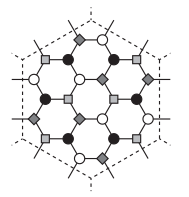}
\caption{The supercell of the four- sublattice system enabling the transformation of the model ~\cite{Chaloupka10}.
}
\label{fig:Chaloupka10_1b}
\end{figure}

In the rotated basis, 
the Hamiltonian of Eq. (\ref{HKI}) reads
\begin{eqnarray}
\tilde{H}_{HK} = C \sum_{\langle ij \rangle || \gamma} [(\alpha -1) {\bf{\tilde{{\tau}}}}_{i} \cdot {\bf{\tilde{{\tau}}}}_{j} 
+(2 - 4 \alpha) {\tilde{\tau}}_{i}^{\gamma} {\tilde{\tau}}_{j}^{\gamma}].
\label{hk_rotl}
\end{eqnarray}
That is, the unitary transformation maps the Hamiltonian back to itself yet with new parameters; as such, this sublattice dependent site dependent rotation
operation realizes a self-duality transformation. Written in terms of the original parameters in Eq. (\ref{HKII}), one sees that Eq. (\ref{hk_rotl}) implies a transformation
\begin{eqnarray}
\alpha \to (2 \alpha -1)/(3 \alpha -2), \nonumber
\\ C \to C (3 \alpha -2).
\label{alpC}
\end{eqnarray}
 Alternatively, under this  self-duality transformation, in the top line of Eq. (\ref{HKI}), $J_{1} \to (J_{1} - 2J_{2})$ while $J_{2} \to - J_{2}$. 
In the $(A, \varphi)$ parameterization of Eq. (\ref{HKI}), $\tan \varphi \to (-\tan \varphi -1)$ \cite{Chaloupka13}
and $A \to A \sqrt{(2+ 2 \tan \varphi + \tan^{2} \varphi)/(1+ \tan^{2} \varphi)}$. 
As in all self-dualities, a repeated application of the transformation twice restores the original 
coupling constants (as is evident in this case from a repeated application of the sublattice dependent 180$^{\circ}$ rotations). 
A schematic of the self-duality in this parameterization and the associated phases and transitions which will shortly be elaborated on later is provided in Fig. (\ref{fig:Chaloupka13_2}).
Fig. \ref{fig:1chal12} reproduced from \cite{Chaloupka13} further provides a schematic of the types of
ground states found for different values of $\varphi$; ground states related to each other by the self-duality transformation
can be formed from one another by applying the sublattice dependent 180$^{\circ}$ rotation discussed above. We next turn to particular values of the parameters for
which an exact characterization of the system or ground states is possible.  

\begin{figure}
\centering
\includegraphics[width=\columnwidth]{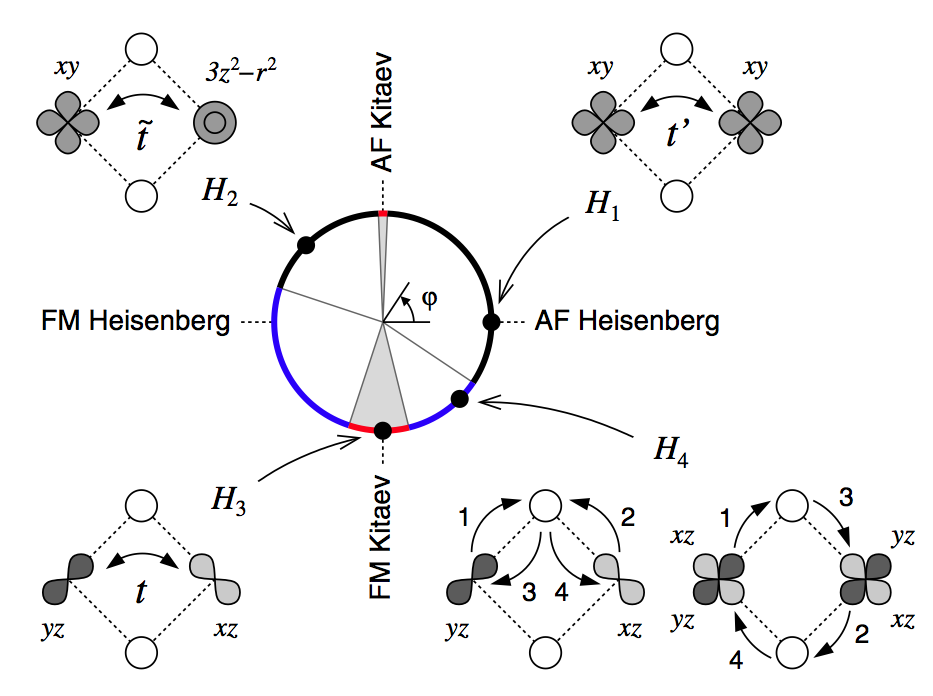}
\caption{(a) Phase diagram of the Kitaev-Heisenberg model containing 2 spin-liquid and 4 spin-ordered phases. The transition points (open dots on the $\phi$ circle) are obtained by an exact diagonalization. The gray lines inside the circle connect the points related by the exact mapping (see text). Open and solid circles in the insets indicate up and down spins. The rectangular box in the zigzag pattern (top-left) shows the magnetic unit cell. (b) Ground-state energy and its second derivative revealing the phase transitions~\cite{Chaloupka13}.}
\label{fig:1chal12}
\end{figure}

 {\it Exact orders of the quantum Heisenberg-Kitaev model at special points:} \\
The quantum system of Eq. (\ref{HKI}) can be illustrated to exactly harbor 
six disparate types of low temperature orders \cite{Chaloupka10,khalS2005,Chaloupka13,Reuther11,Schaffer2012}
at different couplings. Each of the points below lies within a phase that extends over a
finite range of parameters. Earlier works focused on three of the phases present for $C>0$ and $0 \le \alpha \le 1$ (or, equivalently, those in the fourth quadrant, $- \pi/2 \le \varphi \le 0$).
The full totality of the six phases discussed below was first enumerated in \cite{Chaloupka13}.  The full spectrum of phases and their transitions are evident 
in Fig. (\ref{fig:Chaloupka13_2}); the lower panel in this figure provides the results of exact diagonalization. Further numerical results
will be reviewed later on. The character of the transitions (whether continuous or discontinuous) between 
any two given phases must be of the same type as that between the two other phases to which these given phases are dual to. 
In the discussion below, we further invoke the self-duality of the Heisenberg-Kitaev model to illustrate that one may consistently infer exact statements about the system and its low temperature orders.

 (i).  {{$\alpha =0$ and $C>0$ (or, equivalently, $\varphi =0$):}}  \newline
{\bf{Heisenberg antiferromagnet}} \newline
 At this point, the model of Eq. (\ref{HKI}) reduces to the Heisenberg antiferromagnetic model
($J_{1}=0, J_{2}>0$) on the honeycomb lattice exhibiting global SU(2) rotational invariance. 
This Heisenberg model on the honeycomb lattice \cite{Betts78,Reger89,Oitmaa92,Mattsson,Fouet2001}
exhibits Neel order yet  with a notably reduced magnetization as a consequence of quantum fluctuations. As a symmetry analysis shows \cite{Jiang11}, the system is six-fold
degenerate. 

 (ii). {{$\alpha=1$ and $C>0$ (or $\varphi = - \pi/2$):}} \newline
{\bf{Ferromagnetic Kitaev model}} \newline
Here, Eq. (\ref{HKI}) reduces to the ferromagnetic variant of Kitaev's honeycomb model \cite{Kitaev06}. The system exhibits no broken symmetries and a wealth of fascinating characteristics \cite{Kitaev06}. All two point spin correlations vanish for distances beyond 
one lattice constant \cite{shankar,Chen08}. 

 (iii). {{$\alpha = 0$ and $C<0$ (or $\varphi = \pi$)}}: \newline
{\bf{Heisenberg ferromagnet}} \newline
 When $J_{2} <0$ and $J_{1}=0$, the Heisenberg-Kitaev model
of Eq. (\ref{HKI}) becomes that of an SU(2) invariant ferromagnetic system. 
At these parameters, the system is dual to the $\alpha =1/2$ and $C>0$ case 
discussed above.

 (iv). {{$\alpha =1$ and $C<0$}}: \newline
{\bf{Antiferromagnetic Kitaev model}} \newline
When $J_{1} <0$ and $J_{2}=0$ (or, equivalently, $\phi = \pi/2$),
the model reduces to the antiferromagnetic variant of 
Kitaev's honeycomb model. At this point, the system
is, by another duality transformation (as may, e.g., be seen by the ``bond algebraic'' technique
to be reviewed and applied to the Kitaev model in subsection \ref{solnus}) trivially dual 
to that of the ferromagnetic model. As such, 
here the system is a spin liquid. However, by contrast
to the ferromagnetic Kitaev model,
the system is less stable to Heisenberg type perturbations 
which more readily alter it and promote various orders. 

 (v). {{$\alpha=1/2$ and $C>0$ (or $\varphi = - \pi/4$): }} \newline
{\bf{Fluctuation free quantum stripe antiferromagnet}} \newline
At this point the system is exactly solvable due an elegant consideration 
of \cite{Chaloupka10}.  

When $\alpha =1/2$ (or , equivalently, $J_{1} = 2 J_{2}$ or when $\tan \varphi = -1$), 
this spatially dependent spin rotation of Eq. (\ref{uical}) maps this system
to a ferromagnetic Heisenberg model, i.e., to the Hamiltonian
\begin{eqnarray}
\tilde{H}_{HK, ~\alpha = 1/2} = - \frac{1}{2} \sum_{\langle i j \rangle} {\bm{\tilde{\tau}}}_{i}
\cdot {\bm{\tilde{\tau}}}_{j}.
\label{Hese}
\end{eqnarray}
Thus, when $\alpha =1/2$, the system harbors the continuous global SU(2) symmetry
of the ferromagnet.

Moreover, when $\alpha =1/2$, the ground states of the full quantum problem of 
Eqs.(\ref{HKI},\ref{HKII}) are thus
uniform ferromagnetic states \cite{Chaloupka10} in the rotated basis ${\bm{\tilde{\tau}}}$. 
Applying the inverse transformation, ${\bm{\tilde{\tau}}} \to  {\bm{\tau}}$,
a uniform ferromagnetic state in the ${\bm{\tilde{\tau}}}$ basis is seen to correspond
to a  {\em stripe-like antiferromagnet} in the original ${\bm{\tau}}$ basis. 
A cartoon is shown in Fig. \ref{fig:Chaloupka10_1c}.


\begin{figure}
\centering
\includegraphics[width=\columnwidth]{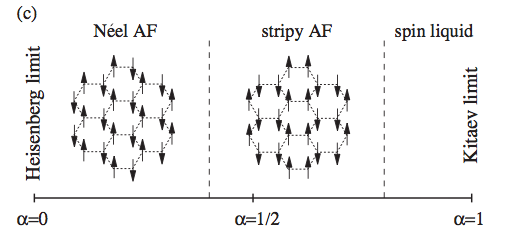}
\caption{Schematic phase diagram for $C >0$ in Eq. (\ref{HKI}): With increasing $\alpha$, the ground state changes from the Ne\`el AF order to the stripy AF state (being a fluctuation-free exact solution at $\alpha=1/2$) and to the Kitaev spin liquid. See the text for the critical values of $\alpha$.~\cite{Chaloupka10}.}
\label{fig:Chaloupka10_1c}
\end{figure}

\onlinecite{Chaloupka10} noted that this stripy antiferromagnet state is an exact eigenstate at $\alpha =1/2$ and is thus {\it fluctuation free}. When $\alpha \neq 1/2$, symmetry considerations indicate that a six-fold
degeneracy persists over the entire extent of the stripy ferromagnetic state \cite{Jiang11}. 

 (vi): {{$\alpha = 1/2$ and $C<0$ (or $\varphi = 3 \pi/4$):}} \newline
{\bf{Zig-zag phase}} \newline 
This point with, in the original Hamiltonian of Eq. (\ref{HKI}), $J_{1} = 2J_{2} >0$ is dual to the point at $\varphi =0$ ($\alpha =1/2, ~C>0$) discussed above. 
Thus, the ($\varphi=0$) antiferromagnetic states on the honeycomb model become transformed by the sublattice dependent 
180$^{\circ}$ rotation that realizes the self-duality transformation into those present when $\varphi = 3 \pi/4$. 
As antiferromagnetic Neel order (such that for $\varphi=0$) is transformed, via this sublattice dependent rotation, into
a ``{\it zig-zag}'' order (i.e., one with ferromagnetic zigzag chains of the pseudo-spins which are antiferromagnetically
coupled to each other), it follows that when $\varphi = 3 \pi/4$, the system may exhibit such zig-zag ordering. 
As the zig-zag state is, at this point, an exact (stationary) eigenstate of the Hamiltonian, this eigenstate is, similarly to the $\alpha = 1/2$ and $C>0$ point above, also {\it fluctuation free}.   
As such, associated with spin-wave fluctuations, gapless spin-wave modes (i.e., Goldstone modes) appear at this point. 
Such zig-zag ordering is of possible pertinence as several experiments have reported observations
consistent with precisely such a zig-zag ordering  \cite{Liu11,Choi2012,Singh10,Sing11}.

\subsubsection{Spin waves and Exact Transition Points}

Similar to the results reviewed for the 120$^{\circ}$ compass model in Section \ref{sec:NBCB}, a spin-wave type analysis may be performed about, for instance,
a zig-zag type ground state. The bare spin-wave dispersion for such fluctuations was reported in \cite{Chaloupka13}. Employing an $(a \times b)= (\sqrt{3} a_{0} \times a_{0})$ 
rectangular unit cell
similar to that in \cite{Choi2012} where $a_{0}$ is the length of the hexagon.
Writing the reciprocal lattice vector as ${\bf{k}} = (\frac{2 \pi}{a} h, \frac{2 \pi}{b} k)$, the resulting four dispersive branches may be exactly computed \cite{Chaloupka13}.



For a choice of parameters $(J_{1}, J_{2}) = -(20.9, 4.01)$ in Eq. (\ref{HKI}) [or, equivalently, $\alpha = 0.723$, $C=-14.46 meV$ (corresponding
to an angular parameterization with $\varphi$ in the second quadrant with $\tan \varphi \simeq 2.61$)]; these are parameters estimated for Na$_{2}$IrO$_{3}$ where there are experimental indications
of a viable zig-zag phase  \cite{Liu11,Choi2012,Singh10,Sing11}. At the solvable fluctuation free $alpha = 1/2$ and $C<0$ point (or, equivalently, $\varphi = 3 \pi /4$) 
where an exact zig-zag state may be proven as discussed above, the four spin-wave mode become two pairs of degenerate modes. The lower energy set of these modes realizes the Goldstone mode behavior. 
For parameter values away from this point a magnon gap is expected by quantum fluctuations \cite{Chaloupka13}. 


Below it is illustrated that, in a somewhat similar vein, exact statements can be made concerning 
transitions between classical orders in this system.  

(i)  {{$\alpha = 1/3$ and $C>0$}} (or $\varphi = \tan^{-1}(-1/2) \simeq - 26.563^{\circ}$): 
{\it Transition between the Neel and stripe antiferromagnetic ground states.}
This point has been examined earlier \cite{Chaloupka10}. 
Examining Eq.(\ref{HKII}), one sees that associated with each 
nearest neighbor link of the lattice along the $\gamma=x,y$ or $z$ direction only the exchange interactions amongst two components of the pseudo-spin $\gamma' \neq \gamma$ appear. That is,
e.g., associated with a link along the $\gamma =x$ direction, one explicitly has
\begin{eqnarray}
H_{HK; ~ (i,i+{\bf{e}}_{x})} = \frac{2}{3} (\tau^{y}_{i} \tau^{y}_{i+ {\bf{e}}_{y}} +
\tau^{z}_{i} \tau^{z}_{i+ {\bf{e}}_{z}}).
\end{eqnarray}  
The appearance of only two spin components suggests that symmetries may appear.
Unlike the situation for $\alpha =0,1/2$ (or, equivalently, $\varphi = -\pi/4,-\pi/2$), however, these
are only emergent symmetries within the ground state (i.e., zero temperature) 
sector of the classical model.  Specifically, {\it Ising type $d=1$ gauge like symmetries} are present
(corresponding to a flip of all spins along chains) and these
lead to an infinite degeneracy of the ground state sector. When acting on a classical
Neel ground state, such $d=1$ Ising type operations lead to the stripy antiferromagnet state. 
Thus, classically, the boundary 
between the Neel and stripy antiferromagnet ground states lies, exactly, at 
$\alpha = 1/3$.  

(ii) {{$\alpha = 1/3$ and $C<0$:}} 
{\it Transition between the ferromagnetic and zigzag ground states}
This point has not been examined in earlier works. We wish to note that the system at this point ($\alpha = 1/3$ and $C<0$) is dual to that at $\alpha = 1/3$ and $C>0$ (and for these values of the parameters related to it
in the angular parameterization via the transformation $\varphi \to (\pi - \varphi)$). As
can be established by incorporating the sublattice dependent rotation, 
similar to its dual point at positive $C$, the $(\alpha = 1/3, C<0)$ point lies at the boundary
between classical ferromagnetic and classical zigzag order.

The two semi-classical points above $(\alpha = 1, C>0$) and $(\alpha = -1, C<0$) are self-dual (i.e., map onto themselves) under the self-duality transformation.
Similar to its counterpart at positive $C$, the $\alpha =1$ system at negative $C$ is a spin liquid \cite{Chaloupka13}.
Both systems at these points correspond to Kitaev model for ferromagnetic ($C>0$) or antiferromagnetic ($C<0$) couplings respectively.
Although the ferromagnetic and antiferromagnetic Kitaev models are identical to each other and share an identical spectrum, 
as stated above,when Heisenberg type perturbations are introduced it is seen the ferromagnetic spin liquid is more
stable and extends over a far larger range of $\varphi$ values than its antiferromagnetic counterpart \cite{Chaloupka13}. 
This is readily seen in Fig. \ref{fig:1chal12}. 
Related phases (including a spiral phase) may appear in the model of Eq. (\ref{LRHK}) \cite{Kargarian2012}. 


\subsubsection{Order by disorder}
Order by disorder occurs at two different situations. One encounters:

(i)  {\em Global rotational symmetry}
As discussed above, when $\alpha =0,1/2$ (for both signs of the constant $C$ in Eq. (\ref{HKI}))
the system rigorously exhibits a global SU(2) symmetry. This symmetry appears as either as a standard global uniform rotational symmetry
in the case of the ferromagnet ($C<0$) or antiferromagnetic ($C>0$) Heisenberg model when $\alpha =0$ 
or to such a global rotational symmetry following a four sublattice dependent 180$^{\circ}$ rotation in
the case of the $\alpha =1/2$  to which these systems are dual. 
Similar to the discussion of the 
120$^{\circ}$ and $90^{\circ}$ compass models, within a spin-wave approximation, 
the classical rendition of Eq.(\ref{HKI}) exhibits, 
also for $\alpha \neq 0, 1/2$, an emergent continuous symmetry
within its ground state sector. Similar to the situation for
the 120$^{\circ}$ and $90^{\circ}$ compass models, this symmetry is lifted
by an order out of disorder mechanism. Physically, a spin gap opens and 
the system preferentially orders along one of the crystalline 
axes of IO$_{6}$ octahedra. The energy penalty for orienting
the pseudo-spins away from the crystalline axes scales, for $\alpha$
close to 1/2, as $\Delta \simeq \frac{2}{\alpha}(\alpha - \frac{1}{2})^{2}$ \cite{Chaloupka10}. 

(ii) {\em $d=1$ Ising type symmetries}
Quantum fluctuations lift the above discussed emergent Ising degeneracy of the ground state sector
of the classical rendition of the $\alpha =1/3$ system \cite{Chaloupka10} and move the transition point
between the two phases to a larger value of $\alpha$.

\subsubsection{Phase Transitions}
There are several zero temperature transition points at values of the coupling constants that are not exactly determined.

(i) {\em Transition between the Neel phase and the stripe antiferromagnet}
The boundary between the ground states of the Neel and stripy antiferromagnet
ground states (a discontinuous transition by symmetry considerations)
lies at $C>0$ near $\alpha \simeq 0.4$ (or $\tan \varphi \simeq -2/3$). Within second order perturbation theory 
\cite{Chaloupka10}, the energies of the Neel and stripy antiferromagnet states 
on a honeycomb lattice of $N$ sites (and $3N/2$ links) are given by 
\begin{eqnarray}
E_{HK, ~ Neel} \simeq - \frac{3}{16} N (3- 5 \alpha), \nonumber
\\ E_{HK, ~ stripy~ AF} \simeq - \frac{1}{8} N (5 \alpha - 3 + \frac{1}{\alpha}).
\end{eqnarray}
Setting $E_{HK, ~ Neel} = E_{HK, ~ stripy ~ AF}$ leads to $\alpha =0.4$ (and a physically irrelevant solution of $\alpha =0.2$). Numerical analysis further corroborate the existence of
{\em a first order transition} between the Neel and stripy antiferromagnetic phases at $\alpha =0.4$.  

(ii) {\em Transition between the stripe antiferromagnet and the ferromagnetic Kitaev 
spin liquid phase} In the Kitaev model (i.e., in the $\alpha =1$ limit of Eq.(\ref{HKI}) with $C=1$),
 the sole nearest neighbor correlations between two spins are
those associated with the iso-spin component $\gamma$ that lies 
along the link direction $\langle i j \rangle$ (i.e., $\gamma || \langle i j \rangle$) \cite{shankar,Chen08}. 
An estimate employing these nearest neighbor correlation suggests that
the spin liquid phase may extend beyond the Kitaev point ($\alpha =1$)
up to a value of $\alpha \simeq 0.86$. This value is close to the numerically
ascertained value of $\alpha \simeq 0.8$ (or $\tan \varphi \simeq -4$) for a critical point
between the spin liquid and stripy antiferromagnetic phase.  

Relying on the above results, in the below,  additional approximate transition points are computed. The
values of the transition points that one finds are consistent with those found by the exact diagonalization  
in~\cite{Chaloupka13}.

(iii) {\em Transition between the ferromagnetic Kitaev spin liquid and the spin ferromagnet} 
Using the self-duality of the system one can compute the boundaries between the Kitaev spin liquid and the spin ferromagnet by relying on earlier results concerning
the transition point between the stripe ferromagnet (which is dual to the stripe antiferromagnet) and the ferromagnetic Kitaev spin liquid. This approach which is now
applied using the self-duality to compute the phase boundaries has not been invoked in earlier works. Setting $\alpha \simeq 0.8$ ($C>0$) to mark the transition point between
the stripe antiferromagnet and the ferromagnetic Kitaev model, the self-duality illustrates that the transition
between the ferromagnetic Kitaev spin liquid and the spin ferromagnet occurs at $\alpha \simeq 1.5$ and $C>0$ (or $\tan \varphi \simeq 3$ with $\varphi$ in the third quadrant). 
The character of the transition between the ferromagnetic Kitaev spin
liquid and the ferromagnet is 
identical to that between the stripe antiferromagnet and the ferromagnetic Kitaev spin
liquid.

(iv) {\em Transition between the ferromagnet and the zig-zag phase}
Once again, one may employ the self-duality of the system to ascertain the transition point (and the character
of the transition) between the ferromagnetic phase (dual to the stripe antiferromagnet) and the zig-zag phase 
(dual to the Neel phase). This transition occurs at $\alpha \simeq 0.25$ for $C<0$ (or, equivalently, when $\varphi$
lies in the second quadrant with $\tan \varphi = -1/3$. 

\subsubsection{Overview of Numerical results}

Numerical validation of the phase diagram comprised of the three phases was first obtained by \cite{Chaloupka10}. In Fig. \ref{fig:Chaloupka10_2}, different measures are presented.  Numerical results \cite{Reuther11}, \cite{Chaloupka10} suggest that the transition between the spin liquid phase and the stripy antiferromagnetic phase is a continuous transition. On their own, direct slave mean field approaches \cite{Schaffer2012} lead to the conclusion that the zero temperature transition between the spin-liquid and stripy antiferromagnetic phase is a discontinuous transition.  However, \cite{Schaffer2012} quantum fluctuations may render the transition continuous in accord with numerical calculations.  Beyond its application to the transition point, the work of \cite{Schaffer2012} nicely illustrates, amongst others, how energy band may become increasingly dispersive once larger Heisenberg terms are introduced, the nature of a slave fermion treatment of this problem, and the Ising character of the Kitaev spin liquid state as gleaned by this method.   

\begin{figure}
\centering
\includegraphics[width=.8\columnwidth]{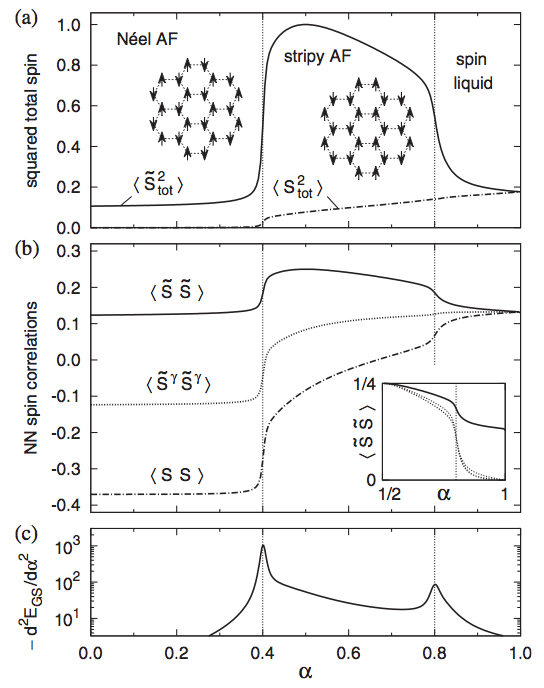}
\caption{(a) Squared total spin of the 24-site cluster, normalized to its value in the fully polarized FM state, as a function of $0 \le \alpha \le 1$ for $C>0$. The solid (dot-dashed) line corresponds to the rotated (original) spin basis. (b) The NN spin correlations: The solid (dot-dashed) line corresponds to a scalar product of the rotated (original) spins. The component of the correlation function matching the bond direction is indicated by a dotted line. This quantity is the same in both bases. The inset compares NN spin correlations (solid line) above $\alpha=0.5$ with longer range spin correlations up to third-nearest neighbors (dotted lines). (c) Negatively taken second derivative of the ground state energy with respect to $\alpha$. Its maxima indicate the phase transitions at $\alpha \approx 0.4$ and 0.8.~\cite{Chaloupka10}.}
\label{fig:Chaloupka10_2}
\end{figure}

It is further noteworthy that an applied uniform field in the spatially dependent  rotated basis discussed above leads to a near saturation  of the rotated ${\bf{\tilde{{\tau}}}}$ moments within the stripy antiferromagnetic phase \cite{Chaloupka10}. The phase diagram of this system has been recently explored in the presence of an external magnetic field \cite{Jiang11} and at finite temperatures \cite{Reuther11}. We next survey the  results found in those cases. 

{\em External Magnetic Field}  
As was surveyed earlier, in Kitaev's honeycomb model, the application of a magnetic field ${\bf{h}}$
along the $\langle 111 \rangle$ direction leads to the appearance of a non-Abelian topological
phase \cite{Kitaev06}. Inspired by this, the authors of \cite{Jiang11} investigated the phase diagram of the amended Hamiltonian 
\begin{eqnarray}
H_{HK+h} =  H_{HK} - {\bf{h}} \cdot \sum_{i} {\bf{\tau}}_{i} 
\label{HK+h}
\end{eqnarray}
Physically such an effective magnetic field coupling to the pseudo-spins as in Eq. (\ref{low_so}),  can be brought about by, e.g., a magnetic field that couples to the original spin degrees of freedom. 
The found phase diagram is depicted in Fig. \ref{fig:Jiang11_1}. As seen in Fig. \ref{fig:Jiang11_1}, the transitions between  the three zero-field phases present when $C>0$ and $0 \le \alpha \le 1$ (Neel, stripy ferromagnet, and ferromagnetic spin liquid) persist yet with additional rich features.  We comment on these below. 

\begin{figure}
\centering
\includegraphics[width=.8\columnwidth]{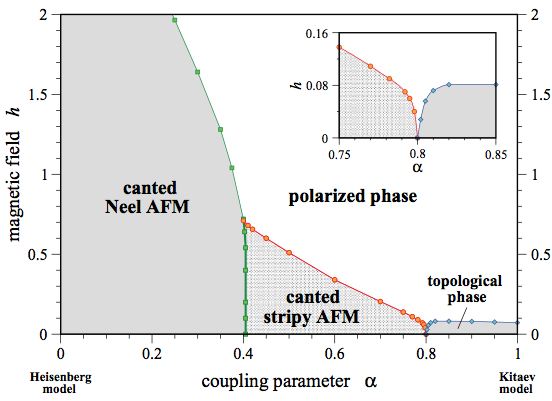}
\caption{Ground-state phase diagram of the Heisenberg-Kitaev model with $C=1$ in Eq. (\ref{HKI}). 1) in a  magnetic field of strength $h$ along the direction $[111]$. Interpolating from the Heisenberg ($\alpha = 0$) to Kitaev ($\alpha = 1$) limit for small field strength, a sequence of three ordered phases is observed: a canted Ne\`el state for $\alpha < 0.4$, a canted stripy Ne\`el state for $0.4 < \alpha < 0.8$, and a topologically ordered state for nonvanishing field around the Kitaev limit. All ordered phases are destroyed for sufficiently large magnetic field giving way to a polarized state.~\cite{Jiang11}.}
\label{fig:Jiang11_1}
\end{figure}

{\em (1)~ The canted and polarized phases}
At sufficiently large $h=|{\bf{h}}|$, the pseudo-spins become 
polarized along the applied field $\langle 111 \rangle$ direction. For $h>0$, the Neel and 
stripy ferromagnetic states both become canted along this direction as well. 
As symmetry considerations indicate, and is numerically seen, 
the six-fold degeneracy of both the canted Neel state and the canted
stripy antiferromagnet persist as $h$ is varied. 

The transition between the canted Neel and polarized state
as well as the transition between the canted stripy antiferromagnet and 
the polarized state are both continuous.  By contrast, the transition between the 
canted Neel state and the canted stripy ferromagnet is a first order
transition. Thus, the point of merger of the two critical lines
separating the canted phases from the polarized states
and the first order line separating the two canted phases
from one another constitutes a viable tricritical point. 

{\em (2)~ The topological spin liquid phase}
The Iridates offer the exciting prospect of potentially realizing
a non-Abelian topological phase. As discussed earlier, 
in Eqs. (\ref{fieldkit}, \ref{hhh}, \ref{HAK1})
in his seminal work \cite{Kitaev06}, Kitaev argued,
via perturbative considerations, that
the application of a magnetic field may lead to
a gapfull phase with non-Abelian statistics.

Specifically, as reviewed in Section \ref{nonabp},  a field in the $[111]$ direction,
leads to a third order perturbative term of the form of Eq.(\ref{hhh})  \cite{Kitaev06}.
This additional three spin term in turn translates, in the bond algebraic technique which we will review in subsection \ref{solnus}, 
to the Hamiltonian of Eq.(\ref{HAK1})
with additional terms that link next nearest neighbor Majorana fermions
($A_{ij}$ is non-zero for sites $i$ and $j$ that 
are next nearest neighbors). The model of Eq. (\ref{HAK1})
was quadratic and solvable. The additional 
bilinear resulting from the field led to the appearance
of a mass and non-Abelian statistics. 

To test this perturbative argument and to better
understand the character of the non-Abelian 
phase and its potential realization in the Iridates, 
the authors of \cite{Jiang11} numerically
probed an even more general Hamiltonian that
than of Eq.(\ref{HK+h}) that is given
by 
\begin{eqnarray}
H_{HK+h+\kappa}=(H_{HK+h} - \kappa \sum_{ijk} \tau^{x}_{i} \tau^{y}_{j} \tau^{z}_{k}).
\label{HK+h+kappa}
\end{eqnarray}
The $\kappa$ term in Eq.(\ref{HK+h+kappa}) extends over spin triads as soon
in Fig. \ref{fig2a:Jiang11_1}.

\begin{figure}
\centering
\includegraphics[width=.4\columnwidth]{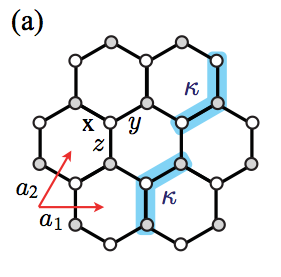}
\caption{The honeycomb lattice spanned by unit vectors ${\bm a}_1 = (1,0)$ and ${\bm a}_2 = (1/2, \sqrt{3}/2)$~\cite{Jiang11}.}
\label{fig2a:Jiang11_1}
\end{figure}

It was numerically established \cite{Jiang11} (see Fig. \ref{fig:Jiang11_4}))
that the non-Abelian phase obtained
by setting $h=0$ (and $\alpha =1$) in the Kitaev model
for $\kappa >0$ is {\em adiabatically connected} to the 
physical system of the Iridates in a magnetic field
given by the Hamiltonian of $H_{HK+h}$
with $\kappa =0$.  
The critical field required to go to
the polarized phase is seen to be monotonic and saturate
when $\kappa$ is large. This is in accord
with theoretical expectations as the Majorana
fermion gap in the non-Abelian phase is monotonic
in $\kappa$ and asymptotically tends to a constant for large $\kappa$ \cite{Jiang11}. 

\begin{figure}
\centering
\includegraphics[width=.8\columnwidth]{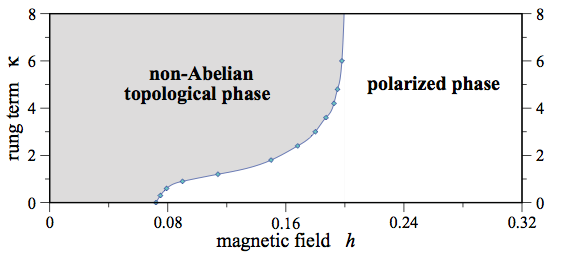}
\caption{
The constant $C=1$ in Eq. (\ref{HKI}). Ground-state phase diagram of the Kitaev model ($\alpha$= 1) in the $h - \kappa$ plane, where $h$ is the strength of a magnetic field pointing in the 111 direction and $\kappa$ is the strength of a time-reversal symmetry breaking three-site term~\cite{Jiang11}.}
\label{fig:Jiang11_4}
\end{figure}

The transition between the non-Abelian phase and the 
polarized phase in the system of $H_{HK+h}$ appears
to be continuous. Moreover, for any finite field $h \neq 0$ 
there is no direct transition between the canted stripy
antiferromagnetic phase and the non-Abelian phase. It seems
that within the $h \to 0$ limit, the two critical lines marking the (i) boundary between the 
stripy antiferromagnet and the polarized phase and (ii) the boundary between the topological spin
liquid and the polarized phase may merge at a single multi-critical point near $\alpha \simeq 0.8$
(in agreement with the value found by \cite{Chaloupka10}).  

{\em Finite temperature transitions} 
\cite{Reuther11} employed the pseudo-fermion renormalization group (PF-FRG) approach
\cite{FRG1,FRG2}
to assess viable ordering at finite temperatures. The pseudo-spins are expressed as
(pseudo-) fermi bi-linears and the resulting system is investigated perturbatively.
The phase diagram resulting from their analysis is presented in Fig. \ref{fig:Reuther11_4}.
As we will discuss below, the critical cutoff scale $\Lambda_{c}$ was argued
to emulate the ordering temperature.
  
\begin{figure}
\centering
\includegraphics[width=.8\columnwidth]{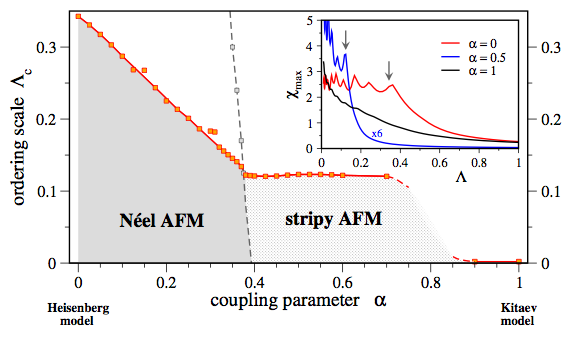}
\caption{Ordering scale $\Lambda_c$ obtained from the FRG calculations for various coupling parameters. The constant $C$ in 
Eq. (\ref{HKI}) was set here to unity. The dashed line indicates the crossover from dominant AFM to dominant s-AFM fluctuations as well as an extrapolation below the ordering transition down to $T = 0$. A regime of enhanced numerical uncertainties is encountered near $\alpha \approx 0.8$. The inset shows the RG flow of the magnetic susceptibility versus frequency cutoff $\Lambda$. The arrows indicate the estimated ordering temperatures $\Lambda_c$ where the RG flow breaks down~\cite{Reuther11}.}
\label{fig:Reuther11_4}
\end{figure}

More generally, following \cite{Honerkamp2001},
who noted that both the frequency cutoff $\Lambda$ and the temperature $T$ play the rule of infra-red cutoffs
the key notion was to identify various
frequency cutoff  $\Lambda$ scales in the PF-FRG approach
to correspond to physical temperature scales. This qualitative identification was motivated by direct 
finite $T$ calculations: this general correspondence was validated
in the analysis of the high temperature susceptibility. 

{\em Curie-Weiss temperature} 
In what follows $C=1$ in Eq. (\ref{HKI}). It was found \cite{Reuther11} that, at high temperatures, the ferromagnetic susceptibility adhered to
the Curie-Weiss law
\begin{eqnarray}
\chi  \sim 1/(\Lambda - \Lambda_{CW}).
\end{eqnarray}
 
The thus extracted Curie-Weiss scale $\Lambda_{CW}>0$  for $\alpha \gtrsim  0.68$
(indicating an overall effective ferromagnetic coupling (as consistent with the ferromagnetic 
couplings in the Kitaev ($\alpha =1$ and $C>0$) limit)). Similarly, the observation of 
$\Lambda_{CW} <0$ for $\alpha \lesssim 0.68$ is consistent with the Heisenberg ($\alpha =0$) limit
of Eq.(\ref{HKI}).  Such a change in the sign of the dominant exchange, as
adduced from the dependence of the susceptibility $\chi$ on the cutoff $\Lambda$, 
is consistent with semi-classical analysis for which the Curie-Weiss temperature is given 
by \cite{Reuther11}
\begin{eqnarray}
\Theta_{CW} = - \frac{3}{4} + \frac{5}{4} \alpha.
\label{semiCW}
\end{eqnarray}
In Eq. (\ref{semiCW}), a crossover from an overall ferromagnetic to antiferromagnetic
behavior occurs at $\alpha = 0.6$ (proximate to the value of $\alpha \simeq 0.68$ suggested
by the dependence of $\chi$ on the cutoff).
The variation of the effective Curie-Weiss cutoff scale $\Lambda_{CW}$ 
over the entire range of $\alpha$ values is provided in Fig. \ref{fig:Reuther11_3}.
Qualitatively, the dependence of $\Lambda_{CW}$ 
on $\alpha$ is similar to that of the semi-classical
approximation of Eq. (\ref{semiCW}). 

\begin{figure}
\centering
\includegraphics[width=.8\columnwidth]{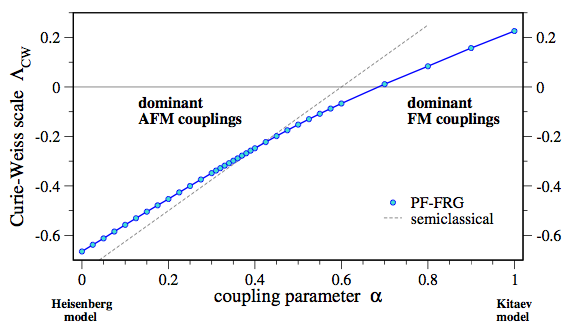}
\caption{The constant $C=1$ in Eq. (\ref{HKI}). The Curie-Weiss scale $\Lambda_{CW}$ obtained from fitting the inverse susceptibilities to a Curie-Weiss law for varying coupling parameter $\alpha$. Around $\alpha \approx 0.68$ the Curie-Weiss scale switches sign indicating a transition of the dominant exchange from antiferromagnetic ($\Lambda_{CW} < 0$) to ferromagnetic ($\Lambda_{CW} > 0$).~\cite{Reuther11}.}
\label{fig:Reuther11_3}
\end{figure}

{\em Ordering temperatures}
The scale of the ordering temperatures $T_{o}$ was identified
with the critical cutoff $\Lambda_{c}$ beyond which 
the Renormalization Group flow breaks down in
the PF-FRG method \cite{Reuther11}, see Fig. \ref{fig:Reuther11_4} \cite{Reuther11}.  

It is seen that within the Neel phase $(\alpha \simeq 0.4$) both the 
inferred (i) ordering temperature $T_{o}$ and (ii) the Curie-Weiss temperature $\Theta_{CW}$
scales are nearly linear in $\alpha$. By contrast, within the stripy ferromagnetic phase apart
from the region near the transition into the spin liquid phase, the relevant ordering temperature $T_{o}$ (as suggested by the value of the critical cutoff scale $\Lambda_{c}$)
is nearly $\alpha$ independent \cite{Reuther11}. Unlike the ordering temperature itself, $\Theta_{CW}(\alpha)$ varied
approximately linearly in this range of $\alpha$ values of the stripy antiferromagnet
phase. In the spin-liquid phase ($\alpha \gtrsim 0.8$), the ordering temperature
$T_{o} =0$. 

Assuming that both the Curie Weiss and ordering temperatures ($\Theta_{CW}$ and $T_{o}$)
are related by an identical multiplicative scale factor to the cutoff scales $\Lambda_{CW}$ and $
\Lambda_{c}$, the ratio $(|\Lambda_{CW}|/\Lambda_{c})$ is equal to the experimentally 
measured ``frustration parameter'' \cite{Art} ratio of $f=(|\Theta_{CW}|/T_{o}$) \cite{Reuther11}. 
This ratio is nearly constant ($f \approx 2$) in the Neel phase and monotonically diminishes 
in size as $\alpha$ is increased beyond the transition point ($\alpha \approx 0.4)$ into
and throughout most of the stripy antiferromagnetic regime. Within the 
spin-liquid phase, $f$ diverges. In comparison to experiments \cite{Reuther11},
 the measured frustration parameter 
 for the iridate Na$_{2}$IrO$_{3}$
 \cite{Singh10} ($f \approx 8$) 
is far larger than the values adduced for the 
Heisenberg-Kitaev model in these phases.
Earlier work has not examined finite temperature
behavior the zig-zag or other
phases which lie outside the $C>0$ and $0 \le \alpha \le 1$
parameter regime. By {\it invoking the self-duality} transformation
of Eq. (\ref{alpC}) to the results of \cite{Reuther11} and its
adduced phase diagram above, one may ascertain 
the extended {\it finite temperature phase diagram} including the zig-zag phase
(this is so as zig-zag phase is dual to the stripe antiferromagnet). 
Thus as the transition temperature
$T_{o}$ in the stripy antiferromagnetic phase, as determined by thermodynamic measurements, 
is nearly constant as a function of the parameter $\alpha$ of Eq.(\ref{HKI}), invoking the 
self-duality relations of Eqs. (\ref{alpC}), one finds that {\it the ordering
temperature of the zig-zag phase is also nearly constant.} 

\subsubsection{Heisenberg-Ising Hamiltonians}

\cite{Bhattacharjee2011} noted that trigonal distortions around
the Ir$^{4+}$ ions can lead to significant crystal field effects. These crystal fields 
may splinter the three $t_{2g}$ orbital states into a degenerate doublet
($e_{1g}^{\prime}$ and $e_{2g}^{\prime}$) and a non-degenerate $a_{1g}$ state.
Spin-orbit coupling may lead to a locking of the spin and orbital degrees of freedom
and thus to the appearance of only two pertinent states, e.g., $|e_{1g}^{\prime} \downarrow \rangle$
and $|e_{2g}^{\prime} \uparrow \rangle$ where $\downarrow$ and $\uparrow$ denote the two
possible values of the electronic spin along the axis of trigonal distortion;  these
replace the two states of Eq. (\ref{low_so}). As before, these two states can be regarded
as eigenstates of a pseudo-spin 1/2 operator. 

 \onlinecite{Bhattacharjee2011} report that a uniform trigonal distortion is most consistent with the experimental results. 
With the aid of the pseudo-spin 1/2 operator discussed above, 
an effective model with Heisenberg (${\bf \tau}_{i} \cdot {\bf \tau}_{j}$)
and Ising ($\tau^{z}_{i} \tau^{z}_{j}$) type interactions each of which 
of which contains nearest, next nearest, and next-next nearest 
couplings is constructed (i.e., a $\tilde{J}_{2}-\tilde{J}_{3}$ Heisenberg-Ising model). Resultant ground states, on the mean field level,
include the zig-zag state. 

Note that the models of  \cite{Bhattacharjee2011} and \cite{Kimchi2011} which report to fit
the experimental data (as well as, of course, the bare Heisenberg-Kitaev model
that was reviewed earlier) can all be written in terms of a generalized 
$\tilde{J}_{2}-\tilde{J}_{3}-J_{2}^{I}-J_{3}^{I}$ type Heisenberg-Ising-Kitaev model
which takes the form 
\begin{eqnarray}
H_{HIK} = H_{HK} - \tilde{J}_{2} \sum_{\langle \langle i k \rangle \rangle} {\bf \tau}_{i} \cdot {\bf \tau}_{k} 
- J_{3} \sum_{\langle \langle \langle i l \rangle \rangle \rangle} {\bf \tau}_{i} \cdot {\bf \tau}_{l} \nonumber
\\ - J_{2}^{I} \sum_{\langle \langle i k \rangle \rangle} \tau_{i}^{z} \tau_{k}^{z} 
- J_{3}^{I} \sum_{\langle \langle \langle i l \rangle \rangle \rangle} \tau_{i}^{z} \tau_{l}^{z}.
\end{eqnarray}

\begin{figure}
\centering
\includegraphics[width=.8\columnwidth]{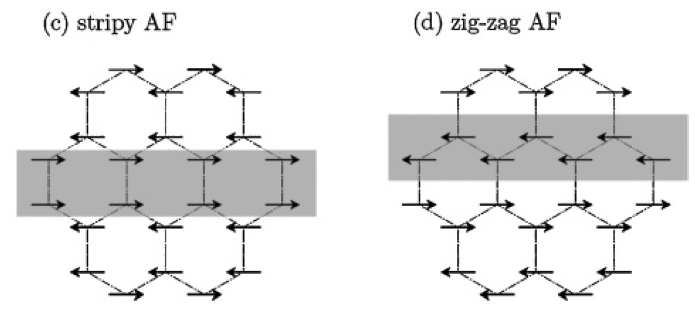}
\caption{Crystal structure and possible antiferromagnetic-ordering patterns of the Ir sublattice. (c), and (d): three possible magnetic structures. In each case, the magnetic unit cell is the same as the structural unit cell. The shaded boxes highlight the stripy and zig-zag chain elements.~\cite{Liu11}.}
\label{fig:Liu11_3cd}
\end{figure}

\subsubsection{Heisenberg-Kitaev $\tilde{J}_{2}-\tilde{J}_{3}$ Model}

In \cite{Kimchi2011}, an extension of the bare Heisenberg-Kitaev Hamiltonian $H_{HK}$ [Eq. (\ref{HKI})] was introduced wherein additional {\it next nearest neighbor} ($\langle i k \rangle$) and {\it next-next nearest neighbor} ($\langle \langle i l \rangle \rangle$) Heisenberg terms appear, 
\begin{eqnarray}
H_{\tilde{J}2J\tilde{3}} &=&
C \Big[ -2 \alpha \sum_{\langle ij \rangle_{\gamma}} \tau_{i}^{\gamma} \tau_{j}^{\gamma} 
+ (1- \alpha) \sum_{\langle i j \rangle} {\bm \tau}_{i} \cdot {\bm \tau}_{j} \Big] \nonumber
\\ &+&  (1-\alpha) [ \tilde{J}_{2} \sum_{\langle \langle i k \rangle \rangle}
{\bf \tau}_{i} \cdot {\bf \tau}_{k}  + \tilde{J}_{3} \sum_{\langle \langle il \rangle \rangle} {\bf \tau}_{i} \cdot {\bf \tau}_{l}].
\label{j2j3ham}
\end{eqnarray}
As $\alpha$ is varied from zero to unity, this extended system interpolates between  an SU(2) symmetric limit to Kitaev's honeycomb model.  Heisenberg exchange type terms between next nearest neighbor and further away sites may arise, in the standard way, from high order processes  including hopping processes and orbital overlaps \cite{Kimchi2011}. Indeed, \cite{Bhattacharjee2011} found $\tilde{J}_{2}$ to be comparable to  the nearest neighbor exchange $C$ (which following \cite{Kimchi2011} has been typically set to unity in Eq. (\ref{j2j3ham})). 

An analysis of Eq. (\ref{j2j3ham}) was performed for the classical ground states \cite{Kimchi2011} along similar lines to that of \cite{Nussinov2001} to find that for general minimizing wave-vectors 
of the interaction kernel $\hat{V}({\bf k})$ of Eqs. (\ref{Fourier_space}, \ref{long_ebasis}),  spirals constitute the sole ground states. For commensurate minimizing wave-vectors, additional Neel, zig-zag type, and stripy ground states were found. This append the  earlier discussed ground states found for the Heisenberg-Kitaev system at the point $\tilde{J}_{2}= \tilde{J}_{3}=0$.  Notably, the zig-zag type phase in this model was found already when $C>0$ and $0 \le \alpha \le 1$. 

Exact diagonalization of the Hamiltonian of Eq. (\ref{j2j3ham}) was further performed  and by employing all of the found eigenstates, the extended phase diagram mapped at finite temperatures. The authors of \cite{Kimchi2011} suggest that the current experimental data may for both Na$_{2}$IrO$_{3}$ and Li$_{2}$IrO$_{3}$ when $\tilde{J}_{2}$ and $\tilde{J}_{3}$ as well as the Kitaev term may be significant (see table I in \cite{Kimchi2011}). A moderate trigonal coupling enhanced both Heisenberg and Kitaev type terms as well as lead to additional (small) Ising interaction terms (i.e., $\tau ^{z}_{i} \tau^{z}_{j}$) and Ising-Kitaev terms ($\tau^{\gamma}_{i} \tau^{z}_{j}$ with $\gamma$ set by the direction $(ij)$) and were not suspected to radically change the obtained results \cite{Kimchi2011}. 

\subsubsection{Spin density functional theory calculations}
Thus far, we reviewed the results as they pertain to the Heisenberg-Kitaev model of Eq. (\ref{HKI}). Also first principle type  spin density functional theory computations have been carried out, sans the reduction to a Heisenberg-Kitaev model, to directly ascertain possible orders of Na$_{2}$IrO$_{3}$~\cite{Liu11}. These calculations further suggested  the existence of non-colinear (``zig-zag'') states different
from the stripy antiferromagnet and that are lower in energy. The zig-zag states are found as ground staters for $C<0$; these states are promoted by inter-orbital exchange (see Section \ref{sec:sporb} and Eq. (\ref{interorbit}) in particular)  which is a dominant contribution \cite{Chaloupka13}. 

To clarify the  structure of the zig-zag states  and their difference as compared to the  stripy antiferromagnetic state, a comparison of these states is provided in Fig. \ref{fig:Liu11_3cd}. Aside from their presence in the Heisenberg-Kitaev model for $C<0$ in Eq. (\ref{HKI}), as briefly alluded to earlier, such zig-zag states
were also found in the phase diagram of the  Heisenberg-Kitaev $\tilde{J}_{2}-\tilde{J}_{3}$ Model \cite{Kimchi2011}
also for $C>0$ and $0 \le \alpha \le 1$. 

\subsubsection{Experimental Results}

Although, as we will now review, some preliminary results are suggestive, 
determining whether exotic aspects of the physics of the Heisenberg-Kitaev model of Eq.(\ref{HKI}) are indeed materialized
in the iridates of the A$_{2}$IrO$_{3}$ type still 
requires more effort. 

We have by now alluded several times to one of the currently most promising maxims in that regard.
Experiments suggest the presence of the zig-zag phase in the iridates \cite{Liu11,Choi2012}. As we
reviewed above, this phase may be expected to be the pertinent one in the iridates \cite{Chaloupka13}
given the typical parameters that characterize the effective Heisenberg-Kitaev model
and the dominance of Eq. (\ref{interorbit}). Furthermore, the dispersion 
computed for spin waves within the zig-zag phase \cite{Chaloupka13} 
and the magnetic susceptibility are consistent
with neutron scattering \cite{Choi2012} and magnetic susceptibility data \cite{Singh10,Sing11}. 
These calculations for predictions concerning the Heisenberg-Kitaev
 model in its zig-zag phase \cite{Chaloupka13} are also consistent with magnetic susceptibility measurements 
 of both Na$_{2}$IrO$_{3}$ and Li$_{2}$IrO$_{3}$ \cite{Sing11}. Fig. (\ref{fig:Chaloupka13_4}) from \cite{Chaloupka13},
 shows a comparison between the experimentally measured susceptibility as a function of temperature and 
 that fitted
 with empirically suggested parameter values for Na$_{2}$IrO$_{3}$ (as well as parameters for Li$_{2}$IrO$_{3}$).
 
\begin{figure}
\centering
\caption{ INSERT FIG. 4 FROM
~\cite{Chaloupka13}.
The parameters are for Na$_{2}$IrO$_{3}$ are given by $(J_{1}, J_{2}) = (-20.9, -4.01)$ in Eq. (\ref{HKI}) [or, equivalently, $\alpha = 0.723$, $C=-14.46 meV$ (corresponding
to an angular parameterization with $\varphi$ in the second quadrant with $\tan \varphi \simeq 2.61$)] . 
The parameters for Li$_{2}$IrO$_{3}$ are given by $(J_{1}, J_{2}) = (-15.7,-5.3)$, or equivalently, $C=-12.15 meV $and $\alpha = 0.564$ (corresponding to 
$\tan \varphi \simeq -1.48$ with $\varphi$ in the second quadrant).}
\label{fig:Chaloupka13_4}
\end{figure}

As discussed above, one of the key results of the finite temperature analysis of 
\cite{Reuther11} was that, within the stripy antiferromagnetic phase, the transition temperature
$T_{o}$, as determined by thermodynamic measurements, 
is nearly constant as a function of the parameter $\alpha$ of Eq.(\ref{HKI})
[see  Fig. \ref{fig:Reuther11_4}]
while the Curie-Weiss temperature varied smoothly with $\alpha$ [Fig. \ref{fig:Reuther11_3}]. 
As first remarked above, invoking the 
self-duality relations of Eqs. (\ref{alpC}), one sees that the same follows for the ordering
temperature of the zig-zag phase. 

An interesting set of experiments \cite{Singh10,Liu11,Sing11} 
on Na$_{2}$IrO$_{3}$ and Li$_{2}$IrO$_{3}$ offers a consistent 
realization of this trend.  In particular, in \cite{Sing11} it
was found that both Na$_{2}$IrO$_{3}$ and Li$_{2}$IrO$_{3}$
exhibit the thermodynamic signatures of a transition at $T_{o} \simeq 15 K$.
(Scattering measurements indicate that long range magnetic ordering,
in Na$_{2}$IrO$_{3}$ is, as to be expected, 
nearly coincident and appears for $T \lesssim  13.3 K$. \cite{Liu11};
magnetic susceptibility measurements (which have been fitted in Fig. (\ref{fig:Chaloupka13_4})) 
similarly show a precipitous drop at $T \simeq 15 K$ \cite{Singh10}.) 
[see Fig. \ref{fig:Sing11_3}
from \cite{Sing11}].

\begin{figure}
\centering
\includegraphics[width=.8\columnwidth]{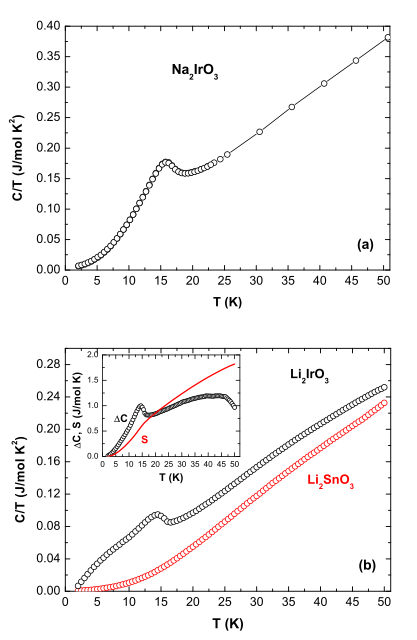}
\caption{ (a) The heat capacity divided by temperature $(C/T)$ versus the temperature $T$ data between $T$ = 1.8 K and 40 K for single crystals of Na$_2$IrO$_3$ and the heat capacity of Na$_2$SnO$_3$ as the lattice contribution $C_{lattice}/T$ versus $T$. The inset shows the $C/T$ versus $T$ data in $H=0$ and 7T applied magnetic field. (b) The difference heat capacity $\Delta C$ and difference entropy $\Delta S$ versus $T$ data between $T$ = 1.8 K and 40 K ~\cite{Sing11}.}
\label{fig:Sing11_3}
\end{figure}

However, while Na$_{2}$IrO$_{3}$ has a Curie-Weiss temperature
$\Theta_{CW} \simeq -125 K$, its cousin Li$_{2}$IrO$_{3}$, with the lighter
Li ion replacing Na,
exhibited a far reduced Curie Weiss temperature of $\Theta_{CW} \simeq -33K$
[see Fig. \ref{fig:Sing11_2} from \cite{Sing11}].

\begin{figure}
\centering
\includegraphics[width=.8\columnwidth]{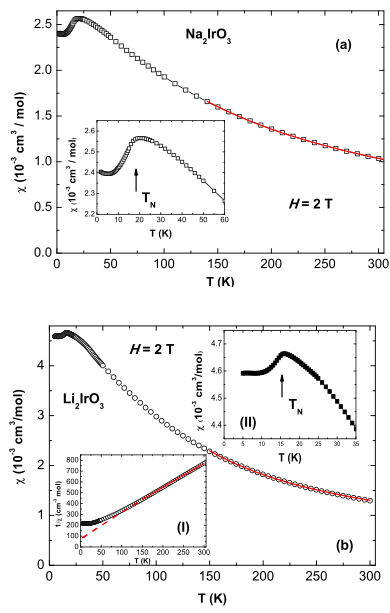}
\caption{ (a) Magnetic susceptibility $\chi$ versus temperature $T$ for Na$_2$IrO$_3$. The fit by the Curie-Weiss (CW) expression $\chi = chi_0 + C/(T - \theta)$ is shown as the curve through the data. The inset shows the anomaly at the antiferromagnetic ordering.	(b) $\chi$ versus $T$ for	Li$_2$IrO$_3$. The solid curve through the data is a fit by the CW expression. The inset (I) shows the $1/\chi(T)$ data for Li$_2$IrO$_3$. The solid curve through the data is a fit by the CW expression and the dashed curve is an extrapolation of the fit to lower $T$ . The inset (II) shows the anomaly at the onset temperature of antiferromagnetic ordering~\cite{Sing11}.}
\label{fig:Sing11_2}
\end{figure}

An initial possible interpretation of these results was that both Na$_{2}$IrO$_{3}$ and Li$_{2}$IrO$_{3}$ correspond
(possibly with some additional terms) to a realization of
Eq. (\ref{HKI}) within the stripy antiferromagnet phase 
yet with two different values $\alpha$. This is 
further bolstered by the observation that the
magnetic susceptibility $\chi(T)$ saturates
at low temperatures \cite{Singh10} to a
large finite value: such a saturation
is consistent with stripe type ordering. 
Resonant X-ray scattering measurements
\cite{Liu11} further indicate Bragg peaks that
are inconsistent with a Neel state. These peaks 
may, however, be accounted for by structures similar to
the stripy antiferromagnet or the somewhat similar zig-zag states.
As noted earlier,
from spin density functional calculations \cite{Liu11}
 the zig-zag states were suggested to be lower in energy 
 than the stripy antiferromagnetic states. Interestingly, 
 as we reviewed in some detail above,
 the zig-zag states are ground states of the Heisenberg-Kitaev model
 within an extensive parameter regime. Notably, the inter-orbital $t_{2g}$-$e_g$ exchange processes [leading to Eq. (\ref{interorbit})]
 may promote precisely the zig-zag phase of very likely
experimental pertinence in the iridates \cite{Chaloupka13}.
 
The experimental results are thus consistent with some of the 
predictions of the Heisenberg-Kitaev model
but, currently, more work (both theoretical and 
experimental) may be required to attain a more
comprehensive picture. Accounts
including (i) effective $C-J_{2}-J_{3}$ Heisenberg-Ising type models 
with interactions that may result from large diagonal couplings
\cite{Bhattacharjee2011}
 as well as 
 (ii) $\tilde{J}_{2}-\tilde{J}_{3}$ Heisenberg-Kitaev models 
\cite{Kimchi2011} 
may well better explain key features of the data.

In closing, we should note  
more rudimentary aspects that have been established in these materials. 
From the Curie-Weiss tails, it was determined that 
both Na$_{2}$IrO$_{3}$ and Li$_{2}$IrO$_{3}$
have an effective spin of $S_{eff} =1/2$ \cite{Singh10,Sing11}. 
Both of these Iridates are Mott insulators. In \cite{Singh10}, it
was found that, for $100K<T<300K$, the electrical
resistivity data for Na$_{2}$IrO$_{3}$ is of the variable range
hopping type as expected for localization by disorder;
there is no sign of activated Arrhenius behavior.
It was further seen that aside from
a pronounced drop of the susceptibility at $T \simeq 15K$,
there is a broad global maximum of the susceptibility
at $T \simeq 23K$.  This broad maximum in
the susceptibility, the high temperature tail 
of the specific heat change (after lattice
subtraction), and the entropy adduced
with this specific heat data all suggest
that short order persists beyond $T_{o} \simeq 15K$
up to higher temperatures. Some early, non thermodynamic,
measurements of Li$_{2}$IrO$_{3}$ \cite{felner2002,kobayahsi2003}.
suggested the non-existence of magnetic order (for measured temperatures of $T>5K$)
\cite{felner2002} and glassy behavior \cite{kobayahsi2003}. 

\subsection{Compass Heisenberg Models}
\label{compass_heisen}

The Heisenberg-Kitaev models studied in subsection \ref{HKCS} constitute a very special
case of the more general Compass-Heisenberg models of Eq. (\ref{hybrid_compass}). 
Non Heisenberg-Kitaev realizations of the Compass-Heisenberg models have not been less
explored. Some key features of the 90$^{\circ}$ compass Heisenberg  model have been reported in \cite{Trousselet10,Trousselet12}.
As reviewed in section \ref{VCsection}, this model and its descendants may describe a placement of rectangular superlattice of vacancy centers on a diamond grid. 
The prominent effect of adding Heisenberg interactions was highlighted in \onlinecite{Trousselet12}. Introducing arbitrarily weak Heisenberg interactions lifts the exponential degeneracy of the compass system and favor a particular 
ordering between the compass model rows (or columns). In the presence of Heisenberg interactions, only a two-fold degeneracy remains. 
The anisotropic compass interactions lead to a gap to spin wave excitations. In small clusters, the lowest energy excitations are those 
in which entire individual rows (or columns) are flipped. These excitations and related issues have been extensively investigated \cite{Trousselet12}.

\section{Kitaev Models \&  Quantum Computing} 
\label{sec:kit}

Recently, there has been much activity in the study of compass-related models as candidates for {\it topological quantum memories} \cite{Kitaev03,Kitaev06}. In such a topological memory, quantum states can in principle be encoded fault tolerantly -- i.e., be {\it protected} against decoherence \cite{Shor96,Kitaev03,Kitaev06,Dennis2002}. This aspect of fault tolerance motivates this activity. Assuming that errors are of a {\it local} nature, topological quantum memories are intrinsically stable because of  physical fault-tolerance to weak, quasi-local, perturbations.

In this section, we review two of the prototypical models which were invented to describe the basic principles of topological quantum computing \cite{Kitaev03,Kitaev06}. As we will see, one of these models, Kitaev's {\it toric code} model, is essentially a pure Ising gauge theory. Its excitations are {\it anyons} that obey Abelian statistics. It appears as a limiting form of the Kitaev {\it honeycomb} model which is far richer and enables non-Abelian quantum computing. Kitaev's honeycomb model is exactly the compass model on a hexagonal lattice, introduced in Sec.~\ref{sec:basic} in Eq. (\ref{eq:HKT}), where the non-Abelian excitations arise in presence of an additional magnetic field.

The key feature of these models is that possess these very interesting physical properties while still being exactly solvable, as will be explained in Sec.~\ref{kith} for the bare Kitaev model [and in Section \ref{nonabp} for an extended variant in the presence of a magnetic field (or, more precisely, additional term involving the products of three neighboring pseudo-spins that supplant the two-spin interactions in Eq. (\ref{eq:HKT})]. 
This sets these Kitaev models apart from the generic compass-type model. 
To introduce the general notions and to put the exciting features of the Kitaev models in perspective, we first briefly review anyonic statistics, braiding and fusion rules while keeping the focus on compass-Kitaev models. For an in-depth review of topological quantum computing we refer to \cite{Preskill,Nayak2008}.

\subsection{Basic Notions of Statistics}
\label{sec:notions}

A potent prediction of quantum mechanics is that identical particles must generally be fermions or bosons. In its simplest form one can consider a many body wavefunction $\psi(r_{1}, ..., r_{N})$ and the effect of permutations on it. If a permutation operator $P_{12}$ that permutes two particles (particles number 1 and 2)  commutes with the Hamiltonian ($[H,P_{12}]=0$) then we can simultaneously diagonalize both the Hamiltonian and $P_{12}$. Such a relation is guaranteed when particles 1 and 2 are identical to one another and thus appear symmetrically in the Hamiltonian. Formally, one may view the operation of $P_{12}$  (and its square)  as the spatial rotation of one particle around the other or the as the {\it braiding} of the world line of one particle around another, see Fig~\ref{fig:braid}.  

Conceptually, $P_{12}^{2}$ can be emulated by the rotation of one particle (say particle 1) about the other  (particle 2) by 360$^ {o}$.  In three and higher spatial dimensions, the permutation operator is its own inverse ($P_{12}^{2} =1$) and consequently these are the only two generic possibilities of bosonic ($P_{12} | \psi \rangle = | \psi \rangle$) or fermionic statistics ($P_{12} | \psi \rangle = - |\psi \rangle$. Similarly, for general pairwise permutations $P$, one has that $P^{2} \psi(r_{1}, r_{2}, ..., r_{N}) = \psi(r_{1}, r_{2}, ..., r_{N})$ and consequently $P = \pm 1$. 
 
\subsubsection{Anyons and Braiding}
\label{sec:anyons}

\begin{figure}
\centering
\includegraphics[width=\columnwidth]{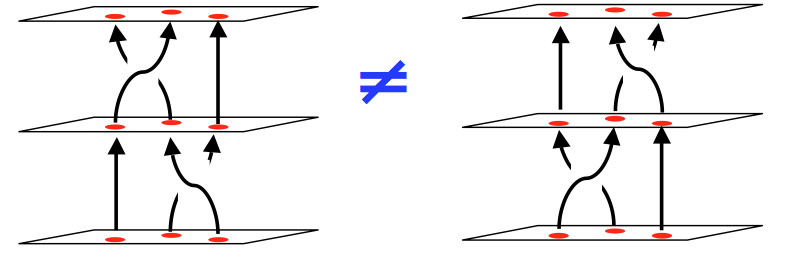}
\caption{Braiding in two-spatial dimensions. If a braid group is non-Abelian (non-commutative) then the order of the braiding operations is important. The statistics in such cases are encoded in matrix representations of braid group.
(Schematic courtesy of S. Simon.)}
\label{fig:braid}
\end{figure}

As first realized by Leinass and Myrheim, \cite{Leinaas77} in two spatial dimensions (or 2+1 space-time dimensions), however, the situation is considerably richer. In two spatial dimensions, the rotation of one particle around the other (emulating the permutation) need not be the same for clock-wise and counter-clockwise directions. Unlike the case in three spatial dimensions when a clockwise rotation can be converted into a counter-clockwise rotation depending on whether we are looking at the rotation from above or below the plane where the rotation takes place, in two spatial dimensions, a directionality can be associated with such a braiding operation, see Fig~\ref{fig:braid}. In higher dimensions (e.g., three spatial or 3+1 space-time dimensions), a double exchange of two particles along a path is topologically equivalent to an exchange along a topologically equivalent path which is shrunk to the origin and thus the wavefunction must be unaltered.
 
Thought about equivalently, in 2+1 space-time dimensions braiding operations can have non-trivial topologies just as shoe string can be tied in three dimensions (and only in three dimensions).  In higher space-time dimensions, just as shoe laces can always untied, particle world lines cannot lead to non-trivial topologies and the only possibilities are (especially so for gapped systems)  those of fermionic and bosonic statistics. 

{\it Anyons} are particles or excitations that are neither bosons nor fermions -- hence the name {\it any}-ons \cite{Leinaas77,Wilczek82a,Wilczek82b}. It is useful to define them formally by considering the configuration space ${\cal{C}}_{N}$ associated with $N$ identical particles that reside on the spatial manifold ${\cal{M}}_{N}$ \cite{Leinaas77}. This space is given by 
\begin{eqnarray}
{\cal{C}}_{N} = \frac{ 
{\cal{M}}_{N} / {\cal{D}}
}
{S_{N}},
\label{cmd}
\end{eqnarray}
where ${\cal{M}}_{N} / {\cal{D}}$ refers to the removal of the space of singular configurations ${\cal{D}}$ (where two or more particles coincide) from the space ${\cal{C}}_{N}$. The division of the quotient group by the permutation $S_{N}$ group of $N$ particles is carried out in order to reflect the indistinguishability of the identical particles. Within the path integral formulation, a correspondence exists between the possible unitary irreducible representations of the first homotopy groups $\pi_{1}({\cal{C}}_{N})$ of the configuration space ${\cal{C}}_{N}$ and the viable statistics is seen. Specifically in two spatial dimensions, the first homotopy is given by the {\it braid group} $\pi_{1}( {\cal{C}}_{N}) = B_{N}$ of $N$ particles whereas in higher dimensions it is equal to the permutation group, $\pi_{1}({\cal{C}}_{N}) = S_{N}$. \cite{WuPRL84}

The generators ${\cal{T}}_{i}$ of all of these homotopy groups (both braiding and permutations) that interchange particles at sites $i$ and $(i+1)$ satisfy the relations
\begin{eqnarray}
{\cal{T}}_{i} {\cal{T}}_{j} = {\cal{T}}_{j} {\cal{T}}_{i},   ~~~~ |i-j| \ge 2, \nonumber
\\ {\cal{T}}_{i} {\cal{T}}_{i+1} {\cal{T}}_{i} = {\cal{T}}_{i+1} {\cal{T}}_{i} {\cal{T}}_{i+1}.
\label{braid}
\end{eqnarray}
For the permutation group, a double interchange corresponds to the identity ${\cal{T}}_{i}^{2} =1$; this additional constraint leads to a subgroup of the braiding group. Specifically, this constraint enables the permutation group to only have the well known $N!$ elements.  By contrast, the braiding group is continuous. 

It is instructive to consider one-dimensional irreducible (or scalar) representations of these groups. Whereas there are only two possible irreducible scalar representations of the permutation group $S_{N}$ (namely the trivial bosonic representation (where it is everywhere the identity 1) and the fermionic representation where it is given by $(-1)^{P} = \pm 1$ for even/odd permutations $P$), there is a continuous set of possible representations for the braid group $B_{N}$ wherein ${\cal{T}}_{j} = e^{i \theta}$. As it must, the braiding group representation $B_{N}$ includes ($\theta = 0, \pi)$ the fermionic and bosonic statistics of the permutation $P_{N}$ representation as a subgroup. For such scalar representations, the ordering of the braiding operations is clearly unimportant. These braiding representations thus correspond to {\it Abelian} anyons. 

One of the simplest realizations of such anyonic statistics is for (Abelian) Fractional Quantum Hall systems. In, e.g., the $\nu =1/3$ state, under the exchange of two excitations the wavefunction can acquire  a phase of $\theta= \pm 2 \pi/3$. That is, revolving one quasi-particle around another leads to the change of phase $| \psi \rangle \to e^{\pm 2 \pi i/3} | \psi \rangle$. The fractional phase here of $2\pi /3$ may be seen to reflect a statistical Aharonov-Bohm effect \cite{Aharonov1959,Kivelson_Rocek} associated with the fractional quasi-particle charge $q = \pm e/3$ with $e$ the electron charge \cite{Laughlin83,Goldman95}.

\subsubsection{Non-Abelian Anyons}
\label{sec:non-Abelian}

Non-Abelian topological excitations display non-trivial anyonic statistics \cite{Moore1991,Nayak2008,Stern2010,Ahibrecht2009}. The end result following a rotation of one anyon around another generally depends on the order in which the rotations have been done, see, e.g., Fig~\ref{fig:braid}. In such a case, the braid group is non-commutative. Such anyons lie at the heart of quantum topological computing schemes: in a remarkable work \cite{Freedman2002}, it was illustrated that particular sorts of non-Abelian anyons allow universal quantum computation. 

For non-Abelian anyons, a rotation of one particle around another (or braiding of their world lines) does not merely lead a change of the wavefunction by a phase. Rather, more generally, when 
non-Abelian anyons are wound around one another there is a unitary operator $U$ that leaves the system with the same energy as it must for identical particles but, however, leads to another physical state, $|\psi \rangle \to U| \psi \rangle$. If the unitary operators $U_{ab}$ corresponding to different exchanges of particle $a$ about particle $b$ do not commute  (i.e., $[U_{ab}, U_{cd}] \neq 0$ or $[U_{ab}, U_{ac}] \neq 0$) then the system exhibits non-Abelian statistics. Formally, in Eq. (\ref{braid}), in such cases, the elements ${\cal{T}}_{i}$ are non-commuting unitary matrices that act on a degenerate space of states.  Non-local operators are associated with the braiding of such anyons. As we will briefly elaborate on later, the corresponding braiding rules describe the different ways in which anyons can behave collectively, that are yet locally indistinguishable.
For the anyonic character of these topological excitations (or defects) to be unambiguous, it is important that they can be {\em localized} so that braiding operations (and thus statistics) are well defined. Such a localization of anyonic excitations generally appears in gapped systems. Thus, the typical size of the anyons ($l_{anyon}$), set by the inverse of the requisite energy gap for their creation, must be far smaller than the scale of the their separation ($R$) during braiding operations, $l_{anyon} \ll R$. 

To briefly make a connection with our discussions thus far in earlier sections, we remark that defects associated with the restoration of intermediate (or $d-$ dimensional) symmetries similar to those discussed in Section \ref{sec:sym} can exhibit anyonic statistics and non-trivial topological conservation laws. As elaborated elsewhere \cite{PNAS}, in systems including many compass models (including Kitaev's), $d=1$ symmetry operations which link different ground states to another can be viewed as a process involving the creation and transport of virtual anyons (more precisely, members of pair formed by an anyon and  an antianyon) around $d=1$ dimensional loops followed by an annihilation. On a finite size system, these degeneracy between states related to one another by operations involving such a transport of anyons  can be lifted by corrections that are exponentially small in the system size (i.e., scale as $O(e^{-cL})$ with the constant $c>0$ and $L$ the linear system size along which the anyon and anti-anyon tunnel and recombine).

\subsubsection{Fusion of Anyons}

Besides braiding, {\it fusion} is the other key process for anyons. A well-known example of fusion is that bringing together two fermions gives a boson. The generalization of this notion to anyons leads to ``fusion rules'' that are of the form
\begin{eqnarray}
a \times b = \sum_{c} N_{ab}^{c} c,
\label{general_fuse}
\end{eqnarray}
where the non-negative integer $N_{ab}^{c}$ denotes the number of distinct ways in which the anyons $a$ and $b$ may be fused together to form the anyon $c$. 
A notable property of fusion is its associativity,
\begin{eqnarray}
(a \times b) \times c = a \times (b \times c).
\label{axbxc}
\end{eqnarray}
When these anyons are Abelian, the fusion outcome is unique. For non-Abelian anyons there are multiple fusion channels: if there exists a pair $(a,b)$ for which   $\sum_c N_{ab}^{c} >1 $ then the anyons $a$ and $b$ are non-Abelian. Therefore the fusion outcome of two non-Abelian anyons is non-unique.

Hilbert spaces can be encoded with anyons whose number increases with the Hilbert space dimension. The fusion rules and associated {\it quantum dimensions} $\{d_{q}\}$ of quasi-particles of type $q$, determine the total dimension of the Hilbert space that can be encoded with a given set of anyons. The ground state degeneracy for $n_{q}$ quasi-particles scales as $d_{q}^{n_{q}}$. This exponential degeneracy factor in non-Abelian systems (wherein $d_{q}>1$) leads to an additive contribution to the entropy that scales as 
$
S_{anyon} = n_{q} k_{B} \ln d_{q}, 
$
 with $k_{B}$ the Boltzmann constant. This entropy carried by the individual quasi-particles may allow for  anyonic adiabatic cooling -- an effect predicted in 
\cite{Gervais2010}. 

\subsubsection{Majorana Fermions}
\label{sec:Majorana}

Majorana fermions \cite{Majorana1937} appear in many systems. Formally, similar to the real ($a$) and imaginary ($b$) parts of a complex number $(z=a_{1}+ia_{2})$ for which $ a_{1} = (z + z^{*})/2$ and $a_{2} = i(z^{*} -z)/2$),  the two operators $c_{1,2}$ being the "real" and "imaginary" parts of a fermionic operator (i.e., $c_{1}=(d+d^{\dagger})/2$ and $c_{2}= i(d^{\dagger} - d)/2$, with $d^{\dagger}$ and $d$ being Fermi creation and annihilation operators) satisfy the Majorana algebra. By that, we mean that the operators $\{c_{j}\}$ satisfy the following  relations that actually define a {\it Majorana-Fermi algebra}:
\begin{eqnarray}
\label{majorana}
\{c_{j}, c_{p}\} &=&0 ~ \mbox{for} ~ j \neq p \nonumber
\\ c_{j}^{2} &=& \frac{1}{2}, \nonumber
\\ c_{j}^{\dagger} &=& c_{j},
\end{eqnarray}
where $\{c_{j}, c_{p}\}$ denotes the anticommutator of two Majoranas, labeled by $j$ and $p$. One of their key physically defining features is that (as $c_{j}^{2}$ is a constant) the parity of number of fermions is important yet not the actual number of particles itself.  Similarly, the creation or annihilation of a Majorana fermion amount to the same operation ($c_{j} = c_{j}^{\dagger}$) -- or stated more colloquially, a Majorana fermion is its own anti-particle. 

From this it is seen that one may represent ($2n)$ Majorana fermions in terms of $n$ fermions or vice versa. Majorana fermions offer one of the simplest realizations of the particles that have non-Abelian statistics 
\cite{Nayak2008,Rowell2009}, as will become explicit in Section~\ref{nonabp} where we review the non-Abelian phase of the Kitaev model.

Such statistics are thought to occur in the $\nu = 5/2$ fractional Quantum Hall state, as first suggested by Moore and Read \cite{Moore1991} by investigating their conjectured $\nu =5/2$ wavefunction. Nayak and Wilczek later illustrated that each quasi-particle in this state carries a zero energy Majorana fermion \cite{Nayak1996}. Currently, it is still not experimentally known whether non-Abelian statistics indeed occurs in this state. 

While, currently, it is not clear if Majorana fermions exist as fundamental particles --neutrinos might possibly offer such a realization-- there is an increasing number of condensed matter systems in which Majorana fermions appear as excitations \cite{Wilczek2009,Franz2010,Ivanov,read,Lee2007,Schnyder2008,Alicea2010}.  

Similar Majorana fermion quasi-particles were theoretically found in superconductors with a $p_{x}+ip_{y}$  gap function by Read and Green \cite{Read1999}. Ivanov \cite{Ivanov} investigated the quasi-particle statistics by examining adiabatic change in these superconductors. This enabled an explicit matrix representation of braid group element. Kitaev illustrated that non-Abelian anyons with zero energy Majorana modes appear in the Kitaev honeycomb model when the Chern number is odd \cite{Kitaev06}. All in all, an extraordinary amount of work was devoted to these viable fractional Quantum Hall  and other states that may exhibit non-Abelian statistics \cite{Willet1987,Xia2004,Eisenstein2002,Moore1991,Read2009,Nayak1996,Ivanov,Stern2004,read,Read1999,LevinWen,Wen1993,Greiter1991,Morf1998,Storni2010,Toke2007,Rezayi00,Wojs06,Feiguin2008}.  As alluded to above, $p-wave$ superconductors \cite{Ivanov,Nayak2008} may display non-Abelian statistics. Non-Abelian anyons might potentially occur also in cold-atom systems \cite{Cooper2001,Gurarie2007}, topological insulator or superconductor based systems \cite{Hasan2010,Fu2008,Fu2009, Nilsson2008,Beenakker2011} and, notably, possibly also in semi-conductor wires \cite{Alicea2010, Alicea2011} and semi-conductor/(s-wave) superconductor hybrids \cite{Sau2010}. 

\subsubsection{Fused Magetic and Electric Charges -- Dyons}
\label{sec:dyon}

Armed with two different types of particles, for instance a particle with {\em electric} charge $q$ and another particle with a {\em magnetic} flux $\phi$ and their relative statistics, see Fig. \ref{fig:em_stat}, one can discuss all possible composites of these basic electric and magnetic particles. In one of the simplest instances of an electromagnetic type theory (or $U(1)$ theory) with magnetic and electric charges, the Aharonov-Bohm phase associated with the rotation of a particle of charge $q$ about a particle carrying a flux $\phi$, wherein the system state $|\psi \rangle \to \exp[2 \pi i q \phi] | \psi \rangle$, leads to a non-trivial statistics when $q \phi$ is not an integer.  

\begin{figure}
\centering
\includegraphics[width=.5\columnwidth]{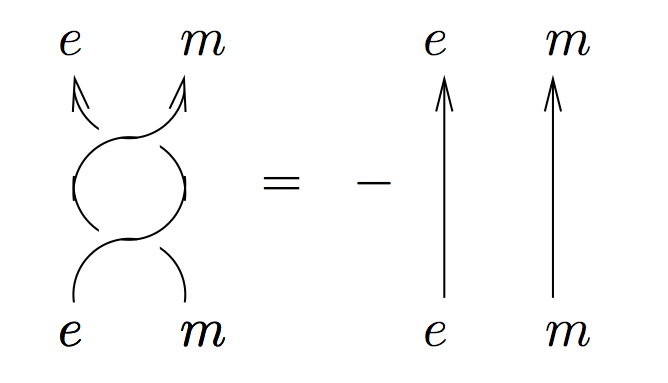}
\caption{Relative statistics between electric and magnetic charges. 
}
\label{fig:em_stat}
\end{figure}

The most basic composite to consider in this context is a {\it dyon} -- a "molecule'' composed of an electrical and a magnetic particle that have been brought together and {\it fused}. As the magnetic particles have trivial statistics amongst  themselves -- the Aharonov Bohm phase associated with transporting one magnetic particle around another is zero -- and similarly all electrically charged particles  have trivial mutual statistics, the only complication that can arise when consider the mutual statistics of two dyons is that  arising from revolving the charge $q$ about the flux $\phi$. That leads to a uniform phase factor of $\exp[ 2 \pi i q \phi]$. That is, the dyons can have a non-trivial fractional statistics among themselves.

Perhaps the simplest anyons are those that arise in an Ising (or $Z_{2}$) theory. Here, there are only two possible values for the electrical and magnetic charges. The electrical charge, which is henceforth denoted by $e$, can assume a value of $\pm 1$ and similarly the magnetic charge $m$ can assume be $\pm 1$.  Revolving an electrical particle around a magnetic particle entails an Aharonov Bohm phase of $\exp[i \pi] = -1$, see Fig. \ref{fig:em_stat}.

In this case, there are only 4 possible basic particle type sectors, which sometimes are referred to as {\it superselection} sectors. These are given by (i) $e$ (the electric charge),  (ii) $m$ (the magnetic charge),  (iii)  a dyon $\epsilon$ composed of a hybrid of an electric and a magnetic charge, and (iv) the vacuum (devoid of particles) which is denoted by $I$. 

The magnetic particles are bosons relative to one another -- there is no phase change in revolving a magnetic particle around another magnetic particle. Similarly, the electrical charges are bosons, the dyons are fermions. Formally one also associate vacuum "particles" to $I$, in which case they are bosonic. The mutual statistics of these four particle types relative to one another is very simple. As stated earlier, revolving an electrical charge around a magnetic charge and also the converse (a magnetic charge around an electrical charge) entail a relative phase of $-1$.  Similarly, revolving a magnetic particle around a dyon or an electrical charge around a dyon both involve a phase factor of $-1$.  

One may similarly ask what occurs when we fuse two particles together -- this is after all what the rather formal sounding name of  "fusion rules" aims to convey. One can consider for example how two electric charges behave when they are fused into a hybrid unit. In an Ising theory, two identical charges (no matter what charge it were $e = \pm 1$) would cancel each other and behave like no charge at all (the vacuum). This is formally encapsulated by the fusion rule
 \begin{eqnarray}
 e \times e = I.
 \label{eeI}
 \end{eqnarray}
 In a similar manner,
 \begin{eqnarray}
 m \times m =I, \nonumber
 \\ e \times m = \epsilon, \nonumber
 \\ m \times e = \epsilon.
 \label{fusion}
 \end{eqnarray}
 We note, for completeness, that the above fusion rules are augmented the universal trivial relations,
  \begin{eqnarray}
  I \times I = I, ~I \times e =e, ~ I \times m=m,~ I \times \epsilon = \epsilon,
  \label{triv_fusi}
  \end{eqnarray}
 (and any other fusion with the identity matrix in other systems). We will not repeat again
 the trivial fusion rule with the identity
 in our future discussions; these are always to be understood.
 
As noted earlier, and is fleshed out in the particular example of the fusion rules above, for {\em Abelian anyons}, the product of the fusion of any pair of particles (on the lefthand side of Eqs. (\ref{eeI},\ref{fusion})) leads to a {\em single unique particle} (on the righthand side of these equations). In terms of Eq. (\ref{general_fuse}), in Eq. (\ref{eeI},\ref{fusion}) there always exists only a single channel for all possible ways of fusion. 
 
As we will review in the upcoming sections, the fusion rules of Eqs. (\ref{eeI},\ref{fusion}) precisely appear in the so-called Toric Code Model \cite{Kitaev03} and, by extension, in  Kitaev's honeycomb model in its Abelian phase \cite{Kitaev06}. Such fusion rules- in particular those for richer {\em non-Abelian counterparts}  (as in, e.g., the non-Abelian phase of Kitaev's honeycomb model) that may lead to several possible fusion products and enable universal computation -- form a cornerstone of various topological quantum computing schemes. The basic idea underlying topological quantum computing is that of preparing particular initial states with such particles (invoking, in effect, the fusion rules), performing calculations via unitary gates that employ braiding of these particles, and performing measurements by fusion \cite{Kitaev06}.  As alluded to above, such particles (both Abelian and non-Abelian)  appear in Kitaev's model which we now discuss.

\subsection{Kitaev-Compass Model -- Features}
\label{kith}

The 120$^\circ$ compass model on the two-dimensional honeycomb lattice was introduced by \onlinecite{Kitaev06} and is often simply referred to as the {\it Kitaev Honeycomb Model}. Because of its central relevance to all that follows, we reiterate here that the Kitaev-compass Hamiltonian $H_{\varhexagon}^{K}$, as defined in Sec.\ref{sec:defineKHM} and illustrated in Fig.~\ref{fig:KitaevH} is given by Eq. (\ref{eq:HKT}) which we write anew,
\begin{eqnarray}
\label{HKF}
H_{\varhexagon}^{K}=
&-&J_x\sum_{ \begin{smallmatrix} \bm{e}_1-&\\ {\sf bonds}& \end{smallmatrix}}\tau^x_{i}\tau^x_{j} 
-J_y \sum_{\begin{smallmatrix} \bm{e}_2-&\\ {\sf bonds}& \end{smallmatrix}}\tau^y_{i}\tau^y_{j} \nonumber \\
&-&J_z \sum_{\begin{smallmatrix} \bm{e}_3-&\\ {\sf bonds}& \end{smallmatrix}}\tau^z_{i}\tau^z_{j},
\end{eqnarray}
where the operators ${\bm \tau}=(\tau^x,\tau^y,\tau^z)$ represent  (pseudo-) spin 1/2 degrees of freedom on neighboring vertices of a honeycomb lattice, labeled by $i$ and $j$. The lattice links may point along three different directions, labeled by $\bm{e}_1$, $\bm{e}_2$ and $\bm{e}_3$, where the angle between the three unit lattice vectors is $120^\circ$. Re-expressed in the explicit form of a compass model
\begin{eqnarray}
H_{\varhexagon}^{K}&=& - \sum_{i,\gamma} J_\gamma \tau^\gamma_i \tau^\gamma_{i+\bm{e}_\gamma} \nonumber \\
&&with
\left\{ 
\begin{array}{l}
  \{\tau^\gamma\}=\{\tau^x,\tau^y,\tau^z\} \\
    \{J_\gamma\}=\{J_x,J_y,J_z\} \\
  \bm{e}_\gamma=\bm{e}_x\cos{\theta_\gamma}  +  \bm{e}_y \sin{\theta_\gamma} \\
  \{\theta_\gamma\}=\{0,2\pi/3,4\pi/3\}.
\end{array} 
\right.
\label{eq:Kitaev_K}
\end{eqnarray}
As we will review in Section \ref{corrkk}, in the limit of strong anisotropy, the Kitaev compass model on the honeycomb lattice (panel (A) of Fig. \ref{fig:honeycomb}) reduces to another well known model in topological 
quantum computing- the ``Toric Code'' model (panel (B) of Fig. \ref{fig:honeycomb}). 

By its very nature, Kitaev's honeycomb model is very similar to the $90^\circ$ compass models and other $120^\circ$ models. However, the Kitaev-compass system has a number of very remarkable properties. These can be assessed in a crisp manner because the model is exactly solvable: it can be mapped exactly onto a system of non-interacting Majorana (as well as Dirac) fermions, as will be detailed in Sec.~\ref{majof}. This allows the derivation of all of the beautiful topological characteristics -- its gapped bulk states, computable Chern numbers and Majorana excitations. Moreover, it will make evident that these Majorana excitations are coupled to a gauge field which embodies the topological charges, i.e., magnetic and electric like charges as introduced in Sec.~\ref{sec:notions}. 

For future purposes it is useful to define an extension to this Hamiltonian $H_{\varhexagon}^{h}$, which actually becomes relevant if the model is studied in an external field $h$. This term involves three pseudo-spins on sites $i$, $j$ and $k$, and is the of form
\begin{eqnarray}
H_{\varhexagon}^{h}&=& - \kappa \sum_{ijk} \tau^x_i \tau^y_j \tau^z_k
\label{eq:Kitaev_h}
\end{eqnarray}
where the sum over $ijk$ is a sum over {\it all} sites connected by the two links $\langle ij \rangle$ and $\langle jk \rangle$. So here the link $\langle ij \rangle$ connects neighboring sites $i$ and $j$, similarly for $\langle jk \rangle$, but sites $i$ and $k$ are {\it next} nearest neighbors. This form of the Hamiltonian might seem rather particular at this point, but when adding it the model will stay exactly solvable. This term is essential to drive the Kitaev-compass Hamiltonian from a ground state with Abelian excitations to a state with non-Abelian ones, as will be discussed in Sec.~\ref{nonabp}

The Kitaev-compass model reduces to the {\it toric code model} in the limit in which one coupling constant is far larger than all of the rest, e.g., $|J_z| \gg |J_{x,y}|$. The excitations in the toric code model, reviewed in Sec.~\ref{corrkk} precisely have magnetic and electric charges introduced in Sec.~\ref{sec:notions}. 


As we reviewed earlier, compass systems such as Kitaev's (and its extensions) may, e.g., be implemented by atoms in optical lattices \cite{Lewenstein07,Duan03} and cavity and ion trap systems  \cite{Kay2008,Trousselet12,Schmeid11}. By focusing on the low energy subspace (in a spirit somewhat similar to that of Section \ref{sec:chiral}) of magnetic clusters, \cite{Wang2010} suggested that Kitaev's model may be constructed via magnetic clusters on a honeycomb lattice.  A proposal for experimentally constructing this system via superconducting quantum circuit elements was advanced in \cite{You2010}. As we discussed in  Section \ref{sec:sporb}, Kitaev-Heisenberg systems might describe the Iridates.

\subsubsection{Relation to Topological Insulators}
\label{generalk2}

\begin{figure}
\centering
\includegraphics[width=\columnwidth]{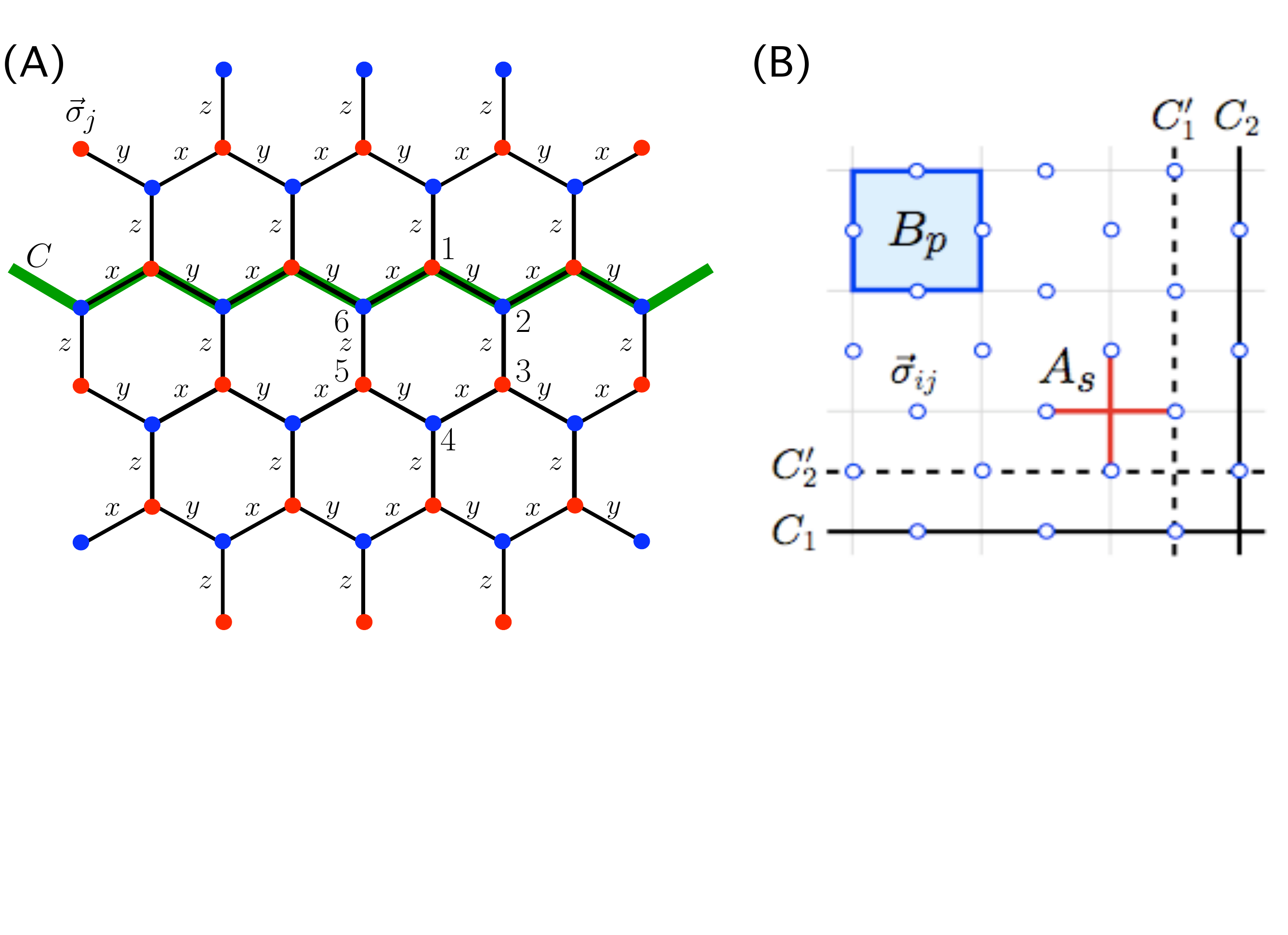}
\vspace{-2.5cm}
\caption{(A) Kitaev's model on a honeycomb lattice and three types of bonds. On each vertex there is an $S=1/2$ degree of freedom indicated by a Pauli matrix $\vec{\sigma}_{j}$ (see text). (B) Elementary plaquette $B_p$ and star $A_s$ interaction terms in Kitaev's Toric code model. Hollow circles in the bonds (links) represent an $S=1/2$ degree of freedom, while thick (dashed or solid) lines represent topological ($d=1$) symmetry operators (see text).
\cite{Nussinov08a}.}
\label{fig:honeycomb}
\end{figure}

In many regards, Kitaev's model furnishes an elegant and exactly solvable realization of a {\it topological insulator} \cite{top_ins,SHE1,SHE2,SHE3,SHE4,SHE5,SHE6,Fu07a,Fu07b}. Topological insulators are systems that are gapped and insulating in the bulk yet due to topological characteristics harbor metallic, zero-energy, edge-states at the system's boundaries. Due to topology, these edge-modes are robust and may retain their metallic character notwithstanding the introduction of disorder.  

The class of topological insulators should include the integer quantum Hall states, which occur in 2D systems in which time-reversal symmetry is broken due to the presence of a magnetic field. 
Their quantum spin-Hall counterparts --the topological insulators mentioned above which can be realized in semiconductors with large spin-orbit coupling-- time reversal symmetry remains unbroken  
\cite{SHE1,SHE2,SHE3,SHE4,SHE5,SHE6,Fu07a,Fu07b}. Similar to the integer quantum Hall systems the Kitaev-compass model exhibits sharp topological quantities such as the Chern number \cite{avron03,gwchern10}, which characterize two dimensional systems of free fermions with an energy gap. In integer quantum Hall systems, the Chern number is just the magnetic filling factor. 
Similar to other topological insulators, the Kitaev-compass model exhibits gapped phases in the bulk with concomitant gapless chiral modes:  In the presence of a magnetic field, the Kitaev-compass model exhibits chiral edge-modes of a Majorana fermion character.

\subsubsection{Majorana Excitations}
The existence edge-states in the Kitaev model constitutes an analogue to quantum Hall systems and other topological insulators. However, in integer quantum Hall systems, the edge-modes are of bona fide fermions and not Majorana fermions. Its is the Majorana character of the excitations that in principle enables the aforementioned fault tolerance relative to {\em all} local fluctuations -- "errors'' in the setting of quantum computing.  The excitations of the Kitaev model flesh out the notions of anyonic statistics introduced in section 
\ref{sec:notions} and afford very crisp realizations of non-trivial topology. Kitaev's model realizes fusion rules such as those of Eqs. (\ref{eeI},\ref{fusion}). 

The system also realizes one of the simplest examples of exotic ideas concerning fractionalization in strongly correlated electronic and spin systems. In its Abelian phase, the {\it magnetic} and {\it electric} excitations in the model may, respectively, be viewed \cite{Sachdev09} as counterparts of  {\it vison} and {\it spinon} excitations in theories of doped quantum antiferromagnets \cite{senthil} with relative "semionic" statistics which requires that when an excitation of one type is moved around another it picks up a phase factor of $(-1)$. 

It should be stressed that while the existence of excitations of Majorana-type is a special feature of the Kitaev-compass model, it is not necessarily an unique feature. In special situations three dimensional topological insulators may also exhibit Majorana fermion type of excitations, for instance ton their surface when placed at an interface with a superconductor~\cite{linder10}. Majorana fermions may also manifest in some of the systems that we earlier referred to in the context of non-trivial statistics: the fractional quantum Hall systems such that of the state of filling fraction $\nu = 5/2$ \cite{read}, at cores of half-vortices in $p$-wave superconductors \cite{Ivanov} and in semi-conductor \cite{Alicea2010, Alicea2011} and semi-conductor/($s$-wave) superconductor systems  \cite{Sau2010}. 

\subsection{Kitaev-Compass Model -- Abelian Phases}
\label{solnus}

As was emphasized earlier, the Kitaev-compass model is exactly solvable in its ground state sector, for any set of coupling constants $J_x$, $J_y$ and $J_z$.  The original solution in \onlinecite{Kitaev06} hinged on introducing several Majorana fermion degrees of freedom per site and making a projection on to a physical Hilbert space and symmetrization. Later approaches invoked a Jordan-Wigner (JW) transformation in two dimensions \cite{Chen07a,Chen08,Feng07,Kells2009}, perturbative methods, e.g., \cite{Vidal2008} and slave fermion methods \cite{Burnell2011,Schaffer2012}. Another approach, which will be followed here, is based on the direct use of  a {\it bond algebra} \cite{nussinov-bond}. It is rather straight-forward and keeps directly track of the local symmetries that the Hamiltonian harbors, which are crucial to the solutions of $H_{\varhexagon}^{K}$ (and the same model augmented by $H_{\varhexagon}^{h}$). The explicit solution via the JW transformation \cite{Chen08} largely inspired the bond algebraic approach, but it is not as direct. The advantage of the bond algebraic method is that it enables the solution without enlarging the Hilbert space and making subsequent projections. Nor does it use at intermediate steps non-local string operators as in the Jordan-Wigner transformation. 

\subsubsection{Bond Algebra, Symmetries, and Anyonic Charge}
\label{bondalgebrasym}

In the Kitaev-compass Hamiltonian $H_{\varhexagon}^{K}$ three types of bonds $\{b_{jk}\}$ appear
\begin{eqnarray}
\tau^{x}_{j} \tau^{x}_{j+\bm{e}_1}, \ \tau^{y}_{j} \tau^{y}_{j+\bm{e}_2}  \  \ and \ \  \tau^{z}_{j} \tau^{z}_{j+\bm{e}_3},
\label{eq:longbonds}
\end{eqnarray}
where $\{\bm{e}_1,\bm{e}_2,\bm{e}_3\}$ are unit vectors along the three directions of the hexagonal lattice. In terms of bond operators the Hamiltonian is
\begin{eqnarray}
H_{\varhexagon}^{K} = \sum_{\langle j k \rangle} J_{jk} b_{jk},
\label{eq:Kitaev_bond}
\end{eqnarray}
with, as in Eq.~(\ref{eq:Kitaev_K}), $J_{jk} = J_{x}, J_{y}$ or $J_{z}$ depending on the orientation of bond $\langle j k \rangle$ along one of the three directions.
One usually supplements this definition of the bond-Hamiltonian with an ordering convention of the bonds, the simplest one being that site $j$ always lies below site $k$ in the honeycomb lattice as for instance shown in Fig.~\ref{fig:honeycomb}.
The pseudo-spin operators anticommute at any given site $j$, e.g., $\{\tau_{j}^{x}, \tau_{j}^{z}\}=0$,  and commute at different sites, e.g.,  $[\tau_{j}^{x}, \tau_{p}^{z}]=0$ for any two sites $j \neq p$. The bonds therefore satisfy an extraordinarily simple algebra \cite{nussinov-bond}:
\begin{enumerate}[(i)]
\item The square of each bond is one.
\item Two bonds that do not share any common site commute.  
\item Two bonds that share one common site anti-commute.
\end{enumerate}
There are no additional algebraic relations that the bonds that appear in the Hamiltonian $H_{\varhexagon}^{K}$ need to satisfy. This set of all algebraic relations between the bonds in a general Hamiltonian is termed the {\it bond algebra} \cite{cobanera,ADP,Nussinov08b}. 
If we can write down another representation of the bonds in Eq. (\ref{eq:longbonds}) for which all of the above algebraic relations are the same,  then the Hamiltonian in the new representation and the original one will share the same spectrum and are thus related by a unitary transformation (and are thus dual to one another). 
Precisely such a change of representation underlies the exact solution of  $H_{\varhexagon}^{K}$ (as further elaborated on in subsections~\ref{majof},\ref{gsfm}). Similar dualities (including those that lead to an effective dimensional reduction) can be established in numerous other compass models, e.g., \cite{Brzezicki08, Eriksson09, cobanera,ADP,Nussinov08b,Vidal09,Karimipour2009,AOP,holography}.

We now pause to examine the symmetries of the Hamiltonian $H_{\varhexagon}^{K}$ of Eq. (\ref{eq:Kitaev_K}). Exact local ($d=0$) gauge symmetries are given by {\it products of pseudo-spins around each hexagon} \cite{Kitaev06},. For each hexagon $i$ labeled by $\varhexagon i$ as in Fig.~\ref{fig:symm_K}, such a symmetry is given by
\begin{eqnarray}
\hat{O}_{\varhexagon i} = \tau^{z}_{1} \tau^{x}_{2} \tau^{y}_{3} \tau^{z}_{4} \tau^{x}_{5} \tau^{y}_{6}.
\label{eq:symkit6}
\end{eqnarray}

\begin{figure}
\centering
\includegraphics[width=.4\columnwidth]{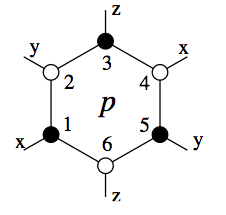}
\caption{Pictorial rendition of the local symmetry of Eq. (\ref{eq:symkit6}) associated with every hexagon~\cite{Kitaev06}.
}
\label{fig:symm_K}
\end{figure}

Each of the six sites of the hexagon contributes only one component $\tau^\gamma$ of its pseudo-spin operator  to the product $\hat{O}_{\varhexagon i}$, where $\gamma$ is either $x$, $y$ or $z$. Precisely which component of these three depends on the type of link that is {\it not} part of the hexagon -- if on site $j$ the bond operator on the "non-hexagon link" is of type $\tau^{\gamma}_{j} \tau^{\gamma}_{j+\bm{e}}$ (thus with $j \in  \varhexagon i$ and $j+\bm{e} \not\in  \varhexagon i$), the pseudo-spin component appearing in $\hat{O}_{\varhexagon i}$ is $\tau^{\gamma}_{j}$.

It can readily be verified that $\hat{O}_{\varhexagon i}$ commutes with {\it any} bond-operator $b_{jk}$ of Eq.~(\ref{eq:longbonds}) and consequently $[H_{\varhexagon}^K, \hat{O}_{\varhexagon i}]=0$. These operators also mutually commute with one another: $[\hat{O}_{\varhexagon i},\hat{O}_{\varhexagon j}]=0$. Moreover the square of each such symmetry operator is one: $\hat{O}_{\varhexagon i}^2=1$.  When it attains a non-trivial eigenvalue, i.e., $\hat{O}_{\varhexagon i} = -1$, the operator $\hat{O}_{\varhexagon i}$ is said to depict an {\em anyonic charge} or {\em vorticity} on hexagon $i$, for reasons which will become clear later.

From the above follows that the system is composed of $2^{N_{h}}$ sectors with $N_{h}=N/2$ being the number of hexagons. Each sector is specified by the the set of eigenvalues of the operators $\{\hat{O}_{\varhexagon i} \}$, where $i=1,..., N_h$:  $| O_{\varhexagon 1} = \pm 1,  O_{\varhexagon 2} = \pm 1, ...,  O_{\varhexagon N_h} \pm 1 \rangle$. 

The model has more symmetries. When the system is placed on a torus, $H_{\varhexagon}^{K}$ also has $d=1$ symmetries, using the classification of symmetries of Section \ref{global_inter}. For any loop $C$ that spans the entire system the symmetry given by  $\prod_{j\in C} \tau^{\gamma}_{j}$, where on each site $j$ the component $\gamma$ is determined by the character of one bond of site $j$ that is not on $C$ (i.e., the bond $\tau^{\gamma}_{j} \tau^{\gamma}_{j+\bm{e}}$ with $j \in  C$ and $j+\bm{e} \not\in  C$). When $C$ is for instance taken to be the zig-zag contour shown in Fig.~\ref{fig:honeycomb} this symmetry is $\prod_{j\in C} \tau^{z}_{j}$, but actually any {\em closed} loop $C$ represents a symmetry. 

\subsubsection{Majorana Representation and Fermionization}
\label{majof}

The relations (i)-(iii) of the previous section define the bond algebra of  $H_{\varhexagon}^{K}$ and it can readily be checked that they are also satisfied by the following substitution for the bonds in Eq.(\ref{eq:longbonds}):
\begin{eqnarray}
b_{jk} = 2i \eta_{jk} c_{j} c_{k},
\label{eq:Majorana_bond}
\end{eqnarray}
where the operators $c_{j}$ represent {\it Majorana fermions}, obeying the Majorana algebra as defined in Eq.~(\ref{majorana}) and $\eta_{jk}$ are Ising-type gauge links: a number that is either $+1 $ or $-1$ on any given link $\langle jk \rangle$.  With the ordering convention of bonds being that site $j$ always lies below site $k$, one has that exchanging sites $j$ and $k$ results in $\eta_{jk} = - \eta_{kj}$. The set $\{\eta_{jk}\}$ encompassing all bonds constitutes a sector of gauge links.
In any given sector $\{\eta_{jk}\}$, the Hamiltonian of Eq.(\ref{eq:Kitaev_bond}) is quadratic in the Majorana fermions $\{c_{i}\}$ and {\em thus exactly solvable}. \cite{Kitaev06,nussinov-bond}

The local ($d=0$) symmetries of Eq.(\ref{eq:symkit6}) can be expressed in terms the bonds as
\begin{eqnarray}
\hat{O}_{\varhexagon i} = \prod_{jk \in \varhexagon i} \eta_{jk}.
\label{eq:O_bond}
\end{eqnarray}
That is, each sector of fixed $\{\eta_{jk}\}$ is an eigenstate of the symmetry operators of Eq.(\ref{eq:symkit6}) with an eigenvalue that is determined only by $\{\eta_{jk}\}$. In $\hat{O}_{\varhexagon i}$, as one multiplies $\eta_{jk}$ for all links $\langle jk \rangle$ that are in the hexagon $i$, one keeps the bond indices $j$ and $k$ ordered with the previously chosen convention of $j$ being below $k$.

The expression for $\hat{O}_{\varhexagon i}$ above highlights the similarity between the local gauge symmetries in this system and such general symmetry (and fluxes) elsewhere. For instance, in a lattice version of electromagnetism, Eq. (\ref{eq:O_bond}) relates to an Aharonov Bohm like phase.  In the current context, Eq. (\ref{eq:O_bond}) relates to the Ising version of such a phase ($O_{\varhexagon i} = \pm 1$). 

As each site belongs to three hexagons and each hexagon contains six sites, the number of hexagons is half the number of lattice sites ($N_{h} = N/2$). Thus to account for all eigenvalues of the operators $\{O_{\varhexagon i}\}$, it suffices to allow the $N/2$ degrees of freedom $\eta_{jk}$ on, for instance, all {\it vertical} bonds along ${\bm e}_3$ to attain a value of $\pm 1$ and to pin $\eta_{jk}$ on all other bonds (those along the ${\bm e}_1$ or ${\bm e}_2$ axis) to be 1. With this particular choice of the local gauge fields $\eta$,  $\hat{O}_{\varhexagon i}$ in Eq.(\ref{eq:O_bond}) reduces to the product of $\eta_{jk}$ on the two vertical links that belong to each hexagon $i$. 

The dimensionality of the original Hilbert space of $N$ pseudo-spins is $2^{N}$. Thus in each of the $2^{N/2}$ sectors of $\eta_{jk}$, there is a remaining Hilbert space of size $2^{N/2}$ on which the Majorana fermions are defined. As explained in Sec.~\ref{sec:Majorana}, one representation for the $N$ Majorana fermions is in terms of $N/2$ spinless Dirac fermions. This may be explicitly done here by setting
\begin{eqnarray}
\label{cdrep}
c_{j} = d_{jk} + d_{jk}^{\dagger}, \nonumber
\\ c_{k} = - i (d_{jk} - d_{jk}^{\dagger}),
\end{eqnarray}
with $d_{jk}$ a spinless Dirac Fermi operator on the vertical link $\langle jk \rangle$ (that is, $k = j + {\bm e}_3$) \cite{Chen08}. The centers of the vertical links of the honeycomb lattice form a square lattice. It is therefore convenient to place the Fermi operators $d_{jk}$ and $d^{\dagger}_{jk}$) at the centers of the vertical links $\langle jk \rangle$) and henceforth denote these by $r$, leaving us with the operators $d_{r}^{\dagger}$, $d_{r}$ and the Ising degrees of freedom ${\eta_r}$. Denoting the unit vectors of the resulting square lattice by ${\bm e}_{x}$ and ${\bm e}_{y}$, the Kitaev-compass Hamiltonian reduces to 
\begin{eqnarray}
H_{\varhexagon}^{K} = J_{x} \sum_{r} ( d_{r}^{\dagger} + d_{r}) (d_{r+ {\bm e}_{x}} ^{\dagger} - d_{r+ \hat{e}_{x}}) \nonumber
\\ + J_{y} \sum_{r} ( d_{r}^{\dagger} +d_{r} ) (d_{r+\hat{e}_{y}}^{\dagger} - d_{r+\hat{e}_{y}}) \nonumber
\\ +  J_{z} \sum_{r} \eta_{r} (2 d_{r}^{\dagger} d_{r} -1).
\label{eq:Kitaev_F}
\end{eqnarray}

The last term constitutes an analogue of a "minimal coupling" term between gauge and matter degrees of freedom that is familiar from electromagnetism  -- in this specific case, an analogue of a coupling between the charge (or matter) density and an electrostatic type potential. 

An advantage of the fermionization procedure employed above is that it does not require the use of elaborate non-local JW transformations. That the representation in terms of spinless fermions is $2^{N/2}$ dimensional can be checked by realizing that there are $N/2$ vertical links $\langle jk \rangle$ and the dimensionality of each spinless Fermion operator is two -- the bond $\langle jk \rangle$ can be either occupied or un-occupied by a fermion. 

Putting all of the pieces together, one sees that the problem of solving $H_{\varhexagon}^{K}$ has now been reduced to a problem involving solely fermions and Ising gauge degrees of freedom $\eta_{r}$,  which at each site $r$ can only attain the value $\pm1$. All excitations that appear in this system can be expressed in terms of the original spin variables ${\bf{\tau}}_{j}$ or, equivalently, in terms of fermions and Ising gauge fields.

The {\it fusion rules} that will appear both in this system and its non-Abelian extension that we will review in Sec.~\ref{nanp}  must relate to {\em fermionic} and {\em Ising gauge} type basic degrees of freedom. 

\subsubsection{Ground State of Fermionized Model}
\label{gsfm}

Within the ground state sector, for all hexagons all $\hat{O}_{\varhexagon i} =1$, or equivalently on the square lattice $\eta_{r} =1$ for all sites $r$. That the ground state must be vortex free is ensured by a corollary of a theorem due to \onlinecite{Lieb1994} and has also been established numerically \cite{Kitaev06}. In momentum space, the fermionized Hamiltonian of Eq.(\ref{eq:Kitaev_F}) assumes the form
\begin{eqnarray}
H_{\varhexagon}^{K} = \sum_{\bm q} \epsilon_{q} d_{\bm q}^{\dagger} d_{\bm q} + i \frac{\Delta_{\bm q}}{2} \left(d_{\bm q}^{\dagger}
d_{-\bm q}^{\dagger} + d_{\bm q} d_{-\bm q} \right),
\label{fermi-pair}
\end{eqnarray}
where ${\bm q}= (q_{x},q_{y})$ and
\begin{eqnarray}
\epsilon_{\bm q} &=& 2J_{z} - 2J_{x} \cos q_{x} - 2 J_{y} \cos q_y, \nonumber \\
\Delta_{\bm q} &=& 2J_{x} \sin q_{x} + 2J_{y} \sin q_{y}.
\label{epD}
\end{eqnarray}
Interestingly, this Hamiltonian has the form of a {\it p-wave BCS type Hamiltonian on the square lattice} \cite{Chen08}, which becomes explicit when the Hamiltonian is cast in the form of a Bogoliubov - De Gennes Hamiltonian
\begin{eqnarray}
H_{\varhexagon}^{K}= 
\left( \begin{array}{cc} d_{\bm q}^{\dagger}  & d_{-\bm q}  \end{array} \right)  H_{BdG}^{K}  (\bm q)
\left( \begin{array}{c} d_{\bm q}  \\ d_{-\bm q}^{\dagger}  \end{array} \right), 
\label{eq:BdG}
\end{eqnarray}
where $ H_{BdG}^{K}  (\bm q)$ is a $2 \times 2$ matrix. It can be cast in the slightly more general form
\begin{eqnarray}
H_{BdG}  (\bm q) =  h_{\bm q} \sigma_x+ \Delta_{\bm q}  \sigma_y + \epsilon_{\bm q}   \sigma_z = {\bm d} ({\bm q}) \cdot {\bm \sigma},
\label{eq:BdG}
\end{eqnarray}
where ${\bm \sigma} = (\sigma_x,\sigma_y,\sigma_z)$ with Pauli matrices $\sigma_{x,y,z}$ and the last line defines the three-component vector ${\bm d}({\bm q})$. Here an extra coupling $h_{\bm q}$ has been introduced for future reference. This coupling is not present within the pure honeycomb Kitaev-compass model.  Thus, $H_{BdG}^{K}= \lim_{h_{\bm q} \rightarrow 0} H_{BdG}  (\bm q)$. 

In the Hamiltonian $H_{BdG}$, the vector ${\bm d}({\bm q})$ acts as a "Zeeman field" applied to the "spin" ${\bm \sigma}$ of a two-level system. All its eigenvalues come in pairs, corresponding to the energies
\begin{eqnarray}
E_{\bm q} = \pm d({\bm q}) = \pm |{\bm d}({\bm q})| = \pm\sqrt{{\bm d}({\bm q}) \cdot {\bm d}({\bm q})}.
\label{BdG_energies}
\end{eqnarray}
and eigenvectors

Diagonalizing the Hamiltonian by a Bogoliubov transformation 
\begin{eqnarray}
\gamma_{\bm q} = u_{\bm q} d_{\bm q}  + v_{\bm q} d_{-\bm q} ^{\dagger}
\label{eq:Bbov_transformation}
\end{eqnarray}
with $|u_{\bm q}|^2+|v_{\bm q}|^2=1$ and $|u_{\bm q}|^2 = \frac{1}{2}\sqrt{1+\frac{\epsilon_{\bm q}}{E_{\bm q}^2 }}$ gives the energy spectrum
No effective chemical potential appears in this problem (i.e., $\mu =0$) so that within the ground state, all fermionic states of negative energy ($E_{q} <0$) are occupied while all states of positive energies are empty. The corresponding ground state wavefunction is
\begin{eqnarray}
|g \rangle= \prod_{\bm q} (u_{\bm q} + v_{\bm q}  d_{\bm q} ^{\dagger} d_{-\bm q} ^{\dagger})|0 \rangle
\label{BCS-p}
\end{eqnarray}

Eigenvalues of the Bogoliubov - de Gennes come in {\it pairs}, at energies $\pm \epsilon$ with $\gamma_{-\epsilon}^\dagger = \gamma_{\epsilon}$, because of particle-hole symmetry. One fermion excitation eigenstate gives two solutions of the BdG equations. At $\epsilon=0$ is a Majorana fermion, as $\gamma_{0}^\dagger = \gamma_{0}$

In the vicinity of band extrema, the dispersions of Eq. (\ref{BCS-p}) is of a parabolic when a gap appears between the two bands in the problem
(i.e., $\min\{|E_{q}|\} >0$) and is linear near the zeros of $E_{q}$ when the system is gapless. 

The ground state corresponds to a BCS condensate.  In \cite{Chen08}, 
real space ground state wave-functions $| \Psi_{0} \rangle$ were explicitly constructed
in the original spin representation in closed forms that do not require 
any implicit projections. These Explicit ground states may be determined 
by writing the BCS wavefunction and undoing all of the steps employed 
here to map the spin problem onto a fermionic one.  Time reversal invariance along with 
Kramers' theorem ensures ground state degeneracy in systems with an odd number
of spins. Explicit ground state wave-functions on a torus were constructed by \cite{Kells2009};
these exhibit a four-fold  topological degeneracy. 

\subsubsection{Gapless and Gapped Phases}

\begin{figure}
\centering
\includegraphics[width=\columnwidth]{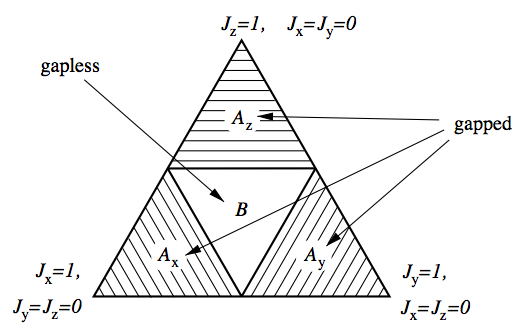}
\caption{Phase diagram of the honeycomb Kitaev-compass model. The triangle is the section of the positive octant $(J_x,J_y,J_z) \geq 0$ by the plane $Jx+Jy+Jz=1$.The diagrams for the other octants are similar. \cite{Kitaev06}}
\label{fig:Kitaev_phases}
\end{figure}

To provide a better understanding of the spectrum, we focus on a 
particular set of couplings. At the symmetric system ($J_{x}=J_{y}=J_{z}$),
the dispersion of Eq. (\ref{BCS-p}) is, within the first Brillouin zone, 
zero at $ {\bf q}^{(+)}  \equiv  {\bf K}_{1}/3 +  2 {\bf K}_{2}/3, {\bf q}^{(-)} = 2 {\bf K}_{1}/3 + {\bf K}_{2}/3$
where ${\bf K}_{1,2}$ denote the reciprocal lattice vectors along 1 and 2 directions \cite{Kitaev06}
(and the equality holds modulo the addition of any reciprocal vectors). 
As the anisotropy of the coupling constants is increased (e.g., setting $|J_{z}|$ fixed and decreasing
$|J_{x,y}|$), the two points ${\bf q}^(\pm)$ veer towards one another until they merge at
the boundary between the gapped and gapless phases \cite{Kitaev06}. Beyond this point, as $|J_{x,y}|$
are further decreased, and the system is in its gapped phase there are no real 
vectors ${\bf q}$ for which $E_{q}$ is zero. 
 
In  a similar fashion,  the spectrum may be computed in other sectors.
By looking at the algebra of the bonds appearing in the Hamiltonian, it is immediately clear that the spectrum is invariant
under a change of sign of any of the exchange constants $J_{\alpha} \to - J_{\alpha}$. 
This, along with an overall global scale invariance of the gapless/gapfull parameter regions under
a uniform scaling of all coupling parameters $J_{x,y,z} \to c J_{x,y,z}$ with $c$ a constant
enables us to delineate the boundaries of the gapless and gapped regions
of the model. Such a phase diagram of the system is provided in Fig. \ref{fig:Kitaev_phases} \cite{Kitaev06}. 
The existence of transitions between these phases are ``topological'' and as such cannot be discerned by any standard local
measurements. The inability of local measurements to discern between different phases underlies
systems with topological order \cite{Kitaev03,PNAS,AOP}.  When expressed in terms of the basic spin degrees of freedom,
anyons involve extended non-local lines. Amongst other probes,
an interesting signature of the topological transitions between the
Abelian phases in Kitaev's honeycomb model is afforded by quantum information
theory measures, in particular the mutual information \cite{Cui2010}. 

The condition for a gapless phase is tantamount to the triangle inequality via Eqs. \ref{epD}. 
This is so as the gapless phase implies $\epsilon_{q} = \Delta_{q} =0$ which
in turn implies from Eqs. \ref{epD} (via the law of cosines) that one can view $q_{x}$ and $q_{y}$ as angles
in the triangle formed by the sides $\{J_{x}, J_{y}, J_{z}\}$ \cite{Chen08}.  
The relation between Kitaev's honeycomb model and the $p-$ wave type pairing
in Eqs. (\ref{eq:Kitaev_F},\ref{fermi-pair},\ref{epD}) was further 
elucidated in several insightful works, \cite{Yue2008,YueNPB}.

\subsubsection{Braiding Statistics}

The Majorana fermion representation of Eq.(\ref{eq:Kitaev_bond}) highlights another
important property of this system-  the braiding statistics formed by
displacing one string of bonds around a closed loop.
The product of bonds along any contour
(open or closed) commutes with 
all other string products of the same form,
including the symmetries $O_{h}$ of Eqs. (\ref{eq:symkit6},\ref{eq:O_bond}).
That is, for any contour $C$ drawn on the 
lattice, the operators
\begin{eqnarray}
O_{C} = \prod_{ij \in C} b_{ij} 
\label{Oc}
\end{eqnarray}
commute 
amongst themselves.
For closed contours $C$, one has a sort of Stokes'
theorem. That is, the symmetries of Eq. \ref{Oc} can be written as \cite{Chen08}
\begin{eqnarray}
O_{C} = \prod_{h \in C} O_{h},
\label{stokes}
\end{eqnarray}
with the product taken over all hexagons $h$ that are enclosed by the loop $C$.
The right-hand side of Eq.(\ref{stokes}) corresponds to the 
total anyonic charge enclosed by $C$.  If an odd number
of anyons (hexagons $h$ for which $O_{h} =-1$) is circumscribed by $C$
then $O_{C} =-1$.   {\em This minus sign is the origin of the anyonic nature} of
the braiding operations in Kitaev's model of Eq. \ref{eq:Kitaev_K}, e.g., \cite{Chen08, Kells08},
in its gapped phase (known as the ``A phase'').  
In the gapless phase of couplings of Kitaev's model (``B phase'') of Eq. \ref{eq:Kitaev_K}, 
the statistics of the vortices is ill defined. However, as will be elaborated on later,
augmenting Eq. \ref{eq:Kitaev_K} by an additional external magnetic field term
leads to the opening of a gap in the B phase. Within this gapped regime, 
the vortices exhibit well defined non-Abelian statistics. 

An immediate corollary of the local ($d=0$) symmetries $O_{h}$ of Eqs. (\ref{eq:symkit6},\ref{eq:O_bond}),
is that, by Elitzur's theorem, only correlation functions that are invariant under all of these symmetries
may attain a non-zero expectation value at finite temperatures. \cite{Chen08}
Thus, all non-zero {\em correlation functions are composed of products 
of bonds (along closed or open contours) as in Eqs. (\ref{Oc},\ref{o_gamma})}.
Similar considerations also apply at zero-temperature. [This is so as within the ground
state $O_{h}=1$ for all $h$ while $O_{h}$, for a particular hexagon $h$, may be chosen to reverse the sign 
of a correlation function unless the correlation function is of the form of 
Eqs. (\ref{Oc},\ref{o_gamma}.) \cite{Chen08}]
These considerations generally lead to string and ``brane'' type correlation functions.  
[In fact, two-body and similar fermionic correlation functions of the local quadratic 
Fermi system defined by Eq. (\ref{eq:Kitaev_F}) become,
upon the use of the JW transformation, such non-local string and brane type
spin correlation functions \cite{Chen08}.]
While the above considerations revolve around symmetries of the spins alone, an
earlier work illustrated, by the use of Majorana fermions, that all 
two-point correlation functions apart from those that form bonds vanish within
the ground state and further related interesting consequences \cite{shankar}.

In the gapped phase, these string correlation functions are 
exponentially damped in spatial distance between the endpoints
of the string (with a similar behavior concerning 
dynamic correlations). Within the gapless phase, the string correlation functions 
decay algebraically in the distance. \\

\subsubsection{Fermion Excitation and Translation}

The fermionization procedure discussed above enables the construction of anyons out of string operators \cite{Chen08}.
A feature directly related to the symmetries of Eqs. (\ref{Oc},\ref{stokes}) is that it is possible to create ``fermionic'' excitations alone sans anyons. One way to see this is  
by invoking symmetry and bond algebra arguments once again. Towards this end, one may 
consider the string product of bonds of the form of Eq. (\ref{Oc}) yet now for 
{\em an open contour} $\Gamma$ (as opposed to the closed contour $C$).  
That is, we may define the operator 
\begin{eqnarray}
O_{\Gamma} = \prod_{ij \in \Gamma} b_{ij}
\label{o_gamma}
\end{eqnarray}
along an open contour $\Gamma$. For the purposes of what follows, let us label the end 
points of $\Gamma$ by $U$ and $V$. Unlike $O_{c}$ of Eq. (\ref{Oc}), the operator of Eq. (\ref{o_gamma})
is not a symmetry. That is, the operator $O_{\Gamma}$ serves as a trivial symmetry for 
all bonds $b_{kl}$ for which (1) $k , l \neq U$ or $V$
and/or (2) lie along $\Gamma:  k,l \in \Gamma$. For all such bonds, 
$[b_{kl}, O_{\Gamma}] =0$. The above includes all bonds 
$b_{kl}$ that have any number of sites along $\Gamma$ (i.e., 0,1, or 2) such that
the bonds do not touch $\Gamma$ only at one point with that point being one of the endpoints
$U$ or $V$. However, if $k=U$ or $V$ and $l \not \in \Gamma$ 
or vice versa (i.e., $l= U$ or $V$ and $k \not \in \Gamma$) 
then $b_{kl}$ will anti-commute with $O_{\Gamma}$:  ~ 
\begin{eqnarray}
\{O_{\Gamma},  b_{kl} \}=0.
\label{ogkl}
\end{eqnarray}
There are four such bonds $b_{kl}$. All other bonds commute with the operator
of Eq. (\ref{o_gamma}), $[O_{\Gamma}, b_{mn}] = 0$. 
As the exact solution that was outlined earlier [Eqs. (\ref{epD}, \ref{BCS-p})]
shows, the ground state sector of Kitaev's model
is not highly degenerate. As $O_{\Gamma}$ flips the energetic contributions of
the four of the bonds $b_{kU}$ and $b_{Vl}$ that touch the endpoints $U$ and $V$, 
all of this suggests that the application
of general $O_{\Gamma}$ (there is an exponentially large number of contours $\Gamma$)
on a ground state cannot give back a ground state but rather
must excite the system. The bonds at the end points of the contour $\Gamma$ have
been modified (by a change of sign) as a result of the anticommutation relation of Eq. (\ref{ogkl})
and together the four disrupted bonds at the two endpoints $U$ and $V$ that
do not lie along $\Gamma$ sum to yield a higher energy state. 
Thus, it is natural to associate defects created by the string operator of Eq. (\ref{o_gamma})
at the endpoints $U$ and $V$ of the contour $\Gamma$.
As seen from Eqs. (\ref{eq:Kitaev_bond}, \ref{o_gamma}), the string operator $O_{\Gamma}$ 
involves only the Majorana fermions
and not the anyons of Eqs. (\ref{eq:symkit6},\ref{eq:O_bond}) and their composites
(Eq. \ref{stokes}). Indeed, it is possible to verify that as each of the bonds of 
the lattice, $b_{ij}$ of Eqs. (\ref{eq:longbonds}, \ref{eq:Kitaev_bond})
commutes with all anyonic charges $O_{h}$ of Eqs. (\ref{eq:symkit6},\ref{eq:O_bond}),
the operator of Eq.( \ref{o_gamma}) does not create [nor, in general, remove or displace) any
anyons]: ~$[O_{\Gamma}, O_{h}]=1$. 
For a closed loop however, the closed string operator
of Eqs. (\ref{Oc}, \ref{stokes}) is a symmetry. 
The existence of general $d=1$ symmetry operators that are products
of defect creation operators along loops \cite{PNAS} has similar incarnations elsewhere (e.g., in Quantum
Hall systems with the creation of quasi-particle/quasi-particle pairs). 
By creating defects
and moving these defects
along entire closed cycles, the defects annihilate and the system returns 
to its low energy (ground state) sector. Putting all of the pieces together, one sees that
it is possible to have fermionic excitations (generated by Eq. (\ref{o_gamma})) alone. 
It is possible to express all of 
these results in terms of the fermions directly similar to \cite{Chen08}. 

When a fermion is transported around a closed loop that encircles a single Ising
vortex (for which $O_{h} =-1$), 
we see that Eqs.(\ref{Oc}, \ref{stokes}) reduce to an overall phase factor of $(-1)$. Thus, in
such an instance the quantum state is multiplied by this overall phase
factor \cite{Kitaev06}.

\subsubsection{Vortex Pair Creation and Translation}

It is common to think about excitations formed by the application of 
single spin operators (i.e., by a rotation of a single spin) or by a product of
two on the ground state. As pointed out by \cite{Dusuel08}, there are subtleties associated
with simple interpretation of  the action of these operations within the low 
energy sector. In what follows, we will focus on such an excitation 
via general symmetry and 
bond algebraic considerations. Towards this end, we consider a single vertical link
$(ij)$. We define, similar to \cite{Kitaev06,Pachos06,Pachos07}, the three operators
$X =  \tau_{i}^{x} \tau_{j}^{x}$, $Y = \tau_{i}^{x} \tau_{j}^{y}$, and $Z= \tau_{j}^{z}$.
These operators are different from those Eq. (\ref{o_gamma}) (including the case of a single two site bond). Each of these three operators anti-commutes with two bond operators. For instance, 
$Z$ anticommutes with the two bonds (other than $b_{ij}$) 
that have $j$ as one of their endpoints. Similarly, the operators $X$ and $Y$ each anti-commute
with exactly two bonds. When acting on the ground state, the flipping operations incurred
by any of the operators $X,Y$ or $Z$ may  increase the system energy. It is furthermore readily
verified that $Y$ and $Z$ may each flip the anyonic charges $O_{h}$ of two hexagonal plaquettes
while $X$ flips the anyonic charges of all four hexagonal plaquettes that contain either the 
site $i$ or $j$ (or both). The flipping of any of the bonds
generated by each of these three operators  can be accounted for by inverting the sign of $\eta$
field along the corresponding link following Eq. \ref{eq:Kitaev_bond}. 
The three operators satisfy 
$S=1/2$ spin algebra:
\begin{eqnarray}
\{X, Y\}= \{X,Z\}=\{Y,Z\}=0, \nonumber
\\ X^{2} = Y^{2} = Z^{2} =1, \nonumber
\\ XY = iZ,~ YZ = iX, ~ZX = iY.
\label{XYZ_kit}
\end{eqnarray}
It is natural to associate ``particles'' $X| \psi \rangle, Y| \psi \rangle, Z| \psi \rangle$
created by the application of the operators $X,Y$ or $Z$ on the ground state wavefunction.  
The last line of Eq. \ref{XYZ_kit} generally suggests that a fusion of two particles into a third might be possible.  This is indeed the case as it has been worked out in some detail in various approaches and 
limits (especially that of $J_{z} \gg J_{x,y}$ lying within the $A$ phases of the system, see Fig. \ref{fig:Kitaev_phases} \cite{Kitaev06, Pachos07}. In that limit,
the energy of the the excitation $X| \psi \rangle$ is nearly equal to that of the sum of
energies corresponding to $Y| \psi \rangle$ and $Z| \psi \rangle$.

 As the anyonic charges of Eq.\ref{eq:symkit6} are symmetries, anyonic excitations
are massive. That is, an anyonic excitation is stationary as it is an 
eigenstate of the Hamiltonian. As discussed in \cite{Dusuel08}, it is possible to create anyons without  fermions by the combined use of one and three spin operations on the ground state.
We now extend the discussion of the single bond operators above and present the general vortex translation 
(or anyon) operator.  An approach related to ours, along with a detailed analysis of energies, is given 
in \cite{Kells08}. An insightful analysis is also provided in \cite{Dusuel08}.
In order to analyze the Ising vortex translation operators, 
we introduce an operator that is identical to that of 
Eq. (\ref{o_gamma}) apart from all important end point corrections that allow it to be expressed
as $O_{\Gamma}$ multiplied by two operators corresponding to the two endpoints. Specifically, 
we consider an open contour $\Gamma$. For each non endpoint vertex $i \in \Gamma$, there is only a single neighbor $l$ that is not on $\Gamma$. For the two end points of $\Gamma$, 
($i_{1} = U$ and $i_{2} = V$), there are two neighbors $l$ that do not lie
on $\Gamma$. One may choose any of these neighbors for the two endpoints in what follows. 
(We will mark the chosen neighbors for the endpoints by $l_{1}$ and $l_{2}$ respectively.)
We denote the direction of a ray parallel to the nearest neighbor link ($\langle il \rangle$) by $\gamma$ (that may be x, y, or z).
We then construct the open contour operator
\begin{eqnarray}
{\cal{T}}_\Gamma = \prod_{i \in \Gamma} \tau_i^\gamma.
\label{vortex-contour}
\end{eqnarray}
Eq. \ref{vortex-contour} is nearly of an identical form to  Eq. \ref{o_gamma} for all points non-boundary
points $i$. However, in Eq. \ref{o_gamma}, the component of the boundary spin operators
that appear in the string operator are such set equal to the two directions $\gamma_{1,2} 
= \langle i_{1,2} j_{1,2} 
\rangle$ with $j_{1,2}$ being the nearest neighbors of $i_{1,2}$ that {\it lie on $\Gamma$}
(i.e., ``going backwards'' away from the endpoints $i_{1,2}$).  By contrast,
in Eq. (\ref{vortex-contour}),  the components of the spins at the two endpoints that appear in the string
product are set by the two directions $\gamma = \langle i_{1,2} l_{1,2} \rangle$ (with $l_{1,2}$ not 
on $\Gamma$).  

For the two hexagonal plaquettes $h^*=h_{1,2}$ that have a single vertex at one of the endpoints of $i_{1}$ or $i_{2}$ of  $\Gamma$ and that furthermore include one of the vertices $l_{1}$ 
or $l_{2}$, we have that 

\begin{eqnarray}
{\cal{T}}_{\Gamma} O_{h*} {\cal{T}}  = -O_{h*}.
\label{charge-inv}
\end{eqnarray}

In eq. (\ref{charge-inv}),$O_{h*}$ denotes the vortex charge of a plaquette $h*$ that lies at an endpoint
of $\Gamma$.

Similar to the operator of Eq. (\ref{o_gamma}), for all other plaquettes $h \neq h*$,
we have that $
{\cal{T}}_{\Gamma} O_{h} {\cal{T}}_{\Gamma} = O_h$
(with no change in the vortex charge).

It is readily verified that the operator $
{\cal{T}}_\Gamma$, albeit flipping the sign of two bonds attached to the endpoints of $\Gamma$, 
does {\em not alter} the bond algebra of all bonds (all non- neighboring bonds commute, neighboring bonds anticommute, and the square of any bond is 1). The sole change triggered by the application of 
${\cal{T}}_{\Gamma}$ is that 
two bond pre-factors $\eta$ are multiplied by a factor of (-1)).
and correspondingly two vortex charges are flipped. 
Thus, the effect of ${\cal{T}}_{\Gamma}$ is to flip the sign of the two vortices at its endpoints.

If the system has a single vortex
$O_{h_{1}}=-1$ at plaquette $h_{1}$
that has only one (endpoint) on $\Gamma$
and furthermore contains one of the two points $l_{1.2}$, then the application 
of ${\cal{T}}_{\Gamma}$ with the contour $\Gamma$ having a single point in the plaquette $h_{1}$ 
(the latter plaquette also containing the point $l_{1}$) as one of its endpoints will move the vortex to another plaquette $h_{2}$  that lies at the other end of the contour $\Gamma$
(and contains the point $l_{2}$).

That is, ${\cal{T}}_\Gamma$ is a {\em vortex translation operator}.
If $\Gamma$ forms a complete closed contour  $C$ 
along a toric cycle (when $h_{1}$ and $h_{2}$ are identified as the same point on the torus) 
then, similar to $O_{\Gamma}$ of Eq. (\ref{o_gamma}), 
${\cal{T}}_{\Gamma}$ veers towards the $d=1$ dimensional symmetry of Eq. (\ref{Oc}). 
In the above, we established that the sole effect of ${\cal{T}}_\Gamma$ 
is to displace a vortex without influencing the system energy from any
of the bonds that do not touch that endpoints of the contour $\Gamma$.

Although trivial, it may be noted that as pair permutations can be written in an SU(2) symmetric
form as
\begin{eqnarray}
{\cal{P}}_{ij}  = \frac{1}{2}(1+  \bf{\tau}_i \cdot \bf{\tau}_j)
\label{pij}
\end{eqnarray}
and as any translation may be expressed as a product of pair permutations, 
general Majorana fermion and vortex translation operators may be expressed
in an SU(2) symmetric form as a product of operators of the form of 
Eq. (\ref{pij}).

\subsection{Kitaev-Compass Model -- non-Abelian Phase}
\label{nonabp}

\subsubsection{Definition of Extended Model}

Kitaev's model for a wide range of couplings 
corresponds, as earlier discussed
(see Eq. (\ref{epD}, \ref{BCS-p})
 to a gapless phase. This region is th so-called ``B'' phase
 of Kitaev's model. It is only in the ``corners''
of the phase diagram of Fig. \ref{fig:Kitaev_phases} ( the so-called ``A'' phase where the $\{J_{x}, J_{y},J{z}\}$ differ substantially from
one another and cannot form the sides of a triangle) that a gap opens up.
As will be elaborated later on (sections (\ref{corrkk}, \ref{kkht}, \ref{abst}), 
in the A phase, gapped Abelian anyons are present. 
Our focus in this section will be on the B phase where
gapless excitations of Eq. (\ref{BCS-p}) were found.
By a modification of Kitaev's honeycomb model, gapped non-Abelian
excitations can arise.
There are various ways in which such excitations can arise. For instance, these
may be triggered by the geometry of the lattice (via, e.g.,
a decoration of the lattice wherein each vertex
of the hexagonal lattice is replaced by a triangle
\cite{Yao2007}). 
In what follows, we review the original investigation
of \cite{Kitaev06} in which a gapped phase with non-Abelian 
excitations originates from the application of an external magnetic
field to a point $(J_{x}, J_{y}, J_{z})$ in the space
of coupling constants for for which the system
would have been gapless if no field were applied.
[In this phase, the ``B phase'', the couplings $J_{x}, J_{y},$ and $J_{z}$ 
satisfy the triangle inequalities.] The B phase has made an appearance in
studies quite removed from Kitaev's model. Interestingly, in the quantum
Hall arena, for half-filled Landau levels, this phase has also been
suggested \cite{Barkeshli2012}
to appear to lie in the interface between 
a $\nu=1/2$ Moore-Read type state and a topological 
superconducting state as a periodic potential is tuned. 

 As we now review, such a field gives rise
to an effective {\em next nearest neighbor coupling} between
Majorana fermions. This additional hopping leads to a gapped
spectrum with non-Abelian chiral modes.
When a magnetic
field ${\bf h}$ is applied along the $[111]$ direction, 
i..e., when Eq. \ref{eq:HKT} is augmented by a Zeeman coupling
\begin{eqnarray}
H' = H - \sum_{i} {\bf h} \cdot {\bf \tau}_{i},
\label{fieldkit}
\end{eqnarray} 
a gap opens up in in the core region (B phase) of the phase diagram of Fig. \ref{fig:Kitaev_phases}.
The (time reversal broken) phase that arises from the application
of this field is very interesting.

In particular, non-Abelian anyons appear 
in the former gapless phase (which includes
the symmetric point $J_{x} = J_{y}= J_{z}$).
The Hamiltonian of Eq. \ref{fieldkit} is not exactly solvable.
It can, however, be treated perturbatively and (ignoring unimportant corrections)
reduced to an exactly solvable system \cite{Kitaev06}. That is,
the magnetic field term in Eq. \ref{fieldkit} gives rise
(with $\kappa \sim  h_{x} h_{y} h_{z}/J^{2}$ in the symmetric point $J_{x} = J_{y} = J_{z} = J$)
to a  (time reversal symmetry breaking) term of the form of Eq. (\ref{eq:Kitaev_h}) which we write here anew,
\begin{eqnarray}
\label{hhh}
H_{h} = - \kappa \sum_{ijk}  \tau_{i}^{x} \tau_{j}^{y} \tau_{k}^{z},
\end{eqnarray}
for all triplets of sites  $(i,j,k)$ formed by the union of two bonds ($(ij)$ and $(jk)$) 
that impinge on site $j$. 
The {\em product of the three spin operators} of Eq. (\ref{hhh}) can be expressed as a {\it product of two neighboring bonds}
of Eqs. (\ref{eq:longbonds},\ref{eq:Majorana_bond}) by use of the relation $c_{j}^{2} =1/2$. 
For instance, for (oriented) links $(ij)$ and $(jk)$ along
the x and z directions respectively (with $i_{z} < j_{z}$ abd $j_{z} < k_{z}$), the product of the bonds of 
Eq. (\ref{eq:Majorana_bond}) reads $b_{ij} b_{jk} = - 2 \eta_{ij} \eta_{jk} c_{i} c_{k}$. 
Eq. (\ref{hhh})  is seen to reduce to a {\it Majorana fermion bi-linear linking (all) next nearest neighbor
sites}. The  Majorana fermion bi-linear ($c_{i} c_{k}$)  resulting from the product of
two bonds has a real prefactor ($-2 \eta_{ij} \eta_{jk}$) as
opposed to the imaginary prefactors that are associated with
single nearest neighbor bonds in Eq. (\ref{eq:Majorana_bond}).  This
relative phase factor of $i$ reflects the time reversal symmetry
breaking of the perturbation. Time reversal symmetry breaking also allows for
the existence of chiral modes wherein fermionic modes may preferentially propagate
in one (clockwise or anti-clockwise) direction. For any pair of next nearest neighbor sites $(ik)$ on the
honeycomb lattice, there is a unique 3 site path (and two bond 
product) that leads to the bi-linear form $c_{i} c_{k}$.
The quadratic character of 
these three-spin perturbations of Eq. (\ref{hhh})in the Majorana fermions 
(and similarly also in the fermions following, e.g., Eq. (\ref{cdrep}))
ensures that even when the
system is augmented by these perturbations, the total Hamiltonian
the final Hamiltonian 
\begin{eqnarray}
\label{khh}
H_{K_{h};h} \equiv H_{K_h} + H_{h}
\end{eqnarray}
formed by the sum of Eqs. (\ref{eq:HKT}, \ref{hhh}) is {\it still exactly solvable}.

\subsubsection{Solution of Extended Model}
\label{solnusnon}
The solution to the problem is of a similar character to the one that earlier led to 
Eqs. (\ref{epD}, \ref{BCS-p}).  As each spin product of
the type $ \sigma_{i}^{x} \sigma_{j}^{y} \sigma_{k}^{z}$ 
is given by a product of two bonds
(each of which commutes with all of the symmetries of Eq. (\ref{eq:symkit6}),
it follows that the perturbation of Eq. (\ref{hhh}) commutes with the 
operators $O_{h}$. As before, in any given sector one can employ the 
representation of Eqs. (\ref{eq:longbonds}) with $\eta$ related to the flux
via the condition of Eq.(\ref{eq:O_bond}). All of the earlier steps taken in Eqs. (\ref{epD}- \ref{fermi-pair})
can thus be exactly reproduced. However, unlike the nearest neighbor Hamiltonian that
we we studied earlier in the absence of an applied external field $h$ (or an effective), 
the next nearest neighbor Fermi interactions lead new non-trivial results.
In particular, the perturbation set non-zero ${\bf{h}}$ allows the earlier gapless phase in the absence of a field to become gapped and thus to support anyons which within
this phase are non-Abelian \cite{Kitaev06}. 
The spectrum of $H_{K_{h};h}$, in the vortex free sector ($O_{h}=1$ for all $h$) 
is then seen to be given by  
\begin{eqnarray}
E_{q} = \pm  \sqrt{\epsilon_{q}^{2} +|\tilde{\Delta}_{q}|^{2}}
\label{energy_eqq}
\end{eqnarray}
where the real p-wave type gap $\Delta_{q}$ \cite{Chen08} of Eq. (\ref{epD}) is now replaced by the complex
\begin{eqnarray}
\tilde{\Delta}_{q} = \Delta_{q} +  4i \kappa [\sin q_{1} - \sin q_{2}  + \sin(q_{2} - q_{1})].
\end{eqnarray}
As can be seen by some simple analysis, the former gapless points  ${\bf q}^{(\pm)}$
of Eq.(\ref{BCS-p}) now acquire a gap when $\kappa \neq 0$. The 
p-wave type gap function \cite{Chen08} $\tilde{\Delta}$ now becomes complex. This suggests that the physics will essentially
be the same as that for $``(p+ip)''$ superconductors \cite{Ivanov}. This is indeed the case
as we will briefly reiterate later on. It is noteworthy that even when the 
Hamiltonian is time reversal invariant (as that of the system without perturbations-
that of Eq. (\ref{eq:HKT})), the ground states may spontaneously break time reversal. 
Indeed, by Kramers' theorem, this must occur whenever the system is defined on a hexagonal 
lattice with an odd number of spins \cite{Chen08}.
In the B phase of Kitaev's model wherein the gap was borne by 
the perturbation, the associated Chern number $\nu = \pm 1$
and the aforementioned non-trivial statistics \cite{Kitaev06}
with non-Abelian topological anyons. We elaborate on 
these anyons and their features next.

\subsubsection{Non-Abelian anyons and their properties}
\label{nanp}

To conform with standard practice, we use $\sigma$ to denote a vortex
(defined, similar, to the Abelian phase by having 
the plaquette product $O_{h}$ of Eqs. (\ref{eq:symkit6},\ref{eq:O_bond})
be $(-1)$, $O_{h}=-1$)
and $\epsilon$ is mark a fermionic mode. The
fusion rules are then of the form
\begin{eqnarray}
\epsilon \times \epsilon =I, \nonumber
\\ \sigma \times \epsilon = \sigma, \nonumber
\\ \sigma \times \sigma = I + \epsilon,
\label{fus_nab}
\end{eqnarray}
augmented by the trivial statement that the fusion of any particle
with the identity operator leads back to that particle (as in, e.g., Eq. (\ref{triv_fusi}) for the Abelian anyons).
As in the case of the Abelian anyons of Eqs. (\ref{eeI}, \ref{fusion}),
each particle is its own anti-particle. The non-trivial
character of the non-Abelian anyons rears its head
in the last line of Eq. \ref{fus_nab}. Two vortices ($\sigma$) may fuse in two
different channels to either annihilate ($I$) each other
or to form a fermion ($\epsilon$). The vortex operators of Eq. (\ref{eq:symkit6}) have, as always, Ising eigenvalues $O_{h}= \pm 1$.
Anyons that satisfy the relations of Eq. (\ref{fus_nab}) are called ``Ising anyons''.
Unlike the case of Abelian anyons
(e.g., Eqs. (\ref{eeI}, \ref{fusion})), {\em fusing two different
non-Abelian may lead to different outcomes} (the vacuum ($I$)
or an $\epsilon$ particle). In terms of Eq. (\ref{general_fuse}), 
when choosing $a=b=\sigma$, there are two different anyonic outcomes
for these non-Abelian anyons. 
In the limit of spatially infinitely distant vortices, 
the fermionic spectrum as adduced from the square lattice
Hamiltonian of Eq. (\ref{HKF}) with $\eta_{r}$ on the vertical links of original 
the honeycomb lattice set by the vortices $O_{h}$ of Eq. (\ref{eq:O_bond}),
exhibits a multitude of fermionic {\it zero modes} \cite{Lahtinen}.  
Thus the hybrid of two well separated vortices ($\sigma$) may lead to
a state in which the vortices annihilate to form the vacuum ($I$) 
or a ``zero energy'' fermionic state $(\epsilon$) 
This degeneracy is lifted once the vortices become close to one another
wherein the fermionic modes $\epsilon$ attain a finite energy cost (or ``mass'').
Repeated
applications of the last of Eqs. (\ref{fus_nab}) 
rationalizes the $2^{n_{\sigma}/2-1}$ fold degeneracy that is present in a system of $n_{\sigma})$ (with this number being an even integer) well separated vortices \cite{Nayak1996}. 
In formal terms, the quantum dimension of the vortices $\sigma$ is $d_{\sigma} = \sqrt{2}$; the 
system degeneracy for $n_{\sigma}$ vortices scales $d_{\sigma}^{n_{\sigma}}$. Due to the unique
outcome of all of the other fusion rules in Eq. (\ref{fus_nab}), the quantum dimensions
of $\epsilon$ and $I$ are $d_{\epsilon}= d_{I} =1$.
The authors of \cite{Lahtinen} studied, in detail, the spectrum
of Kitaev's model and, in particular, the resulting spectrum for different vortex configurations
in the non-Abelian phase.  \cite{Lahtinen2011a} further illustrated how the fusion
rules of Eq. (\ref{fus_nab}) can be made evident by carefully studying the spectrum
as the vortices were made to move towards one another with a distance
that could be made continuous and examining the levels that appears in
the limit of zero spatial separation between two vortices. The effective
interactions between vortices that are held a finite distance
apart  (that trigger the aforementioned lifting of the topological degeneracy) 
exhibit an oscillatory character as a function
of separation. The characteristic modulation length of 
these oscillations is set by the inverse Fermi momentum \cite{Lahtinen2011a} akin
to that associated with vortices in $(p+ip)$ superconductors \cite{Cheng2009}.  
Similar oscillations \cite{Baraban2009} appear in the Moore-Read Pfaffian wavefunction
proposed for fractional Quantum Hall states \cite{Moore1991}. 
We will review these and related properties in Section \ref{vfl}.

Reflecting more on the relations of Eq. (\ref{fus_nab}), one may consider a system with four well separated anyons $\sigma$
(denoted as $1,2,3,$ and $4$) which together fuse to
form the vacuum $I$. In such an instance (similar to EPR type experiments \cite{EPR}), the fusion 
outcome of any two $\sigma$ particles (e.g., $1$ and $2$)
uniquely determines the particular fusion outcome of the other two $\sigma$ particles ($3$ and $4$). 
Namely, the outcome of the fusion of particles $a$ and $b$ is the same as that of fusing particles $3$ and $4$.
As particles $1$ and $2$ can fuse in two different ways, there is a two dimensional space that is related
to these four $\sigma$ particles. Focusing on the $\sigma$ particles $1$ and $2$ is, of course,
artificial. One could have, e.g., focused on the basis set by the fusion result of particles $2$ and $3$. 
The matrix relating to such a change of basis is termed the ``F-matrix''. More generally,
in general systems, the matrix element $[F^{d}_{abc}]_{ef}$ is the amplitude that quasi-particles $a,b,$ and $c$ fuse and create quasi-particle $d$ in a specific fusion channel
$e$. A schematic is shown in Fig. (\ref{fig:F-matrix-explained}). 
\begin{figure}
\centering
\includegraphics[width=.6\columnwidth]{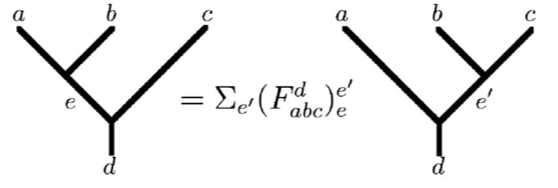}
\caption{
F-matrices relate the results of a fusion processes in which the order of
events is changed. On the left, particles $a$ and $b$ are first fused together
to form particle $e$; this particle is, in turn, then fused with particle $c$ to form particle $d$.
The diagram on the right represents a process in which particles $b$ and $c$ are first fused 
to form a particle $e^{\prime}$ which is then fused with particle $a$ to form $d$. The outcomes
of the two fusion processes are the same by the associativity property of fusion
(Eq. (\ref{axbxc})). The two orthonormal bases defined by $(e,c)$ and $(e^{\prime}, c)$
are related to one another via the $F-$ matrix~\cite{Wootton2008}.}
\label{fig:F-matrix-explained}
\end{figure}
In the basis spanned by $(1, \psi)$, the model has the non-trivial $F$ matrix 
\begin{eqnarray}
F^{\sigma}_{\sigma \sigma \sigma}=   \frac{1}{\sqrt{2}}
\begin{pmatrix}
1 & 1 \\
1  & -1 \\
\end{pmatrix},
\end{eqnarray}
and $F^{\sigma}_{\psi \sigma \psi} = F^{\psi}_{\sigma \psi \sigma} = - 1$. 

Braiding any two such particles with each other cannot change their fusion channel. This is so as their fusion channel or total charge may be determined along a distant path that circumscribes both particles. When two particles fuse in a definite channel,
the outcome of revolving one of these particles around the other can only lead to a phase factor as diagrammatically depicted
in Fig. (\ref{fig:R-matrix-explained}). For the Ising anyons
of Eq. (\ref{fus_nab}), these phase factors are given by
\begin{eqnarray}
R_{\sigma \sigma}^{I} = e^{- i \pi/8} , ~~~ R_{\sigma \sigma}^{\epsilon} = e^{3 \pi i/8}  \nonumber
\\  R_{\epsilon \epsilon}^{I} = 1, ~~~ R_{\sigma \epsilon}^{\sigma} = i.
\label{R-matrix} 
\end{eqnarray}
In Eq. (\ref{R-matrix}), $R^{c}_{ab}$ is the phase factor ratio between the (i) state that results 
by fusing two particles ($a$ and $b$)
to a particle $c$ in a particular fusion channel of Eq.(\ref{general_fuse}) following  
a counter-clockwise exchange of particles $a$ and $b$ to  (ii) the fusion of particles $a$ and $b$ to form particle $c$
without such an exchange. 
\begin{figure}
\centering
\includegraphics[width=.5\columnwidth]{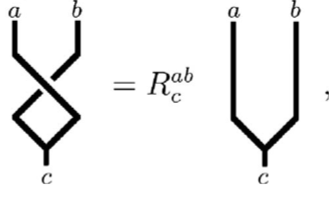}
\caption{The R-matrix describes the effect of a counter-clockwise exchange of 
two particles on their fusion product. On the lefthand side, particles are exchanged in a counter-clockwise manner 
prior to their fusion. The resulting fusion product is related by the matrix element $R^{ab}_{c}$
to that without such a prior exchange~\cite{Wootton2008}.}
\label{fig:R-matrix-explained}
\end{figure}

Similarly, for the Abelian anyons of Eqs.(\ref{eeI}, \ref{fusion}), which are pertinent to the Abelian phase of 
Kitaev's model, the elements of the $R$ matrix are given by
\begin{eqnarray}
R_{I}^{\epsilon \epsilon} = (R_{\epsilon}^{e m})^{2} = -1, ~~ R_{I}^{ee} = R_{I}^{mm} = 1.
\label{R-abelian}
\end{eqnarray}

We briefly comment on general {\em topological quantum computing} aspects \cite{Kitaev06,Nayak2008} which
become most potent in systems with non-Abelian anyons such as those of Eq. (\ref{fus_nab}). 
Within this general framework, an initialization proceeds by the creation of anyon pairs
from the vacuum; this enables a knowledge of the particles and their fusion channels.  
Separating these anyons from one another gives rise to a degenerate manifold on which computations may be done.
For the particular case of the vortex ($\sigma$) anyons of Eq. (\ref{fus_nab}), as alluded to above, 
an $(n-1)$ bit space is generated from $(2n)$ far separated anyons. Braiding operations of anyons 
correspond to unitary gates. Measurements of the ``calculation output'' 
may be performed via the fusion of anyons and measuring
energy to ascertain the fusion channel and/or by an interference experiment. Interference can be performed by
creating a pair of anyons and, similar to the discussion in Section \ref{sec:notions} 
moving each of the members of the pair in different directions around another
anyon along a $d=1$ loop. The final outcome following a recombination and fusion of the anyon pair members with each other
depends on the state of measured anyon \cite{Kitaev06,Nayak2008}. 

As remarked earlier, fusion rules similar to those of Kitaev's
model appear also in $p-$ wave superconductors \cite{Ivanov,Stone2006}.
In the latter arena, $\epsilon$ represent Bogoliubov quasiparticles, $\sigma$
a vortex, and $I$ represents the Cooper pair ground state.
The relation between the two systems might not be that surprising given
the mapping employed here to relate Kitaev's honeycomb model
to a $p-$ wave type superconductor in the form of Eq. (\ref{HKF})
which has a similar form yet with additional next nearest neighbor 
pairing and hoping terms once the interactions of Eq. (\ref{hhh}) are included
in the discussed form of Majorana fermion (and thus according to Eq. (\ref{cdrep}) also fermion)
bi-linears. The non-Abelian nature of this theory
may enable an anyonic quantum computer on a honeycomb model \cite{Kitaev06,Freedman2002}.
{\em An edge mode is associated with each of the various excitations in Eq. (\ref{fus_nab})}. 

Formally, the relations of Eq. \ref{fus_nab} realize 
``SU(2)$_{k=2}$ algebra'' (termed``level two SU(2)'' algebra). This fancy name simply 
means that we may consider the basic objects
to be usual (pseudo-)spins with the identity $I$ representing a trivial $S=0$ spin,
particle $\sigma$ corresponding to a spin $S=1/2$, and $\epsilon$ corresponding to spin $S=1$.
The fusion rules of Eq. \ref{fus_nab} are then what is expected when adding two spins 
$S_{1}$ and $S_{2}$ for which there is the well known total spin decomposition
\begin{eqnarray}
S_{1} \otimes S_{2} = |S_{1}-S_{2}| \oplus (|S_{1}-S_{2}|)+1) \nonumber
\\ \oplus .... \oplus (S_{1} + S_{2}),
\label{addrule}
\end{eqnarray}
with the additional requirement that any total spin $S_{tot} >(k/2)$ on the righthand side of
Eq. \ref{addrule} is to be dropped (this is what the ``level $k$'' qualifier means). As in standard quantum mechanics, when we add two $S=1/2$ spins, there are 
two possible total total spin outcomes- i.e., $S_{tot}=0$ or $1$; this leads to the last line of Eq. \ref{fus_nab}.
Similarly, when adding two $S=1$ spins (two $\epsilon$ particles)
we have $S_{tot}=0,1,2$. However, as all but the $S_{tot}=0$ outcome have a total spin $S_{tot}> k/2$ with $k=2$,
we have the single possible outcome on the first line of Eq. \ref{fus_nab}. 

\subsubsection{Berry, Wilczek-Zee phases \& Relation to Anyonic Statistics}
As is well known, when a system evolves adiabatically along a closed cycle in some general parameter space, it may 
accumulate a geometrical or ``Berry'' phase \cite{Berry,Pancharatnam}.  In quantum computation schemes, {\em anyonic statistics is determined by the Berry phase associated with the motion of
one anyon taken around another}. Non-Abelian statistics may arise when unitary matrices $U_{ab}$ implementing 
different exchanges of  different particles about one another do not commute. As alluded to earlier, in 
 Eq. (\ref{braid}), in tese cases, ${\cal{T}}_{i}$ are non-commuting unitary matrices 
 (of dimension $n={\cal{D}}_{U} \times {\cal{D}}_{U}$ with ${\cal{D}}_{U} >1$) acting on degenerate states. 
 In the Hilbert space spanned by $n$ degenerate eigenstates, the Berry phase becomes a unitary matrix \cite{Wilczek1984},
 \begin{eqnarray}
 U = {\cal{P}} \exp \{i \oint {\cal{A}}(\lambda) d\lambda \},
 \label{UP}
 \end{eqnarray}
 where ${\cal{P}}$ denotes a path-ordering in the parameter space $\lambda$ and 
 \begin{eqnarray}
 {\cal{A}}_{\ell m} (\lambda) = i \langle \Phi^{\ell} (\lambda)| \frac{d}{d \lambda} \Phi^{m}(\lambda) \rangle,
 \label{WZ}
 \end{eqnarray}
 with $1 \le \ell, m \le n$.
 An adiabatic motion of fused vortices leads to Berry phases that tend to the braiding 
 matrices associated with non-Abelian statistics as verified numerically by \cite{Bolukbasi2011}. 
 
 Berry phases may also be examined in the Abelian phase of Kitaev's honeycomb model, e.g., \cite{Lian2011},
 and examine situations (other those of anyons) wherein the phase factors of Eqs.(\ref{UP},\ref{WZ}) 
 may also be scalar (instead of more general matrices). 
 Owing to the effective $d=1$ dimensional character
of the generated excitations in Kitaev's honeycomb model when spins are, e.g., rotated about the $z$ axis, 
the authors of \cite{Lian2011} sought to investigate the system under a ``correlated rotation'' 
corresponding to the evolution of the ground state wavefunction
 $|\Psi_{0}^{\prime}(\phi) \rangle = U(\phi)  | \Psi_{0} \rangle$ where 
\begin{eqnarray}
U(\phi) = \exp[i \phi R],  \\
\nonumber \\  R = \sum_{z-bonds} \tau^{z}_{j} \tau^{z}_{k}
\label{RBerry}
\end{eqnarray} 
with similar forms for $R$ associated with rotations about the $x$ or $y$ spin axes. 
The Berry phase is, in this case, set by 
$ \Psi_{0} | R| \Psi_{0} \rangle$.
We note this Berry phase of \cite{Lian2011} is simply the sum of the energies associated
with the ``z-bonds'' of Kitaev's honeycomb model. The authors of \cite{Lian2011} found that this phase
factor (or bond energies) is, as is to be expected, non-analytic at the transition between the A and B phases. 
Specifically, \cite{Lian2011} examined the system behavior along the isotropic line $J_{x} = J_{y}$ as 
the coupling $J_{z}$ was varied. 
For the Berry phase of Eq. (\ref{RBerry}), second order derivatives
of $\gamma$ were found to be divergent at the transition between 
the two phases. Such may be expected if the bond energies (and their differences as encountered
in the 90$^\circ$ compass model) play a role similar to that of
an order parameter in a continuous second order transition. Numerically, oscillations in $\gamma$ were found 
in the ``B phase'' of Kitaev's honeycomb model while those were absent in the gapped $A$ phase \cite{Lian2011}.

\subsubsection{General Features of (Majorana) Fermi Forms}

As we reviewed above, in principle, Kitaev's 
model can be solved (both without and with 
a perturbatively small magnetic field), within any vortex configuration ($\{O_{h}\}$)
 by mapping to a Fermi bi-linear. Specifically, we may find the exact spectrum of
 the Majorana Fermi bi-linear
 \begin{eqnarray}
H = \frac{i}{4} \sum_{jk}  A_{jk} c_{j} c_{k}
\label{HAK1}
\end{eqnarray}
  with the fermionic substitution of Eq. (\ref{cdrep})
where $\{O_{h}\}$ determine $\eta_{jk}$ by Eq. (\ref{eq:O_bond}) where we set $\eta_{jk}$ on
all vertical links to be $(+1)$ in any sector of the topological charges
$\{ O{h} \}$.  Such general (essentially free fermion systems in the diagonal basis) 
systems are completely characterized by
their band structures $E_{q}$. The exact ground states of this system
corresponds to occupying all states below the Fermi 
energy (whose value is $\mu(T=0) = \epsilon_{F} =0$). 
The pair correlator between any two single particle fermionic
states is one if both states are occupied (have negative energy)
and zero otherwise. Topological features such as 
{\em Chern numbers} are associated with 
the projection operators to the negative energy states for 
these dispersions in momentum space.  
In his seminal work, Kitaev \cite{Kitaev06}
analyzed the resulting Fermi surface topology
using these. 

As noted in section \ref{nonabp}, 
an explicit diagonalization of the general bi-linear of Eq. \ref{HAK1} \cite{Kitaev06,Lahtinen} 
as it pertains to Kitaev's honeycomb model in the presence of the next nearest neighbor fermion interactions
stemming from the perturbation of Eq. (\ref{hhh}) reveals that
the system features a spectrum containing a continuum of positive and negative energy states with
a gap in between them with, as in many other systems (such as integer quantum Hall systems 
and polyacetelyne chains), discrete {\em additional mid-gap (or fermionic 
zero mode) states} whose energy
vanishes for some momenta. In real space, these zero energy states 
correspond to states that are localized near the boundary. As a function
of momentum, the energies of the states changes from positive to negative
to the projection operators onto the negative energy states (that are all occupied
at zero temperature) changes abruptly at these wave-vectors. These discontinuities
are topologically distinguished by the so-called Chern number alluded to earlier.
The general form of the bi-linear of Eq. \ref{HAK1} enables a simple and exact form for the spectral
Chern number $\nu$ \cite{Kitaev06}. A physical consequence of this topological invariant is that energy spectra corresponding to 
different different Chern numbers cannot be adiabatically
connected- a quantum phase transition must intervene as $\nu$ is changed. 

Similar to electronic systems (with a chemical potential 
$\mu =0$), in our discussion of the single particle spectra of Eq. (\ref{BCS-p}),
within the ground state $|\psi \rangle$ all single particle fermionic 
state that are of negative energy are occupied  and those with energy $E_{q} > \mu =0$ are unoccupied. 
The projection operator to the negative energy states $E_q<0$ can be expressed in the original Majorana Fermi basis 
as 
\begin{eqnarray}
P_{jk}= \langle \psi | c_{k} c_{j} | \psi \rangle.
\end{eqnarray}
Formally, this projection depends the wave-vector parameter index and 
thus defines a  (one-dimensional) vector bundle which corresponds to a non-trivial
Chern number. 
Specifically,
\begin{eqnarray}
\nu = \frac{1}{2 \pi i} \int dq_{x} dq_{y} Tr \Big[ P(\vec{q}) \Big( \frac{\partial P}{\partial q_{x}} \frac{\partial P}{\partial q_{y}}  - \frac{\partial P}{\partial q_{y}} \frac{\partial P}{\partial q_{x}} \Big) \Big].
\end{eqnarray}
As briefly noted in section \ref{generalk2}, the Chern number ($\nu$) 
characterizes the difference between the number of different ``left'' and ``right'' movers along the boundary of the system (if such a system
occupies a half plane) \cite{Kitaev06}.  We will elaborate and briefly re-iterate some aspects concerning Chern numbers
below. The Chern number assumes integer values. The values of the Chern numbers aid in identifying 
different phases in Kitaev's honeycomb model. Even values of $\nu$ correspond to Abelian phases 
whereas odd Chern numbers
appear in non-Abelian phases. A non-zero value of the Chern number, $\nu \neq 0$, indicates
a chirality. 

The edge modes carry thermal energy and thus as in 
Quantum Hall systems  \cite{Cappelli2002} (where the Chern number is equal to the 
filling fraction), the Chern number can be adduced by thermal transport measurements \cite{Kitaev06}.
Specifically, at a temperature $T$, the edge energy current is given by
\begin{eqnarray}
J= \frac{\pi \nu}{24} T^{2}.
\end{eqnarray}
The physics of the Fermi gas (that of decoupled Fermi modes) largely underlies the results that we presented for the Kitaev model.
In this case, when considering the edge current,  the physics is that of decoupled 
fermi modes on the system boundary- i.e., that of a one dimensional fermionic 
system \cite{Kitaev06}.  These considerations appear elsewhere and are very powerful.

For multi-band systems, we can project the system onto each band separately and compute the Chern number that corresponds to each such band. The total Chern number associated with the negative bands is opposite in sign and equal in magnitude to the total Chern number associated with the positive energy bands.

Whenever $\nu = \pm 1$, every vortex must carry an unpaired Majorana mode. 
Unfortunately, translational symmetry
is broken by vortices ($O_{h} \neq 1$) and the system in a general sector is
diagonalized in a sector other than that of momentum space. Rather, we need
to diagonalize another general structural matrix defined by $A$. In the general
case, the Chern number is defined modulu 16. This relates to a topological phase
of the vortices given by $\exp[2 \pi i \nu/16]$. Only three of these 
phases are of relevance (the Abelian case of $\nu=0$ and the 
non-Abelian phase with $\nu = \pm 1$ that is induced by
an external magnetic field. In particular, when $\nu$ is odd, it is not
possible to define a structural matrix in the presence of vortices $O_{h} =-1$.
A Chern number of $\nu=0$ corresponds to the corners of the phase diagram of 
Fig. \ref{fig:Kitaev_phases} in the Abelian phase of the model
where the triangle inequality relating $J_{x}, J_{y}$ and $J_{z}$ cannot be satisfied. 

Similar to the general discussion of subsection \ref{sec:notions}, we can define magnetic and electrical topological charges here. In this case, these correspond to the anyon charges $O_{h} =-1$
on alternating horizontal rows of hexagonal plaquettes of the honeycomb lattice. It is possible to change
a magnetic vortex into an electrical one by the creation/absorption of a fermion $\epsilon$.
Denoting by $m$ and $e$ magnetic and electrical charges and by $I$ the vacuum ($O_{h}=1$),
there are, in this (Abelian) phase 
fusion rules identical to Eqs. (\ref{eeI}, \ref{fusion}). 

\subsubsection{Ground States Properties in Vortex-full Sectors}
\label{vfl}
As briefly alluded to earlier, the fusion rules of Eq. (\ref{fus_nab}) 
and their consequences may be understood by examining the 
system in the presence of vortices. The calculations within
the vortex full sectors proceed along the lines identical of subsections \ref{solnus},\ref{nonabp}.
In the bare Kitaev model Hamiltonian of Eq. (\ref{eq:HKT}) [which may be augmented by  
the perturbation of Eq. (\ref{eq:Kitaev_h})], we set $\eta_{r}$ on the vertical links to be such that the vorticity of Eq. (\ref{eq:O_bond})
satisfies $O_{h} = \eta_{r} \eta_{r'}$ where $\eta_r$ and $\eta_r'$ denote the Ising degrees of freedom on 
the two vertical links of the plaquette $h$ of the original hexagonal lattice. As we reviewed earlier, the resulting Fermi form [also
in the presence of the perturbation of Eq (\ref{hhh})] is a bi-linear with on-site and nearest neighbor (resulting
from the Hamiltonian of Eq. (\ref{eq:HKT})) and next nearest neighbor (resulting from Eq. (\ref{hhh}) 
(hopping and pairing) terms on the square lattice which may be exactly solved. 
The Fermi spectrum may, specifically, be exactly determined within each of the $2^{N/2}$ sectors (i.e., 
values $O_{h} = \pm 1$ on each of the $N_{h}=N/2$ hexagons). Practically, it is possible to 
exactly (numerically) solve for the spectrum of the system in simple insightful sectors. 
These include a system with a single pair of vortices, a superlattice of uniformly spaced vortices,
and other intermediate and other related regimes. General trends are seen and we elaborate
on those below. 

{\underline{{\it{Vortex pairs: oscillations, midgap states, and fusion:}}}} 
We first briefly review several key aspects of a system with a single pair of vortices \cite{Lahtinen2011a}
in the presence of a field (Eq. (\ref{hhh})).  When two non-Abelian anyons are held at a finite distance 
$d_{s}$ apart and the system eigenstates are computed, it is found that in addition to the gapped fermionic
modes there appear two additional modes. The continuous particle-hole symmetric spectrum 
is augmented by an additional pair of energies \cite{Lahtinen2011a}
\begin{eqnarray}
\pm \Upsilon^{d_{s}} =  \pm \Delta_{f} \cos(2 \pi d_{s}/\lambda) e^{-d_{s}/\xi}.
\label{lahit}
\end{eqnarray}
These new vortex states of low energy states (i.e., low $\Upsilon$) augment the continuous bands of higher energy Fermi
bands (of energies $\pm E$, ~ $E> \Upsilon^{d_{s}}$) that appear here similarly to the case
of the vortex free lattice (Eq. (\ref{energy_eqq})).
As the separation $d_{s} \to \infty$, these additional modes veer towards zero energy and become {\it midgap states}
between the positive and negative fermionic energy bands $ \pm E_{q}$ of the vortex free system.
By continuously changing the separation $d_{s}$ to zero, \cite{Lahtinen2011a} elegantly illustrated how
the Majorana Fermi fusion mode $\epsilon$ of Eq. (\ref{fus_nab}) is explicitly realized. 
The gap $\Delta_{f}$ of Eq. (\ref{lahit}) sets the minimal energy of these fermionic excitations. 
As is manifest in Eq. (\ref{lahit}), at 
finite separation $d_{S}$, the virtual exchange of these Fermi fusion modes lifts the degeneracy.
The size of the gap $\Delta_{f}$ is an effective confining barrier height that needs to be surmounted
for the fermionic fusion modes to become delocalized \cite{Lahtinen2011a}.  Thus, the picture
that emerges at finite inter-vortex separation $d_{s}$ is 
that of Majorana fusion modes that are bound to the vortices. The modulation length
$\lambda$ in Eq. (\ref{lahit}) is set by the reciprocal of the difference between the wave vectors (${\bf{q}}^{(\pm)}$)
that minimize the energy of Eq. (\ref{energy_eqq}),  \cite{Lahtinen2011a}
\begin{eqnarray}
\lambda \simeq \frac{2 \pi}{|
{\bf{q}}^{(+)}
- 
{\bf{q}}^{(-)}|}.
\end{eqnarray} 
Similar oscillatory decay appears in other systems that harbor Majorana excitations \cite{Cheng2009,Baraban2009}.
More broadly, effective interactions on non-uniform sign such as those of Eq. (\ref{lahit}) also appear in numerous classical
systems including RKKY interactions in spin glasses, theories of structural glasses, and other systems with competing interactions  \cite{Yosida, Kasuya,Ruderman,Chakrabarty2011,Nussinov4,Nussinov1999,Tarjus2005}. 
In a many particle system, such an oscillatory character of the effective pair interactions may 
reflect frustration and consequently lead to multiple inhomogeneous states 
and viable slow dynamics. 

{\underline{{\it{Multiple vortices and vortex lattices}}}}
The understanding of the single vortex pair problem and 
the bound states that form therein allows a qualitative understanding
of the $n$ vortex pair problem. For well separated vortices, there
is a $2^{n}$ fold degeneracy of the ground state. Majorana fermion modes may
tunnel and realize the mode $\epsilon$ associated with the
fusion channel of Eq. (\ref{fus_nab}). Vortices partially bind
 Majorana modes and lead to a lifting of the degeneracy when 
the inter-vortex separation is finite and lead to a localization
of the Majorana zero modes. 

As the number of vortex pairs $n$ is increased yet is far smaller that the number
of sites on the lattice, additional (fusion borne) midgap states appear that are well separated from 
the continuum of modes at positive and negative energies \cite{Lahtinen2011a,Lahtinen2011b}. 
The wave-functions corresponding to these midgap states preferentially have their spatial support
localized on the vortices. By contrast, the real space wave-functions corresponding to 
the fermionic states at the continuum of energies $\pm E$ are uniformly delocalized on the lattice. 
We remark that the physics that emerges as the number of vortices is increased 
is qualitatively similar to that in other systems (e.g., doped semiconductors (or superconductors)) 
with an increased number of dopants that modify the local hopping (and pairing) 
amplitudes. In lattices with a dense set of vortices, 
the new Majorana states that appear (vis a vis those in the vortex free problem) may lead 
continuous bands. These new modes may overlap with those borne out of
the continuous modes \cite{Lahtinen2011a,Lahtinen2011b}. Interactions
may hybridize low-energy bands from the multitude of these new
fusion borne modes. We remark that when the vortices are arranged in a periodic
lattice, the fermionic system is periodic and should
exhibit multiple momentum space bands (which may be regarded as a hybridization of
the individual fusion type modes) associated with an enlarged unit cell (set by the 
period of the vortex lattice). The emergence
and evolution of these new bands as a function of vortex lattice spacing 
was carefully studied in \cite{Lahtinen2011b}.  It is possible to account for the 
Chern number and other characteristics of the different phases as seeing
from as originating from the sum of two different contributions:
(i) continuum Fermi bands and the (ii) new Majorana modes 
that appear in the vortex lattice and may be approximately
viewed as hopping between the vortices (to which they
are preferentially bound)- this may be emulated by a kinetic
term with complex short range hopping amplitudes between Majorana fermions
on the vortex lattice sites \cite{Lahtinen2011b}. The extreme case of the
fully dense vortex lattice (i.e., a vortex on every plaquette) was investigated
in \cite{Lahtinen2010}. The Fermi surface associated with this theory
undergoes a quantum phase transition from the usual 
non-Abelian phase (with Chern number $\nu =-1$) with a single pair of continuous fermionic 
modes to an Abelian chiral phase (with Chern number $\nu =- 2$) that supports the above 
mentioned new additional bands \cite{Kitaev06,Lahtinen2010}. As re-iterated above, these new bands 
in vortex lattices (including the maximal dense vortex lattice) 
may be seen as brought about by the merger
of the Majorana fusion modes that appear in
the dilute vortex pair problem.

\subsection{Classical Ground States \& Dimer Coverings}
\label{classkit}
Similar to the compass models that we studied in earlier sections, fluctuations (both zero-point quantum fluctuations as well
as (classical) thermal fluctuations) can stabilize order in Kitaev's honeycomb model.

Following \cite{Baskaran08}
we review features of the large $S$ rendition of the Kitaev model. 
Similar to the large $S$ extension of the 120$^{\circ}$ and 90$^{\circ}$ compass
models \cite{Nussinov04, Biskup05} in Eq. \ref{H120} and the discussion in 
Section \ref{sec:diso}, we may replace each spin operator in Eq. \ref{eq:HKT}
by a classical vector to obtain the classical ($S \to \infty$) limit. 
We may similarly define the Kitaev model in this fashion
for arbitrary $S$ \cite{Baskaran08}. 

Building on the general discussion of classical ground states
in Section(\ref{sec:diso}),
we note that employing the same
counting arguments that we invoked earlier
shows that there are, on the honeycomb lattice,
$(3N/2)$ independent conditions stemming
from Eq. (\ref{TTGS}). On the other hand, $N$ classical
three component pseudo-spins have $(2N)$ degree of
freedom (i.e., two angles specify the orientation
of a pseudo-spin on the unit sphere for each of
the $N$ pseudo-spins). Thus, by naive counting
there remain $(N/2)$ redundant degrees of 
freedom suggesting a local emergent gauge
symmetry. (This number is equal to the number
of hexagons: $N_{h} = N/2$.)

The system has a rich structure at its symmetric point: $J_{x}=J_{y}=J_{z}$ in
the classical rendition of Eq. \ref{eq:HKT}. Similar to the classical
compass models discussed in Section \ref{sec:diso} \cite{Nussinov04,Biskup05} there is an
{\em emergent continuous symmetry} in the ground state
sector. To elucidate these, we note that the honeycomb is bipartite-- it
is possible to partition of the sites of the lattice into two 
sublattices (say, A and B) such that points in the A sublattices have
all of their neighbors in the B sublattice (and vice versa). 
With this in mind, we note that we may generate ground state if
we orient {\em all} of the spins
in the A sublattice along the same direction and then orienting
all of the spins on the B sublattice along the opposite direction. \cite{Baskaran08}

The above set of ground states harboring emergent symmetries is augmented
by a discrete set of states. \cite{Baskaran08} In these, we set two spins at the endpoints
of a link to be anti-parallel to each other and be oriented along the spatial direction 
of the link. For instance, we may set the two spins along a link that is parallel to the spatial $x$ direction
to be anti-parallel to each other and point along the $\hat{e}_{x}$ and $(-\hat{e}_{x}$) directions. 
These ground states thus reduce to dimer coverings and are of a similar form
to the suggested low energy 
states that we encountered in the 120$^{\circ}$ model on the honeycomb lattice \cite{Nasu08}.
[See section \ref{120qhn}.]
The problem of enumerating these discrete states may be mapped onto that of examining dimer coverings of the lattice.
That is, we may ask in how many ways we can place dimers on the links of the lattice.
Once a dimer is placed, we orient the spins at the endpoints to be parallel/anti-parallel 
to the spatial direction of the dimer. The number of dimer coverings of the 
honeycomb lattice with N sites is equal to 
\begin{eqnarray}
{\cal{N}}_{dimer} \simeq 1.175^{N}.
\end{eqnarray}
If we take into account the possibility of placing the spins parallel to anti-parallel to the spatial direction of the link, we find that the total number of discrete classical ground states is 
\begin{eqnarray}
{\cal{N}}_{discrete~classical} = {\cal{N}}_{dimer} \times 2^{N/2} \simeq 1.662^{N}.
\end{eqnarray}

\subsubsection{Symmetries of Spin-S Systems}
\label{symmetries_SSK}
For general (pseudo-spin) $S$ renditions of Kitaev's honeycomb model, 
the conserved $S=1/2$ quantities of Eq. (\ref{eq:symkit6}) 
are replaced by \cite{Baskaran08}
\begin{eqnarray}
O_{h} = \prod_{j=1}^{6} e^{i \pi T_{j}^{a_{j}}}
\end{eqnarray}
where $a_{j}$ denotes the direction of an outward ``spoke'' at site $j$ that is normal to the
hexagon $h$ and $T_{j}$ is the $j$th pseudo-spin compoent.  Similar to the (pseudo-)spin one half case, generalizations of
Eqs. (\ref{Oc},\ref{stokes}) associated with any closed contour $\Gamma$ are conserved quantities
(symmetries of the model). Rather explicitly,
these are \cite{Baskaran08}
\begin{eqnarray}
O_{\Gamma} = \prod_{j \in \Gamma} e^{i \pi T_{j}^{a_{j}}}.
\end{eqnarray}

\subsubsection{Spin-wave Expansion}
The degeneracy of the classical ground states described in Section \ref{classkit} is partially 
lifted by quantum 
fluctuations as manifest in $(1/S)$ corrections to the energy. For a one-dimensional version of the 
general spin S Kitaev model, it was found that incorporating ($1/S)$ corrections made 
the discrete set of classical ground states lower in  
energy than their continuous counterparts. \cite{Baskaran08}.

With each dimer covering that represents a discrete classical ground state,
there is an associated set of self avoiding walks. These walks are formed by
walks such that no link that lies on the self avoiding walk is a dimer
that represents the spin orientation in the ground state. 

\begin{figure}
\centering
\includegraphics[width=.4\columnwidth]{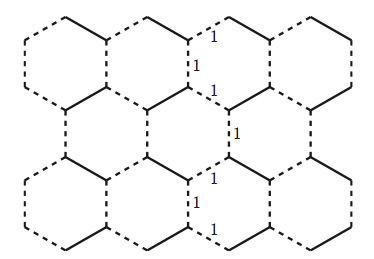}
\includegraphics[width=.4\columnwidth]{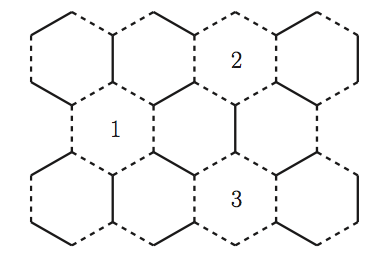}
\caption{Mappings between dimer configurations (solid lines) and self-avoiding random walks (dashed lines), see text.~\cite{Baskaran08}.}
\label{fig:baskaran_dimer}
\end{figure}

\subsubsection{Quantum Ground States and Dimer Coverings}
\label{qgsd}
In a very interesting work, \cite{Nash2009} illustrated that,
in the absence of vortices, the zero temperature phases of the (pseudo-spin 1/2) Kitaev model
on general lattices are in a one to one correspondence with those 
of classical dimer coverings \cite{Fisher1966,Friedland2008} 
on the same lattice. This result may be viewed
as complimentary to the results of \cite{Baskaran08} as they pertain
to the classical, large $S$, Kitaev model on the honeycomb lattice.
[See, subsection \ref{classkit}.]  On sufficiently complicated lattice realizations of 
the pseudo-spin 1/2 Kitaev mode
wherein the couplings are orientation dependent and furthermore 
are varied to become spatially non-uniform e.g., a honeycomb lattice with 
an eighteen site basic periodic
block, a new gapped (``C'') phase was seen 
to emerge inside the ``B'' phase.  

Viable lower energy states for general pseudo-spin $S$ were provided by \cite{Mandal2010}. 
These are constructed out of superpositions of the product states. The idea underlying these
states is elegant and we briefly review it below within the general 
approach employed in this review. The honeycomb lattice can be decomposed
into disjoint hexagons and a set of links between these
on which the different hexagons interact: $\sum_{hex} H_{hex} + \sum_{int} H_{int}$ \cite{Mandal2010}.
The interactions around a hexagonal plaquette (those in $H_{hex}$) satisfy,
obviously, the bond algebra earlier discussed  Eq. (\ref{eq:longbonds}). For a single hexagon,
the algebra is that of bonds around a six site chain that anticommute if they are nearest neighbors,
commute otherwise, and always square to 1. Thus, by a trivial change of basis , all bonds
may be made to be of the xx or yy type. 
The resulting six site Hamiltonian is then 
\begin{eqnarray}
H_{hex} = - J(T_{1}^{x} T_{2}^{x} + T_{2}^{y} T_{3}^{y} + T_{3}^{x} T_{4}^{x} \nonumber
\\ + T_{4}^{y} T_{5}^{y} + T_{5}^{x} T_{6}^{x} + T_{6}^{y} T_{1}^{y}).
\end{eqnarray}
The classical ground state of this Hamiltonian would be the uniform state wherein all spins
are polarized along some direction $\phi$ in the xy plane. \cite{Mandal2010}
then consider the quantum product state analogue of the corresponding coherent states
at the six sites $|\phi ... \phi \rangle = | \phi \rangle_{1} \otimes ... |\phi \rangle_{6}$.
The spin state at site $i$ can be expressed as $|\phi\rangle_{j}  = (\frac{1}{2})^{S} \exp[i \phi T_{j}^{-}]
|T_{z}=S \rangle_{j}$ with $T_j^{-} = (T_{j}^{x} - i T_{j}^{y})$.  A candidate product state over the entire
system of all hexagons is then suggested to be $ \otimes |\psi \rangle_{h}$
with, at each hexagon $h$, 
\begin{eqnarray}
|\psi_{h} \rangle = \frac{1}{2 \pi} \int_{0}^{2 \pi} e^{-6S \phi i} |\phi \phi ... \phi \rangle.
\end{eqnarray}
Computing the energy of this trial state with the full Hamiltonian (involving also the interactions
between hexagons) leads to reasonable results. This procedure and be improved in a variety
of ways (e.g., employing exact quantum ground states on large units and optimizing with respect
to the interactions between those units). 

\subsubsection{Bounds on Ground State Energies}

In an elegant work \cite{Mandal2010}, exact bounds on the ground state energies of the
pseudo-spin $S$ rendition of Kitaev's honeycomb model were provided
for the system at its symmetric point ($J_{x}=J_{y}=J_{z}$) in the absence of 
perturbations.  A lower bound for the energy was obtained by determining the minimal 
possible energy for a small unit of a site on one sublattice with all of its (three) nearest neighbors
(on the other sublattice); the Hamiltonian can be written as a sum of the interactions
found in these chosen units (or bonds). Such a decomposition has 
afforded bounds on ground state energies of,
e.g., bipartite quantum antiferromagnets \cite{Anderson1951,Carlson1998}. 
Upper bounds on the ground state energy were obtained via a variational argument
using a product of wave-functions on two-sites (extensions include larger size units as discussed
in subsection \ref{qgsd} above).  Putting all of the pieces together, the ground state energy was found to satisfy inequalities for all $S$ systems. In the $S \to \infty$ limit, these read
\begin{eqnarray}
-\frac{1}{2} + \frac{1}{4S} \le \frac{E_{0}}{JN} \le - \frac{1}{2} + 0.374 \frac{1}{S}
\label{bounds_S_E}
\end{eqnarray}
A related yet slightly different lower upper quadratic estimate was obtained
by \cite{Baskaran08} wherein the $1/S$ correction in the bound of 
Eq. (\ref{bounds_S_E}) was replaced by $(0.289/S)$.

\subsubsection{Classical Order out of Disorder}

The authors of \cite{Mandal2010} report that 
unlike most other compass models, Kitaev's model
in its classical version does not exhibit a finite temperature order out of disorder
effect and that quantum fluctuations are required to attain order. 
This conclusion relates to the general character of
quantum and classical order out of disorder mechanisms, see section \ref{dcq}.

\subsection{Toric Code Model}
\label{corrkk}

\subsubsection{Relation between Kitaev Compass \& Toric Code Models}

In its low energy sector, the gapped phase of Kitaev's model of Eq. \ref{eq:HKT} is {\em adiabatically connected} to Kitaev's toric code model which we will study in subsection \ref{KTCM}.  In the limit  of extreme 
anisotropy (e.g., $|J_{z}|\gg |J_{x,y}|$), 
Kitaev's honeycomb model of Eq. (\ref{eq:HKT})
reduces to the Toric code model of Eqs. 
(\ref{kitaevmodel},\ref{AB_defn}) \cite{Kitaev06}. 
In this limit, all of the bonds along the $z$ direction are 
strongly correlated and effectively the strong correlated spins 
along the two sides of a given $z$ bond (a ``vertical dimer'') may be replaced
by one spin. Geometrically, the replacement of vertical dimers of
the honeycomb lattice by singlet sites is
similar to what we have earlier encountered in the solution of
Kitaev's honeycomb model by replacing
it by a fermionic system on a square lattice
[Eq.(\ref{cdrep})]. Analogously, the resulting geometry 
in the case of extreme anisotropy  (e.g., $|J_{z}|\gg |J_{x,y}|$) is that 
of the square lattice of the Kitaev Toric code model
on the square lattice which we will next discuss. Within the low energy subspace
of Kitaev's honeycomb model, the Toric model below 
results from the lowest non-trivial order  (a fourth order contribution) in the perturbative expansion in
$(J_{x,y}/J_{z})$ about the $J_{z} \to \infty$ limit of decoupled vertical dimers.  
When Kitaev's honeycomb model is not projected onto the low
energy subspace (that of no fermionic excitations), 
there are additional terms that augment the correspondence to 
Kitaev's toric code model \cite{Dusuel08}.
As we will discuss in the section \ref{KTCM}, Kitaev's toric code 
model exhibits an Abelian (``$Z_{2} \times Z_{2}$
type'') statistics identical to that of Eqs. (\ref{eeI},\ref{fusion}). 
As the gapped phase of Kitaev's honeycomb model
is adiabatically linked to the toric code model, the honeycomb model
also exhibits, everywhere within its gapped phase, this $Z_{2} \times Z_{2}$ symmetry.

\subsubsection{Kitaev's Toric Code model: Definition, Symmetries \& Ground States}
\label{KTCM}

Kitaev's Toric code model \cite{Kitaev03} in $D=2$ spatial dimensions 
is defined on a square lattice with $L \times L=N_s$ sites, where on each
link $\langle ij\rangle$ there is a $S=1/2$ degree of freedom, see Fig. \ref{fig:honeycomb}(B). The Hamiltonian of this model is given by
\begin{eqnarray}
H_{K} = -\sum_{s} A_{s} -\sum_{p} B_{p}\;,
\label{kitaevmodel}
\end{eqnarray}
with the operators 
\begin{eqnarray}
A_{s} = \prod_{\langle ij\rangle\in {\sf star}(s)} \sigma_{ij}^{x},\hskip 1cm
B_{p} = \prod_{\langle ij\rangle\in {\sf plaquette}(p)} \sigma_{ij}^{z} ,
\label{AB_defn}
\end{eqnarray}
and $\sigma_{ij}^{\kappa}$ ($\kappa=x,y,z$) representing Pauli
matrices. $B_{p}$ and $A_{s}$ describe the plaquette (or face) and star
(or vertex) operators, respectively, with ($\forall s,s',p,p'$)
\begin{eqnarray}
[A_s,A_{s'}]=[B_p,B_{p'}]=[A_s,B_p]=0 ,
\end{eqnarray} 
generating an Abelian group which is known as the {\it code's stabilizer\/}
\cite{Kitaev03},
and the two $d=1$ (Ising type (${\mathbb Z}_{2}$)) symmetries 
are given by \cite{Kitaev03}
\begin{eqnarray}
Z_{1,2} = \prod_{\langle ij\rangle\in C_{1,2}} \sigma_{ij}^{z}, \hskip 1cm
X_{1,2} = \prod_{\langle ij\rangle\in C^{\prime}_{1,2}} \sigma^{x}_{ij},  
\nonumber \\ 
\{X_{\mu}, Z_{\mu} \} = 0 \ ,\hskip .5cm 
 [X_{\mu}, Z_{\nu}]   = 0 \ ,\hskip .5cm \ \mu\neq\nu ,
\label{kit}
\end{eqnarray}
where $C_1(C_2')$ are horizontal and $C_2(C_1')$ vertical closed
contours [i.e., loops on the lattice (dual lattice)] that span an entire cycle of the torus. The non-commutativity
of the $d=1$ symmetry operators $X_{\mu}$ and $Z_{\mu}$
implies that the ground state sector is degenerate. It is
on these ground state basis states, that topological
quantum computing schemes were devised. The logical
operators $Z_{1,2}$ and $X_{1,2}$ commute with the code's stabilizer
but are not part of it, thus acting non-trivially on the two 
{\it encoded\/} Toric code qubits. 

The ground states of the toric code model are trivial and embody a very beautiful
topological structure akin to that found in some other systems with exactly solvable ground
states such as the Quantum Dimer Model \cite{RK,RK1}  that may be fleshed out
by examining loop coverings. As these states are extremely easy to ascertain we will do so below. First, we note that within the ground state of Eqs. \ref{kitaevmodel},\ref{AB_defn}, all the plaquette and star operators 
may be set to $A_{s}= B_{p}=1$.  The ground states can also be characterized by the two
additional commuting quantum numbers $Z_{1,2}$ (or, similarly, by $X_{1,2}$). 
To bring to the fore standard loop representations of such states,
we investigate Kitaev's toric code model in the $|\sigma^{x} \rangle$ 
basis of all spins and denote each ``right pointing''
spin $\sigma_{si}^{x}=+1$ by a solid line on the edge $(si)$.
The condition $A_{s}=1$ at all lattice sites $s$
will hold if and only if at all such 
lattice sites, the solid lines composed of
these links form (any number of continuous) 
closed loops. This also allows solid loops to
share common sites. This is so as for any continuous
solid loop, the number of edges formed by putting four ``spokes''
at lattice sites, is even within any 
plaquette (either none, two, or four). 
Each of these ``loop states'' $|\phi \rangle$ is a ground state of
the first term in Eq. \ref{kitaevmodel}. In 
order to form simultaneous eigenstates
of the second (plaquette term) in Eq. \ref{kitaevmodel},
we need to form a linear combination $| \psi \rangle$ of these
``loop states'' such that we have that for all plaquettes $p$, 
$B_{p}|\psi \rangle = | \psi \rangle$. That, however, is easy to achieve
by a projection onto the sector of $B_{p} =1$ for all plaquettes $p$
(for any plaquette $p$, the plaquette operator $B_{p}$ has two eigenvalues (which are
$\pm 1)$). That is, if we set 
\begin{eqnarray} 
| \psi \rangle = \Big( \prod_{p} [\frac{1}{2} (1+ B_{p})] \Big) | \phi \rangle,
\label{ksgs}
\end{eqnarray}
where the product in Eq. \ref{ksgs} is over all plaquettes $p$, 
then $| \psi \rangle$ will constitute a projected state of $| \phi \rangle$
where, within each plaquette $p$, we have that 
$B_{p} | \psi \rangle = | \psi \rangle$. 
Now, the action of $B_{p}$ on any loop state is trivial-
$B_{p}$ flips the location of the solid state within plaquette $p$.
To see this, we recall that on any link $(ij)$, we have that 
$\sigma^{z}_{ij} \sigma^{x}_{ij} \sigma^{z}_{ij} = - \sigma^{x}_{ij}$
and that $B_{p}$ is defined by Eq. \ref{AB_defn}. Together, these two
relations imply that within any plaquette $p$, for all solid links $(ij)$ that
initially corresponded $\sigma^{x}_{ij} = +1$ in the state $|\phi \rangle$,
$B_{p} | \phi \rangle$ leads to a state that within the plaquette
$p$ has the values of $\sigma^{x}$ reversed. That
is, for all $(ij) \in p$, we have that $\sigma^{x}_{ij} \to - \sigma^{x}_{ij}$.  
Thus, in the loop representation, the solid lines (denoting those
links for which $\sigma^{x}_{ij} =1$) invert their location:
all links that were not part of a solid loop in the plaquette $p$,
become part of the loop and vice versa. Thus the operators $B_{p}$
create and reconnect loops. From Eq. \ref{ksgs}, we see
that the terms $``1''$ and $``B_{p}''$ have equal amplitudes in the product
over all plaquettes $p$. This implies that we have a very simple algorithm for
generating ground states. First we focus on any loop state $|\phi \rangle$
and then we create an equal amplitude superposition of all loop states formed
by flipping the links in {\em all subsets} of plaquettes $p$ that appear
in the lattice- this corresponds to expanding the product in Eq. \ref{ksgs} to
contain all terms: $\{1, \{B_{p}\}, \{B_{p} B_{p'} \}, ... \}$. The local loop reconnection operators
$B_{p}$ cannot link states that belong to different topological sectors.
As, by Eqs. (\ref{AB_defn},\ref{kit}),  $[X_{1,2}, B_{p}]=0$ 
(and, similarly, $[Z_{1,2}, A_{s}]= [Z_{1,2},B_{p}]=[Z_{1,2},A_{s}]=0$),
the $B_{p}$ can only connect state that lie in the same topological sector. There
are four such topological sectors $(X_{1} = \pm 1, X_{2} = \pm 1)$. Thus, there
are four independent ground states. These four ground states embody
the ${\mathbb Z}_{2}\times {\mathbb Z}_{2}$ symmetries of Eqs. (\ref{kit})). 
This number of ground states may also 
be arrived at by noting that there are, on a torus of size $L \times L$, 
$(2L^{2})$ spin 1/2 degrees of freedom sitting on each link.
The ground state are specified by the conditions $A_{s}=1$ for all $s$
and $B_{p}=1$ for all plaquettes $p$. There are $L^{2}$ sites $s$ and $L^{2}$ plaquettes
$p$. On a torus, however, due to periodic boundary conditions, there are two additional
constraints of the form of $\prod_{s} A_{s} = \prod_{p} B_{p} =1$. 
These two constraints lead to $(L^{2}-1)$ independent star operators $A_{s}$
and a similar number of independent plaquette operators $B_{p}$.  
Thus, the conditions $A_{s}=1$ and $B_{p}=1$ encompass $(2L^{2}-2)$ independent ($S=1/2$) degrees
of freedom. As there $(2L^{2})$ spin-1/2 operators in the entire lattice, there are $2$ independent 
$S=1/2$ degrees of freedom left (these correspond to the two degrees of freedom spanned by 
the topological operators $X_{1}$ and $X_{2}$) or four ground states. 
The operators $X_{1,2}$ have a very simple geometrical interpretation:
they monitor the even/odd parity of the number of times
loops cross a given cycle ($C_{1}^{\prime}$ or $C_{2}^{\prime}$) of the torus. 
The same considerations apply for the $Z_{1,2}$ operators. 
The four ground states of the Kitaev toric model on the simple torus are
the equal amplitude superpositions of all states (in the
$\sigma^{z}_{i}$ product basis)
that have $B_{p} =1$ and lie in each of the 4 sectors
$Z_{1} =  \pm 1$ and $Z_{2} = \pm 1$.
On a general manifold having $g$ handles (the ``genus'' number)- instead
of the simple case of a torus with $g=1$, similar
considerations lead to operators of the form of Eq.(\ref{kit})
associated with each of the non-trivial cycles of the system $C_{1}, ..., C_{2g}$ (and $C_{1}^{\prime}, ..., C_{2g}^{\prime}$).  
The eigenvalues of $Z_{1}, Z_{2}, ..., Z_{2g}$ (or of $X_{1}, ..., X_{2g}$) may label the ground states.
In the aftermath, it is seen that there are $4^{g}$ independent ground states that
correspond to equal amplitudes superpositions of loops
within each of the $4^{g}$ topological sectors. This dependence
of the degeneracy solely on the topology of the system (and not on any
local property) exemplifies
the topological character of the ground states 
and was often stated to reflect the {\em topological order} that is present in this
system. On a sphere (and all other simply connected manifolds ($g=0$)), any loop operator may be contracted
to a single point; as there are no non-trivial loops on such manifolds, all loop products may be expressed in terms
of plaquette (or star) loops. As $A_{s}=B_{p}=1$ within the ground state and there are no additional
symmetries (of the form of Eq. (\ref{kit})) there is no non-trivial topological degeneracy.

Similar to ``spin-liquids'', these equal amplitude superpositions of closed loops on non-trivial manifolds
are topological- it is impossible between the four ground states by making 
any local measurements. Rather, the toric codes states are characterized by topological quantities
such as the eigenvalues of the $d=1$ symmetry operators $X_{1}$ and $X_2$ around closed loops of the torus. 
The ground states are formed by equal amplitude superpositions of all closed loops
that belong to a given topological sector (characterized, in this case, by a set of eigenvalues
of $X_1$ and $X_2$ that denote the parity of the loop covering).
An exactly analogous situation occurs in the Quantum Dimer Model \cite{RK,RK1,PNAS}.

\subsubsection{Excitations in Kitaev's Toric Code Model}
\label{EKTCM}
The elementary excitations of $H_K$ are of two types \cite{Kitaev03}
\begin{eqnarray}
\ket{\Psi_z(\Gamma)}&=&\prod_{\langle ij\rangle\in\Gamma} \sigma_{ij}^z 
\ket{\Psi_0}\equiv S^{z}(\Gamma) \ket{\Psi_0}, \nonumber \\  
\ket{\Psi_x(\Gamma')}&=&\prod_{\langle ij\rangle\in\Gamma'} \sigma_{ij}^x 
\ket{\Psi_0}\equiv S^{x}(\Gamma') \ket{\Psi_0} ,
\label{QQbar}
\end{eqnarray}
where $\Gamma(\Gamma')$ is an open string on the lattice (dual lattice)
and $\ket{\Psi_0}$ is a ground state. [If $\Gamma(\Gamma')$ denotes
closed contours that circumscribe an entire Toric cycle then the string
operators $S^{x,z}$ will become the Toric symmetries of Eq.
(\ref{kit})].  In the case of the open contours of Eq. (\ref{QQbar}),
the operators $S^{x,z}$ generate excitations at the end points of
these strings (thus always coming in pairs) with abelian fractional
statistics (anyons). Excitations living on the vertices represent {\it
electric charges} while the ones living on the plaquettes are {\it
magnetic vortices}. These {\em magnetic} and {\em electric} type excitations obey
the fusion rules of Eqs. (\ref{eeI}, \ref{fusion}). 
We elaborate on these below.

There are many degenerate low energy excitations (hence the lack of robustness
to thermal excitations as we will later describe). The excitations may be viewed as
being of two types: those corresponding to ``faulty'' stars: $A_{s'} =-1$ 
(electric charges $e$) and to faulty
plaquettes: $B_{p'}=-1$ (magnetic charges $m$). 

We focus first on faulty stars. In order to have $A_{s'}=-1$ within the loop representation,
we see that we must have {\em an odd number} of links emanating from the site $s'$ for which 
$\sigma_{is'}^{x} = +1$. Geometrically, this corresponds to ``broken bonds'' \cite{Sachdev09} 
for which the solid line of bonds starts/terminates at the site $s'$. Thus, ``star defects'' correspond to
sources/sinks of sold lines that supplant the closed loops of solid lines that appear within the ground
states. In the Kitaev model representation, these sources of lines correspond to ``electric charges''.
It is notable that as a state with $A_{s'}=-1$ at a particular site $s'$ is an eigenstate of the Hamiltonian of 
Eq. \ref{kitaevmodel} (all states can be written in the complete orthogonal eigen-basis spanned by 
$| \{A_{s}\}, \{B_{p}\}, \{X_{1,2}\} \rangle$). 
Thus, such a defect corresponds to a quantum mechanical stationary state. In order words, formally, we can think
of a defect with $A_{s}=-1$ as an infinitely massive object that does not move. 

In a similar manner, we can analyze ``plaquette defects'' with $B_{p'}=-1$. 
A state with a single defect at plaquette $p'$ is given by 
\begin{equation}
| \psi' \rangle = 
\Big[ \prod_{p \neq p'} \Big(\frac{1+ B_{p}}{2} \Big) \Big] 
\frac{1 - B_{p'}}{2} | \phi \rangle.
\label{plaquette_defect_state}
\end{equation}
In Eq. \ref{plaquette_defect_state}, one still creates an equal amplitude superposition of loop states 
for all states related by the application of the plaquette inversion operator $B_{p \neq p'}$ as 
in Eq. \ref{ksgs}. However, loops related by reconnection on the plaquette $p'$, lead to
an opposite sign superposition in Eq. \ref{plaquette_defect_state}. This relative change
of phase (of $(-1)$) in the amplitude of the superposed loop states can be thought
of as that stemming from an Aharonov-Bohm (AB) phase associated with half a fluxon.
Thus, plaquette defects can be viewed as magnetic charges. Similar to the
electric charges, magnetic charges are also infinitely heavy
as states with plaquette defects are eigenstates of the Hamiltonian.

When an electric charge (a star defect) encircles the half fluxon (the magnetic
charge), the associated AB phase factor is that of $(-1)$.
Thus, the relative statistics of the two types of defects relative
to one another is semionic. Longhand, the state excited state with an
electric charge defect (and a magnetic defect) is the sum of loops with
one end loop at the location of the electric charge. Moving the electric
charge, on its own, around the magnetic defect leads to a phase factor of
$(-1)$. This phase factor multiplies the entire state in which all of the other closed
loops suffer no change in the phase factor as a result of parallel transport
around the magnetic charge (the half fluxon associated with the plaquette defect). 

As already observed in 1884 by Heavyside \cite{Heavyside}, 
the electric and magnetic fields are dual to each other in vacuum electromagnetism. With the above definitions of 
electric and magnetic excitations in Kitaev's model in tow, an ``electric-magnetic''
duality also rears its head in Kitaev's toric code model. Specifically, one may replace
the original lattice by its dual lattice (whose vertices are the center of the plaquettes
of the original square lattice) and replace all $\sigma^x \leftrightarrow \sigma^z$.
Under such a transformation, on the dual lattice following the change of basis, 
all star operators $A_{s}$ turn into plaquette operators $B_{p}$ and vice versa, the closed loop
string operators $X \leftrightarrow Z$, and the electric and magnetic field excitations or 
string operators of Eq. (\ref{QQbar}) map onto each other ($S^{x}(\Gamma') \leftrightarrow S^{z}(\Gamma)$). 
Kitaev's toric code model is self-dual under this mapping (i.e., it maps onto itself).
In Kitaev's model and others, it has recently become popular to consider ``quantum double models'' in which
two dual replicas of the same system are considered in unison. Kitaev's model \cite{Kitaev03} may realize
such quantum double systems wherein there are as many magnetic and electric excitations 
when considering two such dual copies of the same system. There are extensive studies of boundary effects and generalizations of
the excitations and structures e.g., \cite{Buerschaper,Beigi2011} that appear in Kitaev's toric code model to related 
models with other groups. In such cases \cite{Buerschaper} moving dyons around
each other leads to a general unitary map. Generalizations of the basic ideas embodied within 
Kitaev's toric code model to systems with non-Abelian groups enable universal quantum computation, 
\cite{Bravyi2005,Nayak2008,Bravyi2006,Kauffman2004,Mochon2003,Freedman2002}.

\subsubsection{Relation between Vortex Excitations in Kitaev-Compass \& Toric Code Models}
\label{kkht}

As noted earlier, within the limit of large anisotropy (e.g., $|J_{z}| \gg |J_{x,y}|$), 
Kitaev's honeycomb model reduces to the toric code system \cite{Kitaev06}. 
In this limit,  two spins on each link along, e.g., the z direction are parallel (or anti-parallel) to one another and the hexagonal lattice reduces to a square lattice. As was just 
discussed in section(\ref{KTCM}), in Kitaev's toric code model,
there are two types of charges that correspond to faulty plaquettes
($B_{p}=-1$ -- magnetic charges $m$) and 
stars with $A_{s}=-1$ (electric charges $e$).
Following \cite{Kitaev06}, we next examine the (vertical dimer contraction) transformation that 
relates the honeycomb lattice to the square lattice. In doing so, it is seen \cite{Kitaev06} 
that magnetic ($m$) and electric charges ($e$) in Kitaev's toric code model on the
square lattice are none other than vortices $O_{h}=-1$ in Kitaev's
honeycomb lattices as they appear on alternating rows, see Fig. \ref{fig:em_vortices_K}.

\begin{figure}
\centering
\includegraphics[width=.6\columnwidth]{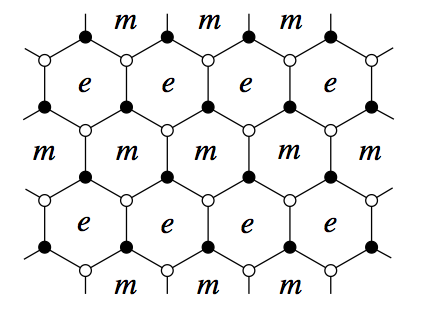}
\caption{Magnetic ($m$) and electric ($e$) charges live on alternating rows of hexagons~\cite{Kitaev06}.}
\label{fig:em_vortices_K}
\end{figure}

As the limit of extreme anisotropy is adiabatically connected, by continuously 
varying the couplings $J_{x,y,z}$,  to all points of the ``$A_{z}$ phase'' of Kitaev's model
(we elaborate on the designation of the $A_{z}$ phase below),
one may infer that the same identification of magnetic and electric
 charges holds for all other points in the $A_{z}$ phase.  
 We now briefly define the $A_{z}$ phase, see Fig. \ref{fig:Kitaev_phases}.
 
Earlier, following Kitaev, 
we denoted  the A phase as the one in which the couplings $J_{x}, J_{y}$ and $J_{z}$
did not satisfy the triangle inequality (and for which the system was found to 
gapped even in the absence of a perturbation (Eqs. (\ref{epD}, \ref{BCS-p})). 
This region can be further subdivided. 
The $A_{z}$ phase denotes that region of parameter space for which 
$|J_{z}| > |J_{x}| + |J_{y}|$ (i.e., the region of the $A$ phase in which the $J_{z}$ couplings
are most dominant).  

In a similar vein, replicating the above considerations to the cases in which $|J_{x}| \gg |J_{y,z}|$
(or $|J_{y}|  \gg |J_{x,z}|$) we will find that in the $A_{x}$ ($A_{y}$) phase of Kitaev's honeycomb model, vortices on alternating rows orthogonal to the $x$ (or $y$) direction will correspond to 
electric and magnetic charges. These charges obey the fusion rules of Eqs.(\ref{eeI}, \ref{fusion}).

We now turn to explicit forms for the operators that translate vortices in Kitaev's honeycomb model.
These operators explicitly enable the braiding operations of excitations around each another.  
In Kitaev's toric code model, there are two different types of vortex translation operators (related to the $
X$ and $
Z$ qubit operators).
These are given by Eq. (\ref{QQbar}).
\begin{eqnarray}
{\cal{T}}_{cx}^x = \prod_{ij \in cx} \sigma^x_{ij},
\label{tcx}
\end{eqnarray}
With $cx$ a contour that bisects edges (ij).

When acting on the ground state,
${\cal{T}}_{cx}^x$ generates two plaquettes p with negative vorticity,
$
B_p=-1$.
Similarly, when acting on a plaquette with a non-trivial vortex charge,
${\cal{T}}_{cx}^x$ shifts a vortex at plaquette $p_{1}$ to one at a plaquette $p_2$ if the plaquettes $p_1$ 
and $p_2$ lie at the endpoints of the contour $cx$ (have one side of the plaquette belonging to $cx$).
When $cx$ forms a closed loop along one of the toric cycles,
${\cal{T}}_{cx}^x$ become one of the two symmetry operators $
X_1$ or $
X_2$.

Similarly, one may construct another type of vortex translation operator given by
\begin{eqnarray}
{\cal{T}}_{cz}^z = \prod_{ij \in cz} \sigma^z_{ij}.
\label{tcz}
\end{eqnarray}
In eq.(\ref{tcz}), the edges (ij) lie on the contour $cz$.
The operators ${\cal{T}}_{cz}^z$ shift or create vortices of the "star" type:
$
A_s=-1$.

Analogous to the case of the operators 
${\cal{T}}_{cx}^x$, the operators of eq.(\ref{tcz}), become toric code symmetries when $cz$ forms a toric cycle. These are identified with $
Z_{1,2}$.

Depending on the direction of the contour $\Gamma$ in eq.(\ref{vortex-contour}), it may become one of the two types of Eqs.(\ref{tcx},\ref{tcz}).

For instance, if one considers a contour composed of xx and yy bonds on the hexagonal lattice then $
{\cal{T}}_{\Gamma}$ is a product of $\sigma^z$ operators along the contour. When we shrink each zz bond to a single point in the strong coupling limit, the resulting zig-zag chain of ${\cal{T}}_{\Gamma}$ 
is of the same type as that of eq.(\ref{tcz}).

Similarly, considering the same xx yy zig-zag contour c in the strong coupling limit of $
| J_x|\gg | J_{y,z}|$,
each xx type link reduces to a point and the geometry becomes akin to that of eq.(\ref{tcx}). 

\subsubsection{Abelian Braiding statistics}
\label{abst}

We now explain the braiding rules of Kitaev's toric code model and directly relate these
to the operator representations of the translation operators of Eqs. (\ref{QQbar}) 
[or Eqs.(\ref{tcx},\ref{tcz})]. One may introduce an electric charge at site $s$: $A_{s}=-1$
and then apply the translation of Eq. (\ref{QQbar}) [or Eq. (\ref{tcz})] to translate along a 
contour $\Gamma$ the encircles a magnetic vortex (one with $B_{p}=-1$). In doing this operation,
the translation operator becomes, similar to the Stokes' type relation of Eq. (\ref{stokes}),
a phase factor of $(-1)$. To see this, we note that 
\begin{eqnarray}
\prod_{ij \in \Gamma} \sigma^{z}_{ij} = \prod_{p \in {\cal{A}}} B_{p},
\label{stokes_kit}
\end{eqnarray}
where ${\cal{A}}$ is the region bounded by $\Gamma$. If there is a single magnetic
vortex ($B_{p}=-1$) that is in ${\cal{A}}$ then Eq. (\ref{stokes_kit}) reduces to a factor
of $(-1)$. We symbolically denote this result as in Fig. \ref{fig:braid_Kitaev_view}.

\begin{figure}
\centering
\includegraphics[width=.8\columnwidth]{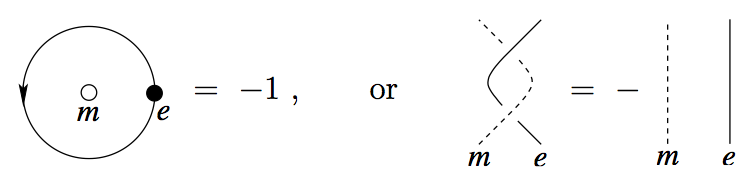}
\caption{A pictorial view of the braiding rules between electric ($e$) and magnetic ($m$) charges~\cite{Kitaev06}.}
\label{fig:braid_Kitaev_view}
\end{figure}

Thus, one may regard the two quasi-particles of an ``electric charge'' ($A_{s}=-1$)
and a ``magnetic charge'' as having a non-trivial statistics (leading to a factor of $(-1)$)
when one particle is encircled around the other. 

Along similar lines it is seen that taking an electric charge along a contour $\Gamma$ 
around another single electric
charge (with no magnetic particles present) will lead to a trivial phase factor of $(+1)$.
In this case,  as no magnetic particles are present, $B_{p}=1$
everywhere within the contour $\Gamma$ and by Eq. (\ref{stokes_kit}), 
there is a trivial accumulated phase factor. Thus, the electric particles behave as 
bosons relative to themselves. The same similarly holds for magnetic particles amongst
themselves. The fusion of an electric and magnetic particle- the dyon $\epsilon$ of Eq. (\ref{fusion})-
is a fermionic particle \cite{Kitaev06}. To see this, one may think about the motion of one dyon around the other. Naively, when considering the motion of one dyon (say, dyon number 1) around the other (dyon number 2)
we might, at first, anticipate to have two factors of $(-1)$ as one the electric charge of dyon number 1 encircles the magnetic charge of dyon number 2 and similarly the electric charge of dyon number 1 revolves around the magnetic charge of dyon number 2. However, it should be kept in mind that dyon number 1 is  rigid entity in which the magnetic
charge in that dyon is held at a fixed relative position ${\bf r}$ relative to the electric charge of dyon number1. 
When the internal structure of the dyon is taken into account it is seen that while revolving 
dyon number 1 in a closed loop around dyon number 2, the world line of the magnetic particle in dyon number 
1 winds around that of the 
electric particle of this dyon.

This additional internal winding leads to a third factor of $(-1)$
associated with the winding of dyon number 1 around dyon number 2. Thus, putting all of the pieces together,
when one dyon is revolved around the other it accumulates a factor of $(-1)$. Thus, dyons do behave as fermions amongst themselves. 
 
\subsubsection{Relation between non-Abelian Ising Statistics \& Toric Code Abelian Statistics}

In subsections \ref{EKTCM},\ref{kkht}, we discussed the explicit construction of defects in the Kitaev honeycomb 
model and their relation to those in the toric code model.  \cite{Wootton2008} studied the
general question as to how to construct non-Abelian anyons out of strings in the Abelian toric code model.
Their elegant construct may be viewed as a discrete lattice variant of similar continuum constructions, e.g., 
 
\subsubsection{Finite Temperature Behavior \& Dynamics}

As discussed in subsection \ref{KTCM}, the ground state (protected subspace of the code) is 4-fold 
degenerate (with Abelian ${\mathbb Z}_{2}\times {\mathbb Z}_{2}$ symmetry of Eqs. (\ref{kit})). There is, 
furthermore, an energy gap to excitations. 

Thermal effects in Kitaev's toric code model (including those related to {\em thermal fragility}-- 
the appearance of finite autocorrelation (or memory) times at all positive temperatures) and generalizations thereof 
were investigated by several groups \cite{Nussinov08a,Nussinov08c,PNAS,AOP,Alicki2009,Iblisdir2009,Iblisdir2010,Yoshida2011}.
In simple physical terms, as a consequence of the $d=1$ symmetries that this system harbors and the generalized Elitzur theorem \cite{Batista05,holography} which we discussed, 
precisely as in the Ising chain, two defects can be created with a finite energy penalty and thereafter separated arbitrarily far apart from one another with no energy cost. Thus, as a consequence
of this dimensional reduction, precisely as in the one-dimensional Ising mode, 
information initially stored in the qubits $Z_{i}$ (or $X_{i}$) will, at finite temperatures, be erased in a finite (i.e., system size independent) time scale. 
Beyond the specific information stored in these qubits, Kitaev's toric code model genuinely has a 1D zero temperature Ising type transition and is disordered
at any finite temperature \cite{PNAS, AOP}. 
That this is so is even more evident as the spectrum is that of two uncoupled circular
Ising chains with each chain being of length $N_{S}$. [$2N_s$ is the total number of links of the original $D=2$
lattice.] This mapping has consequences for the system dynamics and thermodynamics. As Ising chains are disordered
at any finite temperature, it is 
seen as stated above that, in thermal equilibrium state, 
this model is disordered at all non-zero temperatures
($\langle X_{\mu} \rangle = \langle Z_{\nu} \rangle =0$)
and that there is a trivial critical point at  temperature of
\begin{eqnarray}
T_{c}^{Kitaev~Toric~Code} = 0.
\end{eqnarray}
Due to the duality between Kitaev's model and the Ising
chains \cite{PNAS, AOP}, no non-trivial finite temperature spontaneous symmetry breaking
or any other transition can take place. At any non-vanishing temperature, no matter how small, 
entropic contributions to the free energy overwhelm energy penalties 
and lead to a free energy which is everywhere analytic. 
Nevertheless, if temperatures far below the spectral gap may
be achieved, the autocorrelation time $\tau$ associated with the topological
invariants (the $d=1$ symmetries) of
Eq. (\ref{kit}) can be made very large.
That is, e.g., the autocorrelation function of the toric
operators of Eq.(\ref{kit}),
\begin{eqnarray}
\langle Z_{1} (0) Z_{1}(t) \rangle \sim \exp(-|t|/\tau),
\label{autoZ}
\end{eqnarray} 
with, at asymptotically low temperatures, $\tau = \exp(\beta \Delta)$
where $\Delta$ the spectral gap.
Similar effects were found in some three dimensional variants of the 
Toric code model including those which further exhibit
finite temperature transitions \cite{Nussinov08a,Nussinov08c}.
Several high dimensional systems reduce to Ising chains following
a similar bond-algebraic mapping \cite{holography}.  
Bounds of the form of Eq. (\ref{autoZ}) are not specific to
the Kitaev model but are rather generic to any system with $d=1$ gauge
like symmetries. The autocorrelation functions in these systems
have a canonical one dimensional (chain type) character.
In finite size systems, whenever the associated 
correlation length is larger than the system size, the system
may appear ordered and a finite temperature crossover occurs
when the correlation length is comparable to the system size \cite{PNAS}.
The existence of these finite size effects is evidenced in the topological
entanglement entropy \cite{Castelnovo2007}. Quench dynamics of the entanglement entropy in Kitaev's toric code
model has been investigated \cite{Rahmani2010}. Aspects of general topological quantum orders, associated symmetries, 
thermal effects (and their manifestations in correlations and stability of quantum information storage to perturbations)
in Kitaev's toric code and Kitaev's honeycomb compass model and related models have been extensively
discussed elsewhere, e.g., \cite{Dennis2002,Nussinov08a,Nussinov08c,PNAS,AOP,Michnicki12,Bravyi08,cheng12,Mazac12,son12,spyros11,Schuch2010,Kargarian2009,Abasto2009}.

\subsubsection{External fields, Disorder, Dilution, Coupling to Phonons \& Photons}

The effect of external fields was investigated in 
\cite{Trebst2007,Hamma2008,Tupitsyn2010,Vidal09,Vidal09b,cobanera,ADP,holography}
via numerous approaches (including, in certain cases, those of dualities and self-dualities). 
A very rich phase diagram harboring continuous and discontinuous transitions, multi-criticality,
and dimensional reduction has been found. Recent works have also investigated 
the effects of disorder (realized as a distribution in
the coupling constants) \cite{Wootton2011,Stark2011}
and how such disorder may lead to localization of errors
in a toric code quantum computing based schemes. 
The key steps in the solution of Kitaev's honeycomb model
that were outlined in section \ref{solnus} \cite{Chen08} do not change in such 
a case.

The relation between Kitaev's honeycomb model and p-wave type BCS superconductors 
suggested in \cite{Chen08} that impurity bound states may appear. The effect of site dilution in 
Kitaev's model was investigated in detail in \cite{Willans2010,Willans2011}.  
Within the 90$^{\circ}$ compass model wherein dilution trivially lowers the coordination number
and removes bonds leading to higher degeneracy \cite{Nussinov08b}. Similarly,
dilution of the Kitaev model leads to extended zero energy states near each vacancy \cite{Willans2010,Willans2011}. 
It is readily seen by, e.g., bond algebra considerations that were discussed earlier for the (undiluted) Kitaev model [as well
as those employed for the diluted 90$^{\circ}$ model  \cite{Nussinov08b}], that Kitaev's model will remain exactly
solvable if the number of bonds is reduced due to dilution. 

In the gapless phase, vacancies were shown to bind to plaquette fluxes. In the presence of
an external magnetic field of strength $h$, interactions between induced moments lead to an attraction between 
the vacancies. In the absence of a vacancy in the bulk, the magnetic susceptibility $\chi \sim const$. A single vacancy leads
to a susceptibility that scales as $\chi \sim |\ln h|$. For two nearby vacancies on the same sublattice, at low temperatures,
$\chi \sim 1/(|h| |\ln h|^{3/2})$ \cite{Willans2010,Willans2011}. In the gapped phase, zero modes are localized by the vacancies. 
%

Lastly, we briefly note the interesting work of \cite{Bonderson2012} where the stability of the Kitaev model and its
degeneracy were illustrated for both (acoustic) phonon coupling as well as coupling with photons (via additional general
Zeeman couplings in the low energy limit). 

\subsubsection{Doping with Kinetic Vacancies}

Kitaev honeycomb models and Kitaev Heisenberg models doped with kinetic vacancies were explored 
in \cite{Mei2011,You2011,Hyart2011}.
These systems describe doped systems with spin-1/2 fermions that hop on the lattice and that 
additionally exhibit a spin exchange terms similar to those of  the Kitaev or Kitaev-Heisenberg models. 
Specifically, the spin Hamiltonians of the Kitaev and Kitaev-Heisenberg models 
of Eqs.(\ref{eq:HKT}, \ref{HKI}) respectively were, similar to $t-J$ models \cite{Eduardo_book}, 
augmented by a kinetic hopping term 
\begin{eqnarray}
H_{t} = -t \sum_{\langle ij \rangle \sigma} {\cal{P}} c_{i \sigma}^{\dagger} c_{j \sigma} {\cal{P}}.
\label{tJK}
\end{eqnarray}
A few standard words regarding the notation and $t-J$ type physics of Eq.(\ref{tJK}) are due. 
The operator ${\cal{P}}$ in Eq. (\ref{tJK}) projects out doubly occupied sites (lattice sites that
harbor fermions of both spin $S=1/2$ up and down flavors). By virtue of this ``no double occupancy''
constraint, a fermion can only (with an amplitude $(-t)$) to an empty site (or, equivalently, a hole can kinetically 
hop to occupied neighboring sites). The spin exchange between sites
that are occupied by fermions is given by the Kitaev or Kitaev-Heisenberg models
of Eqs.(\ref{eq:HKT},\ref{HKI}). [In the standard $t-J$ models, the spin exchange between occupied sites
is that of the rotationally invariant Heisenberg form (the scalar product between nearest neighbor spins).]  
In disparate analysis using the so-called ``SU(2)'' and ``U(1) slave boson'' methods respectively
(in which the electronic creation operators $c_{i \sigma}^{\dagger}$ are decomposed into
products of bosonic charge (holon) operators that carry SU(2) or U(1) degrees of freedom
and fermionic ``spinon'' operators) , two 
works \cite{You2011,Hyart2011} suggested the emergence of spin triplet 
superconductivity with $p$ wave type pairing as the Kitaev type spin systems were doped when the Kitaev couplings
were far stronger than the uniform spin exchange couplings in Eq. (\ref{HKI}). Transitions
separate the topologically trivial and non-trivial superconducting states at high and low dopings. 
At high ratios of the Heisenberg exchange couplings relative to the Kitaev couplings in Eq (\ref{HKI}), 
the dominant Heisenberg exchange may, similar to the standard $t-J$ model, 
lead to spin singlet $d-$ wave type superconductivity.  As highlighted by \cite{Hyart2011}, the topological superconducting phase (in which pairs of counter-propagating Majorana modes  appear) may appear over a much wider region of the phase diagram of the Kitaev-Heisenberg model than the Kitaev spin liquid phase itself in the absence of doping. If this model and its analysis indeed capture the quintessential features of doped Iridates, it might be easier to observe these non-trivial doped phase than the Kitaev type spin liquid of the Kitaev-Heisenberg model. The authors of \cite{You2011}
suggest the appearance of a time reversal symmetry breaking sate (with $p_{x} + ip_{y}$ type pairing)
while \cite{Hyart2011} arrive, within their calculations, at a lower energy time reversal invariant p-wave type
pairing with Ising topological superconductivity.  It remains to be seen which state is indeed favored by the 
complete energetic contributions.  Another work \cite{Mei2011} examined the Kitaev model at low dopings 
using a ``dopon'' representation that attempts to protects the $Z_{2}$ gauge structure of the Kitaev model 
and concluded that the doped Kitaev model is a Fermi liquid albeit with a temperature dependent Lorentz number
(the ratio $L = \kappa/(\sigma T)$  where $\sigma$ is the thermal conductivity, $\sigma$ is the electric conductivity and $T$ is the temperature)) 
and a large Wilson ratio (the ratio between the zero-temperature magnetic susceptibility $\xi$ and the coefficient of the linear temperature term in the specific heat $\gamma$ ($C_{v}= \gamma T$) of  a Fermi liquid, e. g., \cite{Varma2002}) .

\subsubsection{Generalizations of Kitaev's Models}
Since Kitaev's original works \cite{Kitaev06}, many related models (in both two and three spatial dimensions) building on his original construct of his model have been advanced
\cite{gwchern10,Yao2007,nussinov-bond,YangS,Si2007, Si2008npb,Yao2009,Wu2009, Ryu2009,Mandal2009,Levin2011,Yao2011,Nakai2011,Biswas2011,Lai2011a,Lai2011b,Lai2011c,Wang2010,Tikhonov2010,
Tikhonov2011,Chua2011,Baskaran2009} 
most of which are exactly solvable. 
Many of these exactly solvable models flesh out
rich phase diagrams and non-trivial excitations (and symmetries).  One key principle in which the exact solvability of these models (and their equivalence (or duality) to far simple quadratic forms in various guises) becomes apparent is that of the bond algebras \cite{nussinov-bond,cobanera,ADP,Nussinov08b} 
which were employed in the solution of Kitaev's model presented in the preceding Section.

\section{Conclusions}
Complementing more standard theories with isotropic interactions between various fundamental fields (such as spin, charge, color, or more general ``pseudo-spin''), there exists a plethora of physical systems in which the couplings between the pertinent internal degrees of freedom are direction dependent. The couplings in these ``compass models'' depend on the direction of the vectors connecting the interacting sites relative to a lattice (or continuum Cartesian or other directions). Such anisotropic direction dependent interactions are ubiquitous. Indeed, the anisotropic components of the interactions between dipoles when these are placed on lattices have precisely such a form. 
In compass models, external lattice (or other) directions lift the standard rotational invariance of the interactions. As we reviewed, in recent decades, numerous condensed matter systems have been discovered
to host precisely such compass type interactions. The paradigmatic class of physical systems described by compass interactions is afforded by transition metal materials where the real space form of the pertinent electronic orbitals lead to exactly such direction dependent interactions. The associated orbital ordered have been observed to persist, in some materials, up to temperatures (that can range up to ${\cal{O}}(10^{3}$K))
 that may significantly exceed magnetic ordering temperatures (when these are present) in these materials. Other primary examples of compass type interactions include diverse spin systems on frustrated lattices, bosonic and fermionic gases on optical lattices, materials with strong spin-orbit interactions, and other systems. Due to the anisotropic character of the interactions, the study of these systems is, by comparison to more standard rotationally invariant systems, a supremely interesting and challenging problem. Notably, as we reviewed for many particular compass Hamiltonians, some of these systems may be quantum liquids or, conversely, 
 may lead low temperature phases of matter in which order is triggered by fluctuation effects. Rich phenomena, such as dimensional reduction and holography spawned by the unusual (exact or emergent) symmetries that
 these systems typically have, appear in these systems. The rich states of matter that compass models exhibit still remain largely unexplored. Items that have only started to being examined in recent years include 
 the precise understanding of the nature of the phase transitions that they exhibit. To date, no effective field theories of these systems have been studied (nor even written down- our review includes one of the first general
 forms of these unusual anisotropic field theories). The intimate connections between compass models and topological quantum information (such as that exemplified by Kitaev's model) and, in particular, topological states of
 matter (e.g., those displayed by recently discovered topological insulators) are likely to lead to new insights. 

\section{Acknowledgments}
We have been very fortunate to closely interact with numerous colleagues on some of the systems described in this work. These include, amongst many others, 
C. Batista, M. Biskup, L. Chayes, H.-D. Chen, E. Cobaenra, M. Daghofer, T. P. Devereaux, E. Fradkin,  P. Horsch,  G. Khaliullin, D. Khomskii, S. Kumar, F. Nogueira, F. Mack, M. Mostovoy. A. Oles, K. Shtengel, K. Wohlfeld, J. Zaanen and P. Zoller. 
In particular, ZN is grateful to nearly a decade long interaction with G. Ortiz on many aspects of compass models. We are thankful for the understanding of these colleagues for not promptly writing other papers that need to be finished. We would like to thank our spouses for their patience and encouragement during this work.  Work at Washington University (WU) in St Louis was partially supported by the National Science Foundation under NSF Grant number DMR- 1106293 (ZN) as well as the CMI of WU at the initial stages of this work. ZN is grateful to the hospitality of the IFW, Dresden during which central parts of this work were done.

\section{Appendix A: The bond algebra of the plaquette orbital model}
\label{Aplaq}

In Eq. (\ref{plaq_ocm}), following \cite{Wenzel09,Biskup10}, we introduced the ``plaquette orbital model''. We remarked therein that its local algebraic structure is similar to that of 90$^\circ$ compass model on the square lattice
[Eq. (\ref{2dpocm})].  In this very brief appendix, we clarify this observation and invoke the bond algebraic structure \cite{cobanera,ADP,clock,holography,Nussinov2012,nussinov-bond,Nussinov08b} which we have earlier used in 
subsection \ref{solnus} in describing the exact solvability of Kitaev's model on the honeycomb lattice

The Hamiltonian defining the plaquette orbital model is a sum of two types of terms (or ``bonds''): \newline
(A) $\tau^{x}_{i} \tau^{x}_{j}$ for all links that belong to the $A$ plaquette sublattice,$ \langle i j \rangle \in A$.  \newline
(B)  $\tau^{y}_{i} \tau^{y}_{j}$ for all links  $\langle i j \rangle$  that belong to the $B$ plaquette sublattice.   \newline
The decomposition into the two plaquette ($A$ and $B$ sublattices) is shown in Fig. (\ref{fig:Biskup10_1}) \cite{Biskup10}.

The algebra satisfied by these bonds is very simple and is encapsulated by the following relations

\noindent (i) The square of each bond is one. \newline
(ii) Any two bonds that are of different type (i.e., one bond is of type A and the other is of type B) that share 
one common site anticommute: $\{\tau^{x}_{i} \tau^{x}_{j}, \tau^{y}_{i} \tau^{y}_{k} \} =0$ with the curly brackets denoting the anticommutator. (By fiat, given the type of the interactions,
$\langle ij \rangle \in A$ and $\langle ik \rangle \in B$.)  \newline
(iii) Bonds of different type commute if they share no common site: $[\tau^{x}_{i} \tau^{x}_{j}, \tau^{y}_{k} \tau^{y}_{l}]=0$ (with $i,j,k,$ and $l$ corresponding to four different sites).  \newline
(iv) Any bond of type A commutes with any other bond of type A and, similarly,
any bond of type B commutes with all bonds of the B type. 

Thus, locally, each bond (having a square that is unity) anticommutes with two other neighboring bonds and commutes with the two other nearest neighbor bonds  (as well all other bonds on the lattice). 

The bond algebra associated with the 90$^\circ$ compass model of Eq. (\ref{2dpocm}) is very much like that of the plaquette orbital model. This system has a decomposition into
two types of similar bonds: 

\noindent
(a) $\tau^{x}_{i} \tau^{x}_{j}$ on all horizontal links.  \newline
(b) $\tau^{y}_{i} \tau^{y}_{j}$ on all vertical links. 

The algebra satisfied by these bonds is specified by a similar list:

\noindent
(i) The square of each bond is one. \newline
(ii) Any two bonds that are of different type that share 
one common site anticommute: $\{\tau^{x}_{i} \tau^{x}_{j}, \tau^{y}_{i} \tau^{y}_{k} \} =0$. newline
(iii) Bonds of different type commute if they share no common site: $[\tau^{x}_{i} \tau^{x}_{j}, \tau^{y}_{k} \tau^{y}_{l}]=0$ (with $i,j,k,$ and $l$ corresponding to four different sites).  \newline
(iv) Any horizontal bond commutes with any other horizontal bond and, analogously,
any vertical bond commutes with all vertical bonds. 

The local algebra is congruent to that of the plaquette orbital model: each bond anticommutes with two out of its four nearest neighbors. 
This equal structure implies that in their Cayley tree (or Bethe lattice) approximations, the 90$^{\circ}$ and the plaquette compass model are identical.

\section{Appendix B: Gell-Mann Matrices\label{sec:Gell-man}}
\label{sec:appendix}

The Gell-Mann matrices are a representation of the infinitesimal generators of the special unitary group SU(3). This group has dimension eight and therefore it has a set with eight linearly independent generators, which can be written as $\lambda_i$, with $i$ taking values from 1 to 8. They obey the commutation relations
\begin{equation}
[\lambda_i,\lambda_j] = \frac{i}{2} f^{ijk} \lambda_k,
\end{equation}
where a sum over the index $k$ is implied. The constants $f^{ijk}$ are $ f^{123}=1$, $ f^{147}= f^{165}=f^{246}= f^{257}=f^{345}= f^{376}=1/2$ and $ f^{458}= f^{678}=\sqrt{3}/2$ and are antisymmetric in the three indices.
The Gell-Mann matrices representations involving $3 \times 3$ matrices, that act on complex vectors with 3 entries. They have the additional properties that are traceless, Hermitian, and obey the relation Tr$(\lambda_i \lambda_j)$ $= 2\delta_{ij}$. 

\begin{gather*} 
\lambda_1=  \begin{pmatrix}
0 & 1 & 0\\
1  & 0 & 0 \\
0  & 0 & 0 \\
\end{pmatrix}, \quad 
\lambda_2 = \begin{pmatrix}
0  & -i & 0\\
i  & 0 & 0 \\
0  & 0 & 0 \\
\end{pmatrix}, \quad 
\lambda_3 =  \begin{pmatrix}
1  & 0 & 0\\
0  & -1 & 0 \\
0  & 0 & 0\\
\end{pmatrix}
\end{gather*}

\begin{gather*} 
\lambda_4=  \begin{pmatrix}
0 & 0 & 1\\
0  & 0 & 0 \\
1  & 0 & 0 \\
\end{pmatrix}, \quad 
\lambda_5 = \begin{pmatrix}
0  & 0 & -i\\
0  & 0 & 0 \\
i  & 0 & 0 \\
\end{pmatrix}, \quad 
\lambda_6 =  \begin{pmatrix}
0  & 0 & 0\\
0  & 0 & 1 \\
0  & 1 & 0\\
\end{pmatrix}
\end{gather*}

\begin{gather*} 
\lambda_7=  \begin{pmatrix}
0 & 0 & 0\\
0  & 0 & -i \\
0  & i & 0 \\
\end{pmatrix}, \quad 
\lambda_8 = \frac{1}{\sqrt{3}}\begin{pmatrix}
1  & 0 & 0\\
0  & 1 & 0 \\
0  & 0 & -2 \\
\end{pmatrix}
\end{gather*}

The matrices $\lambda_3$ and  $\lambda_8$ commute. Three independent SU(2) subgroups are formed by the elements of vectors ${\bm \mu}^1$, ${\bm \mu}^2$ and ${\bm \mu}^3$, where ${\bm \mu}^1= \frac{1}{2}( \lambda_1,  \lambda_2, \lambda_3 )$, ${\bm \mu}^2 = \frac{1}{2}( \lambda_4,  \lambda_5, \lambda_+ )$, and ${\bm \mu}^3=\frac{1}{2}( \lambda_6,  \lambda_7, \lambda_- )$. Here the $\lambda_+$, $\lambda_-$ are linear combinations of  $\lambda_3$ and  $\lambda_8$: $\lambda_{\pm}= \lambda_3\cos \left( \frac{2\pi}{3} \right) \pm \lambda_8 \sin \left( \frac{2\pi}{3}  \right)$, so that, as is expected for a SU(2) spin $1/2$, the commutator $[{\bm \mu}_1^{\gamma},{\bm \mu}_2^{\gamma}]= \frac{i}{2}{\bm \mu}_3^{\gamma}$, for each $\gamma=1, 2, 3$.

The operators 
\begin{eqnarray}
\hat{R}^+=  \begin{pmatrix}  0 & 0 & 1 \\  1 & 0 &0  \\  0 & 1 & 0 \end{pmatrix}; \hat{R}^-=  \begin{pmatrix}  0 & 1 & 0 \\  0 & 0 &1  \\  1 & 0 & 0 \end{pmatrix}
\end{eqnarray}
rotate the vectors $\bm \mu$ onto eachother: ${\bm \mu^2} = \hat{R}^- {\bm \mu^1} \hat{R}^+$

\section{Appendix C: Classical \&  Quantum fluctuations in the large $n$ limit}
\label{appendixB}

The large $n$ limit (see section \ref{sec:clas}) of the theory of Eq. (\ref{long_ebasis}) is exactly solvable. 
As such, it allows us to easily point out to a difference between the classical and quantum theories. 
In the large $n$ limit of the classical system, order appears in $D$ dimensional system appears if 
and only if the classical (lowest order  (${\cal{O}}(1/n)^{0}$))
self-energy diagram stemming from the 
Boltzmann distribution of harmonic modes (and its related equipartition theorem) 
\begin{eqnarray}
\label{cse}
\Sigma^{(0)}_{cl} = \sum_{\alpha} \int \frac{d^{D}k}{(2 \pi)^{D}}  \frac{1}{v_{\alpha}({\bf k}) + \mu}
\end{eqnarray}
does not diverge as the ``mass'' $\mu$ veers towards $[-\min_{k,\alpha} \{v_{\alpha}({\bf k})]$.
The integration in Eq. (\ref{cse}) is performed over the first Brillouin zone- a region of finite volume.
Thus, $\Sigma^{(0)}_{cl}$ can diverge only from infra-red contributions. 
%
In systems in which the mode spectra $v_{\alpha}({\bf k})$
disperse quadratically about their minimum,
the relevant integral converges in dimensions $D>2$ but fails to converge in low dimensions
due to the large relative phase space volume of low energy modes.
Quantum mechanically, in large $n$ systems, 
see, e.g., \cite{Nussinov4,Serral2004}, the corresponding self-energy 
is governed by the Bose function set by the modes $\omega_{k}$.  The pertinent zero temperature
dispersion of $v_{\alpha}({\bf k})$ in the argument of the integrand governing
the convergence or divergence of \ref{cse} in the classical case is replaced
in quantum case by the square root forms $\sqrt{v_{\alpha}({\bf k})}$.  Qualitatively
similar Bose type distributions and dispersions are found in 1/S calculations.
As power counting suggests,
the convergence of the integral and thus the character of the fluctuations arising 
from classical and quantum effects are different.
It is possible to have ordering of the quantum system at zero temperature
while the classical counterpart 
of Eq. (\ref{cse}) exhibits an infra-red divergence.  Such a case arises
in two-dimensional ferromagnets
Precisely this sort of situation arises in the 120$^{\circ}$ compass model-
the large $n$ quantum version of the model exhibits low temperature order (quantum order
out of disorder) yet its classical counterpart exhibits no finite temperature order.
As it will turn out, however, once the 120$^{\circ}$ system is constrained 
to its original $n=2$ component version, both classical thermal fluctuations 
and quantum effects lead to similar sorts of ordering.




\end{document}